\documentclass[10pt,paper]{Jnosso}
\pdfoutput=1
\usepackage{graphicx}
\usepackage{amsmath}

\skip\footins = 1\bigskipamount plus 2pt minus 4pt

\title{RNS derivation of $N$-point disk amplitudes from the revisited S-matrix approach}

\author{Luiz Antonio Barreiro$^{1}$ and  Ricardo Medina$^{2}$\\
$^{1}$Departamento de F{\'i}sica, UNESP\\
\hphantom{$^{1}$}Rio Claro, S{\~a}o Paulo, Brazil\\
$^{2}$Instituto de Matem\'atica e Computa\c{c}\~ao, Universidade Federal de Itajub\'{a}\\
\hphantom{$^{1}$}Itajub\'a, Minas Gerais, Brazil\\
\hphantom{$^{1}$}E-mails: \email{luiz.a.barreiro@gmail.com,
rmedina50@gmail.com}.}

\abstract{In the past year, in arXiv:1208.6066 we proposed a revisited S-matrix approach to efficiently find the bosonic terms of the open superstring low energy effective lagrangian (OSLEEL). This approach allows to compute the ${\alpha'}^N$ terms of the OSLEEL using open superstring $n$-point amplitudes in which $n$ is very much lower than $(N+2)$ (which is the order of the required amplitude to obtain those ${\alpha'}^N$ terms by means of the conventional S-matrix approach). In this work we use our revisited S-matrix approach to examine the structure of the scattering amplitudes, arriving at a closed form for them. This is a RNS derivation of the formula first found by Mafra, Schlotterer and Stieberger in arXiv:1106.2645, using the Pure Spinor formalism. We have succeeded doing this for the 5, 6 and 7-point amplitudes. In order to achieve these results we have done a careful analysis of the kinematical structure of the amplitudes, finding as a by-product a purely kinematical derivation of the BCJ relations (for N=4, 5, 6 and 7). Also,
following the spirit of the revisited S-matrix approach, we have found the $\alpha'$ expansions for these amplitudes up to ${\alpha'}^6$ order in some cases, by only using the well known open superstring 4-point amplitude, cyclic symmetry and tree level unitarity: we have not needed to compute any numerical series or any integral involving polylogarithms, at any moment.}

\begin{document}
\maketitle
\flushbottom

\section{Introduction}

\label{Introduction}

\noindent In the recent years there has been a considerable progress in perturbative String Theory (considering D-brane systems as well). Acheivements have been going on in S-matrix calculations involving Ramond-Ramond massless states and open superstrings \cite{Ehsan1, Garousi1, Ehsan2, Ehsan3, Garousi2, Ehsan4}\footnote{There have been various recent results about mixed open/closed superstring computations, to all $\alpha'$ order, by E. Hatefi \cite{Ehsan5}.}, the closed and the open sector of the superstring low energy effective lagrangian \cite{Garousi3, Garousi4, Barreiro0}, higher loop calculations of closed and open superstring scattering \cite{Gomez1, Green0, Mafra0}, Mellin correspondence between supergravity and superstring amplitudes \cite{Stieberger1, Stieberger2} and a deeper understanding of the $\alpha'$ expansion of tree level open and closed superstrings \cite{Schlotterer1, Broedel2, Broedel1, Stieberger4}, among other things.  \\
\noindent One of the very important results that has been acheived is Mafra-Schlotterer-Stieberger (MSS) formula for the (tree level) $N$-point scattering of nonabelian massless open superstrings \cite{Mafra1}:
\begin{eqnarray}
A(1, \ldots , N) & = & \sum_{\sigma_N  \varepsilon  S_{N-3}}  F^{\{\sigma_N\}}(\alpha') \ A_{SYM}(1,\{ 2_{\sigma}, 3_{\sigma}, \ldots, (N-2)_{\sigma} \}, N-1, N) \ ,
\label{MSS}
\end{eqnarray}
where $A(1, \ldots , N)$ is the subamplitude and
\begin{eqnarray}
{\cal A}_N & = & i (2 \pi)^{10} \delta^{(10)} (k_1 + \ldots +k_N ) \ \biggl[ \ \mbox{tr}(\lambda^{a_1} \lambda^{a_2} \ldots \lambda^{a_N}) \ A(1, 2, \ldots, N) +
\left( \begin{array}{c}
          \mbox{non-cyclic} \\
          \mbox{permutations}
          \end{array}
\right)  \ \biggr]    \ \ \ \
\label{N-point}
\end{eqnarray}
is the complete $N$-point open superstring (tree level) scattering amplitude (where $N \geq 3$).\\
\noindent In (\ref{MSS}) $\sigma_N=\{ 2_{\sigma}, 3_{\sigma}, \ldots, (N-2)_{\sigma} \}$ denotes a permutation of  indexes $\{ 2,3, \ldots, (N-2) \}$ and the $F^{\{\sigma_N\}}(\alpha')$'s are the momentum factors which contain the $\alpha'$ information of the scattering amplitude\footnote{Besides $\alpha'$, the $F^{\{\sigma_N\}}(\alpha')$'s depend on the $k_i \cdot k_j$ scalar products, which can be written in terms of the independent Mandelstam variables of this $N$-point process.} (see eqs.(\ref{MSS2}) and (\ref{MSS3})) . There are $(N-3)!$ terms in the sum in (\ref{MSS}). \\
\noindent Formula (\ref{MSS}) considers all possible scattering processes involving external gauge bosons and their fermionic massless superpartners. In it, $A_{SYM}$ denotes the tree level scattering subamplitude of this process in $D=10$ Super Yang-Mills theory and $A$ (on the left hand-side) is the corresponding scattering subamplitude in Open Superstring Theory.  In (\ref{N-point}) the $\lambda^a$'s are the gauge group matrices in the adjoint representation (see eq. (\ref{adjointm})).   \\
\noindent Formula (\ref{MSS}) has the merit of clarifying that the kinematics of the open superstring $N$-point amplitude, at any $\alpha'$ order, is governed by the Super Yang-Mills kinematics, that is, from the low energy theory. It also has the virtue of identifying an explicit general integral formula, valid for any $N \geq 3$, for all the $F^{\{\sigma_N\}}(\alpha')$ momentum factors (see eq.(\ref{MSS2})).\\
\noindent During the past and the present year quite nontrivial results have been obtained for the $\alpha'$ expansion of the $F^{\{\sigma_N\}}(\alpha')$  momentum factors, in eq.(\ref{MSS}). These factors are disk worldsheet integrals  which can be found, but in terms of non elementary functions so, besides the cases of $N=3$ and $N=4$, their $\alpha'$ expansion is a non trivial thing to compute\footnote{They are $(N-3)$ multiple integrals (see eq.(\ref{MSS2})). In the case of the 5-point amplitude one finds the $_{3}F_{2}$  Hypergeometric function (see, for example, \cite{Kitazawa1} and \cite{Brandt1});  in the case of $6$-point amplitudes one finds a double series of $_{4}F_{3}$ Hypergeometric functions (see, for example, \cite{Oprisa1} and \cite{Hemily1}); and for $N \geq 7$ one has to deal with even more complicated expressions. In all these cases the coefficients of the $\alpha'$ expansion are given in terms of Harmonic (or Euler-Zagier) sums and/or Polylogarithmic integrals: all of them are nowadays known how to be calculated (see, for example, refs. \cite{Vermaseren1} and \cite{Remiddi1}), but the required calculations to find them grow enormously with the $\alpha'$ order.}. \\
\noindent Besides the fact of improving the $\alpha'$ order of expansions of the past \cite{Paiva1, Barreiro1, Oprisa1} in various orders \cite{Boels1, Schlotterer1, Broedel2},  a conjecture has been established for the general form of the $\alpha'$ expansion of the $F^{\{\sigma_N\}}(\alpha')$'s (for arbitrary $N$ and arbitrarily high order in $\alpha'$\cite{Schlotterer1})\footnote{See Appendix \ref{N5 and higher} for this explicit form.}'\footnote{The mentioned conjecture of ref. \cite{Schlotterer1} has been checked in \cite{Broedel2} up to ${\alpha'}^{21}$ order for $N=5$,  up to ${\alpha'}^{9}$ order for $N=6$ and up to ${\alpha'}^{7}$ order for $N=7$.} and, also, a recursive formula (in $N$) for these $\alpha'$ expansions has been proved in \cite{Broedel1} (by means of a generalization of the work in ref.\cite{Drummond1})\footnote{The mathematical framework used in refs. \cite{Drummond1, Broedel2} is related, but is not directly the same one of ref. \cite{Schlotterer1}.}.\\
\noindent All these results have as a common starting point the MSS formula, given in eq.(\ref{MSS}). This formula was derived using the Pure Spinor formalism \cite{Berkovits1}, which is manifestly supersymmetric right from the beginning. Formula (\ref{MSS}) is the final (and simple) result of an ellaborated study involving  pure spinor superspace and its cohomology structure \cite{Mafra2}, first applied in the calculation of Super Yang-Mills amplitudes and afterwards extended to the corresponding calculations in Open Superstring Theory \cite{Mafra1}. \\
\noindent The purpose of the present work is two-fold. On one side, we show that it is possible to arrive to MSS's formula in (\ref{MSS}) working only in the Ramond-Neveu-Schwarz (RNS) formalism \cite{Friedan1}\footnote{The subtlety is that we do not deal with fermion vertex operators at all. We only work with the $N$-point gauge boson amplitude, $A_b(1, \ldots, N)$, given in eq.(\ref{ANfermionic}), which comes from only gauge boson vertex operators in the RNS formalism \cite{Schwarz1}. See section \ref{For gauge bosons and massless fermions} for more details about this.}. For the moment, we only have a proof for $3 \leq N \leq 7$ within this approach, but we think that a deeper understanding of our procedure can, eventually, lead to the proof for arbitrary $N$. On the other side, we shed light in how the $\alpha'$ expansion of the $F^{\{\sigma_N\}}(\alpha')$ momentum factors in (\ref{MSS}) can be obtained, order by order in $\alpha'$, not by doing the explicit computations of the coefficients of its expansion (which are given in terms of multiple zeta values (MZV's)\cite{Brown1, Mafra3, Boels1, Schlotterer1}), but by demanding tree level unitarity of the amplitudes and (presumably) using only the $\alpha'$ expansion of the $5$-point amplitude\footnote{This statement seems to be in contradiction with the final part of the abstract of this work, where we sayed that we would only use the $\alpha'$ information from the $4$-point amplitude.\\
The clarifying statement is that, for the calculation purposes of this work, in which at most we have obtained ${\alpha'}^6$ order results, it will be enough to use the $4$-point amplitude information. As explained in section \ref{Using}, the $5$-point amplitude information will become important from ${\alpha'}^8$ order onwards and, as argued in that section, we claim that the $pure$ $5$-point amplitude is enough to find the whole $\alpha'$ terms of the $F^{\{\sigma_N\}}(\alpha')$ momentum factors.}.\\
\noindent The basic tool in which our findings are supported is the `revisited S-matrix approach', found by us in the past year \cite{Barreiro0}. This method was initially proposed as an efficient tool to determine, order by order in $\alpha'$, the bosonic part of the open superstring low energy effective lagrangian (OSLEEL) but, as we will see in the present work, it has a direct counterpart in the determination of the scattering amplitudes of the theory, allowing us to arrive to (\ref{MSS}) and also to the $\alpha'$ expansion version of it. This method is intrinsically kinematic and supersymmetric, although it is not {\it manifestly} supersymmetric. It deals, first, with the pure external gauge boson interactions and only at the end, it incorporates the interactions between external gauge bosons and its fermionic superpartners.\\
\noindent The kinematics is present right from the beginning since the main statement of the method has to do with the kinematical structure of the $N$-point amplitudes of gauge bosons: in Open Superstring Theory they should not contain $(\zeta \cdot k)^N$ terms, at any $\alpha'$ order \cite{Barreiro0}. With respect to supersymmetry, it is also present right from the beginning since (we believe that) it is the reason for the absence of those kinematical terms in the amplitudes\footnote{It is well known that the $(\zeta \cdot k)^N$ terms are indeed present in the case of $3$ and $4$-point amplitudes of massless states in Open Bosonic String Theory \cite{Schwarz1, Kawai1} and, from the general integral formula for the $N$-point amplitude \cite{Schwarz1}, it is also believed that they are present in this general case. Therefore, it is quite reasonable to conjecture that the absence of the $(\zeta \cdot k)^N$ terms, in the case of supersymmetric open strings, is a consequence of Spacetime Supersymmetry.}.\\
\noindent The structure of this work is as follows. In section \ref{Brief review} we give a brief review of the revisited S-matrix method. We explain in it why we claim that demanding the absence of the $(\zeta \cdot k)^N$ terms in the $N$-point amplitude of gauge bosons and, based on the conjecture of ref. \cite{Schlotterer1} and the main result of ref.\cite{Broedel1}, using only the $\alpha'$ expansion of the $5$-point amplitude, that is enough information to find the complete (bosonic part of the) OSLEEL. We claim there, then, that using similar arguments there should be a direct analog of this situation from the perspective of the scattering amplitudes, that is, at a given order in $\alpha'$, knowing the $5$-point amplitude is enough information to know any higher $N$-point amplitude\footnote{See footnote $8$ in the previous page.}.\\
\noindent We begin the ellaboration of these ideas in section \ref{Kinematical} by examining the space of $N$-point kinematic expressions which are on-shell gauge invariant and which do not contain $(\zeta \cdot k)^N$ terms. We find that this space is $(N-3)!$ dimensional (at least for $3 \leq N \leq 7$). We then check that a BCJ basis for Yang-Mills subamplitudes (see eq.(\ref{basis})) \cite{Bern1} can indeed be chosen as a basis for this space. In light of this important kinematical result, the determination of the explicit expressions of  the open superstring subamplitudes (for gauge bosons) and of the BCJ relations themselves become simply a linear algebra problem: we know a vector of the space (that is, a given subamplitude for which we want an expression) and what is left to do is to find the components of this vector with respect to the basis of the vector space. We do this calculation  in section \ref{Closed} for the open superstring subamplitudes (for gauge bosons), arriving precisely to the bosonic part of (\ref{MSS}), and in Appendix \ref{BCJ}, for the BCJ relations themselves (arriving to the same result of refs. \cite{Bjerrum1, Stieberger3}). Then, in the last part of section \ref{Closed}, we briefly explain why once we have found the gauge boson amplitudes in this manifestly gauge invariant way, the corresponding amplitudes involving fermions are immediate, and thus leading to MSS result in eq. (\ref{MSS}) (this time considering there all possible scattering processes involving external gauge bosons and their supersymmetric partners).\\
\noindent In section \ref{Finding the} we apply our revisited S-matrix method to find the $\alpha'$ expansion of the $F^{\{\sigma_N\}}(\alpha')$ momentum factors of (\ref{MSS}), in the case of $N=5,6,7$ \footnote{Because of computer limitations, we have done the explicit calculations only up to ${\alpha'}^6$ order in the first two cases and up to ${\alpha'}^4$ order, in the last case. But it is clear that the method can be used to obtain higher order $\alpha'$ contributions. See more details in section \ref{Finding the}.}. In order to acheive this, besides the requirements of the revisited S-matrix method, we demand cyclicity and tree level unitarity to be obeyed by the subamplitudes.\\
\noindent We end in section \ref{Summary} by summarizing our results and conclusions.\\
\noindent Throughout this work all scattering amplitudes are tree level ones.\\
\noindent Since, at some points, we have needed to deal with huge calculations and formulas, we have considered only the simplest ones in the main body of this work and we have left the more extensive ones for the appendices\footnote{Moreover, there are some extremely long expressions that we have preferred not to include them in the text of this work and only to attach them as `txt' files, in the version that we have submitted to the hep-th arXiv.}. These last ones usually do not offer any new conceptual insight, but they have played an important role checking our main statements.

\section{Review of the revisited S-matrix method}

\label{Brief review}

\subsection{Finding terms of the OSLEEL in an efficient way}

\label{Finding terms}

\noindent Let ${\cal L}_{\mbox{eff}}$ be the general low energy effective lagrangian (LEEL) for nonabelian gauge bosons in (either bosonic or supersymmetric) Open String Theory . It has the following form:
\begin{eqnarray}
{\cal L}_{\mbox{eff}} & = & \frac{1}{g^2} \ \mbox{tr} \biggl[ \  -\frac{1}{4} F^2 + (2 \alpha') F^3 + (2 \alpha')^2 F^4 + (2 \alpha')^3 ( F^5 + D^2 F^4) +
  \nonumber \\
&& \hphantom{  \frac{1}{g^2} \ \mbox{tr} \biggl[ \ \ } (2 \alpha')^4 ( F^6 + D^2 F^5 + D^4 F^4) + {\cal O}\big( (2 \alpha')^5 \big) \ \biggr] \ .
\label{Leff}
\end{eqnarray}
Each of the $F^n$ and the $D^{2p}F^q$ terms in (\ref{Leff}) is an abbreviation of the sum of different contractions of Lorentz indexes for those sort of terms. For example,``$F^4 \ $'' really denotes an abbreviation of $ ( b_{1}F^{\mu\lambda}F^{\nu}_{\ \lambda}F_{\mu}^{\ \rho}F_{\nu\rho}  +b_{2}F^{\mu}_{\ \lambda}F_{\nu}^{\ \lambda}F^{\nu\rho}F_{\mu\rho} +b_{3}F^{\mu\nu}F_{\mu\nu}F^{\lambda\rho}F_{\lambda\rho}
+b_{4}F^{\mu\nu}F^{\lambda\rho}F_{\mu\nu}F_{\lambda\rho} )$ \cite{Tseytlin1}, where $\{ b_1, b_2, b_3, b_4\}$ are the coefficients to be determined. \\
\noindent In the second column of the table in (\ref{table2})\footnote{This table has been taken from eq.(3.4) of ref. \cite{Barreiro0}.} we have written the number of coefficients that the general LEEL contains at the first orders in ${\alpha'}$ \footnote{The terms which are being taken into account in the LEEL are only the ones which remain invariant under field redefinitions.}. These coefficients are the ones that the conventional S-matrix approach usually finds by computing the open string $N$-point amplitudes (from $N=4$ up to $N=p+2$, at least) at ${\alpha'}^p$ order.
\begin{eqnarray}
\begin{tabular}{c|c|c}
$p$ & Dimension of the general basis & Dimension of the constrained basis\\
      & at order ${\alpha'}^p$              & at order  ${\alpha'}^p$ \\
\hline
1 & 1 & 0 \\
2 & 4 & 1 \\
3 & 13 & 1 \\
4 & 96 & 0 \\
\vdots & \vdots & \vdots
\label{table2}
\end{tabular}
\end{eqnarray}
\noindent In the third column of table (\ref{table2}) we have written the number of coefficients (which is extremely small!) that the revisited S-matrix approach really needs to find in order to determine the OSLEEL at a given $\alpha'$ order. The reason for the smallness of these numbers (in relation to the corresponding ones in the second column) is that in the revisited S-matrix method (only applicable to the case of the supersymmetric string) the $N$-point amplitudes satisfy the constraint \cite{Barreiro0}
\begin{eqnarray}
\mbox{absence of} \ (\zeta \cdot k)^N \ \mbox{terms} \ .
\label{absence}
\end{eqnarray}
This constraint implies further linear restrictions that the $b_j$ coefficients of the LEEL in (\ref{Leff}) should satisfy (see section 4 of \cite{Barreiro0} for more detalis about these restrictions). These restrictions are so strong that only a small number of coefficients remain free: this number is precisely the `dimension of the constrained basis' appearing in the third column of table (\ref{table2}). These same type of contraints had correctly been found about ten years before, by Koerber and Sevrin, using the method of BPS configurations (which is not directly a String Theory one) \cite{Koerber1}. In \cite{Barreiro0} we pointed out that the (probable) reason for the constraint in (\ref{absence}) is Spacetime Supersymmetry.  \\
\noindent Due to the highly constrained form that the OSLEEL lagrangian adopts after demanding the requirement in eq. (\ref{absence}), in order to determine the ${\alpha'}^p$ order terms of it, a $(p+2)$-point amplitude  calculation, in Open Superstring Theory, is no longer needed (as in the conventional S-matrix method): very much lower $N$-point amplitudes (expanded at ${\alpha'}^p$ order) are expected to be enough for this purpose. In fact, in \cite{Barreiro0} we saw that the $\alpha'$ expansion of the $4$-point momentum factor (given in eq.(\ref{expansionGamma})) is enough to determine explicitly the OSLEEL, at least up to ${\cal O}({\alpha'}^4)$ terms.\\

\subsection{Using the $4$ and the $5$-point $\alpha'$ information to obtain $N$-point information}

\label{Using}

\noindent So, a main idea that arises naturally from the revisited S-matrix method is the fact that the $\alpha'$ expansion of the $4$-point momentum factor (whose coefficients are completely known in terms of integer zeta values, at any $\alpha'$ order,  see eq. (\ref{formula1})) is enough information to obtain completely the $\alpha'$ expansion of higher $N$-point amplitudes, at least up to a certain order in $\alpha'$. For example, with the calculations that we did in \cite{Barreiro0} it is clear that the $5$ and the $6$-point amplitudes (and $any$ higher $N$-point amplitude) can be completely determined at least up ${\alpha'}^4$ order, because we found the OSLEEL explicitly up to that order, bypassing $5$ and $6$-point worldsheet integral calculations. \\
\noindent In fact, in \cite{Barreiro0} we raised the possibility that the OSLEEL could be determined to any $\alpha'$ order by means of the revisited S-matrix method plus the known $\alpha'$ expansion of the $4$-point momentum factor (see eqs. (\ref{formula1}) and (\ref{expansionGamma})), but this can hardly happen since there are higher order coefficients given in terms of multiple zeta values (MZV's)\footnote{See Appendix \ref{Multiple} for an extremely short review on MZV's.}, like $\zeta(3,5)$, $\zeta(3,7)$, $\zeta(3,3,5)$, etc., which already show up in the $\alpha'$ expansion of the $5$-point amplitude (at ${\alpha'}^8$, ${\alpha'}^{10}$ and ${\alpha'}^{11}$ order, respectively \cite{Schlotterer1, Boels1})\footnote{We thank Rutger Boels for calling our attention to this point.}. These coefficients are not expected to be given by linear combinations of products of $\zeta(n)$'s  (in which the coefficients of these linear combinations are rational numbers)\footnote{In this work we will refer to these peculiar MZV's as $non \ trivial$ MZV's, in opposition to the $trivial$ ones, which are known to be given as rational linear combinations of products of $\zeta(n)$'s \cite{Blumlein1}. The $non \ trivial$ MZV's that we will be referring to are $only$ the ones that appear in the MZV basis of this last reference. A few examples of $trivial$ MZV's can be found in formulas (\ref{zeta12}), (\ref{zeta22}) and (\ref{zeta14}) of Appendix \ref{Momentum factors}.}\cite{Brown0}, so they are not present in the $\alpha'$ expansion of the $4$-point amplitude.\\
\noindent Since ${\alpha'}^8$ is the first order at which these $non \ trivial$ MZV's arise, we do not have a proof, but we believe that up to ${\alpha'}^7$ order any $N$-point amplitude can be found by means of the revisited S-matrix method plus only the known $\alpha'$ expansion of the $4$-point momentum factor, $F^{\{2\}}$,  (see eqs. (\ref{A1234-3}), (\ref{formula1}) and (\ref{expansionGamma})). \\
\noindent From ${\alpha'}^8$ order onwards we expect the $4$-point amplitude to only give a partial (but still important) information for the determination of the OSLEEL terms\footnote{Curiously enough, from eqs.(\ref{F})-(\ref{Qns}) of Appendix \ref{Momentum factors} we see that at ${\alpha'}^9$ order the $5$-point amplitude does not contain any $non \ trivial$ MZV's and, therefore, we suspect that the OSLEEL can eventually be completely determined at this $\alpha'$ order, by only using $4$-point amplitude information.}. For example, the ${\alpha'}^p D^{2p-4}F^4$ terms of the OSLEEL (for $p=2, 3, 4, \ldots $) are still going to be $completely$ determined by the $4$-point amplitude \cite{DeRoo1,Chandia1}.  \\
\noindent One might ask if new MZV's, besides the ones that already appear in the $5$-point amplitude, will eventually appear at a given (highly enough) $\alpha'$ order. This would be a signal, from this $\alpha'$ order onwards, that $\alpha'$ information from a $6$-point (or eventually higher $N$-point) amplitude would be required, in order to determine those $\alpha'$ terms of the OSLEEL. But this does not seem to be the case. A remarkable observation was done in ref.\cite{Schlotterer1}, claiming that the coefficients of the $\alpha'$ expansion of the $F^{\{\sigma_N\}}(\alpha')$ momentum factors, for $N>5$, are always the same as the ones that appear for $N=5$: what changes (from one $N$ to another) is only the kinematic polynomial which is being multiplied by this coefficient (see Appendix \ref{N5 and higher} for more details)\footnote{For increasing $N$, the number of independent Mandelstam variables grows and, therefore, the kinematic polynomial, which depends on them, gets bigger.}. This conjecture is also consistent with the recent discovery that the $F^{\{\sigma_N\}}(\alpha')$'s can be iteratively obtained (from the $N$-point point of view), to any order in $\alpha'$, from a unique and same Drinfeld associator (which is a generating series for the MZV's)\cite{Broedel1}. So, the MZV's that already appear in the $\alpha'$ expansion of the $5$-point amplitude, will be the same ones that will appear for higher $N$-point amplitudes\footnote{Shortly speaking, the reason of why not all the MZV's of the $\alpha'$ expansion of the $5$-point function do appear in the $4$-point case (see eqs.(\ref{formula1}) and (\ref{expansionGamma})), is that the kinematic polynomial that would go multiplying them is $zero$, in the case of $N=4$. See refs. \cite{Schlotterer1} and \cite{Broedel1} for more details of this explanation.}.\\
So we are left with the conjecture that our revisited S-matrix method plus the $\alpha'$ expansion of the $5$-point amplitude, are enough informations to find $all$ the $\alpha'$ corrections coming from Open Superstring Theory to the Yang-Mills lagrangian\footnote{Even the ${\alpha'}^p D^{2p-4}F^4$ terms (for $any$ $p=2, 3, 4, \ldots $), that were found in \cite{Chandia1} using the $4$-point amplitude, could in principle be determined using $only$ the $5$-point amplitude $\alpha'$ expansion.}.\\
\noindent In spite of this conjecture, for the calculations that we do in this work, which at most go to ${\alpha'}^6$ order (see section \ref{Finding the}), we will still use only the known $4$-point $\alpha'$ expansion (see Appendix \ref{Gamma factor}).

\section{Basis for open superstring and Yang-Mills subamplitudes}

\label{Kinematical}

In this section we prove that, from a kinematical structure perspective, open superstring and Yang-Mills (tree level) subamplitudes belong to a same space of kinematical $N$-point expressions (where $N \ge 3$) and that a possible basis for this space is given by the $(N-3)!$ Yang-Mills subamplitudes\footnote{The arguments that we present in this work have been only proved for $N=3,4,5,6,7$, but we suspect that they can be generalized for an arbitrary $N$.}
\begin{eqnarray}
{\cal B}_N & = & \Big \{ \ A_{YM}(1,\{ 2_{\sigma}, 3_{\sigma}, \ldots, (N-2)_{\sigma} \}, N-1, N) \ , \ \ \ \sigma \ \varepsilon \ S_{N-3} \ \Big \} \ ,
\label{basis}
\end{eqnarray}
where $\{ 2_{\sigma}, 3_{\sigma}, \ldots, (N-2)_{\sigma} \}$ denotes a $\sigma$ permutation of  indexes $\{ 2,3, \ldots, (N-2) \}$\footnote{All the kinematical proof that we deal with does not depend on the spacetime dimension D, as it also happens with the BCJ relations \cite{Bern1}, for example. In spite of this, evidence has been found in ref. \cite{Nastase1} that the basis of this space might indeed depend on D.\\
It will happen that this subtlety can be kept aside from the kinematical analysis that we do. The implications of our results will still be valid for $any$ D, as it happens with the the BCJ relations.\\
See the introductory paragraph in Appendix \ref{Linear}, for a few more details about this.}. \\
In order to acheive this, in subsection \ref{Some general} we review some important facts about the structure of gauge boson scattering subamplitudes. Then, based on this previous background, in the following subsections we argue, case by case (from $N=3$ up to $N=7$), that the set of subamplitudes in (\ref{basis}) is indeed a basis for the corresponding space\footnote{The details of the computations that support our claim, when $N=5,6,7$, are given in Appendix \ref{N-point basis}.}.

\subsection{Some general facts about the structure of scattering amplitudes of gauge bosons}

\label{Some general}

If we compare open superstring and Yang-Mills (tree level) $N$-point subamplitudes for gauge bosons, both theories being treated in the Lorentz gauge, from the point of view of the structure of its kinematical terms they have in common the constraint in eq. (\ref{absence}), namely, the absence of $(\zeta \cdot k)^N$ terms. In the first case this has been recently emphasized in \cite{Barreiro0}, together with the strong implications that it has for determining the bosonic terms of the low energy effective lagrangian of the theory in a very simplified way and, in the second case, the claim in (\ref{absence}) can easily be confirmed by considering Feynman rules in the construction of tree level scattering amplitudes.\\
So, let us consider the space of all scalar $N$-point kinematical expressions constructed with the polarizations $\zeta_i$ and the momenta $k_i$  of $N$ external gauge bosons in a nonabelian theory (like Open Superstring Theory or Yang-Mills theory, for example). The momenta and polarizations should satisfy:
\begin{eqnarray}
\label{momentum}
\mbox{Momentum conservation:} & \ \ & k_1^{\mu} + k_2^{\mu} + \ldots  + k_N^{\mu} = 0 \ .\\
\label{mass-shell}
\mbox{Mass-shell condition:} & \ \ & k_1^2 = k_2^2 = \ldots = k_N^2 = 0 \ . \\
\label{transversality}
\mbox{Transversality (Lorentz gauge) condition:} & \ \ & \zeta_i \cdot k_i  =  0 \ , (i = 1, \ldots , N) \ .
\end{eqnarray}
Let us denote this space by ${\cal V}_N$. We further restrict ${\cal V}_N$ such that its elements $T(1,2,\ldots, N)$ obey the following conditions:   \\
\begin{eqnarray}
\begin{array}{rl}
\mbox{1.} & \mbox{They are multilinear in the polarizations $\zeta_i$ ($i=1,2, \ldots, N$).} \\
\mbox{2.} & \mbox{They do not contain $(\zeta \cdot k)^N$ terms.} \\
\mbox{3.} & \mbox{(On-shell) Gauge invariance: whenever any $\zeta_i \rightarrow k_i$ ($i=1,2, \ldots, N$) then $T(1,2,\ldots, N)$} \\
& \mbox{becomes $zero$.}  \\
\end{array}
\label{requirements}
\end{eqnarray}
\noindent These four requirements are simply properties of tree level gauge boson subamplitudes in Open Superstring and Yang-Mills theories and we want the elements of ${\cal V}_N$ to also satisfy them. Since these elements, the $T(1,2,\ldots, N)$'s, are Lorentz scalars, they can only be constructed from linear combinations of \footnote{Some care must be taken with the expression ``linear combination'' because, as we will see immediately after (\ref{general-form}), the coefficients that go multiplying the terms that appear in it will, in general, depend in the momenta $k^{\mu}_i$ .}
\begin{eqnarray}
(\zeta \cdot \zeta)^1 (\zeta \cdot k)^{N-2} ,  (\zeta \cdot \zeta)^2 (\zeta \cdot k)^{N-4}, \ldots ,  (\zeta \cdot \zeta)^{[N/2]} (\zeta \cdot k)^{N - 2 [N/2]} \ ,
\label{general-form}
\end{eqnarray}
where $[p]$ denotes the integer part of $p$ and $N \ge 3$. \\
Besides the terms excluded in (\ref{absence}) and the ones that we have mentioned in (\ref{general-form}), there are no more possibilities of kinematical terms that can be constructed from the polarizations $\zeta_i$ and the momenta $k_i$ of the external gauge bosons. The only place where some extra dependence in the momenta could be considered is in the scalar coefficients which in $T(1,2,\ldots, N)$ go multiplying the kinematical terms in (\ref{general-form}): these coefficients are allowed to be given in the terms of the $k_i \cdot k_j$ factors (or equivalently, in terms of the Mandelstam variables) and, eventually, in terms of a length scale (for example $\sqrt{\alpha'}$, in the case of String Theory) .  \\
For example, in the case of $N=3$, the Yang-Mills subamplitude contains only $ (\zeta \cdot \zeta)^1 (\zeta \cdot k)^1$ terms. It is given by \cite{Schwarz1}
\begin{eqnarray}
A_{YM}(1,2,3) & = & 2 g \ \big[  \  (\zeta_1 \cdot k_2)(\zeta_2
\cdot \zeta_3) + (\zeta_2 \cdot k_3)(\zeta_3 \cdot \zeta_1) + (\zeta_3
\cdot k_1)(\zeta_1 \cdot \zeta_2)  \ \big] \ .
\label{A3YM}
\end{eqnarray}
It is easy to see that it is an element of ${\cal V}_3$. In subsection \ref{N3} we will prove that the only elements of ${\cal V}_3$ are multiples of $A_{YM}(1,2,3)$ (see eq.(\ref{3point3})), so the fact that $A_{YM}(1,2,3) \ \varepsilon \ {\cal V}_3$ will turn out to be an immediate thing. \\
In order to do a counting of all the possible independent kinematical terms that can be taken from the list in (\ref{general-form}) to construct the expression for an element $T(1,2,\ldots , N) \ \varepsilon \ {\cal V}_N$, respecting the kinematic conditions in (\ref{momentum}), (\ref{mass-shell}) and (\ref{transversality}) and also the requirement in (\ref{absence}), we need first to analize the $(\zeta \cdot k)$ terms. In principle, for each $i$ there are $N$ possible $(\zeta_i \cdot k_j)$  terms (because $j$ runs from $1$ to $N$). But taking into account the the transversality condition (\ref{transversality}) and also momentum conservation (\ref{momentum}), this implies that for each $i$ ($=1, 2, \ldots, N$) we have:
\begin{eqnarray}
\sum_{j \neq i}^N (\zeta_i \cdot k_j)  = 0 \ .
\label{restrict1}
\end{eqnarray}
So, at the end, for each $i$ there are only $(N-2)$ independent $(\zeta_i \cdot k_j)$  terms.\\
With this information we are in conditions to do a counting of the different independent terms, specified by structure in eq. (\ref{general-form}),  that in principle are allowed to appear in $T(1,2, \ldots, N)$.  This leads us to the following table, for the number of independent allowed kinematical terms of $T(1,2, \ldots, N)$:
\begin{eqnarray}
    \begin{tabular}{c|c|c}
    \hline
       $N$ & Element of ${\cal V}_N$  & Number of independent allowed kinematical terms \\
      \hline
      3 & $T(1,2,3)$ & 3 $ (\zeta \cdot \zeta)^1 (\zeta \cdot k)^1$ terms \\
      4 &  $T(1,2,3,4)$ &  24 $ (\zeta \cdot \zeta)^1 (\zeta \cdot k)^2 $ , 3 $(\zeta \cdot \zeta)^2$ terms \\
      5 & $T(1,2,3,4,5)$ &  270 $ (\zeta \cdot \zeta)^1 (\zeta \cdot k)^3 $ , 45 $ (\zeta \cdot \zeta)^2 (\zeta \cdot k)^1$ terms \\
      6 & $T(1,2,3,4,5,6)$ &  3840 $ (\zeta \cdot \zeta)^1 (\zeta \cdot k)^4$ , 720 $ (\zeta \cdot \zeta)^2 (\zeta \cdot k)^2$ , 15  $(\zeta \cdot \zeta)^3$ terms \\
      7 & $T(1,2,3,4,5,6,7)$ & 65625 $ (\zeta \cdot \zeta)^1(\zeta \cdot k)^5$ , 13125 $ (\zeta \cdot \zeta)^2 (\zeta \cdot k)^3$ , 525  $ (\zeta \cdot \zeta)^3 (\zeta \cdot k)^1$ terms
      \label{table1}  \\
      \hline
      \end{tabular}
\end{eqnarray}

\vspace{0.5cm}

\noindent It is not difficult to prove that the number of independent $(\zeta \cdot \zeta)^j (\zeta \cdot k)^{N-2j} $ terms (where $j= 1, 2, \ldots, [N/2]$), for an arbitrary $N$, is given by
\begin{eqnarray}
d_{N,j} & = & \frac{N (N-1) (N-2) \ldots (N-(2j-1))}{2^j \ j!} \ (N-2)^{N-2j} \ .
\label{number}
\end{eqnarray}
This is the formula that has been used to compute the corresponding number of kinematical terms in table (\ref{table1})\footnote{The number of independent kinematical terms mentioned in (\ref{table1}) should not be confused with the number of independent \underline{coefficients} that will appear multiplying each of those terms. It might even happen that some of these coefficients will become $zero$ after demanding the requirement of on-shell gauge invariance and, consequently, the corresponding kinematical terms referred to in table (\ref{table1}) will not be present in the final expression of $T(1, \ldots, N)$.}. We have only worried to specify in this table the cases $N=3,4,5,6,7,$ because these are the ones that we will consider in the present work. \\
In the next subsections, when writing all the allowed independent kinematical terms in $T(1,\ldots, N)$, in particular when choosing the $(N-2)$ independent $(\zeta_i \cdot k_j)$ terms, our choice will be the following list of $N (N-2)$ terms :
\begin{eqnarray}
\begin{array}{cccc}
\Big \{ (\zeta_1 \cdot k_2) \ , &  (\zeta_1 \cdot k_3) \ , & \ldots \ , & (\zeta_1 \cdot k_{N-1}) \ ,  \hphantom{ \ \Big \}  }\\
\hphantom{ \Big \{  } (\zeta_2 \cdot k_1) \ , &  (\zeta_2 \cdot k_3) \ , & \ldots \ , & (\zeta_2 \cdot k_{N-1}) \ ,  \hphantom{ \ \Big \}  } \\
\hphantom{ \Big \{  }     \vdots & \vdots & \vdots & \vdots \\
\hphantom{ \Big \{  } (\zeta_{N-1} \cdot k_1) \ , &  (\zeta_{N-1} \cdot k_2) \ , & \ldots \ , & (\zeta_{N-1} \cdot k_{N-2}) \ , \hphantom{ \ \Big \}  } \\
\hphantom{ \Big \{  } (\zeta_{N} \cdot k_1) \ , &  (\zeta_{N} \cdot k_2) \ , & \ldots \ , & (\zeta_{N} \cdot k_{N-2}) \ \Big \} \ ,
\end{array}
\label{zetakterms}
\end{eqnarray}
that is, for each $i$ ($=1, 2, \ldots, N-1$) we have used the restriction in (\ref{restrict1}) to eliminate $(\zeta_i \cdot k_N)$ in terms of the remaing $(N-2)$  $(\zeta_i \cdot k_j)$ terms ($j=1, 2, \ldots, N-1$, with $j \neq i$) and, in the last line of (\ref{zetakterms}), we have eliminated $(\zeta_N \cdot k_{N-1})$ in terms of the remaining $(\zeta_N \cdot k_j)$ terms ($j=1, 2, \ldots, N-2$).\\

\subsection{Finding a basis for ${\cal V}_N$, when $3 \leq N \leq 7$}

\label{Finding a}

\noindent In this subsection we will explicitly present the derivation of a basis for ${\cal V}_N$ only in the case $N=3$ and $N=4$, which involve simple calculations and illustrate the procedure to arrive at that basis. For $N=5,6,7$ we will just mention the final result and we will leave the details of the main calculations to Appendix \ref{N-point basis}.\\

\subsubsection{Case of $N=3$}

\label{N3}

It is well known that in the case of $3$-point amplitudes of massless states, momentum conservation and the mass-shell condition (\ref{mass-shell}) imply that the momenta obey
the relations
\begin{eqnarray}
k_i \cdot k_j & = & 0 \ (i,j = 1,2,3) \ .
\label{kikj0}
\end{eqnarray}
According to the table in (\ref{table1}), $T(1,2,3)$ has only three independent $(\zeta \cdot \zeta)^1 (\zeta \cdot k)^1$ terms, which, following our prescription in (\ref{zetakterms}), leads to:
\begin{eqnarray}
T(1,2,3) & = &   \lambda_1 (\zeta_1 \cdot k_2)(\zeta_2 \cdot \zeta_3) + \lambda_2 (\zeta_2 \cdot k_1)(\zeta_3 \cdot \zeta_1)
+ \lambda_3 (\zeta_3 \cdot k_1)(\zeta_1 \cdot \zeta_2) \ .
\label{3point}
\end{eqnarray}
Demanding on-shell gauge invariance for gauge boson 1, $A(1,2,3)$ should become $0$  when $\zeta_1 \rightarrow k_1$. Using the mass-shell condition (\ref{mass-shell}) for $k_1$  and the condition in (\ref{kikj0}), then
\begin{eqnarray}
T(1,2,3)\Big|_{\zeta_1 = k_1} = 0 \nonumber \\
\Rightarrow  \lambda_2 (\zeta_2 \cdot k_1)(\zeta_3 \cdot k_1) + \lambda_3 (\zeta_3 \cdot k_1)(\zeta_2 \cdot k_1) = 0 \nonumber \\
\Rightarrow  (\lambda_2 + \lambda_3)  (\zeta_2 \cdot k_1)(\zeta_3 \cdot k_1) = 0 \nonumber \\
\Rightarrow  \lambda_2 = -\lambda_3 \ .
\label{3-point-on-shell-1}
\end{eqnarray}
Similarly, demanding on-shell gauge invariance for gauge bosons 2 and 3 we arrive, respectively, to the conditions
\begin{eqnarray}
\label{3-point-on-shell-2}
\lambda_3 = \lambda_1 \ , \\
\label{3-point-on-shell-3}
\lambda_1 = -\lambda_2 \ .
\end{eqnarray}
To arrive to (\ref{3-point-on-shell-2}) and (\ref{3-point-on-shell-3}) we have needed to use, in accordance to eq. (\ref{zetakterms}), that $(\zeta_3 \cdot k_2) = - (\zeta_3 \cdot k_1)$, $(\zeta_1 \cdot k_3) = - (\zeta_1 \cdot k_2)$ and $(\zeta_2 \cdot k_3) = - (\zeta_2 \cdot k_1)$.\\
The three conditions $\{ (\ref{3-point-on-shell-1}),  (\ref{3-point-on-shell-2}),  (\ref{3-point-on-shell-3}) \}$ are linearly dependent. From them we conclude that $\lambda_1 =  - \lambda_2 =  \lambda_3 $ and, for convenience, we choose them to be $2 g \lambda$. So, finally, (\ref{3point}) can be written as\footnote{In eq.(\ref{3point2}) we have substituted back the relation $(\zeta_2 \cdot k_1) = - (\zeta_2 \cdot k_3)$ in order to obtain the familar expression of $A_{YM}(1,2,3)$.}
\begin{eqnarray}
T(1,2,3) & = &   \lambda \cdot 2 g  \Big[  (\zeta_1 \cdot k_2)(\zeta_2 \cdot \zeta_3) +  (\zeta_2 \cdot k_3)(\zeta_3 \cdot \zeta_1)
+  (\zeta_3 \cdot k_1)(\zeta_1 \cdot \zeta_2) \Big] \ ,
\label{3point2}
\end{eqnarray}
where $\lambda$ is an arbitrary dimensionless factor and $g$ is the open string coupling constant (which agrees with the one from the Yang-Mills lagrangian). \\
Eq. (\ref{3point2}) can equivalently be written as
\begin{eqnarray}
T(1,2,3) & = &   \lambda \cdot A_{YM}(1,2,3) \ ,
\label{3point3}
\end{eqnarray}
where $A_{YM}(1,2,3)$ is has been given in eq. (\ref{A3YM}).\\
So ${\cal B}_3= \{  A_{YM}(1,2,3) \}$ is a basis for ${\cal V}_3$ and the dimension of ${\cal V}_3$ is 1 ($dim({\cal V}_3)=1$).  \\
In this case, the constant $\lambda$ cannot have any momentum dependence due to the conditions (\ref{kikj0}), so it can only be a numerical constant. \\
\noindent Two trivial, but immediate applications of (\ref{3point3}) are the case of $T(1,2,3)$ being any of the Yang-Mills 3-point subamplitudes, for which this equation becomes the cyclic or the reflection (or combination of both) properties (where $\lambda=1$ or $\lambda=-1$) and the case of $T(1,2,3)$ being the open superstring subamplitude, $A(1,2,3)$, for which eq.(\ref{3point3}) implies that $A(1,2,3)$ receives no $\alpha'$ corrections and that $\lambda=1$.

\subsubsection{Case of $N=4$}

\label{N4}

In this case, according to the table in (\ref{table1}) the open superstring 4-point subamplitude has a kinematical expression which consists of twenty four $(\zeta \cdot \zeta)^1 (\zeta \cdot k)^{2}$ and three $(\zeta \cdot \zeta)^2$ terms, all of them being independent, so, following the prescrition in (\ref{zetakterms}), leads to:
\begin{eqnarray}
T(1,2,3,4) & = & \lambda_1 (\zeta_1 \cdot \zeta_2) (\zeta_3 \cdot k_1) (\zeta_4 \cdot k_1) + \lambda_2 (\zeta_1 \cdot \zeta_2) (\zeta_3 \cdot k_1) (\zeta_4 \cdot k_2) + \lambda_3 (\zeta_1 \cdot \zeta_2) (\zeta_3 \cdot k_2) (\zeta_4 \cdot k_1) + \nonumber \\
&& \lambda_4 (\zeta_1 \cdot \zeta_2) (\zeta_3 \cdot k_2) (\zeta_4 \cdot k_2) + \lambda_5 (\zeta_1 \cdot \zeta_3) (\zeta_2 \cdot k_1) (\zeta_4 \cdot k_1) + \lambda_6 (\zeta_1 \cdot \zeta_3) (\zeta_2 \cdot k_1) (\zeta_4 \cdot k_2) + \nonumber \\
&& \lambda_7 (\zeta_1 \cdot \zeta_3) (\zeta_2 \cdot k_3) (\zeta_4 \cdot k_1) + \lambda_8 (\zeta_1 \cdot \zeta_3) (\zeta_2 \cdot k_3) (\zeta_4 \cdot k_2) + \lambda_9 (\zeta_1 \cdot \zeta_4) (\zeta_2 \cdot k_1) (\zeta_3 \cdot k_1) + \nonumber \\
&& \lambda_{10} (\zeta_1 \cdot \zeta_4) (\zeta_2 \cdot k_1) (\zeta_3 \cdot k_2) + \lambda_{11} (\zeta_1 \cdot \zeta_4) (\zeta_2 \cdot k_3) (\zeta_3 \cdot k_1) + \lambda_{12} (\zeta_1 \cdot \zeta_4) (\zeta_2 \cdot k_3) (\zeta_3 \cdot k_2) + \nonumber \\
&& \lambda_{13} (\zeta_2 \cdot \zeta_3) (\zeta_1 \cdot k_2) (\zeta_4 \cdot k_1) + \lambda_{14} (\zeta_2 \cdot \zeta_3) (\zeta_1 \cdot k_2) (\zeta_4 \cdot k_2) + \lambda_{15} (\zeta_2 \cdot \zeta_3) (\zeta_1 \cdot k_3) (\zeta_4 \cdot k_1) + \nonumber \\
&& \lambda_{16} (\zeta_2 \cdot \zeta_3) (\zeta_1 \cdot k_3) (\zeta_4 \cdot k_2) + \lambda_{17} (\zeta_2 \cdot \zeta_4) (\zeta_1 \cdot k_2) (\zeta_3 \cdot k_1) + \lambda_{18} (\zeta_2 \cdot \zeta_4) (\zeta_1 \cdot k_2) (\zeta_3 \cdot k_2) + \nonumber \\
&& \lambda_{19} (\zeta_2 \cdot \zeta_4) (\zeta_1 \cdot k_3) (\zeta_3 \cdot k_1) + \lambda_{20} (\zeta_2 \cdot \zeta_4) (\zeta_1 \cdot k_3) (\zeta_3 \cdot k_2) + \lambda_{21} (\zeta_3 \cdot \zeta_4) (\zeta_1 \cdot k_2) (\zeta_2 \cdot k_1) + \nonumber \\
&& \lambda_{22} (\zeta_3 \cdot \zeta_4) (\zeta_1 \cdot k_2) (\zeta_2 \cdot k_3) + \lambda_{23} (\zeta_3 \cdot \zeta_4) (\zeta_1 \cdot k_3) (\zeta_2 \cdot k_1) + \lambda_{24} (\zeta_3 \cdot \zeta_4) (\zeta_1 \cdot k_3) (\zeta_2 \cdot k_3) + \nonumber \\
&& \rho_{1} (\zeta_1 \cdot \zeta_2) (\zeta_3 \cdot \zeta_4)  + \rho_{2} (\zeta_1 \cdot \zeta_3) (\zeta_2 \cdot \zeta_4)  + \rho_{3} (\zeta_1 \cdot \zeta_4) (\zeta_2 \cdot \zeta_3)  \ .
\label{4point}
\end{eqnarray}
In Appendix \ref{Calculations} we find that demanding on-shell gauge invariance of the subamplitude when $\zeta_1 \rightarrow k_1$ we can arrive to the following thirteen linearly independent relations:
\begin{eqnarray}
\lambda_4 = 0 \ , \ \lambda_8=0 \ , \ \lambda_{12}=0 \ , \
\lambda_3 + \lambda_{10}=0 \ ,\ \lambda_{7} +\lambda_{11}=0 \ , \ \lambda_{1}  +\lambda_{5}+\lambda_{9}=0 \ ,\ \lambda_{6} +\lambda_{2}=0 \ , \nonumber \\
\ 2 \rho_{1} +\lambda_{21}s-\lambda_{23}(s+t)=0 \ , \
2 \rho_{2} +\lambda_{17}s-\lambda_{19}(s+t)=0 \ , \ 2 \rho_{3} +\lambda_{13}s-\lambda_{15}(s+t)=0 \ , \ \nonumber \\
 \lambda_{14}s - \lambda_{16} (s+t) =0 \ , \ \lambda_{18} s - \lambda_{20}(s+t) =0 \ ,  \ \lambda_{22} s - \lambda_{24}(s+t) =0 \ , \ \hspace{1cm}
\label{4point1}
\end{eqnarray}
where
\begin{eqnarray}
s =  (k_1 + k_2)^2 = 2 k_1 \cdot k_2 \ \ \ \mbox{and} \ \ \ t =  (k_1 + k_4)^2 = 2 k_1 \cdot k_4
\label{Mandelstam4}
\end{eqnarray}
are two of the three Mandelstam variables that appear in the 4-point scattering\footnote{The other one is $u=(k_1 +k_3)^2 =  2 k_1 \cdot k_3 = -s -t$.}'\footnote{Notice that our convention for the $4$-point Mandelstam variables has a different sign than the common one in very cited references, like \cite{Schwarz1}, \cite{Green1} and \cite{Polchinski1}. We have followed this convention in order to be compatible with the one that we use for higher $N$-point Mandelstam variables. See Appendix \ref{Mandelstam variables}}.\\
The set of equations in (\ref{4point1}) is the 4-point analog to eq.(\ref{3-point-on-shell-1}), found for the 3-point subamplitude when
$\zeta_1 \rightarrow k_1$.\\
In the same way, demanding on-shell gauge invariance of the $T(1,2,3,4)$, when  $\zeta_2 \rightarrow k_2$,  $\zeta_3 \rightarrow k_3$ and $\zeta_4 \rightarrow k_4$, we can arrive to a set of thirteen linearly independent equations in each case. The explicit expression of these additional equations and the details of its solution can be found in Appendix \ref{Calculations}.\\
The important thing is that in the whole set of fifty two equations that come from demanding on-shell gauge invariance (eqs. (\ref{4point1}),  (\ref{4point4}),    (\ref{4point5}) and  (\ref{4point6})) only half of them are linearly independent as a whole. The solution of this system is given in Appendix \ref{Calculations}, in eq. (\ref{4pointsolution}): it consists of 7 null coefficients ($\lambda_1$, $\lambda_4$, $\lambda_8$, $\lambda_{12}$, $\lambda_{15}$, $\lambda_{19}$ and $\lambda_{21}$) and 20 which are given in terms of the Mandelstam variables $s$ and $t$ and a $unique$ arbitrary parameter, which, for convenience, we have chosen to be $\lambda_{24}$, written as $4 g^2 \lambda^{\{2\}} /t$ (where $\lambda^{\{2\}}$ is arbitrary). After substituing this solution in (\ref{4point}) we finally arrive to\footnote{Substituing eqs. (\ref{4point1}),  (\ref{4point4}),    (\ref{4point5}) and  (\ref{4point6}), together with $\lambda_{24}=4 g^2 \lambda^{\{2\}} /t$, in eq. (\ref{4point}), leads to an expression which is only on-shell equivalent to the one in eq.(\ref{A1234-1}): it is necessary to use eqs. (\ref{u}) and (\ref{additional}) in order to check the equivalence between both formulas.}
\begin{eqnarray}
T(1,2,3,4) & = & \lambda^{\{2\}} \cdot 8 g^2 \frac{1}{st}   \Biggl\{ - \frac{1}{4} \Bigl[ ts(\zeta_1 \cdot \zeta_3)(\zeta_2 \cdot
  \zeta_4) + su(\zeta_2 \cdot \zeta_3)(\zeta_1 \cdot \zeta_4) +
  ut(\zeta_1 \cdot \zeta_2)(\zeta_3 \cdot \zeta_4) \Bigr] +
\nonumber
\\&&\hphantom{  \lambda \cdot 8 g^2 \frac{1}{st}   \Biggl\{    } - \frac{1}{2} s \Bigl[ (\zeta_1 \cdot k_4)(\zeta_3 \cdot
  k_2)(\zeta_2 \cdot \zeta_4) + (\zeta_2 \cdot k_3)(\zeta_4 \cdot
  k_1)(\zeta_1 \cdot \zeta_3) +
\nonumber \\&&
\hphantom{ \lambda \cdot 8 g^2 \frac{1}{st}   \Biggl\{  + \frac{1}{2} s \Bigl[}
+(\zeta_1 \cdot k_3)(\zeta_4 \cdot
  k_2)(\zeta_2 \cdot \zeta_3)
+ (\zeta_2 \cdot k_4)(\zeta_3 \cdot k_1)(\zeta_1 \cdot \zeta_4)
  \Bigr] +
\nonumber\\ &&\hphantom{  \lambda \cdot 8 g^2 \frac{1}{st}   \Biggl\{    } - \frac{1}{2} t \Bigl[ (\zeta_2 \cdot k_1)(\zeta_4 \cdot
  k_3)(\zeta_3 \cdot \zeta_1) + (\zeta_3 \cdot k_4)(\zeta_1 \cdot
  k_2)(\zeta_2 \cdot \zeta_4) +
\nonumber \\&& \hphantom{ \lambda \cdot 8 g^2 \frac{1}{st}   \Biggl\{  + \frac{1}{2} s \Bigl[}
+ (\zeta_2 \cdot k_4)(\zeta_1 \cdot k_3)(\zeta_3 \cdot \zeta_4) +
  (\zeta_3 \cdot k_1)(\zeta_4 \cdot k_2)(\zeta_2 \cdot \zeta_1)
  \Bigr] +
\nonumber\\ && \hphantom{  \lambda \cdot 8 g^2 \frac{1}{st}   \Biggl\{    }
-\frac{1}{2} u \Bigl[ (\zeta_1 \cdot k_2)(\zeta_4 \cdot
  k_3)(\zeta_3 \cdot \zeta_2) +
 (\zeta_3
  \cdot k_4)(\zeta_2 \cdot k_1)(\zeta_1 \cdot \zeta_4) +
\nonumber \\&& \hphantom{ \lambda \cdot 8 g^2 \frac{1}{st}   \Biggl\{  + \frac{1}{2} s \Bigl[}
+  (\zeta_1 \cdot k_4)(\zeta_2 \cdot k_3)(\zeta_3 \cdot \zeta_4) +
  (\zeta_3 \cdot k_2)(\zeta_4 \cdot k_1)(\zeta_1 \cdot \zeta_2)
  \Bigr] \ \Biggr\} \ ,
\label{A1234-1}
\end{eqnarray}
or equivalently,
\begin{eqnarray}
T(1,2,3,4) & = & \lambda^{\{2\}} \cdot A_{YM}(1,\{2\},3,4) \ ,
\label{A1234-2}
\end{eqnarray}
where $A_{YM}(1,\{2\},3,4)=A_{YM}(1,2,3,4)$ is the familiar Yang-Mills 4-point subamplitude\footnote{This explicit expression for $A_{YM}(1,2,3,4)$ can be found in many places in the literature, for example, in section 3 of \cite{Brandt1}. But there are some sign differences due to the different convention that we have used for the $4$-point Mandelstam variables in eq.(\ref{Mandelstam4}).}'\footnote{We have used the curly brackets, `$\{ \}$', in the index $2$, just as a remainder of the rule, mentioned in eq.(\ref{basis}), to choose the Yang-Mills subamplitudes of ${\cal B}_N$.}.\\
So ${\cal B}_4=  \{  A_{YM}(1,2,3,4) \}$ is a basis for ${\cal V}_4$ and $dim({\cal V}_4)=1$.  \\
In (\ref{A1234-2}) $\lambda^{\{2\}}$ is a dimensionless factor which may depend in the dimensionless variables $\alpha' s$ and $\alpha' t$, where, at this point, $\alpha'$ could be any squared length scale (that in the case of String Theory would be the string fundamental constant). $\lambda^{\{2\}}$ corresponds to what it is usually called the ``momentum factor''.

\subsubsection{Case of $N=5, 6, 7$}

\label{N567}

The procedure for $N>4$ is exactly the same one presented in subsections \ref{N3} and \ref{N4}, but for the corresponding $N$-point subamplitudes whose general kinematical structure we have mentioned in table (\ref{table1}). Since for these cases the calculations are much more involved than the ones that we have already done for $N=4$, we leave the details of them for Appendix \ref{N-point basis}. In all these cases it is possible to arrive to an $(N-3)!$ dimensional basis for ${\cal V_N}$ ($dim({\cal V}_N)=(N-3)!$) and it is always possible to choose this basis in terms of only Yang-Mills subamplitudes, in particular the one indicated in at the beginning of this section, in eq. (\ref{basis}). That set of subamplitudes precisely contains $(N-3)!$ elements. That set constitutes one of the possible basis that have appeared in the literature as basis for the space of Yang-Mills subamplitudes \cite{Bern1, Bjerrum1, Stieberger1} and also for the space of open superstring subamplitudes \cite{Mafra1}, which is the purpose of this whole section to prove.  \\
In Appendix \ref{N-point basis} we see that in the cases for $N>4$ our calculations have not lead directly to the basis in (\ref{basis}), but only after having verified that the dimension of ${\cal V_N}$ is $(N-3)!$ and that the set of amplitudes mentioned in (\ref{basis}) is linearly independent\footnote{Our proposal of basis in eq. (\ref{basis}) is based in the previously known results of the BCJ relations for $N=5,6,7$ \cite{Bern1, Bjerrum1, Stieberger1}. Except for the case of $N=4$, without these previously known results it would have been quite difficult to guess this basis and, as seen in Appendix \ref{N-point basis}, we would have had only $(N-3)!$ known (and very long) $N$-point kinematical expressions for which we would  have no interpretation at all.}.\\
So, summarizing, with the calculations and proofs that we have done in Appendix \ref{N-point basis} we can write an arbitrary element of ${\cal V}_5$, ${\cal V}_6$ and ${\cal V}_7$, respectively as\footnote{In eqs. (\ref{5point3}), (\ref{6point}) and (\ref{7point}) we are following the same sort of notation that the authors of \cite{Mafra1} used to write the linear combinations of the subamplitudes, that is, the momentum factors are labelled by a superscript denoting the corresponding $\sigma_N$ permuation.}
\begin{eqnarray}
\label{5point3}
T(1,2,3,4,5) & = &\lambda^{\{23\}}  A_{YM}(1,\{2,3\},4,5) + \lambda^{\{32\}}  A_{YM}(1,\{3,2\},4,5) \ ,
\end{eqnarray}
\begin{multline}
T(1,2,3,4,5,6)  = \\
\begin{split}
= \ & \lambda^{\{234\}}  A_{YM}(1,\{2,3,4\},5,6) +  \lambda^{\{324\}}  A_{YM}(1,\{3,2,4\},5,6) + \lambda^{\{243\}}  A_{YM}(1,\{2,4,3\},5,6) +  \\
& \lambda^{\{342\}}  A_{YM}(1,\{3,4,2\},5,6) +  \lambda^{\{423\}}  A_{YM}(1,\{4,2,3\},5,6) + \lambda^{\{432\}}  A_{YM}(1,\{4,3,2\},5,6)
\end{split}
\label{6point}
\end{multline}
and
\begin{eqnarray}
\label{7point}
 T(1,2,3,4,5,6,7)  & = &  \lambda^{\{2345\}}  A_{YM}(1,\{2,3,4,5\},6,7) +  \lambda^{\{2354\}}  A_{YM}(1,\{2,3,5,4\},6,7) +  \\
& &\lambda^{\{2435\}}  A_{YM}(1,\{2,4,3,5\},6,7) +  \ldots + \lambda^{\{5432\}}  A_{YM}(1,\{5,4,3,2\},6,7)
 \ , \nonumber
\end{eqnarray}
\noindent where the $\lambda^{\{\sigma_N\}}$'s are the corresponding momentum factors, which may depend in the dimensionless Mandelstam variables ($\alpha' s_i$ and $\alpha' t_j$) of each $N$-point scattering process of gauge bosons.\\
An independent set of Mandelstam variables for $N=5,6,7$ is given in Appendix \ref{N-point basis}. \\
\noindent In the right-hand side of (\ref{7point}) there are $4!=24$ terms that are being summed, in accordance to eq.(\ref{basis}) for $N=7$. \\
The same as in eq. (\ref{A1234-2}), in eqs. (\ref{5point3}), (\ref{6point}) and (\ref{7point}) we have inserted curly brackets, `$\{ \}$', just as a remainder of the rule, mentioned in eq.(\ref{basis}), to choose the Yang-Mills subamplitudes of ${\cal B}_5$, ${\cal B}_6$ and ${\cal B}_7$, respectively.\\

\subsection{Finding the components of an element of ${\cal V}_N$ with respect to the basis ${\cal B}_N$}

\label{Finding the components}

\noindent Once we have accepted that the set ${\cal B}_N$ is indeed a basis for ${\cal V}_N$, at least for $3 \leq N \leq 7$, the next step consists in finding the components of an element of ${\cal V}_N$ with respect to ${\cal B}_N$. We will present here a procedure which we expect to be valid for any $N \geq 3$ (even for $N > 7$)\footnote{At least in this subsection we will work with the hypothesis that ${\cal B}_N$, as specified in (\ref{basis}), is a basis of ${\cal V}_N$ for $any$ $N \geq 3$.}.\\
\noindent Let $T(1, \ldots, N) \ \varepsilon \ {\cal V}_N$. To find the components of $T(1, \ldots, N)$ with respect to the basis ${\cal B}_N$ we have to find the momentum factors, $\lambda^{\{\sigma_N\}}$'s, such that
\begin{eqnarray}
T(1, \ldots, N) & = & \sum_{ \sigma_N \ \varepsilon \ S_{N-3} } \lambda^{\{\sigma_N\}} \ A_{YM}(1,\{ 2_{\sigma}, 3_{\sigma}, \ldots, (N-2)_{\sigma} \}, N-1, N) \ ,
\label{T}
\end{eqnarray}
where  $\sigma_N=\{ 2_{\sigma}, 3_{\sigma}, \ldots, (N-2)_{\sigma} \}$ denotes the same permutation of  indexes $\{ 2,3, \ldots, (N-2) \}$ that we referred to in eq.(\ref{basis}).\\
\noindent The natural (but tedious!) way to find the $\lambda^{\{\sigma_N\}}$'s is by writing down the expression of $T(1, \ldots, N)$, and of each Yang-Mills submplitude in ${\cal B}_N$, in terms of the kinematical terms listed in eq. (\ref{general-form}), with the convention (\ref{zetakterms}) for the $(\zeta \cdot k)$ terms, for example. Then, a linear system for the $\lambda^{\{\sigma_N\}}$'s arises from demanding that the coefficient of each
$(\zeta \cdot \zeta)^j (\zeta \cdot k)^{N-2j}$ term (where $j=1, \ldots, [N/2]$), in both sides of (\ref{T}), is the same. Since this linear system is overdetermined, it is not necessary to consider $all$ the equations of it in order to find the $(N-3)!$ $\lambda^{\{\sigma_N\}}$'s. \\
\noindent For small values of $N$ it is easy to see that considering in (\ref{T}) the kinematical terms with the smallest number of $(\zeta \cdot k)$ terms (that is, the ones for $j=[N/2]$), that is enough information to find the $\lambda^{\{\sigma_N\}}$'s. Consider, for example, the cases of $N=3,4,5$. From the table in eq.(\ref{table1}) we see that the $3$-point amplitude contains $3$ $(\zeta \cdot \zeta)^1 (\zeta \cdot k)^{1}$ terms, the $4$-point amplitude contains $3$ $(\zeta \cdot \zeta)^2$ terms and the $5$-point amplitude contains $45$ $(\zeta \cdot \zeta)^2 (\zeta \cdot k)^{1}$ terms. It is clear, then, that considering these particular kinematical terms in (\ref{T}), that would be enough information to find the $\lambda^{\{\sigma_N\}}$'s, since for $N=3$ and $N=4$ there is only one component and for $N=5$ there are two components to determine.\\
\noindent But, for sufficiently large $N$ \footnote{Consider $N \geq 12$, for example.}, the number of kinematical terms of the type mentioned above (the ones for $j=[N/2]$) is less than $(N-3)!$ and, then, considering only those type of terms will not be enough information to find $all$ the $\lambda^{\{\sigma_N\}}$'s.\\
\noindent Since, at this point, we are worried about a strategy to find the components of $T(1, \ldots, N)$, for an arbitrary $N$, our proposal will be to consider the kinematical terms for $j=1$, namely, the $(\zeta \cdot \zeta)^1 (\zeta \cdot k)^{N-2}$ terms. Those are the ones, among all the possible types of kinematical terms in (\ref{general-form}), which show up in more amount in the most general expression for $T(1, \ldots, N)$\footnote{It is not difficult to see, by considering formula (\ref{number}), that the number of independent $(\zeta \cdot \zeta)^1 (\zeta \cdot k)^{N-2}$ terms  is bigger than $(N-3)!$.}.\\
\noindent So, in the next two sections, when we will look for the momentum factors that are present in the BCJ relations and in the open superstring formula, we will consider the $(\zeta \cdot \zeta)^1 (\zeta \cdot k)^{N-2}$ terms in our kinematical analysis.

\section{Closed expression for the $N$-point disk amplitude using the RNS formalism}

\label{Closed}

\noindent An immediate and natural application of the analysis done in subsection \ref{Finding the components} is to find the momentum factors in the case that $T(1, \ldots, N)$ is {\it any} of the possible Yang-Mills or open superstring $N$-point subamplitudes, because they are gauge invariant and the $(\zeta \cdot k)^N$ are absent in them\footnote{In the case of Yang-Mills subamplitudes, it was already mentioned in the beginning of subsection \ref{Some general} that they do not contain $(\zeta \cdot k)^N$ terms.}. In the first case, the result becomes one of the BCJ relations and in the second case, the result becomes a closed formula for the open superstring subamplitudes. We do the exercise for the BCJ relations in Appendix \ref{BCJ} and in this section we do the corresponding calculation for the open superstring, arriving to MSS's result, in eq.(\ref{MSS}), by means of a RNS formalism (for $3 \leq N \leq 7$).  \\
\noindent We will first derive MSS formula in the case of pure gauge boson scattering in subsection \ref{For gauge bosons only} and then, in subsection \ref{For gauge bosons and massless fermions}, we will quickly see that there is no need to deal with fermion vertex operators (at least for the tree level amplitudes) in order to find the scattering amplitudes involving fermions (once the closed formula has been found for gauge bosons).\\

\subsection{For gauge bosons only}

\label{For gauge bosons only}

\noindent Using the RNS formalism, it is known that the $N$-point subamplitude for gauge bosons in Open Superstring Theory is given by the following integral formula \cite{Green1}\footnote{In eq.(\ref{ANfermionic}) we have used an index {\it b} in the $N$-point subamplitude as a remainder that it corresponds to the scattering process involving only bosons.}:
\begin{eqnarray}
A_b(1, 2, \ldots,N) & = & 2 \frac{g^{N-2}}{(2 \alpha')^2}
 (x_{N-1}-x_1)(x_{N}-x_1) \
 \int_0^{x_{N-1}} dx_{N-2} \int_0^{x_{N-2}} dx_{N-3} \ \dots  \int_0^{x_3} dx_2 \ \times \nonumber \\
&& {} \times \int d \theta_1 \ldots d \theta_{N-2} \prod_{p<q}^N (
 x_q - x_p - \theta_q \theta_p )^{2 \alpha' k_p \cdot k_q}\times  \int d \phi_1 \ldots d \phi_N
\nonumber \\
& &
\times \ \mbox{exp} \left(   \sum_{i \neq j}^N
\frac{  (2 \alpha')^1  (\theta_j-\theta_i) \phi_j (\zeta_j \cdot k_i) -1/2 \ (2 \alpha')^1  \phi_j \phi_i (\zeta_j \cdot \zeta_i) }{x_j-x_i-\theta_j \theta_i}    \right) \ .
\label{ANfermionic}
\end{eqnarray}
\noindent In this formula, the $\theta_i$'s and the $\phi_j$'s are Grassmann variables, while the $x_k$'s are common real variables, where $x_1 < x_2 < \ldots < x_N$. Briefly speaking, the result in eq.(\ref{ANfermionic}) has been obtained by averaging over the ground state of the theory, the product of $N$ gauge boson vertex operators, localized at positions $x_1$, $x_2$, $\ldots$, $x_N$. The residual symmetry of the integrand can be gauged fixed, for example, by demanding $\{ x_1=0$, $x_{N-1}=1$, $x_N = + \infty\}$ and $\{ \theta_{N-1} =\theta_N=0 \}$ \footnote{In (\ref{ANfermionic}) we have already kept $x_1$, $x_{N-1}$, $x_N$, $\theta_{N-1}$ and $\theta_N$ fixed, but we have not yet chosen the peculiar values mentioned in the text. See section 7.3 of ref. \cite{Green1}, for more details.}.\\
\noindent In this section we will prove that in the case of $T(1, \ldots, N)=A_b(1, 2, \ldots,N)$, given in eq.(\ref{ANfermionic}), then the momentum factors of eq.(\ref{T}) are given precisely by the ones appearing in MSS formula, eq.(\ref{MSS}). The integral formula that Mafra, Schlotterer and Stieberger find for them is \cite{Mafra1}
\begin{eqnarray}
F^{\{23 \ldots N-2\}}(\alpha') & = & \int_0^1 dx_{N-2} \int_0^{x_{N-2}} dx_{N-3} \ \ldots \int_0^{x3} dx_{2} \ \biggl(  \prod_{i>j \geq 1}^{N-1} (x_i - x_j)^{2 \alpha' k_i \cdot k_j} \biggr)  \times \nonumber \\
&&  \hphantom{  \int_0^1 dx_{N-2} \int_0^{x_{N-2}} dx_{N-3} \ \ldots \int_0^{x3} \ \   }      \times \biggl\{ \  \prod_{p=2}^{N-2} \ \sum_{q=1}^{p-1} \Big(\frac{ \ 2 \alpha' k_p \cdot k_q \ }{x_p - x_q}\Big) \  \biggr\} \ ,
\label{MSS2}
\end{eqnarray}
where $x_1=0$ and $x_{N-1}=1$\footnote{In the expression in (\ref{MSS2}) $x_N$ has already be taken to $+\infty$: that is why it does not appear on it.}.\\
\noindent Formula (\ref{MSS2}) is the one for the momentum factor which in eq.(\ref{MSS}) goes multiplying the subamplitude $A_{YM}(1, \{ 2, 3, \ldots, N-2 \}, N-1, N)$. The MSS prescription for the remaining $F^{\{ \sigma_N \}}(\alpha')$ momentum factors consists in interchanging the indices $\{ 2, 3, \ldots, N-2 \}$, according to the $\sigma_N$ permutation, {\it only} in the curly brackets of the right hand-side of eq.(\ref{MSS2}). This interchange of indices should be done in {\it both}, the $k_j$'s momenta and the $x_j$'s variables inside the curly brackets. This will become more clear in the case-by-case study that we will consider in the following subsections.\\
\noindent Before going into the derivation of the momentum factors we have two remarks:
\begin{enumerate}
\item As they stand, formulas (\ref{ANfermionic}) and (\ref{MSS2}) are not applicable to the case of $N=3$. Since it is very well known that in this case $A_b(1,2,3)=A_{YM}(1,2,3)$, the $3$-point momentum factor is simply {\it defined} as being $1$.
\item There is an equivalent expression that Mafra, Schlotterer and Stieberger give for the momentum factor in eq.(\ref{MSS2}), for $N \geq 5$, and it comes by integrating by parts on it \cite{Mafra3}:
\begin{eqnarray}
F^{\{23 \ldots N-2\}}(\alpha') & = & \int_0^1 dx_{N-2} \int_0^{x_{N-2}} dx_{N-3} \ \ldots \int_0^{x3} dx_{2} \ \biggl(  \prod_{i>j \geq 1}^{N-1} (x_i - x_j)^{2 \alpha' k_i \cdot k_j} \biggr)  \times \nonumber \\
&&      \times \biggl\{ \  \prod_{p=2}^{[N/2]} \ \sum_{q=1}^{p-1} \Big(\frac{ \ 2 \alpha' k_p \cdot k_q \ }{x_p - x_q}\Big) \  \biggr\} \
 \biggl\{ \  \prod_{p=[N/2]+1}^{N-2} \ \sum_{q=p+1}^{N-1} \Big(\frac{ \ 2 \alpha' k_p \cdot k_q \ }{x_p - x_q}\Big) \  \biggr\} \ , \ \ \ \
\label{MSS3}
\end{eqnarray}
\end{enumerate}
\noindent In the following subsections we will see that we reproduce formula (\ref{MSS2}) in the case of $N=4$ and formula (\ref{MSS3}) in the cases of $N=5,6,7$.\\
\noindent In the case of $N=4$ we will present the derivation with enough details to see the consistency of the relations that arise when we consider the $(\zeta \cdot \zeta)^1 (\zeta \cdot k)^{N-2}$ terms of the subamplitudes. For $N = 5, 6, 7$ the procedure will be the same one, but we will not present so many details because the open superstring and the Yang-Mills subamplitudes become too large. For $N=5$ we will explain how to arrive at the known expression of the two momentum factors (eq.(\ref{MSS3})), together with some evidence of the self consitency of the calculations, and for $N=6,7$ we will just explain how to arrive to the corresponding momentum factors.

\subsubsection{Case of $N=4$}

\label{N4-3}

\noindent In this case the relation (\ref{A1234-2}) guarantees that choosing $T(1,2,3,4)=A_b(1,2,3,4)$ it is possible to write down
\begin{eqnarray}
A_b(1,2,3,4) & = & F^{\{2\}}(\alpha') \ A_{YM}(1, 2,3,4) \ ,
\label{A1234-3}
\end{eqnarray}
for a certain momentum factor $F^{\{2\}}(\alpha')$ that we want to find.\\
\noindent $A_{YM}(1, 2,3,4)$ has already been given in (\ref{A1234-1}) and the expression for $A_b(1,2,3,4)$ is not difficult to be obtained from (\ref{ANfermionic}) for $N=4$ (after expanding the exponential and integrating over the two Grassmann $\theta_i$'s and the four Grassmann $\phi_j$'s).   \\
\noindent Following the procedure proposed in subsection \ref{Finding the components}, after writing all $(\zeta \cdot k)$ terms in the basis given in (\ref{zetakterms}) and equating the $(\zeta \cdot \zeta)^1 (\zeta \cdot k)^2$ terms of both sides of (\ref{A1234-3}), we arrive to
\begin{multline}
2 g^2 \ \biggl\{  \ (\zeta_1 \cdot \zeta_2)  \Big [ \ - (4 \alpha') I^{[4]}_1 (\zeta_3 \cdot k_1) (\zeta_4 \cdot k_2) + (4 \alpha')  I^{[4]}_3 (\zeta_3 \cdot k_2) (\zeta_4 \cdot k_1) \Big] \ +  \hspace{4cm} \\
(\zeta_1 \cdot \zeta_3)  \Big [   (4 \alpha') I^{[4]}_1 (\zeta_2 \cdot k_1) (\zeta_4 \cdot k_1) + (4 \alpha')  I^{[4]}_1 (\zeta_2 \cdot k_1) (\zeta_4 \cdot k_2) - (4 \alpha')  I^{[4]}_2 (\zeta_2 \cdot k_3) (\zeta_4 \cdot k_1)  \  \Big] \ + \\
(\zeta_1 \cdot \zeta_4)  \Big [  \ - (4 \alpha') I^{[4]}_1 (\zeta_2 \cdot k_1) (\zeta_3 \cdot k_1) - (4 \alpha')  I^{[4]}_3 (\zeta_2 \cdot k_1) (\zeta_3 \cdot k_2) + (4 \alpha')  I^{[4]}_2 (\zeta_2 \cdot k_3) (\zeta_3 \cdot k_1)  \  \Big] \ + \\
(\zeta_2 \cdot \zeta_3)  \Big [  - (4 \alpha') I^{[4]}_3 (\zeta_1 \cdot k_2) (\zeta_4 \cdot k_1) - (4 \alpha')  I^{[4]}_3 (\zeta_1 \cdot k_2) (\zeta_4 \cdot k_2) - (4 \alpha')  I^{[4]}_2 (\zeta_1 \cdot k_3) (\zeta_4 \cdot k_2)  \  \Big] \ + \\
(\zeta_2 \cdot \zeta_4)  \Big [  \  (4 \alpha') I^{[4]}_1 (\zeta_1 \cdot k_2) (\zeta_3 \cdot k_1) + (4 \alpha')  I^{[4]}_3 (\zeta_1 \cdot k_2) (\zeta_3 \cdot k_2) + (4 \alpha')  I^{[4]}_2 (\zeta_1 \cdot k_3) (\zeta_3 \cdot k_2)  \  \Big] \ + \\
(\zeta_3 \cdot \zeta_4)  \Big [  -  (4 \alpha') I^{[4]}_3 (\zeta_1 \cdot k_2) (\zeta_2 \cdot k_3) + (4 \alpha')  I^{[4]}_1 (\zeta_1 \cdot k_3) (\zeta_2 \cdot k_1) - (4 \alpha')  I^{[4]}_2 (\zeta_1 \cdot k_3) (\zeta_2 \cdot k_3)  \  \Big] \  \biggr\} = \\
\begin{split}
&= 2 g^2 \ F^{\{2\}}(\alpha')  \ \biggl\{  \  \ (\zeta_1 \cdot \zeta_2)  \Big [  - \frac{4}{s} (\zeta_3 \cdot k_1) (\zeta_4 \cdot k_2) + \big( \frac{4}{s} + \frac{4}{t} \big) (\zeta_3 \cdot k_2) (\zeta_4 \cdot k_1) \ \Big] \ +   \\
&(\zeta_1 \cdot \zeta_3)  \Big [  \ \frac{4}{s} (\zeta_2 \cdot k_1) (\zeta_4 \cdot k_1) +  \frac{4}{s}  (\zeta_2 \cdot k_1) (\zeta_4 \cdot k_2) -  \frac{4}{t} (\zeta_2 \cdot k_3) (\zeta_4 \cdot k_1)  \  \Big] \ + \\
& (\zeta_1 \cdot \zeta_4)  \Big [  - \frac{4}{s} (\zeta_2 \cdot k_1) (\zeta_3 \cdot k_1) - \big( \frac{4}{s} + \frac{4}{t}  \big) (\zeta_2 \cdot k_1) (\zeta_3 \cdot k_2) + \frac{4}{t} (\zeta_2 \cdot k_3) (\zeta_3 \cdot k_1)  \  \Big] \ + \\
& (\zeta_2 \cdot \zeta_3)  \Big [  - \Big( \frac{4}{s} + \frac{4}{t} \Big)  (\zeta_1 \cdot k_2) (\zeta_4 \cdot k_1) -  \Big( \frac{4}{s} + \frac{4}{t} \Big)       (\zeta_1 \cdot k_2) (\zeta_4 \cdot k_2) -     \frac{4}{t}   (\zeta_1 \cdot k_3) (\zeta_4 \cdot k_2) \ \Big] \ +  \\
&(\zeta_2 \cdot \zeta_4)  \Big [  \ \frac{4}{s} (\zeta_1 \cdot k_2) (\zeta_3 \cdot k_1) + \Big( \frac{4}{s} + \frac{4}{t} \Big)  (\zeta_1 \cdot k_2) (\zeta_3 \cdot k_2) + \frac{4}{t} (\zeta_1 \cdot k_3) (\zeta_3 \cdot k_2)  \  \Big] \ + \\
& (\zeta_3 \cdot \zeta_4)  \Big [  -  ( \frac{4}{s} + \frac{4}{t}  )  (\zeta_1 \cdot k_2) (\zeta_2 \cdot k_3) + \frac{4}{s}  (\zeta_1 \cdot k_3) (\zeta_2 \cdot k_1) - \frac{4}{t} (\zeta_1 \cdot k_3) (\zeta_2 \cdot k_3)  \  \Big] \  \biggr\} \ ,
\end{split}
\label{F2-0}
\end{multline}
\noindent where\footnote{In (\ref{I2}) we have introduced the superscript `[4]' as a remainder that the integrals that have appeared are the ones for the $4$-point case.}
\begin{eqnarray}
I^{[4]}_1 = \int_0^1 dx_2 \ {x_2}^{\alpha' s -1} (1-x_2)^{\alpha' t} \ , \ \ \
I^{[4]}_2 = \int_0^1 dx_2 \ {x_2}^{\alpha' s } (1-x_2)^{\alpha' t -1} \ , \nonumber \\
I^{[4]}_3 = \int_0^1 dx_2 \ {x_2}^{\alpha' s -1} (1-x_2)^{\alpha' t -1} \ .  \hspace{2cm}
\label{I2}
\end{eqnarray}
\noindent In (\ref{F2-0}), when we substituted $A_{YM}(1,2,3,4)$ from eq.(\ref{A1234-1}), we have used that $u=-s-t$.\\
\noindent Comparing the coefficient of the $(\zeta_1 \cdot \zeta_2)(\zeta_3 \cdot k_1)(\zeta_4 \cdot k_2 )$ term in both sides of (\ref{F2-0}) we find that
\begin{eqnarray}
F^{\{2\}}(\alpha') & = & \alpha' s \ I^{[4]}_1 \ ,
\label{F2-1}
\end{eqnarray}
or equivalently,
\begin{eqnarray}
\label{F2-2}
F^{\{2\}}(\alpha') & = & 2 \alpha' k_2 \cdot k_1  \int_0^1 dx_2 \ x_2^{2 \alpha' k_2 \cdot k_1-1} (1-x_2)^{2 \alpha' k_3 \cdot k_2} \\
\label{F2-3}
 & = &  \int_0^1 dx_{2} \ \biggl(  \prod_{i>j \geq 1}^{3} (x_i - x_j)^{2 \alpha' k_i \cdot k_j} \biggr) \frac{ \ 2 \alpha' k_2 \cdot k_1 \ }{x_2 - x_1} \ ,
\end{eqnarray}
where $x_1=0$ and $x_3=1$.\\
\noindent Formula (\ref{F2-3}) corresponds precisely to formula (\ref{MSS2}) for the case of $N=4$, as we had anticipated.\\
\noindent Notice that comparing other $(\zeta \cdot \zeta)^1 (\zeta \cdot k)^2$ terms in eq.(\ref{F2-0}) we may arrive to different expressions for $F^{\{2\}}(\alpha')$. For example, if we compare the coefficient of $(\zeta_1 \cdot \zeta_3)(\zeta_2 \cdot k_3)(\zeta_4 \cdot k_1 )$ and the coefficient of $(\zeta_2 \cdot \zeta_3)(\zeta_1 \cdot k_2)(\zeta_4 \cdot k_1 )$, in both sides of (\ref{F2-0}), we find that, respectively,
\begin{eqnarray}
F^{\{2\}}(\alpha') =  \alpha' t \ I^{[4]}_2 \ , \ \ F^{\{2\}}(\alpha')  =  \alpha' \frac{s \ t}{s+t} \ I^{[4]}_3 \ .
\label{F2-4}
\end{eqnarray}
Integrating by parts in the definition of $I^{[4]}_2$ (and considering the analytic continuation in $\alpha' s$ and $\alpha' t$ of the three integrals in eq.(\ref{I2})), we have that
\begin{eqnarray}
I^{[4]}_2 & = & \frac{s}{t} \ I^{[4]}_1 \ .
\label{I2-2}
\end{eqnarray}
Also, multiplying the identity $1/ \{ x_2 (1-x_2) \} = 1/x_2 + 1/(1-x_2)$, by the term $x_2^{2 \alpha' s} (1-x_2)^{2 \alpha' t}$ and integrating in the interval $[0,1]$, this identity becomes
\begin{eqnarray}
I^{[4]}_3 & = I^{[4]}_1 + I^{[4]}_2 \ .
\label{I2-3}
\end{eqnarray}
Using relations (\ref{I2-2}) and (\ref{I2-3}) it is easy to see that the two alternative expressions for $F^{\{2\}}(\alpha')$, given in eq.(\ref{F2-4}), are equivalent to the one in (\ref{F2-1}) (and, therefore, equivalent to the one in (\ref{F2-3})). \\
\noindent One might even ask what would have happended if we had included the three $(\zeta \cdot \zeta)^2$ terms in both sides of (\ref{F2-0}) and compared their corresponding coefficients. The answer is that we would have found two of the three integral representations that we already have for $F^{\{2\}}(\alpha')$ plus a new one:
\begin{eqnarray}
F^{\{2\}}(\alpha') & = & \frac{s \ (\alpha' s-1)}{s+t} \   \int_0^1 dx_2 \ {x_2}^{\alpha' s -2} (1-x_2)^{\alpha' t} \ .
\label{F2-5}
\end{eqnarray}
It can be easily verified that this expression can be obtained using integration by parts in $I^{[4]}_3$, in the second equation in (\ref{F2-4}).

\subsubsection{Case of $N=5$}

\label{N4-5}

\noindent In this case the relation (\ref{5point3}) guarantees that choosing $T(1,2,3,4,5)=A_b(1,2,3,4,5)$ it is possible to write down
\begin{eqnarray}
A_b(1,2,3,4,5) & = &  F^{\{23\}}(\alpha')  \ A_{YM}(1,2,3,4,5) +    F^{\{32\}}(\alpha')  \ A_{YM}(1,3,2,4,5) \ .
\label{A12345-3}
\end{eqnarray}
\noindent Following the procedure proposed in subsection \ref{Finding the components}, in eq.(\ref{A12345-3}) we consider only the $(\zeta \cdot \zeta)^1 (\zeta \cdot k)^3$ terms,
\begin{multline}
A_b(1,2,3,4,5)\Big|_{(\zeta \cdot \zeta)^1 (\zeta \cdot k)^3}  = \\
\begin{split}
&=  F^{\{23\}}(\alpha')  \ A_{YM}(1,2,3,4,5)\Big|_{(\zeta \cdot \zeta)^1 (\zeta \cdot k)^3} +    F^{\{32\}}(\alpha')  \ A_{YM}(1,3,2,4,5)\Big|_{(\zeta \cdot \zeta)^1 (\zeta \cdot k)^3} \ ,
\end{split}
\label{A12345-4}
\end{multline}
where we are supposed to write all $(\zeta \cdot k)$ terms in the basis given in (\ref{zetakterms}) for $N=5$.\\
On one side, in section $5$ of ref.\cite{Brandt1} it has been explained in detail how the $x_5 \rightarrow + \infty$ limit is taken and how the Grassmann integrations are done in eq.(\ref{ANfermionic}), in the case of $N=5$\footnote{There are three $\theta_i$'s and five $\phi_j$'s in this case.}, so all the terms of the left hand-side of eq.(\ref{A12345-4}) are known.\\
\noindent On the other side, we give the expression for all the  $(\zeta \cdot \zeta)^1 (\zeta \cdot k)^3$ terms of $A_{YM}(1,2,3,4,5)$ (and, therefore, also the corresponding terms of $A_{YM}(1,3,2,4,5)$) in eq.(\ref{A5}), so the right hand-side of eq.(\ref{A12345-4}) is also completely known, except for $ F^{\{23\}}(\alpha')$ and $ F^{\{32\}}(\alpha')$.\\
\noindent If we examine the coefficient of $(\zeta_1 \cdot \zeta_2)(\zeta_3 \cdot k_4)(\zeta_4 \cdot k_1)(\zeta_5 \cdot k_2)$ in both sides of eq.(\ref{A12345-4}) we find that\footnote{In our $N=5$ calculations we have substituted the $k_i \cdot k_j$ invariants in terms of the independent Mandelstam variables that we have chosen. See eqs.(\ref{Mandelstam5}) and (\ref{Mandelstam5-2}).}
\begin{multline}
\frac{8}{s_1 s_3}  \  F^{\{23\}}(\alpha')  = \\
\begin{split}
= 8  {\alpha'}^2  \int_0^1 dx_3  \int_0^{x_3} dx_2 \ {x_2}^{\alpha' s_1-1} {x_3}^{\alpha' ( s_4-s_1-s_2 )} {(x_3-x_2)}^{\alpha' s_2} {(1-x_2)}^{\alpha' ( s_5-s_2-s_3 )}  {(1-x_3)}^{\alpha'  s_3-1} \ .
\end{split}
\label{F23-1}
\end{multline}
\noindent This equation leads directly to an expression for $F^{\{23\}}(\alpha')$, which might be written as\footnote{In (\ref{F23-2}) we have substituted back the explicit expressions for the $N=5$ Mandelstam variables.}
\begin{eqnarray}
F^{\{23\}}(\alpha') &=& \int_0^1 dx_{3}  \int_0^{x_3} dx_{2}      \ \biggl(  \prod_{i>j \geq 1}^{4} (x_i - x_j)^{2 \alpha' k_i \cdot k_j} \biggr) \frac{ \ 2 \alpha' k_2 \cdot k_1 \ }{x_2 - x_1} \cdot   \frac{ \ 2 \alpha' k_4 \cdot k_3 \ }{x_4 - x_3}    \ ,
\label{F23-2}
\end{eqnarray}
where $x_1=0$ and $x_4=1$.\\
\noindent The fact that in eq.(\ref{F23-1}) $F^{\{32\}}(\alpha')$ is not present means that the $(\zeta_1 \cdot \zeta_2)(\zeta_3 \cdot k_4)(\zeta_4 \cdot k_1)(\zeta_5 \cdot k_2)$ term is present in $A_b(1,2,3,4,5)$ and $A_{YM}(1,2,3,4,5)$, but it is not present in $A_{YM}(1,3,2,4,5)$. \\
\noindent Similarly, if we examine the coefficient of  $(\zeta_1 \cdot \zeta_3)(\zeta_2 \cdot k_4)(\zeta_4 \cdot k_1)(\zeta_5 \cdot k_3)$ in both sides of eq.(\ref{A12345-4}) we find that
\begin{multline}
-\frac{8}{ (s_1+s_2-s_4 ) ( s_5-s_2-s_3 ) }  \  F^{\{32\}}(\alpha')  = \\
\begin{split}
= 8  {\alpha'}^2  \int_0^1 dx_3  \int_0^{x_3} dx_2 \ {x_2}^{\alpha' s_1} {x_3}^{\alpha' ( s_4-s_1-s_2 )-1} {(x_3-x_2)}^{\alpha' s_2} {(1-x_2)}^{\alpha' ( s_5-s_2-s_3 )-1}  {(1-x_3)}^{\alpha'  s_3} \ ,
\end{split}
\label{F32-1}
\end{multline}
from which we can arrive at
\begin{eqnarray}
F^{\{32\}}(\alpha') &=& \int_0^1 dx_{3}  \int_0^{x_3} dx_{2}      \ \biggl(  \prod_{i>j \geq 1}^{4} (x_i - x_j)^{2 \alpha' k_i \cdot k_j} \biggr) \frac{ \ 2 \alpha' k_3 \cdot k_1 \ }{x_3 - x_1} \cdot   \frac{ \ 2 \alpha' k_4 \cdot k_2 \ }{x_4 - x_2}    \ .
\label{F32-2}
\end{eqnarray}
Formulas (\ref{F23-2}) and (\ref{F32-2}) correspond to the expressions of the two momentum factors that appear in the case of $N=5$. These two formulas agree completely with the expected expression for $F^{\{23\}}(\alpha')$ and $F^{\{32\}}(\alpha')$, according to formula (\ref{MSS3}).  \\
\noindent The two specific kinematical structures that allowed us to arrive directly to the known expressions of $ F^{\{23\}}(\alpha')$ and $ F^{\{32\}}(\alpha')$ are not the only ones that lead to those expressions. For example, we have checked that examining the coefficient of $(\zeta_1 \cdot \zeta_4)(\zeta_2 \cdot k_1)(\zeta_3 \cdot k_4)(\zeta_5 \cdot k_1)$ and $(\zeta_1 \cdot \zeta_4)(\zeta_2 \cdot k_1)(\zeta_3 \cdot k_4)(\zeta_5 \cdot k_2)$, in both sides of (\ref{A12345-4}), also leads to eq.(\ref{F23-1}). We have also checked that examining the coefficient of $(\zeta_1 \cdot \zeta_4)(\zeta_2 \cdot k_4)(\zeta_3 \cdot k_1)(\zeta_5 \cdot k_1)$ and $(\zeta_1 \cdot \zeta_4)(\zeta_2 \cdot k_4)(\zeta_3 \cdot k_1)(\zeta_5 \cdot k_3)$, in both sides of (\ref{A12345-4}), also leads to eq.(\ref{F32-1}). \\
\noindent For the sake of completeness  we will examine in eq.(\ref{A12345-4}) two other kinematical structures, namely,  $(\zeta_1 \cdot \zeta_2)(\zeta_3 \cdot k_1)(\zeta_4 \cdot k_2)(\zeta_5 \cdot k_1)$ and $(\zeta_1 \cdot \zeta_2)(\zeta_3 \cdot k_4)(\zeta_4 \cdot k_2)(\zeta_5 \cdot k_1)$. This leads, respectively, to the following two relations that $F^{\{23\}}(\alpha')$ and $F^{\{32\}}(\alpha')$ should satisfy:
\begin{eqnarray}
\frac{2}{s_1} \  F^{\{23\}}(\alpha') \ + \ \frac{2 \ (s_2 + s_3 - s_4 -s_5)}{ ( s_5-s_2-s_3 ) (s_1+s_2-s_4  ) } \ F^{\{32\}}(\alpha') \ & = & \ \ \ 2 \ {\alpha'}^2 \ s_4 \ I^{[5]}_1  \ , \nonumber \\
- \frac{2 \ ( s_1 + s_5 )}{s_1 s_3} \  F^{\{23\}}(\alpha') \ + \ \hspace{2.1cm}  \frac{2 }{ ( s_2+s_3-s_5 ) } \ F^{\{32\}}(\alpha') \ & = & - 2 \ {\alpha'}^2 \ s_5 \ I^{[5]}_2 \ ,
\label{system2}
\end{eqnarray}
where
\begin{eqnarray}
I^{[5]}_1 & = &     \int_0^1 dx_3  \int_0^{x_3} dx_2 \ {x_2}^{\alpha' s_1-1} {x_3}^{\alpha' ( s_4-s_1-s_2 )} {(x_3-x_2)}^{\alpha' s_2} {(1-x_2)}^{\alpha' ( s_5-s_2-s_3 )-1}  {(1-x_3)}^{\alpha'  s_3 - 1} \ ,  \nonumber \\
&& \\
I^{[5]}_2 & = &   \int_0^1 dx_3  \int_0^{x_3} dx_2 \ {x_2}^{\alpha' s_1 -1} {x_3}^{\alpha' ( s_4-s_1-s_2 )-1} {(x_3-x_2)}^{\alpha' s_2} {(1-x_2)}^{\alpha' ( s_5-s_2-s_3 )-1}  {(1-x_3)}^{\alpha'  s_3} \ . \nonumber \\
&&
\label{I5}
\end{eqnarray}
\noindent Solving the linear system in (\ref{system2}) leads to integral expressions for $F^{\{23\}}(\alpha')$ and $F^{\{32\}}(\alpha')$ which apparently differ from the ones found in (\ref{F23-1}) and (\ref{F32-1}), respectively. But this is exactly the same thing that we analized in eqs.(\ref{F2-4}) and (\ref{F2-5}), for the case of $N=4$: there are integration by parts and partial fraction relations that allow us to write a same momentum factor in apparently different integral representations. Since the expressions found for the momentum factors this time are double integrals, there is a much rich variety of relations that can be found for them and it is not always an immediate thing to prove the equivalence between this different integral representations. In refs. \cite{Machado1} and \cite{Barreiro1} it was explained how using these relations, $A_b(1,2,3,4,5)$ for the first time could be written in a simple form, as a sum of two contributions (in direct analogy to eq.(\ref{A12345-3})).\\

\subsubsection{Case of $N=6$ and $N=7$}

\label{N67-3}

\noindent In the case of $N=6$ and $N=7$ we will not refer any longer to the alternative integral expressions that show up for the momentum factors (and which can always be proved by combining integration by parts and partial fraction techniques). These expressions will simply have to be equivalent to the ones we are looking for, because of the consistency of our method. \\
\noindent In this case relations (\ref{6point}) and (\ref{7point}) guarantee that choosing $T(1,2,3,4,5,6)=A_b(1,2,3,4,5,6)$ and $T(1,2,3,4,5,6,7)=A_b(1,2,3,4,5,6,7)$, respectively, it is possible to write down similar relations to the ones in those equations, but relabelling the momentum factors as
\begin{eqnarray}
\lambda^{\{ \sigma_6\}} \rightarrow F^{\{ \sigma_6\}}(\alpha') \ , \ \ \ \ \ \lambda^{\{ \sigma_7\}} \rightarrow F^{\{ \sigma_7\}}(\alpha') \ ,
\label{lambdas}
\end{eqnarray}
\noindent where $\sigma_6$ and $\sigma_7$ are all possible $S_{N-3}$ permutations (for $N=6$ and $N=7$) of indices $\{2, 3, 4 \}$ and
$\{2, 3, 4, 5 \}$, respectively.\\
\noindent The final result is that we succeed in arriving at the six and the twenty four momentum factors of formula (\ref{MSS3}) for $N=6$ and $N=7$, respectively, that is:
\begin{eqnarray}
F^{\{234\}}(\alpha') &=& \int_0^1 dx_{4}  \int_0^{x_4} dx_{3}  \int_0^{x_3} dx_{2}   \ \biggl(  \prod_{i>j \geq 1}^{5} (x_i - x_j)^{2 \alpha' k_i \cdot k_j} \biggr)
\biggl\{   \frac{ \ 2 \alpha' k_2 \cdot k_1 \ }{x_2 - x_1} \cdot   \frac{ \ 2 \alpha' k_5 \cdot k_4 \ }{x_5 - x_4} \times \nonumber \\
&&\hspace{7cm}  \Big(  \frac{ \ 2 \alpha' k_3 \cdot k_1 \ }{x_3 - x_1} +  \frac{ \ 2 \alpha' k_3 \cdot k_2 \ }{x_3 - x_2}   \Big) \biggr\} \hspace{1cm}
\label{G1}
\end{eqnarray}
and
\begin{eqnarray}
F^{\{2345\}}(\alpha') &=& \int_0^1 dx_5 \int_0^{x_5} dx_{4}  \int_0^{x_4} dx_{3}  \int_0^{x_3} dx_{2}   \ \biggl(  \prod_{i>j \geq 1}^{6} (x_i - x_j)^{2 \alpha' k_i \cdot k_j} \biggr)  \ \biggl\{   \frac{ \ 2 \alpha' k_2 \cdot k_1 \ }{x_2 - x_1}   \times \nonumber \\
&&\hphantom{  \int_0^1   }    \frac{ \ 2 \alpha' k_6 \cdot k_5 \ }{x_6 - x_5} \cdot  \Big( \frac{ \ 2 \alpha' k_3 \cdot k_1 \ }{x_3 - x_1} +   \frac{ \ 2 \alpha' k_3 \cdot k_2 \ }{x_3 - x_2}  \Big)
 \Big(  \frac{ \ 2 \alpha' k_5 \cdot k_4 \ }{x_5 - x_4} +  \frac{ \ 2 \alpha' k_6 \cdot k_4 \ }{x_6 - x_4}   \Big) \biggr\} \ . \hspace{0.8cm}
\label{H1}
\end{eqnarray}
\noindent In eq.(\ref{G1}) it is understood that $\{x_1=0, x_5=1\}$ while in eq.(\ref{H1}) is is understood that $\{x_1=0, x_6=1\}$.\\
As we mentioned after eq.(\ref{MSS2}), the MSS prescription for finding the remaining momentum factors consists in simply doing a permutation of indices {\it inside} the curly brackets of (\ref{G1}) and (\ref{H1}), according to the $\sigma_N$ permutation that is being considered (where $N=6$ and $N=7$, respectively). For example, doing $2 \leftrightarrow 3$ in (\ref{G1}) and (\ref{H1}) we arrive, respectively, at
\begin{eqnarray}
F^{\{324\}}(\alpha') &=& \int_0^1 dx_{4}  \int_0^{x_4} dx_{3}  \int_0^{x_3} dx_{2}   \ \biggl(  \prod_{i>j \geq 1}^{5} (x_i - x_j)^{2 \alpha' k_i \cdot k_j} \biggr)    \biggl\{  \ \frac{ \ 2 \alpha' k_3 \cdot k_1 \ }{x_3 - x_1} \cdot   \frac{ \ 2 \alpha' k_5 \cdot k_4 \ }{x_5 - x_4}   \times \nonumber \\
&&\hspace{7cm}   \Big(  \frac{ \ 2 \alpha' k_2 \cdot k_1 \ }{x_2 - x_1} +  \frac{ \ 2 \alpha' k_2 \cdot k_3 \ }{x_2 - x_3}   \Big) \  \biggr\}  \hspace{1cm}
\label{G2}
\end{eqnarray}
and
\begin{eqnarray}
F^{\{3245\}}(\alpha') &=& \int_0^1 dx_5 \int_0^{x_5} dx_{4}  \int_0^{x_4} dx_{3}  \int_0^{x_3} dx_{2}   \ \biggl(  \prod_{i>j \geq 1}^{6} (x_i - x_j)^{2 \alpha' k_i \cdot k_j} \biggr)  \ \biggl\{   \frac{ \ 2 \alpha' k_3 \cdot k_1 \ }{x_3 - x_1}   \times \nonumber \\
&&\hphantom{  \int_0^1   }    \frac{ \ 2 \alpha' k_6 \cdot k_5 \ }{x_6 - x_5} \cdot  \Big( \frac{ \ 2 \alpha' k_2 \cdot k_1 \ }{x_2 - x_1} +   \frac{ \ 2 \alpha' k_2 \cdot k_3 \ }{x_2 - x_3}  \Big)
 \Big(  \frac{ \ 2 \alpha' k_5 \cdot k_4 \ }{x_5 - x_4} +  \frac{ \ 2 \alpha' k_6 \cdot k_4 \ }{x_6 - x_4}   \Big) \biggr\} \ . \hspace{0.8cm}
\label{H7}
\end{eqnarray}
In Appendix \ref{Linear system}  we specify the kinematical terms that allow us to write a linear system of equations which has a $unique$ solution for the momentum factors and which is precisely given by the ones in eqs.(\ref{G1}) and (\ref{H1}) (and for the remaining momentum factors, by means of a $\sigma_N$ permutation of the corresponding indices). We also specify, in that appendix, the corresponding linear system of equations for the $F^{\{ \sigma_6\}}(\alpha')$'s and for the  $F^{\{ \sigma_7\}}(\alpha')$'s.

\subsection{For gauge bosons and massless fermions}

\label{For gauge bosons and massless fermions}

\noindent In this section we briefly explain how to find the amplitudes involving fermions directly from the corresponding amplitude involving only gauge bosons. From the point of view of vertex operators, in the RNS formalism, this is a non trivial thing to do since it is known that there arise difficulties when calculating amplitudes of $n$ gauge bosons and $2m$ massless fermions in the case of $m \geq 3$ \footnote{The difficulties have to do with the fact that the total super ghost charge of the involved vertex operators must be $-2$ and fermion vertex operators are generally known in the $+1/2$ and $-1/2$ picture (therefore, four fermions gives $4 \times (-1/2) = -2$ and this case is fine). We thank E. Hatefi for clarifying this issue to us. \\
\noindent In spite of this difficulty in ref. \cite{Kostelecky1} the {\it pure} fermion $6$-point amplitude was obtained using the RNS formalism.}(where $n= 1, 2, \ldots $)  \cite{Friedan1}.\\
\noindent The shortcut to find the amplitudes involving (any number of even) fermions is that we do not need to compute them from the beginning, by considering correlation functions of vertex operators: the closed formula for the $N$-point amplitude of gauge bosons and D=10 Supersymmetry are enough ingredients to find the amplitudes involving fermions. We already saw this in section {\it 2.3} of ref.\cite{Barreiro2}, in the case of the open superstring $5$-point amplitude, where the ansatz that we proposed for the fermion amplitudes consisted in simply changing the original gauge boson subamplitudes of the $2$-dimensional basis ($A_{YM}(1,2,3,4,5)$ and $A_{F^4}(1,2,3,4,5)$, in the case of ref.\cite{Barreiro2}) by the corresponding expression of their superpartner subamplitudes.\\
\noindent In the $N$-point case we do exactly the same thing: we change the
gauge boson subamplitudes of the basis (in this case given by the Yang-Mills subamplitudes in (\ref{basis})) by their superpartner subamplitudes. The resulting formula is precisely the one obtained by Mafra, Schlotterer and Stieberger \cite{Mafra1} (see eq. (\ref{MSS})).\\
\noindent Besides the expression in (\ref{MSS}) there is no other possibility for the scattering amplitudes involving fermions since, as analyzed in \cite{Green1}, the global supersymmetry requirement is sufficient to determine the structure of the boson/fermion interactions $uniquely$\footnote{This analysis can be found, for example, in sections 5.3.1 and 7.4.1 of ref. \cite{Green1}. We thank N. Berkovits for suggesting part of the specific sections of this book to us.}. In formula (\ref{MSS}) this means that the summed variation of all possible boson/fermion subamplitudes, under the supersymmetry transformations ,
\begin{eqnarray}
\delta \zeta^{\mu}_j = \frac{i}{2} ( \bar{\epsilon} \gamma^{\mu} u_j ) \ , \ \ \
\delta u_j =  \frac{i}{2} ( \gamma_{\mu \nu} \epsilon )  \zeta^{\mu}_j k^{\nu}_j   \ , \ \ \
\delta \bar{u}_j =  \frac{i}{2}    (  \bar{\epsilon}    \gamma_{\mu \nu} ) \zeta^{\mu}_j k^{\nu}_j  \ , \ \ \ (j=1, \ldots, N)
\label{susytransf}
\end{eqnarray}
is $zero$, after using the on-shell and the physical state conditions, together with momentum conservation\footnote{The formulas in (\ref{susytransf}) are nothing else than the momentum space version of the supersymmetric transformations of the fields $A^a_{\mu}$ and $\psi^a$ of D=10 Super Yang-Mills theory.}.\\

\section{Finding the $\alpha'$ expansion of the momentum factors}

\label{Finding the}

\noindent It was mentioned in ref. \cite{Mafra3} that the factorization properties of open superstring subamplitudes could be used as a complementary approach to determine the ${\alpha'}$ expansion of the momentum factors $F^{\{\sigma_N\}}(\alpha')$\footnote{See, for example, Appendix C of this reference.}. We agree completely with this observation and, indeed, we will use these sort of factorization properties in this section as part of the tools to find the ${\alpha'}$ expansion of the momentum factors (see eqs.(\ref{unitarity2}) and (\ref{collinear})). Our remark at this point is that there is a more general form for the factorization (or tree-level unitarity) relations than the ones considered in ref. \cite{Mafra3}, which includes as a particular case the collinear limit considered there (see eq.(\ref{collinear})). This form, together with the cyclic property of the subamplitude, allows us to obtain all the $\alpha'$ terms of the momentum factors,  up to quite high ${\alpha'}$ order, by just using the very well known $4$-point (Veneziano-type) momentum factor. \\
This is a very non trivial result that will allow us to bypass $\alpha'$ expansions of worldsheet integrals for $N=5$, $N=6$ and $N=7$, at least up to ${\alpha'}^6$ order in the first two cases and ${\alpha'}^4$ order in the last one\footnote{Our limitation to go to higher orders in $\alpha'$, by just considering the expansion of the $4$-point Gamma factor, is just a computational one since the expressions become extremely huge as the $\alpha'$ order grows. As we explained in subsection \ref{Using}, we believe that using using the 4-point amplitude $\alpha'$ expansion, we are able to obtain completely the $\alpha'$ expansion of any $N$-point momentum factor up to ${\alpha'}^7$ order.}, in the same spirit that the revisited S-matrix method succeeds in finding the corresponding $\alpha'$ terms of the OSLEEL which would commonly be determined by open superstring $N$-point amplitudes (where $N>4$)\cite{Barreiro0}.\\

\vspace{0.5cm}

\subsection{Tree level unitarity of the amplitudes}

\label{Using tree}

\noindent In the case of gauge bosons, the tree level unitarity relations state that the $N$-point subamplitude obeys\footnote{In eq.(\ref{unitarity1}) and the remaing ones that contain expressions involving the subamplitudes, in this subsection, we will not use the index `{\it b}' on them because their dependence in the polarization vectors $\zeta_i$ has been explicited and, therefore, it is clear that they are gauge boson subamplitudes.\\
\noindent As a general rule, we will only use the `{\it b}' index in the subamplitude variable $A$ when it has not been explicited its dependence in the polarizations.}\cite{Polchinski1}
\begin{eqnarray}
{\cal A}(\zeta_1, k_1;\ldots ; \zeta_N, k_N) \sim i  \int \frac{d^D k}{(2 \pi)^D}
\frac{{\cal A}^{\mu } (\zeta _{1},k_{1};\ldots ;\zeta _{m-1},k_{m-1};k) \ {\cal A}_{\mu}(-k;\zeta _{m},k_{m};\ldots ;\zeta _{N},k_{N})}{ -k^2  }  \ \ \ \ \
\label{unitarity1}
\end{eqnarray}
\noindent where ${\cal A}^{\mu }$ and ${\cal A}_{\mu }$ are, respectively, $m$ and $(N+2-m)$-point subamplitudes and where the pole that ${\cal A}(\zeta_1, k_1;\ldots ; \zeta_N, k_N)$ has in the Mandelstam variable $(k_{1}+k_{2}+\cdots +k_{m-1})^{2}$ is being taken to {\it zero}:
\begin{eqnarray}
(k_{1}+k_{2}+\ldots +k_{m-1})^{2} \rightarrow 0 \ .
\label{zero}
\end{eqnarray}
\noindent All ${\cal A}$ subamplitudes  in (\ref{unitarity1}) include the `$i \ \delta$' factor that takes into account momentum conservation (see eq.(\ref{N-point})) and $D$ ($=10$) is the spacetime dimension.\\
\noindent Eq.(\ref{unitarity1}) is applicable for $N \geq 4$ and $m=3, \ldots, N-1$, so there are $(N-3)$ unitarity relations for Mandelstam variables like $(k_{1}+k_{2}+\cdots +k_{m-1})^{2}$. \\
\noindent The remaining unitarity relations arise when the other Mandelstam variables,
\begin{eqnarray}
 (k_{2}+k_{3}+\ldots +k_{m-1})^{2}, (k_{3}+k_{4}+\ldots +k_{m-1})^{2}, \ldots , (k_{m-2}+k_{m-1})^{2}  \ ,
\label{}
\end{eqnarray}
are individually taken to zero. There are $N(N-3)/2$ independent Mandelstam variables and that is the total number of independent unitarity relations.\\
\noindent Substituing the explicit dependence of each ${\cal A}$ subamplitude in the `$i \ \delta$' factor in (\ref{unitarity1}) and calculating the $D$-dimensional integral leads to
\begin{eqnarray}
A(\zeta_1, k_1;\ldots ; \zeta_N, k_N) \sim
\frac{A^{\mu } (\zeta _{1},k_{1};\ldots ;\zeta _{m-1},k_{m-1};k)  \ A_{\mu}(-k;\zeta _{m},k_{m};\ldots ;\zeta _{N},k_{N})   }
 {  (k_{1}+k_{2}+\cdots +k_{m-1})^{2}  }  \ ,
\label{unitarity2}
\end{eqnarray}
\noindent where, now, the $A$, $A^{\mu}$ and $A_{\mu}$ subamplitudes no longer include the $\delta$ dependence and the momentum conservation on each of them must be implicitly assumed. These subamplitudes are the ones that we have been dealing with throughout this work.\\
\noindent $A^{\mu}$ and $A_{\mu}$ are subamplitudes in which the corresponding polarization vector has been taken away:
\begin{eqnarray}
\label{Asuper}
A^{\mu }(\zeta _{1},k_{1};\ldots ;\zeta _{m-1},k_{m-1};k) & = & \frac{\partial }{\partial \zeta_{\mu}} A(\zeta _{1},k_{1};\ldots ;\zeta _{m-1},k_{m-1}; \zeta, k) \ , \\
\label{Asub}
A_{\mu}(-k;\zeta _{m},k_{m};\ldots ;\zeta _{N},k_{N}) & = & \frac{\partial }{\partial \zeta^{\mu}} A(\zeta, -k;\zeta _{m},k_{m};\ldots ;\zeta _{N},k_{N}) \ ,
\end{eqnarray}
where
\begin{eqnarray}
k^{\mu} = - (  {k}^{\mu}_1  + \ldots + {k}^{\mu}_{m-1} ) =   {k}^{\mu}_m  + \ldots + {k}^{\mu}_{N} \
\label{kmu1}
\end{eqnarray}
and where the assymptotic mass-shell condition (\ref{zero}) is being taken into account.\\
\noindent Formula (\ref{unitarity2}) states that the residue that the $N$-point subamplitude has in $(k_{1}+k_{2}+\cdots +k_{m-1})^{2} = 0$ is given by the product of two lower-point subamplitudes (an $m$-point and a $(N+2-m)$-point one). It is valid at any $\alpha'$ order. In particular, if $\alpha' \rightarrow 0$ it means that Yang-Mills subamplitudes also satisfy it.\\
\noindent When eq.(\ref{unitarity2}) is considered for the particular case of $m=3$, it becomes
\begin{eqnarray}
A(\zeta_1, k_1;\ldots ; \zeta_N, k_N) \sim
\frac{A^{\mu }(\zeta _{1},k_{1};\zeta_2, k_2 ;k)  \ A_{\mu}(-k;\zeta _{3},k_{3};\ldots ;\zeta _{N},k_{N})   }
 {  (k_{1}+k_{2})^{2}  }  \ ,
\label{unitarity3}
\end{eqnarray}
or equivalently
\begin{eqnarray}
A(  \zeta_1, k_1;\ldots ; \zeta_N, k_N  ) &\sim &
\frac{1}{2(k_{1}\cdot k_{2})}\  \ V^{\mu }_{(12)}  \ \frac{\partial }{\partial \zeta
^{\mu }}A(\zeta ,k_{1}+k_{2};\zeta_3,k_{3};\ldots;\zeta_{N},k_{N})\ ,
\label{collinear}
\end{eqnarray}
where
\begin{eqnarray}
V^{\mu }_{(12)}=-g \ \Big[ \ (\zeta _{1}\cdot \zeta _{2}) \ (k_{1}-k_{2})^{\mu }-2(\zeta
_{2}\cdot k_{1}) \ \zeta^{\mu }_{1}+2(\zeta _{1}\cdot k_{2}) \  \zeta^{\mu }_{2} \ \Big] \
\label{V12}
\end{eqnarray}
\noindent is the (contracted) Yang-Mills vertex.\\
\noindent For $m=3$, condition (\ref{zero}) implies that $(k_1+k_2)^2 = 2 k_1 \cdot k_2 $ goes to {\it zero}:
\begin{eqnarray}
k_1 \cdot k_2 \rightarrow 0 \ .
\label{parallel}
\end{eqnarray}
\noindent If $k_1$ and $k_2$ are non zero light-like vectors, then condition (\ref{parallel}) implies that these momenta are parallel ($k_1 || k_2$) and, therefore, the unitarity relation (\ref{collinear}) is nothing else than the collinear version of the factorization property of gauge boson subamplitudes \cite{Mangano1}. \\
\noindent In the case of arbitrary $m$ ($=3, \ldots , N-1$), if $\{k_1, \ldots, k_{m-1} \}$ are non zero physical (Minkowski) momenta, then condition (\ref{zero}) implies that {\it all} of them are parallel ($k_1 \ || \ k_2 \ || \ldots || \ k_{m-1}$)\footnote{Due to global momentum conservation, it turns out that {\it all} $k_i$ become parallel, for $i=1, \ldots, N$, not only the first $m$ momenta.}. Subsequently, all Mandelstam variables should tend to zero (because if the momenta tend to be parallel then $k_i \cdot k_j \rightarrow 0$). This is what happens for the Mandelstam variables in the physical region.\\
\noindent So, the word of caution when considering eq.(\ref{unitarity2}) subject to condition (\ref{zero}), is that we are considering there the subamplitude expressions where the Mandelstam variables have been analytically continued in the complex plane. Then, the Mandelstam variables are generally out of the physical region and, therefore, condition (\ref{zero}) does not have any implication for any of the other Mandelstam variables: they remain being independent in spite of one of them (namely, $(k_{1}+k_{2}+\cdots +k_{m-1})^{2}$) tending to zero. This fact will be implicit in the calculations that we will do in subsections \ref{Case of the 5-point}, \ref{Case of the 6-point} and \ref{Case of the 7-point}. In these subsections we will use formula (\ref{collinear}) and its generalization, eq.(\ref{unitarity2}), to find $\alpha'$ terms of the $N$-point amplitude for $N=5,6,7,$ by just using the well known open superstring $4$-point $\alpha'$ expansion (see eqs.(\ref{formula1}) and (\ref{expansionGamma})).\\

\subsection{Analyticity of the momentum factors}

\label{Analyticity}

\noindent Before going to the details of the $\alpha'$ calculations in the $N$-point amplitudes (where $N=5,6,7$) an important remark proceeds.\\
\noindent From the analysis that we did in section \ref{Closed} we have that the $N$-point subamplitude of gauge bosons in Open Superstring Theory is given by
\begin{eqnarray}
A_b(1, \ldots, N) & = & \sum_{ \sigma_N \ \varepsilon \ S_{N-3} } F^{\{\sigma_N\}}(\alpha') \ A_{YM}(1,\{ 2_{\sigma}, 3_{\sigma}, \ldots, (N-2)_{\sigma} \}, N-1, N) \ ,
\label{Ab}
\end{eqnarray}
where the $F^{\{\sigma_N\}}(\alpha')$'s are given by the integral expression (\ref{MSS2}) (or equivalently, if $N \geq 5$, by eq.(\ref{MSS3})).\\
\noindent In any of the two expressions of the momentum factors we see that the $\alpha'$ dependence of them is completely enclosed in the dimensionless Mandelstam variables, $\alpha' s_i$ and $\alpha' t_j$. So, on one side, if these variables are analytically continued in the complex plane, then the $F^{\{\sigma_N\}}(\alpha')$'s will have a well defined Laurent expansion with respect to $\alpha' s_i=\alpha' t_j=0$.\\
\noindent On the other side, it is well known that the low energy limit of the gauge boson (tree level) subamplitudes in Open Superstring Theory gives the corresponding Yang-Mills subamplitudes:
\begin{eqnarray}
\lim_{\alpha' \rightarrow 0} A_b(1, \ldots, N) & = & A_{YM}(1, \ldots, N) \ .
\label{low}
\end{eqnarray}
In eq.(\ref{Ab}) this last condition implies that
\begin{eqnarray}
\lim_{\alpha' \rightarrow 0} F^{\{\sigma_N\}}(\alpha') & = & \left\{ \begin{array}{ccc}
                                                                                                               1 & \mbox{if} & \sigma_N = \{2, 3, \ldots, N-2 \} \\
                                                                                                               0 & \mbox{if} & \sigma_N \neq \{2, 3, \ldots, N-2 \}
                                                                                                              \end{array}  \right.
\label{Fs}
\end{eqnarray}
where $\{2, 3, \ldots, N-2 \}$ is the identity permutation of those indices.\\
\noindent The result in (\ref{Fs}), together with the behaviour of the $F^{\{\sigma_N\}}(\alpha')$'s under an analytical continuation of the dimensionless Mandelstam variables, mentioned before, implies that the momentum factors are analytic functions at the origin of the complex plane ($\alpha' s_i=\alpha' t_j=0$). This implies that the $F^{\{\sigma_N\}}(\alpha')$'s have a well defined $\alpha'$ expansion as a (Maclaurin) power series. \\
\noindent This is a quite non trivial result since it is well known, by explicit computations, that the $\alpha'$ expansion of many of the individual multiple integrals which arise in the expanded version of $A_b(1, \ldots, N)$ (see eq.(\ref{ANfermionic})) do indeed have some negative powers of $\alpha'$ \footnote{See, for example, Appendix {\it A.3} of \cite{Brandt1} for the case of $N=5$, and refs.\cite{Oprisa1, Hemily1} for the case of $N=6$.}. In fact, the explicit expression for the $F^{\{\sigma_N\}}(\alpha')$'s, either in (\ref{MSS2}) or in (\ref{MSS3}), is given by a $(N-3)$ multiple integral multipled by ${\alpha'}^{N-3}$, where $(N-3)$ is a positive integer number. So, at least the first of those multiple integrals (the one corresponding to $\sigma_N = \{2, 3, \ldots, N-2\}$) does have an $\alpha'$ expansion which begins as  ${\alpha'}^{-(N-3)}$.  \\
\noindent The analytic behaviour of the $F^{\{\sigma_N\}}(\alpha')$'s, when $\alpha' \rightarrow 0$, will be an important fact which is behind the $\alpha'$ expansions which we will obtain in the next three subsections.\\

\subsection{Case of the 5-point momentum factors}

\label{Case of the 5-point}

\begin{figure}[th]
\centerline{\includegraphics*[scale=0.1,angle=0]{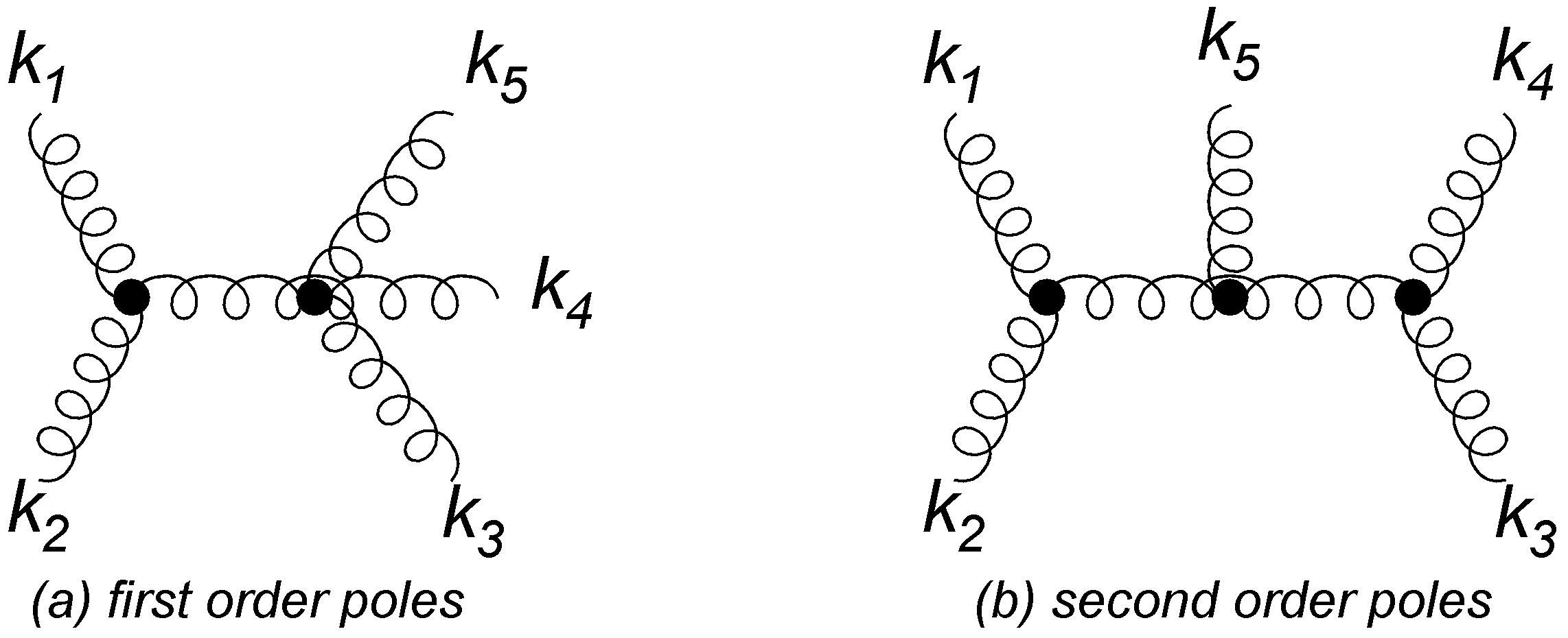}}
\caption{Two type of Feynman diagrams contribute to the YM 5-point amplitude. Permutations of the legs of these diagrams should also be considered in order to account for the full amplitude.}
\label{feynmandiag}
\end{figure}

\noindent In the $N=5$ case, we will give the details of how using tree level unitarity of the subamplitudes we can arrive at the ${\alpha'}^2$ order terms of  $F^{\{23\}}(\alpha')$ and $F^{\{32\}}(\alpha')$. The higher order terms of these two momentum factors will be afterwards written as a consequence of repeating the procedure for the corresponding $\alpha'$ order.\\
\noindent It will turn out to be convenient to introduce here the notation
\begin{eqnarray}
\alpha_{ij} & = & 2 \ k_i \cdot k_j \ ,
\label{alphaij}
\end{eqnarray}
where $k_i$ and $k_j$ are, respectively, the $i$-th and the $j$-th external leg momentum (and $i < j$). This notation will also be useful in the case of $N=6$ (subsection \ref{Case of the 6-point}) and the case of $N=7$ (subsection \ref{Case of the 7-point}).\\
\noindent Now, in order to find the $\alpha'$ expansion of $F^{\{23\}}(\alpha')$ and $F^{\{32\}}(\alpha')$ within our approach, it will be convenient to write, once again, eq.(\ref{A12345-3}) in the present subsection:
\begin{eqnarray}
A_b(1,2,3,4,5) & = &  F^{\{23\}}(\alpha')  \ A_{YM}(1,2,3,4,5) +    F^{\{32\}}(\alpha')  \ A_{YM}(1,3,2,4,5) \ .
\label{A12345-7}
\end{eqnarray}
\noindent We begin by looking for the poles of all three subamplitudes involved in eq.(\ref{A12345-7}). In figure \ref{feynmandiag} we have drawn the two type of Feynman diagrams that contribute to the Yang-Mills $5$-point amplitude. Examining all possible diagrams of this type we can infer that $A_{YM}(1,2,3,4,5)$ and $A_{YM}(1,3,2,4,5)$ have first and second order poles in the $\alpha _{ij}$ variables (one and two propagators, respectively) that we list in table (\ref{poles5p}).\\
\begin{equation}
\begin{tabular}{ccc}
\hline
& first order & second order \\ \hline
$A_{YM}(1,2,3,4,5)$ & $\alpha _{12}$, $\alpha _{23}$, & $\alpha _{12}\alpha
_{34}$, $\alpha _{23}\alpha _{45}$, \\
& $\alpha _{34}$, $\alpha _{45}$, $\alpha _{15}$ & $\alpha _{34}\alpha _{15}$%
, $\alpha _{45}\alpha _{12}$, $\alpha _{15}\alpha _{23}$ \\ \hline
$A_{YM}(1,3,2,4,5)$ & $\alpha _{13}$, ~$\alpha _{23}$, & $\alpha _{13}\alpha
_{24}$, $\alpha _{23}\alpha _{45}$, \\
& $\alpha _{24}$, $\alpha _{45}$, $\alpha _{15}$ & $\alpha _{24}\alpha _{15}$%
, $\alpha _{45}\alpha _{13}$, $\alpha _{15}\alpha _{23}$ \\ \hline
&& \\
\end{tabular}
\label{poles5p}
\end{equation}
\noindent On the other hand, in the left side of equation (\ref{A12345-7}), the
5-point superstring subamplitude, $A_b(1,2,3,4,5)$, presents only first order poles in
the same variables that $A_{YM}(1,2,3,4,5)$ does, namely,
$\{ \alpha _{12}$,~$\alpha _{23}$,~$\alpha _{34}$,~$\alpha _{45}, ~\alpha_{15}\}$ \footnote{Since $A_b(1,2,3,4,5)$ agrees with $A_{YM}(1,2,3,4,5)$ when $\alpha' \rightarrow 0$, and $A_{YM}(1,2,3,4,5)$ does have second order poles, the statement that $A_b(1,2,3,4,5)$ has only first order poles is only valid perturbatively, for any {\it non zero} order in $\alpha'$.}. We will refer to these poles as the {\it physical} poles.\\
\noindent For reasons that will become clear in the next lines, it will be convenient to write $F^{\{23\}}(\alpha')$ and $F^{\{32\}}(\alpha')$ in terms of the simple poles that $A_{YM}(1,2,3,4,5)$ and $A_{YM}(1,3,2,4,5)$ have, respectively:
\begin{eqnarray}
\label{defF23}
F^{\{23\}}(\alpha') & = & F^{\{23\}}[\ \alpha_{12}, \alpha_{23},  \alpha_{34},  \alpha_{45},  \alpha_{15}; \ \alpha'  ] \ , \\
\label{defF32}
F^{\{32\}}(\alpha') & = & F^{\{32\}}[\ \alpha_{13}, \alpha_{23},  \alpha_{24},  \alpha_{45},  \alpha_{15}; \ \alpha' ] \ .
\end{eqnarray}
\noindent Now, since the momentum factors have a well defined $\alpha'$ power series (as it was seen in the previous subsection), we may write that\footnote{In eqs.(\ref{expF23-1}) and (\ref{expF32-1}) we are already using the fact the $F^{\{ \sigma_5\}}(\alpha')$'s satisfy low energy requirement in eq.(\ref{Fs}).}
\begin{eqnarray}
\label{expF23-1}
F^{\{23\}}(\alpha')& = & 1 + \Big( a_1 \ \alpha_{12} + a_2 \ \alpha_{23} + a_3 \ \alpha_{34} + a_4 \ \alpha_{45} + a_5 \ \alpha_{15} \Big) \alpha'  + \Big( b_1 \ \alpha_{12} \alpha_{23} + b_2 \ \alpha_{12} \alpha_{34} + \nonumber \\
&&   \  b_3 \ \alpha_{12} \alpha_{45} + b_4 \ \alpha_{12} \alpha_{15} + b_5 \ \alpha_{23} \alpha_{34} + b_6 \ \alpha_{23} \alpha_{45} +  b_7 \ \alpha_{23} \alpha_{15} + b_8 \ \alpha_{34} \alpha_{45} + \nonumber \\
&&  \ b_9 \ \alpha_{34} \alpha_{15} + b_{10} \ \alpha_{45} \alpha_{15}
+ b_{11} \ \alpha_{12}^2 + b_{12} \ \alpha_{23}^2 + b_{13} \ \alpha_{34}^2  + b_{14} \ \alpha_{45}^2 + + b_{15} \ \alpha_{15}^2  \Big) {\alpha'}^2+ \nonumber \\
 && \ {\cal O}({\alpha'}^3) \ , \\
\label{expF32-1}
F^{\{32\}}(\alpha')& = &  \Big( c_1 \ \alpha_{13} + c_2 \ \alpha_{23} + c_3 \ \alpha_{24} + c_4 \ \alpha_{45} + c_5 \ \alpha_{15} \Big) \alpha' + \Big( d_1 \ \alpha_{13} \alpha_{23} + d_2 \ \alpha_{13} \alpha_{24} + \nonumber \\
&& \ d_3 \ \alpha_{13} \alpha_{45} + d_4 \ \alpha_{13} \alpha_{15} + d_5 \ \alpha_{23} \alpha_{24} + d_6 \ \alpha_{23} \alpha_{45} + d_7 \ \alpha_{23} \alpha_{15} + d_8 \ \alpha_{24} \alpha_{45} + \nonumber \\
&& \ d_9 \ \alpha_{24} \alpha_{15} + d_{10} \ \alpha_{45} \alpha_{15} + d_{11} \ \alpha_{13}^2 + d_{12} \ \alpha_{23}^2 + d_{13} \ \alpha_{24}^2  + d_{14} \ \alpha_{45}^2 + d_{15} \ \alpha_{15}^2 \Big) {\alpha'}^2+ \nonumber \\
&& \ {\cal O}({\alpha'}^3) \ .
\end{eqnarray}
\noindent We will find the value of all the $a_i$'s, $b_j$'s, $c_k$'s and $d_l$'s that participate in these two expansions. We will do this by demanding three requirements:
\begin{enumerate}
\item Absence of {\it unphysical} poles.
\item Tree level unitarity.
\item Cyclic invariance.
\end{enumerate}
All these requirements will be demanded at every {\it non zero} $\alpha'$ order, in $A_b(1,2,3,4,5)$.\\

\noindent \underline{Step 1}: Absence of {\it unphysical} poles in $A_b(1,2,3,4,5)$\\

\noindent According to eq.(\ref{A12345-7}), the only unphysical poles that could arise in $A_b(1,2,3,4,5)$ are the ones that come from $A_{YM}(1,3,2,4,5)$, namely, $\alpha_{13}$ and $\alpha_{24}$. These poles appear always as {\it simple} poles in $A_{YM}(1,3,2,4,5)$, in spite of they participating, also, in the second order poles of this subamplitude (see table(\ref{poles5p})).  \\
\noindent The only possible way to cancel, in the right hand-side of eq.(\ref{A12345-7}), the independent simple poles that $A_{YM}(1,3,2,4,5)$ has at $\alpha_{13}=0$ and $\alpha_{24}=0$, is demanding that $F^{\{32\}}(\alpha')$ should be possible to be written with a common $\alpha_{13} \alpha_{24}$  factor:
\begin{eqnarray}
F^{\{32\}}(\alpha') & = & \alpha_{13} \alpha_{24} \ {\alpha'}^2 \ f^{\{32\}}[\ \alpha_{13}, \alpha_{23},  \alpha_{24},  \alpha_{45},  \alpha_{15}; \ \alpha' ] \ ,
\label{factor1}
\end{eqnarray}
where $f^{\{32\}}[\ \alpha_{13}, \alpha_{23},  \alpha_{24},  \alpha_{45},  \alpha_{15}; \ \alpha' ]$ is analytic in all its five dimensionless $\alpha' \alpha_{ij}$ variables.\\
\noindent Comparing eqs.(\ref{factor1}) and (\ref{expF32-1}) we see that the only possibility is that
\begin{eqnarray}
c_i = 0 \ , (\mbox{for} \ i=1, \ldots, 5) \ \mbox{and} \ d_j =0 \ , (\mbox{for} \ j=1, \ldots, 15; \ \mbox{except for} \ j=2) \ ,
\label{nullcoeffs}
\end{eqnarray}
so, in (\ref{expF32-1}) this automatically leads us to
\begin{eqnarray}
F^{\{32\}}(\alpha')& = &  d_2 \ \alpha_{13} \alpha_{24} \ {\alpha'}^2 + {\cal O}({\alpha'}^3) \ .
\label{simpleF32-1}
 \end{eqnarray}

\vspace{0.5cm}

\noindent \underline{Step 2}: Unitarity relation for $A_b(1,2,3,4,5)$\\

\noindent In Appendix \ref{Unitarity-N5} we prove that demanding unitarity of $A_b(1,2,3,4,5)$  with respect to its $\alpha_{12}$ pole implies that
the $5$-point momentum factor, $F^{\{23\}}(\alpha')$, is related to the $4$-point momentum factor, $F^{\{2\}}(\alpha')$, by the following relation:
\begin{eqnarray}
F^{\{23\}}[  \ 0, \alpha_{23}, \alpha_{34}, \alpha_{45}, \alpha_{15} ;  \alpha ^{\prime }  ]=F^{\{2\}}[ \alpha_{45}, \alpha_{34}; \alpha ^{\prime }]  \ ,
\label{factorization5p1}
\end{eqnarray}
where we have introduced the notation
\begin{eqnarray}
\label{not-F2a}
F^{\{2\}}(\alpha') & = & F^{\{2\}}[ \alpha_{12}, \alpha_{23}; \alpha ^{\prime }]  \ , \\
\label{not-F23}
F^{\{23\}}(\alpha') & = & F^{\{23\}}[  \alpha_{12}, \alpha_{23}, \alpha_{34}, \alpha_{45}, \alpha_{15} ;  \alpha ^{\prime }  ] \ .
\end{eqnarray}
\noindent Notice that the two Mandelstam variables in which $F^{\{2\}}$ is being evaluated in (\ref{factorization5p1}) are not the same two ones that are implicit in the notation (\ref{not-F2a})\footnote{The notation introduced in eq.(\ref{not-F2a}) is not casual: $\alpha_{12}$ and $\alpha_{23}$ are the two Mandelstam variables in which the $4$-point amplitudes in eq.(\ref{A1234-3}) have poles.}.\\
\noindent Relation (\ref{factorization5p1}) is an extremely strong constraint since in the left hand-side $F^{\{23\}}$ is being evaluated at arguments which do not appear in the right hand-side of this relation (namely, $\alpha_{23}$ and $\alpha_{15}$). Here, we will just look for the implications that relation (\ref{factorization5p1}) has in the determination of the coefficients in the $\alpha'$ expansion of $F^{\{23\}}(\alpha')$ only up to ${\alpha'}^2$ order. In fact, after substituing (\ref{expF23-1}) and (\ref{expansionGamma}) in (\ref{factorization5p1}), we can conclude that
\begin{eqnarray}
a_i = 0 \ \ (i=2,3,4,5) \ , \ \ \ b_5 = b_6 = b_7 = 0 \ , \ \ \ b_8 = -\frac{\pi^2}{6} = -\zeta(2) \ , \ \ \
b_j = 0 \ \ (j=9, \ldots, 15)  \hspace{0.7cm}
\label{abs}
\end{eqnarray}
and, therefore, the $\alpha'$ expansion of $F^{\{23\}}(\alpha')$ begins as
\begin{eqnarray}
\label{expF23-2}
F^{\{23\}}(\alpha')& = & 1 + \big( a_1 \ \alpha_{12} \big) \alpha'  + \nonumber \\
&& \big( b_1 \ \alpha_{12} \alpha_{23} + b_2 \ \alpha_{12} \alpha_{34} + b_3 \ \alpha_{12} \alpha_{45} + b_4 \ \alpha_{12} \alpha_{15} -\zeta(2) \ \alpha_{34} \alpha_{45}   \big) {\alpha'}^2+  \ {\cal O}({\alpha'}^3) \ . \hspace{1cm}
\end{eqnarray}

\vspace{0.5cm}

\noindent \underline{Step 3}: Cyclic invariance of $A_b(1,2,3,4,5)$\\

\noindent In Appendix \ref{Cyclicity-N5} we prove that demanding cyclic invariance of $A_b(1,2,3,4,5)$ in (\ref{A12345-7}) implies that
\begin{eqnarray}
\label{F23cycl}
F^{\{23\}}(\alpha ^{\prime }) &=&F_{cycl}^{\{23\}}(\alpha ^{\prime })+\frac{%
\alpha _{13}+\alpha _{23}}{\alpha _{35}}F_{cycl}^{\{32\}}(\alpha ^{\prime })
 \ , \\
\label{F32cycl}
F^{\{32\}}(\alpha ^{\prime }) &=&\frac{\alpha _{13}}{\alpha _{35}}%
F_{cycl}^{\{32\}}(\alpha ^{\prime })  \ ,
\end{eqnarray}%
\noindent where $F_{cycl}^{\{23\}}(\alpha')$ and $F_{cycl}^{\{32\}}(\alpha')$ denote
doing $\{k_1 \rightarrow k_2$, $k_2 \rightarrow k_3$, $\ldots$, $k_5 \rightarrow k_1\}$ in
$F^{\{23\}}(\alpha')$ and $F^{\{32\}}(\alpha')$, respectively.\\
\noindent The $N=5$ BCJ relations (which we have completely found in Appendix \ref{N5-2}) have played an important role in the intermediate steps to arrive to eqs.(\ref{F23cycl}) and (\ref{F32cycl}) (see Appendix \ref{Cyclicity-N5} for more details).\\
\noindent In light of the result in (\ref{expF23-2}), after doing the calculations (\ref{F23cycl}) implies that\footnote{At ${\alpha'}^2$ order, relation (\ref{F32cycl})  does not give any new information: it is simply a consistency condition.}
\begin{eqnarray}
a_1 = 0 \ , \ \ \ b_1 = b_3 = 0 \ , \ \ \ b_2 = - b_4 = d_2 = \zeta(2)  \ ,
\end{eqnarray}
so (\ref{expF23-2}) and (\ref{simpleF32-1}) finally become, respectively,\\
\begin{eqnarray}
\label{}
F^{\{23\}}(\alpha')& = & 1 +  \big(   \alpha_{12} \alpha_{34} - \ \alpha_{12} \alpha_{15} - \ \alpha_{34} \alpha_{45}   \big) \ \zeta(2) \ {\alpha'}^2+  \ {\cal O}({\alpha'}^3) \ , \\
\label{}
F^{\{32\}}(\alpha')& = &  \zeta(2) \ \alpha_{13} \alpha_{24} \ {\alpha'}^2 + {\cal O}({\alpha'}^3) \ .
\end{eqnarray}
We have successfully executed steps $1$, $2$ and $3$, together with the corresponding $N=6$ calculations (see subsection \ref{Case of the 6-point}),  up to ${\alpha'}^6$ order, finding {\it all} the coefficients \footnote{We have also computed ${\alpha'}^7$ and higher order calculations, but some undetermined coefficients arise.\\
This is not in conflict with the revisited S-matrix method since this method only guarantees that, at a given ${\alpha'}^p$ order calculation, the $(\zeta \cdot k)^{p+2}$ terms should be absent in the $(p+2)$-point amplitude. For $p=7$ this means to have demanded a $9$-point calculation, which we have not done in this work.}. The explicit result up to ${\alpha'}^4$ terms is the following:
\begin{eqnarray}
F^{\{23\}}(\alpha ^{\prime }) &=&1+{\alpha ^{\prime }}^{2}\zeta (2)\left(
\alpha _{12}\alpha _{34}-\alpha _{34}\alpha _{45}-\alpha _{12}\alpha
_{15}\right)  + \notag \\
&&{\alpha ^{\prime }}^{3}\text{$\zeta (3)$}\left( -\alpha _{12}^{2}\alpha
_{34}-2\alpha _{12}\alpha _{23}\alpha _{34}-\alpha _{12}\alpha
_{34}^{2}+\alpha _{34}^{2}\alpha _{45}+\alpha _{34}\alpha _{45}^{2}+\alpha
_{12}^{2}\alpha _{15}+\alpha _{12}\alpha _{15}^{2}\right)  + \notag \\
&&\frac{{\alpha ^{\prime }}^{4}}{10}\zeta ^{2}(2)\left( 4\alpha
_{12}^{3}\alpha _{34}+5\alpha _{12}^{2}\alpha _{23}\alpha _{34}+5\alpha
_{12}\alpha _{23}^{2}\alpha _{34}+4\alpha _{12}^{2}\alpha _{34}^{2}+5\alpha
_{12}\alpha _{23}\alpha _{34}^{2}+4\alpha _{12}\alpha _{34}^{3} + \right.
\notag \\
&&7\alpha _{12}\alpha _{23}\alpha _{34}\alpha _{45}-3\alpha _{12}\alpha
_{34}^{2}\alpha _{45}-4\alpha _{34}^{3}\alpha _{45}-\alpha _{34}^{2}\alpha
_{45}^{2}-4\alpha _{34}\alpha _{45}^{3}-4\alpha _{12}^{3}\alpha _{15}  - \notag
\\
&&\left. 3\alpha _{12}^{2}\alpha _{34}\alpha _{15}+7\alpha _{12}\alpha
_{23}\alpha _{34}\alpha _{15}+3\alpha _{12}\alpha _{34}\alpha _{45}\alpha
_{15}-\alpha _{12}^{2}\alpha _{15}^{2}-4\alpha _{12}\alpha _{15}^{3}\right)+\mathcal{O}(\text{$\alpha ^{\prime 5})$} \ , \
\end{eqnarray}
\begin{eqnarray}
F^{\{32\}}(\alpha ^{\prime }) &=&{\alpha ^{\prime }}^{2}\zeta (2)\alpha
_{13}\alpha _{24} +{\alpha ^{\prime }}^{3}\text{$\zeta (3)$}\left( \alpha _{13}^{2}\alpha
_{24}+\alpha _{13}\alpha _{23}\alpha _{24}+\alpha _{13}\alpha
_{24}^{2}-2\alpha _{13}\alpha _{24}\alpha _{45}-2\alpha _{13}\alpha
_{24}\alpha _{15}\right)  + \notag \\
&&\frac{{\alpha ^{\prime }}^{4}}{10}\zeta ^{2}(2)\left( 4\alpha
_{13}^{3}\alpha _{24}+11\alpha _{13}^{2}\alpha _{23}\alpha _{24}+14\alpha
_{13}\alpha _{23}^{2}\alpha _{24}+4\alpha _{13}^{2}\alpha _{24}^{2}+11\alpha
_{13}\alpha _{23}\alpha _{24}^{2} + \right.  \notag \\
&&~~~4\alpha _{13}\alpha _{24}^{3}-12\alpha _{13}^{2}\alpha _{24}\alpha
_{45}-12\alpha _{13}\alpha _{23}\alpha _{24}\alpha _{45}-5\alpha _{13}\alpha
_{24}^{2}\alpha _{45}+12\alpha _{13}\alpha _{24}\alpha _{45}^{2}  - \notag \\
&&\left. 5\alpha _{13}^{2}\alpha _{24}\alpha _{15}-12\alpha _{13}\alpha
_{23}\alpha _{24}\alpha _{15}-12\alpha _{13}\alpha _{24}^{2}\alpha
_{15}+7\alpha _{13}\alpha _{24}\alpha _{45}\alpha _{15}+12\alpha _{13}\alpha
_{24}\alpha _{15}^{2}\right)  \notag \\
&&+\mathcal{O}(\text{$\alpha ^{\prime 5})$}
\end{eqnarray}
\noindent The corresponding full expressions up to ${\alpha'}^6$ terms are given in the text files that we have submitted, attached to this work, to the hep-th arXiv preprint basis.\\
\noindent We have confirmed that our results are in perfect agreement with the ones found previously in \cite{Barreiro1, Boels1, Broedel3}.

\vspace{0.5cm}

\subsection{Case of the 6-point momentum factors}

\label{Case of the 6-point}

\noindent Besides the notation introduced in eq.(\ref{alphaij}), here it
will be convenient to introduce the notation
\begin{eqnarray}
{\beta}_{ijk} & = & \alpha_{ij} + \alpha_{ik} + \alpha_{jk} \ ,
\label{tijk}
\end{eqnarray}
where $i < j < k$.\newline
\noindent The notation in eqs.(\ref{alphaij}) and (\ref{tijk}) will also be
valid for the $N=7$ case (subsection \ref{Case of the 7-point}).\\
\noindent So, we will repeat the procedure that we did in subsection \ref{Case of the 5-point} for $N=5$. For a general $N$ the procedure consists in the following four stages: \\
\\
\noindent {\bf i)} \ \ Identify the poles of each of the $(N-3)!$ Yang-Mills subamplitudes, \\
\hphantom{   \bf i) } $A_{YM}(1,\{ 2_{\sigma}, 3_{\sigma}, \ldots, (N-2)_{\sigma} \}, N-1, N)$, where $\sigma_{N} \ \varepsilon \ S_{N-3}$.\\
\noindent {\bf ii)} \ Define each $F^{\{ \sigma_N\}}(\alpha')$ momentum factor as a function of the $N(N-3)/2$ Mandelstam variables which were found in stage {\bf i)}.\\
\noindent {\bf iii)} Write the $\alpha'$ expansion of each $F^{\{ \sigma_N\}}(\alpha')$ up to the desired $\alpha'$ order, in terms of unknown coefficients.   \\
\noindent {\bf iv)} Determine the coefficients of the previous $\alpha'$ expansions by following the three steps specified in subsection \ref{Case of the 5-point}, namely, demand absence of {\it unphysical} poles, tree level unitarity and cyclic invariance. \\

\noindent Let us do this procedure for the case of $N=6$.\\

\noindent \underline{Stage {\bf i)}:}\\

\begin{figure}[th]
\centerline{\includegraphics*[scale=0.1,angle=0]{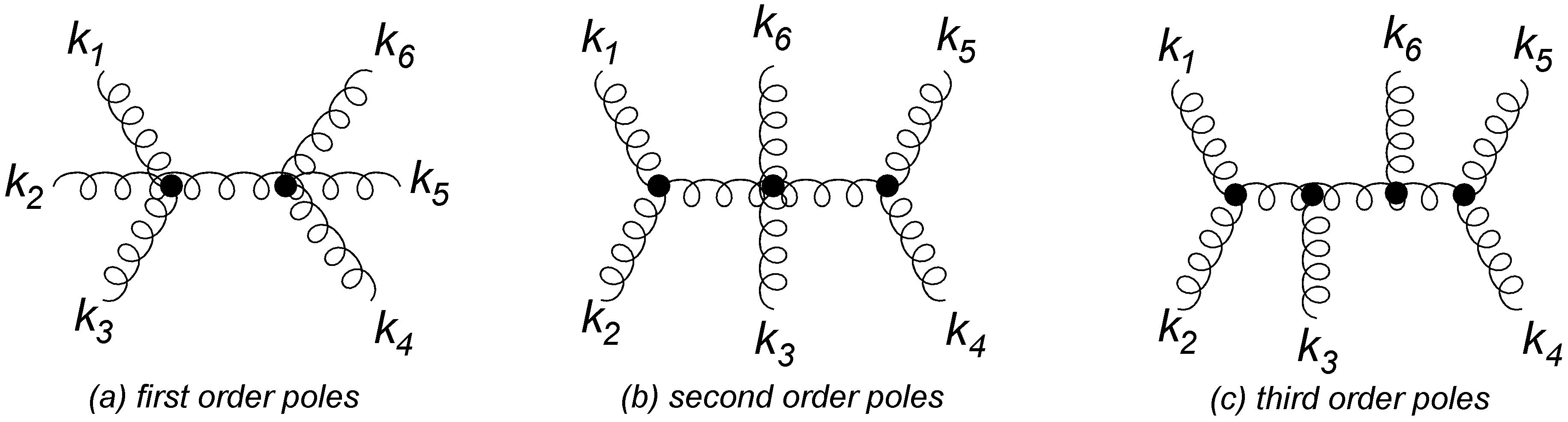}}
\caption{Three type of Feynman diagrams contribute to the YM 6-point amplitude. Permutations of the legs of these diagrams should also be considered in order to account for the full amplitude.}
\label{feynmandiag2}
\end{figure}

\noindent In this case there are first, second and third order poles, as we can see in figure \ref{feynmandiag2}.

\noindent From the Feynman diagrams in figure \ref{feynmandiag2}, and considering
cyclic permutations of the external momenta, we obtain the poles of $A_{YM}(1,2,3,4,5,6)$, shown in table (\ref{poles6p}),
\begin{equation}
\begin{tabular}{cccc}
\hline
& first & second & third \\
& order & order & order \\ \hline
&  & $\alpha _{12}\alpha _{45}$, $\alpha _{23}\alpha _{56}$, $\alpha
_{34}\alpha _{16}$, & $\alpha _{12}{\beta}_{123}\alpha _{45}$, $\alpha _{23}{%
\beta}_{234}\alpha _{56}$, \\
& ${\beta}_{123}$ & $\alpha _{12}{\beta}_{345}$, $\alpha _{23}{\beta}_{123}$%
, $\alpha _{34}{\beta}_{234}$, & $\alpha _{12}{\beta}_{345}\alpha _{45}$, $%
\alpha _{23}{\beta}_{123}\alpha _{56}$, \\
$A_{YM}(1,2,3,4,5,6)$ & ${\beta}_{234}$ & $\alpha _{45}{\beta}_{345}$, $%
\alpha _{56}{\beta}_{123}$, $\alpha _{16}{\beta}_{234}$, & $\alpha _{12}{%
\beta}_{345}\alpha _{34}$, $\alpha _{23}{\beta}_{123}\alpha _{45}$, \\
& ${\beta}_{345}$ & $\alpha _{12}{\beta}_{123}$, $\alpha _{23}{\beta}_{234}$%
, $\alpha _{34}{\beta}_{345}$, & $\alpha _{12}\alpha _{34}\alpha _{56}$, $%
\alpha _{23}\alpha _{45}\alpha _{16}$ \\
&  & $\alpha _{45}{\beta}_{123}$, $\alpha _{56}{\beta}_{234}$, $\alpha _{16}{%
\beta}_{345}$ & $\alpha _{34}{\beta}_{345}\alpha _{16}$, $\alpha _{34}{\beta}%
_{234}\alpha _{16}$, \\
&  &  & $\alpha _{34}{\beta}_{234}\alpha _{56}$ \\ \hline
\end{tabular}
\label{poles6p}
\end{equation}%
\noindent where we are using the notation of eqs.(\ref{alphaij}) and (\ref{tijk}).\\
\noindent The poles for the remaining five Yang-Mills subamplitudes can be obtained by
permutations of the indexes $\{2,3,4\}$. \\

\noindent \underline{Stage {\bf ii)}:}\\

\noindent So, we will work with the six momentum factors as functions of nine independent Mandelstam variables, according to
\begin{eqnarray}
\label{F234-1}
F^{\{234\}}(\alpha') =  F^{\{234\}}[  \alpha_{12},  \alpha_{23},  \alpha_{34},  \alpha_{45},  \alpha_{56},  \alpha_{16}; \beta_{123},  \beta_{234},  \beta_{345}     ; \alpha' \ ] \ , \\
\label{F324-1}
F^{\{324\}}(\alpha') =  F^{\{324\}}[  \alpha_{13},  \alpha_{23},  \alpha_{24},  \alpha_{45},  \alpha_{56},  \alpha_{16}; \beta_{123},  \beta_{234},  \beta_{245}     ; \alpha' \ ] \ , \\
\label{F243-1}
F^{\{243\}}(\alpha') =  F^{\{243\}}[  \alpha_{12},  \alpha_{24},  \alpha_{34},  \alpha_{35},  \alpha_{56},  \alpha_{16}; \beta_{124},  \beta_{234},  \beta_{345}     ; \alpha' \ ] \ , \\
\label{F342-1}
F^{\{342\}}(\alpha') =  F^{\{342\}}[  \alpha_{13},  \alpha_{34},  \alpha_{23},  \alpha_{25},  \alpha_{56},  \alpha_{16}; \beta_{134},  \beta_{234},  \beta_{245}     ; \alpha' \ ] \ , \\
\label{F423-1}
F^{\{423\}}(\alpha') =  F^{\{423\}}[  \alpha_{14},  \alpha_{24},  \alpha_{23},  \alpha_{35},  \alpha_{56},  \alpha_{16}; \beta_{124},  \beta_{234},  \beta_{235}     ; \alpha' \ ] \ , \\
\label{F432-1}
F^{\{432\}}(\alpha') =  F^{\{432\}}[  \alpha_{14},  \alpha_{34},  \alpha_{23},  \alpha_{25},  \alpha_{56},  \alpha_{16}; \beta_{134},  \beta_{234},  \beta_{235}     ; \alpha' \ ] \ .
\end{eqnarray}
\noindent These are the $N=6$ generalization of formulas (\ref{defF23}) and (\ref{defF32}) (seen for the case of $N=5$).\\

\noindent \underline{Stage {\bf iii)}:}\\

\noindent The general form of the $\alpha'$ power series for the momentum factors is the following:
\begin{equation}
\label{expF234}
F^{\{234\}}(\alpha ^{\prime })=1+\sum_{N=1}^{\infty }{\alpha ^{\prime }}^{N}%
\hspace{-0.5cm}\underset{n_{5},n_{6},n_{7},n_{8},n_9=0}%
{\sum_{n_{1},n_{2},n_{3},n_{4}, }^{ \ \ N \ , }}  \hspace{-0.5cm}%
A_{n_{1},n_{2}, \ldots ,n_{9}}^{(N)}\ \alpha _{12}^{n_{1}} \ \alpha
_{23}^{n_{2}} \ \alpha _{34}^{n_{3}} \ \alpha _{45}^{n_{4}} \ \alpha
_{56}^{n_{5}} \ \alpha _{16}^{n_{6}} \ {\beta}_{123}^{n_{7}} \ {\beta}_{234}^{n_{8}} \ {%
\beta}_{345}^{n_{9}} \ ,
\end{equation}%
\begin{equation}
\label{expF324}
F^{\{324\}}(\alpha ^{\prime })=\sum_{N=1}^{\infty }{\alpha ^{\prime }}^{N}%
\hspace{-0.5cm}\underset{n_{5},n_{6},n_{7},n_{8},n_9=0}%
{\sum_{n_{1},n_{2},n_{3},n_{4}, }^{ \ \ N \ , }}  \hspace{-0.5cm}%
B_{n_{1},n_{2}, \ldots ,n_{9}}^{(N)}\ \alpha _{13}^{n_{1}} \ \alpha
_{23}^{n_{2}} \ \alpha _{24}^{n_{3}} \ \alpha _{45}^{n_{4}} \ \alpha
_{56}^{n_{5}} \ \alpha _{16}^{n_{6}} \ {\beta}_{123}^{n_{7}} \ {\beta}_{234}^{n_{8}} \ {%
\beta}_{245}^{n_{9}} \ ,
\end{equation}%
\begin{equation}
\label{expF243}
F^{\{243\}}(\alpha ^{\prime })=\sum_{N=1}^{\infty }{\alpha ^{\prime }}^{N}%
\hspace{-0.5cm}\underset{n_{5},n_{6},n_{7},n_{8},n_9=0}%
{\sum_{n_{1},n_{2},n_{3},n_{4}, }^{ \ \ N \ , }}  \hspace{-0.5cm}%
C_{n_{1},n_{2}, \ldots ,n_{9}}^{(N)}\ \alpha _{12}^{n_{1}} \ \alpha
_{24}^{n_{2}} \ \alpha _{34}^{n_{3}} \ \alpha _{35}^{n_{4}} \ \alpha
_{56}^{n_{5}} \ \alpha _{16}^{n_{6}} \ {\beta}_{124}^{n_{7}} \ {\beta}_{234}^{n_{8}} \ {%
\beta}_{345}^{n_{9}} \ ,
\end{equation}%
\begin{equation}
\label{expF342}
F^{\{342\}}(\alpha ^{\prime })=\sum_{N=1}^{\infty }{\alpha ^{\prime }}^{N}%
\hspace{-0.5cm}\underset{n_{5},n_{6},n_{7},n_{8},n_9=0}%
{\sum_{n_{1},n_{2},n_{3},n_{4}, }^{ \ \ N \ , }}  \hspace{-0.5cm}%
D_{n_{1},n_{2}, \ldots ,n_{9}}^{(N)}\ \alpha _{13}^{n_{1}} \ \alpha
_{34}^{n_{2}} \ \alpha _{23}^{n_{3}} \ \alpha _{25}^{n_{4}} \ \alpha
_{56}^{n_{5}} \ \alpha _{16}^{n_{6}} \ {\beta}_{134}^{n_{7}} \ {\beta}_{234}^{n_{8}} \ {%
\beta}_{245}^{n_{9}} \ ,
\end{equation}%
\begin{equation}
\label{expF423}
F^{\{423\}}(\alpha ^{\prime })=\sum_{N=1}^{\infty }{\alpha ^{\prime }}^{N}%
\hspace{-0.5cm}\underset{n_{5},n_{6},n_{7},n_{8},n_9=0}%
{\sum_{n_{1},n_{2},n_{3},n_{4}, }^{ \ \ N \ , }}  \hspace{-0.5cm}%
E_{n_{1},n_{2}, \ldots ,n_{9}}^{(N)}\ \alpha _{14}^{n_{1}} \ \alpha
_{24}^{n_{2}} \ \alpha _{23}^{n_{3}} \ \alpha _{35}^{n_{4}} \ \alpha
_{56}^{n_{5}} \ \alpha _{16}^{n_{6}} \ {\beta}_{124}^{n_{7}} \ {\beta}_{234}^{n_{8}} \ {%
\beta}_{235}^{n_{9}} \ ,
\end{equation}%
\begin{equation}
\label{expF432}
F^{\{432\}}(\alpha ^{\prime })=\sum_{N=1}^{\infty }{\alpha ^{\prime }}^{N}%
\hspace{-0.5cm}\underset{n_{5},n_{6},n_{7},n_{8},n_9=0}%
{\sum_{n_{1},n_{2},n_{3},n_{4}, }^{ \ \ N \ , }}  \hspace{-0.5cm}%
F_{n_{1},n_{2}, \ldots ,n_{9}}^{(N)}\ \alpha _{14}^{n_{1}} \ \alpha
_{34}^{n_{2}} \ \alpha _{23}^{n_{3}} \ \alpha _{25}^{n_{4}} \ \alpha
_{56}^{n_{5}} \ \alpha _{16}^{n_{6}} \ {\beta}_{134}^{n_{7}} \ {\beta}_{234}^{n_{8}} \ {%
\beta}_{235}^{n_{9}} \ .
\end{equation}%
\noindent In all expansions, from (\ref{expF234}) to (\ref{expF432}), the prime on the internal sum means that the set of indices
$\{  n_1$,  $n_2$,  $n_3$,  $n_4$,  $n_5$,  $n_6$,  $n_7$,  $n_8$,  $n_9 \}$ obeys the constraint
\begin{eqnarray}
n_{1}+n_{2}+n_{3}+n_{4}+n_{5}+n_{6}+n_{7}+n_{8}+n_9=N \ .
\label{constraint-N}
\end{eqnarray}
\noindent In these expansions we have already used the condition of low energy behaviour, eq.(\ref{Fs}), for the momentum factors: that is why the $\alpha'$ series of $F^{\{234\}}(\alpha')$ begins with `$1$' and the five others begin with ${\cal O}({\alpha'}^1)$ order contributions\footnote{Once the unitarity and the cyclicity relations have been demanded it will naturally happen that the order `$1$' coefficients of all expansions will be zero, but as a matter of principles here we are just proposing their general $\alpha'$ expansions in such a way that they respect the low energy requirement in (\ref{Fs}) and the analyticity of the momentum factors.}.

\vspace{0.5cm}

\noindent \underline{Stage {\bf iv)}:}\\

\noindent Here we apply the three steps that will allow us to obtain the coefficients of the $\alpha'$ expansions in eqs.(\ref{expF234})-(\ref{expF432}). We will do our calculations up to ${\alpha'}^6$ order terms.\\

\noindent \underline{Step 1}: Absence of {\it unphysical} poles in $A_b(1,2,3,4,5,6)$\\

\noindent Each of the $\alpha'$ expansions initially has 9 coefficients at ${\alpha'}^1$ order,
 45 coefficients at ${\alpha'}^2$ order, 165 coefficients at ${\alpha'}^3$ order,  495 coefficients at ${\alpha'}^4$ order,  1287 coefficients at ${\alpha'}^5$ order and 3003 coefficients at ${\alpha'}^6$ order\footnote{In this case, the geral formula for the number of coefficients at ${\alpha'}^N$ order is $(N+8)!/(N! \ 8!)$.}. \\
\noindent Demanding absence of unphysical poles in $A_b(1,2,3,4,5,6)$ implies that there should be cancellations between the unphysical simple poles which come in the five Yang-Mills subamplitudes which are different of $A_{YM}(1,2,3,4,5,6)$ (which has only physical poles). After this procedure has been done, the number of independent coefficients reduces considerably according to table (\ref{tableaux}).\\
\noindent There is also the subtlety that $F^{\{234\}}(\alpha')$ and the remaning five momentum factors should be such that they cancel the third order poles that come in $A_{YM}(1,2,3,4,5,6)$. These poles are not present in $A_b(1,2,3,4,5,6)$ (at non zero $\alpha'$ order). \\
\noindent In the case of the ${\alpha'}^1$ order terms, the requirement of absence of unphysical poles is strong enough to forbid them: the coefficients of those terms are all zero and that is the reason of why the first line in table (\ref{tableaux}) contains only $0$'s (zeroes).  \\
\noindent We will not write down here the partial result that we have obtained for the expansions in eqs.(\ref{expF234})-(\ref{expF432}), as we did at the end of Step 1 in subsection \ref{Case of the 5-point}. Instead, for each of the six $F^{\{ \sigma_6\}}(\alpha')$'s we have presented two data at a given order in $\alpha'$, in table (\ref{tableaux}). The first data is the number of undetermined coefficients after the actual Step (Step 1) and the second data is the number of undetermined coefficients after Step 2. We see there, that before demanding cyclic symmetry (in Step 3) there is still a big number of undetermined coefficients. \\

\noindent \underline{Step 2}: Unitarity relations for $A_b(1,2,3,4,5,6)$\\

\noindent Since in the $N=6$ case there are two type of simple poles ($\alpha_{ij}$ and $\beta_{ijk}$, see table (\ref{poles6p})), demanding unitarity of $A_b(1,2,3,4,5,6)$ with respect to each of them will lead to independent unitarity relations. The remaining unitarity relations will be a consequence of the previous ones once cyclic symmetry has been taken into account (in Step 3).\\
\noindent In Appendix \ref{Unitarity-N6} we prove that demanding unitarity of $A_b(1,2,3,4,5,6)$  with respect to its $\alpha_{12}$ pole implies that the $6$-point momentum factors, $F^{\{234\}}(\alpha')$ and $F^{\{243\}}(\alpha')$, are related to the $5$-point momentum factors, $F^{\{23\}}(\alpha')$ and $F^{\{32\}}(\alpha')$, by means of the following two relations:
\begin{eqnarray}
\label{fact-F234}
F^{\{234\}}[ \ 0 ,\alpha _{23},\alpha _{34},\alpha_{45},\alpha _{56},\alpha _{16},{\beta }_{123},{\beta }_{234},{\beta }_{345};\alpha ^{\prime }]&=& F^{\{23\}}[ {\beta }_{123},\alpha _{34},\alpha _{45},\alpha _{56},{\beta }_{345};\alpha^{\prime }] \ , \hspace{1.0cm}\\
\label{fact-F243}
F^{\{243\}}[ \ 0 ,  \alpha_{24},  \alpha_{34},  \alpha_{35},  \alpha_{56},  \alpha_{16}; \beta_{124},  \beta_{234},  \beta_{345}     ; \alpha' ]&=&F^{\{32\}}[ {\beta }_{124},\alpha _{34},\alpha _{35},\alpha _{56},{\beta }_{345};\alpha^{\prime }] \ .
\end{eqnarray}
\noindent In analogy to relation (\ref{factorization5p1}), found after demanding unitarity for the $5$-point amplitude, relations (\ref{fact-F234}) and (\ref{fact-F243}) are strong constraints for the coefficients of the $\alpha'$ expansions of $F^{\{234\}}$ and $F^{\{243\}}$ since there are four arguments (like $\alpha_{23}$, $\alpha_{16}$, $\beta_{123}$ and $\beta_{234}$, in the first case) which are not present in the right hand-side of the corresponding relation.\\
\noindent Also, demanding unitarity of $A_b(1,2,3,4,5,6)$  with respect to its $\beta_{123}$ pole implies that the $6$-point momentum factors,
$F^{\{234\}}(\alpha')$ and $F^{\{324\}}(\alpha')$, are related to the $4$-point momentum factor, $F^{\{2\}}(\alpha')$, by means of the relation:\\
\begin{eqnarray}
F^{\{234\}}[\alpha _{12},\alpha _{23},\alpha _{34},\alpha_{45},\alpha _{56},\alpha _{16}, \ 0 ,{\beta }_{234},{\beta }_{345};\alpha ^{\prime }]  -\frac{\alpha _{12}}{\alpha _{12}+\alpha _{23}} \times \hspace{4.8cm} \nonumber \\
F^{\{324\}}[ - \alpha _{12} - \alpha_{23}    ,\alpha _{23}, \beta_{234} - \alpha_{23} -\alpha_{34}   ,\alpha _{45},\alpha
_{56},\alpha _{16}, \ 0 ,{\beta }_{234},  -\alpha_{23}+\alpha_{45} +\alpha_{16} - \beta_{345}   ;\alpha
^{\prime }]=  \notag \\
\hspace{6cm}=F^{\{2\}}[\alpha _{12},\alpha _{23};\alpha ^{\prime
}] \ F^{\{2\}}[\alpha _{56},\alpha _{45};\alpha ^{\prime }] \ . \nonumber \\
\label{fact-mixed}
\end{eqnarray}
\noindent In this case there are even stronger contraints\footnote{But, in contrast to (\ref{fact-F234}) and (\ref{fact-F243}), there is only one unitarity relation now.} because, besides the fact that there are four arguments which are not present in the right hand-side of (\ref{fact-mixed}), in the left hand-side a miraculous cancelation of the $(\alpha_{12}+\alpha_{23})$ denominator should happen, since in the right hand-side there is only a product of two $\alpha'$ power series (which involve no denominators at all).\\
\noindent The curious non zero arguments in which $F^{\{324\}}$ is being evaluated, in the left hand-side of (\ref{fact-mixed}), are simply the original ones (see eq.(\ref{F324-1})), but written in terms of the basis of Mandelstam variables used for $F^{\{234\}}(\alpha')$ (see eq.(\ref{F234-1})), with $\beta_{123}=0$.\\
\noindent The three relations that we have written in eqs.(\ref{fact-F234}), (\ref{fact-F243}) and (\ref{fact-mixed}), are conditions for only three of the six momentum factors ($F^{\{234\}}$, $F^{\{243\}}$ and $F^{\{324\}}$). For these momentum factors the number of its undetermined coefficients has been reduced as a consequence of Step 2. This is illustrated in table (\ref{tableaux}), in the second data which we have presented for each $F^{\{ \sigma_6\}}(\alpha')$ at a given order in $\alpha'$\footnote{This data corresponds to the number of coefficients which are still undetermined after Step 2, at a given $\alpha'$ order.}. \\
\noindent For the remaining momentum factors, the number of undetermined coefficients has not changed from Step 1 to Step 2. \\
\noindent After demanding cyclic invariance the coefficients of all six momentum factors will be related and it will be possible to find them all, at least up to ${\alpha'}^6$ order.\\
\begin{equation}
\begin{tabular}{|c|c|c|c|c|c|c|c|c|c|c|c|c|}
\hline
Order & \multicolumn{2}{|c|}{$F^{\{234\}}(\alpha ^{\prime })$} &
\multicolumn{2}{|c|}{$F^{\{324\}}(\alpha ^{\prime })$} &
\multicolumn{2}{|c|}{$F^{\{243\}}(\alpha ^{\prime })$} &
\multicolumn{2}{|c|}{$F^{\{342\}}(\alpha ^{\prime })$} &
\multicolumn{2}{|c|}{$F^{\{423\}}(\alpha ^{\prime })$} &
\multicolumn{2}{|c|}{$F^{\{432\}}(\alpha ^{\prime })$} \\ \hline
${\alpha'}^1$ & 0 & 0 & 0 & 0 & 0 & 0 & 0 & 0 & 0 & 0 & 0 & 0 \\ \hline
${\alpha'}^2$ & 44 & 4 & 3 & 3 & 1 & 2 & 3 & 3 & 3 & 3 & 3 & 3 \\ \hline
${\alpha'}^3$ & 164 & 24 & 25 & 10 & 25 & 17 & 25 & 25 & 25 & 25 & 25 & 25 \\ \hline
${\alpha'}^4$ & 494 & 109 & 177 & 53 & 177 & 81 & 177 & 177 & 177 & 177 & 177 & 177 \\
\hline
${\alpha'}^5$ & 1286 & 371 & 405 & 201 & 405 & 287 & 405 & 405 & 405 & 405 & 405 & 405
\\ \hline
${\alpha'}^6$ & 3002 & 1039 & 1155 & 615 & 1155 & 842 & 1155 & 1155 & 1155 & 1155 & 1155
& 1155 \\ \hline
\end{tabular}
\label{tableaux}
\end{equation}

\vspace{0.5cm}

\noindent \underline{Step 3}: Cyclic invariance of $A_b(1,2,3,4,5,6)$\\

\noindent In Appendix \ref{Cyclicity-N6} we prove that demanding cyclic invariance for $A_b(1,2,3,4,5,6)$, in the closed formula that we have for it,
\begin{multline}
A_b(1,2,3,4,5,6)  = \\
\begin{split}
& F^{\{234\}}(\alpha')  A_{YM}(1,2,3,4,5,6) +  F^{\{324\}}(\alpha')  A_{YM}(1,3,2,4,5,6) + F^{\{243\}}(\alpha')  A_{YM}(1,2,4,3,5,6) +  \\
& F^{\{342\}}(\alpha')  A_{YM}(1,3,4,2,5,6) +  F^{\{423\}}(\alpha')  A_{YM}(1,4,2,3,5,6) + F^{\{432\}}(\alpha')  A_{YM}(1,4,3,2,5,6)  \ ,
\end{split}
\label{6point-aux}
\end{multline}
 implies the following six relations for the momentum factors:
\begin{eqnarray}
F^{\{234\}}(\alpha')&=&F_{cycl}^{\{234\}}(\alpha^\prime)+ \frac{\alpha _{56}-{\beta}_{123}}{{\beta}_{123}-\alpha_{45}-\alpha_{56}} F_{cycl}^{\{243\}}(\alpha^\prime)+
\frac{   \alpha _{12}-{\beta}_{123}  }{ {\beta}_{123}+{\beta}_{345}-\alpha _{12}-\alpha _{45} }  F_{cycl}^{\{342\}}(\alpha^\prime) \nonumber \\
&&+\frac{\text{$F_{cycl}^{\{432\}}(\alpha^\prime)$}}{\left( {\beta}_{123}+{\beta}_{345}-\alpha
_{12}-\alpha _{45}\right) \left( -{\beta}_{345}+\alpha _{12}+\alpha _{34}-\alpha
_{56}\right) }  \times \nonumber \\
&&\lbrack \alpha _{12}({\beta}_{123}+{\beta}_{345}-\alpha _{12}-\alpha _{34}+\alpha
_{56}-\alpha _{45})+{\beta}_{123}(\alpha _{34}+\alpha _{45}-{\beta}_{345}-\alpha _{56})] + \nonumber \\
&&\frac{\text{$F_{cycl}^{\{423\}}(\alpha^\prime)$}}{\left( {\beta}_{345}-\alpha
_{12}-\alpha _{34}+\alpha _{56}\right) \left( -{\beta}_{123}+\alpha _{45}+\alpha
_{56}\right) }\alpha _{45}(\alpha _{12}-{\beta}_{123}) \ ,
\label{F234-expansion}
\end{eqnarray}
\begin{multline}
\begin{split}
F^{\{324\}}(\alpha')=&\frac{\text{$F_{cycl}^{\{342\}}(\alpha^\prime)$}}{{\beta}_{123}+{\beta}_{345}-\alpha _{12}-\alpha _{45}}%
(\alpha _{12}+\alpha _{23}-{\beta}_{123})   \\
&+\frac{\text{$F_{cycl}^{\{432\}}(\alpha^\prime)$}}{\left( {\beta}_{123}+{\beta}_{345}-\alpha
_{12}-\alpha _{45}\right) \left( -{\beta}_{345}+\alpha _{12}+\alpha _{34}-\alpha
_{56}\right) }  \\
&\times \lbrack \alpha _{12}({\beta}_{123}+{\beta}_{345}-\alpha _{12}-\alpha _{23}-\alpha
_{34}+\alpha _{56}-\alpha _{45}) \\
&+{\beta}_{123}(\alpha _{34}+\alpha
_{45}-{\beta}_{345}-\alpha _{56})+\alpha _{23}({\beta}_{345}-\alpha _{34}+\alpha
_{56}-\alpha _{45})]   \\
&+\frac{\text{$F_{cycl}^{\{423\}}(\alpha^\prime)$}}{\left( {\beta}_{345}-\alpha
_{12}-\alpha _{34}+\alpha _{56}\right) \left( -{\beta}_{123}+\alpha _{45}+\alpha
_{56}\right) }\alpha _{45}(\alpha _{12}+\alpha _{23}-{\beta}_{123}) \ ,
\end{split}
\end{multline}
\begin{multline}
\begin{split}
F^{\{243\}}(\alpha')=&\text{$F_{cycl}^{\{324\}}(\alpha^\prime)$}+\frac{\text{$F_{cycl}^{\{243\}}(\alpha^\prime)$}}{%
{\beta}_{123}-\alpha _{45}-\alpha _{56}}(\alpha _{56}-\alpha _{34}-{\beta}_{123}) \\
&\frac{%
\text{$F_{cycl}^{\{342\}}(\alpha^\prime)$}}{{\beta}_{123}+{\beta}_{345}-\alpha _{12}-\alpha _{45}}%
(\alpha _{12}-\alpha _{34}-{\beta}_{123})  + \\
&\frac{\text{$F_{cycl}^{\{432\}}(\alpha^\prime)$}}{\left( {\beta}_{123}+{\beta}_{345}-\alpha
_{12}-\alpha _{45}\right) \left( {\beta}_{345}-\alpha _{12}-\alpha _{34}+\alpha
_{56}\right) }  \times \\
&\lbrack \alpha _{34}(-{\beta}_{123}+{\beta}_{345}-\alpha _{34}-\alpha
_{45}+\alpha _{56})+({\beta}_{123}-\alpha _{56})({\beta}_{345}-\alpha _{45})]  + \\
&\frac{\text{$F_{cycl}^{\{423\}}(\alpha^\prime)$}}{\left( -{\beta}_{345}+\alpha _{12}+\alpha
_{34}-\alpha _{56}\right) \left( -{\beta}_{123}+\alpha _{45}+\alpha _{56}\right) }
\times \\
&\lbrack \alpha _{56}({\beta}_{123}+\alpha _{12}+\alpha _{34}-\alpha
_{45}-\alpha _{56})+({\beta}_{123}+\alpha _{34})(\alpha _{45}-\alpha
_{12})] \ ,
\end{split}
\end{multline}
\begin{multline}
\begin{split}
F^{\{342\}}(\alpha^\prime) = &\frac{\text{$F_{cycl}^{\{432\}}(\alpha^\prime)$}}{\left( {\beta}_{123}+{\beta}_{345}-\alpha
_{12}-\alpha _{45}\right) \left( -{\beta}_{345}+\alpha _{12}+\alpha _{34}-\alpha
_{56}\right) }  \\
&\times \lbrack \alpha _{12}({\beta}_{123}-{\beta}_{234}+{\beta}_{345}-\alpha _{12}-\alpha
_{45}+\alpha _{56}) \\
&+{\beta}_{123}({\beta}_{234}-{\beta}_{345}-\alpha _{23}+\alpha _{45}-\alpha
_{56})+\alpha _{23}({\beta}_{345}-{\beta}_{234}+\alpha _{23}-\alpha _{45}+\alpha _{56})]
 \\
&+\frac{\text{$F_{cycl}^{\{423\}}(\alpha^\prime)$}}{\left( {\beta}_{345}-\alpha _{12}-\alpha
_{34}+\alpha _{56}\right) \left( -{\beta}_{123}+\alpha _{45}+\alpha _{56}\right) }
 \\
&\times \lbrack \alpha _{23}({\beta}_{123}+{\beta}_{234}-\alpha _{12}-\alpha _{23}-\alpha
_{34}+\alpha _{45})+({\beta}_{123}-\alpha _{12})(\alpha _{34}-{\beta}_{234}-\alpha
_{45})] \ ,
\end{split}
\end{multline}
\begin{multline}
\begin{split}
F^{\{423\}}(\alpha^\prime)=&\frac{\text{$F_{cycl}^{\{243\}}(\alpha^\prime)$}}{{\beta}_{123}-\alpha _{45}-\alpha _{56}}(\alpha
_{23}+\alpha _{56}-{\beta}_{123}-{\beta}_{234})   \\
&+\frac{\text{$F_{cycl}^{\{432\}}(\alpha^\prime)$}}{\left( {\beta}_{123}+{\beta}_{345}-\alpha
_{12}-\alpha _{45}\right) \left( {\beta}_{345}-\alpha _{12}-\alpha _{34}+\alpha
_{56}\right) }   \\
&\times \lbrack {\beta}_{345}({\beta}_{123}+{\beta}_{234}-\alpha _{23}-\alpha _{56})+(\alpha
_{34}+\alpha _{45})(\alpha _{23}+\alpha _{56}-{\beta}_{123}-{\beta}_{234})]   \\
&+\frac{\text{$F_{cycl}^{\{423\}}(\alpha^\prime)$}}{\left( {\beta}_{345}-\alpha _{12}-\alpha
_{34}+\alpha _{56}\right) \left( -{\beta}_{123}+\alpha _{45}+\alpha _{56}\right) } \\
&\times \lbrack \alpha _{56}(\alpha _{23}+\alpha _{45}-{\beta}_{123}-{\beta}_{234}-\alpha
_{12}+\alpha _{56})+(\alpha _{12}-\alpha _{45})({\beta}_{123}+{\beta}_{234}-\alpha
_{23})] \ ,
\end{split}
\end{multline}
\begin{multline}
F^{\{432\}}(\alpha^\prime)= \\
\begin{split}
=&\frac{\text{$F_{cycl}^{\{432\}}(\alpha^\prime)$}}{\left( {\beta}_{123}+{\beta}_{345}-\alpha
_{12}-\alpha _{45}\right) \left( {\beta}_{345}-\alpha _{12}-\alpha _{34}+\alpha
_{56}\right) } \times \\
&\lbrack \alpha _{23}({\beta}_{123}+{\beta}_{234}-{\beta}_{345}-\alpha _{23}+\alpha
_{34}+\alpha _{45}-\alpha _{56})+(\alpha _{56}-{\beta}_{234}-{\beta}_{123})(\alpha
_{34}-{\beta}_{345}+\alpha _{45})]  \\
&+\frac{\text{$F_{cycl}^{\{423\}}(\alpha^\prime)$}}{\left( {\beta}_{123}-\alpha _{45}-\alpha
_{56}\right) \left( {\beta}_{345}-\alpha _{12}-\alpha _{34}+\alpha _{56}\right) }
\times \lbrack \alpha _{23}(\alpha _{12}-{\beta}_{123}-{\beta}_{234}+\alpha _{23}-\alpha
_{45}) + \\
&\alpha _{56}(\alpha _{12}+{\beta}_{123}+{\beta}_{234}-\alpha _{45}-\alpha
_{56})+({\beta}_{123}+{\beta}_{234})(\alpha _{45}-\alpha _{12})] \ ,
\end{split}
\label{F432-expansion}
\end{multline}
\noindent where each $F_{cycl}^{\{ \sigma_6 \}}(\alpha')$ denotes doing  $\{k_1 \rightarrow k_2$, $k_2 \rightarrow k_3$, $\ldots$, $k_6 \rightarrow k_1\}$ in the corresponding $F^{\{ \sigma_6 \}}(\alpha')$ momentum factor.\\
\noindent The $N=6$ BCJ relations (which we have completely found in Appendix \ref{N6-2})  have played an important role in the intermediate steps to arrive to eqs.(\ref{F234-expansion})-(\ref{F432-expansion}) (see Appendix \ref{Cyclicity-N6} for more details).\\
\noindent So, summarizing, we have successfully executed steps $1$, $2$ and $3$, together with the corresponding $N=5$ calculations (see subsection \ref{Case of the 5-point}),  up to ${\alpha'}^6$ order, finding {\it all} the coefficients. The explicit result up to ${\alpha'}^3$ terms is the following:
\begin{eqnarray}
F^{\{234\}}(\alpha ^{\prime }) &=&1+\text{$\alpha $}^{\prime 2}\zeta
(2)\left( -{\beta }_{123}{\beta }_{345}+{\beta }_{345}\alpha _{12}+{\beta }%
_{123}\alpha _{45}-\alpha _{45}\alpha _{56}-\alpha _{12}\alpha_{16}\right) +
\notag \\
&&\text{$\alpha ^{\prime }$}^{3}\text{$\zeta (3)\left( {\beta }_{123}^{2}{%
\beta }_{345}+{\beta }_{123}{\beta }_{345}^{2}-{\beta }_{345}^{2}\alpha
_{12}-{\beta }_{345}\alpha _{12}^{2}-2{\beta }_{345}\alpha _{12}\alpha
_{23}\right. $} - \notag \\
&&{\beta }_{123}^{2}\alpha _{45}-2{\beta }_{234}\alpha _{12}\alpha
_{45}+2\alpha _{12}\alpha _{23}\alpha _{45}-2{\beta }_{123}\alpha
_{34}\alpha _{45} + \notag \\
&&\left. 2 \alpha _{12}\alpha _{34}\alpha _{45}-{\beta }_{123}\alpha
_{45}^{2}+\alpha _{45}^{2}\alpha _{56}+\alpha _{45}\alpha _{56}^{2}+\alpha
_{12}^{2}\alpha _{16}+\alpha _{12}\alpha _{16}^{2}\right)  + {\cal O}({\alpha'}^4) \ , \hspace{0.5cm}
\end{eqnarray}
\begin{eqnarray}
F^{\{324\}}(\alpha ^{\prime }) &=&{\alpha ^{\prime }}^{2}\zeta (2)\left( {%
\beta }_{245}\alpha _{13}-\alpha _{13}\alpha _{45}\right)  + \notag \\
&&{\alpha ^{\prime }}^{3}\text{$\zeta (3)$}\alpha _{13}\left[ {\beta }%
_{245}^{2}-2{\beta }_{123}\left( {\beta }_{245}-\alpha _{45}\right) +\alpha
_{45}\left( -\alpha _{13}-\alpha _{23}+2\alpha _{24}+\alpha _{45}+2\alpha
_{16}\right) \right. + \notag \\
&&\left. {\beta }_{245}\left( \alpha _{13}+\alpha _{23}-2\left( \alpha
_{45}+\alpha _{16}\right) \right) \right]  +  {\cal O}({\alpha'}^4) \ ,
\end{eqnarray}
\begin{eqnarray}
F^{\{243\}}(\alpha ^{\prime }) &=&{\alpha ^{\prime }}^{2}\zeta (2)\left( {%
\beta}_{124}\alpha _{35}-\alpha _{12}\alpha _{35}\right) +  \notag \\
&&{\alpha ^{\prime }}^{3}\text{$\zeta (3)$}\alpha _{35}\left[ {\beta}%
_{124}^{2}+{\beta}_{124}\left( -2{\beta}_{345}-2\alpha _{12}+\alpha
_{34}+\alpha _{35}-2\alpha _{56}\right) \right.  + \notag \\
&&\left. \alpha _{12}\left( 2{\beta}_{345}+\alpha _{12}+2\alpha
_{24}-\alpha _{34}-\alpha _{35}+2\alpha _{56}\right) \right] +  {\cal O}({\alpha'}^4) \ ,
\end{eqnarray}
\begin{eqnarray}
F^{\{342\}}(\alpha ^{\prime }) &=&{\alpha ^{\prime }}^{2}\text{$\zeta $}(2)\alpha _{13}\alpha _{25}  -\notag \\
&&{\alpha ^{\prime }}^{3}\text{$\zeta $}(3)\alpha _{13}\alpha _{25}\left(
-2{\beta}_{134}-{\beta}_{234}+\alpha _{13}-\alpha _{25}+\alpha _{34}+2\alpha
_{56}+2\alpha _{16}\right)  +  {\cal O}({\alpha'}^4) \ , \hspace{0.7cm}
\end{eqnarray}
\begin{eqnarray}
F^{\{423\}}(\alpha ^{\prime }) &=&{\alpha ^{\prime }}^{2}\zeta (2)\alpha
_{14}\alpha _{35} +  \notag \\
&&{\alpha ^{\prime }}^{3}\text{$\zeta (3)$}\alpha _{14}\alpha _{35}\left( {%
\beta}_{234}+2{\beta}_{235}+\alpha _{14}-\alpha _{23}-\alpha _{35}-2\alpha
_{56}-2\alpha _{16}\right)   +  {\cal O}({\alpha'}^4) \ , \hspace{0.7cm}
\end{eqnarray}
\begin{eqnarray}
F^{\{432\}}(\alpha ^{\prime }) &=&-{\alpha ^{\prime }}^{2}\zeta (2)\alpha
_{14}\alpha _{25}  -{\alpha ^{\prime }}^{3}\text{$\zeta (3)$}\alpha _{14}\alpha _{25}\left( {%
\beta}_{234}+\alpha _{14}+\alpha _{25}-2\alpha _{56}-2\alpha _{16}\right)  +  {\cal O}({\alpha'}^4) \ .  \hspace{0.8cm}
\end{eqnarray}
\noindent The corresponding full expressions up to ${\alpha'}^6$ terms are given in the text files that we have submitted, attached to this work, to the hep-th arXiv preprint basis.\\
\noindent We have confirmed that our results are in perfect agreement with the ones found previously in \cite{Mafra3, Broedel3}.

\vspace{0.5cm}

\subsection{Case of the 7-point momentum factors}

\vspace{0.5cm}

\begin{figure}[th]
\centerline{\includegraphics*[scale=0.1,angle=0]{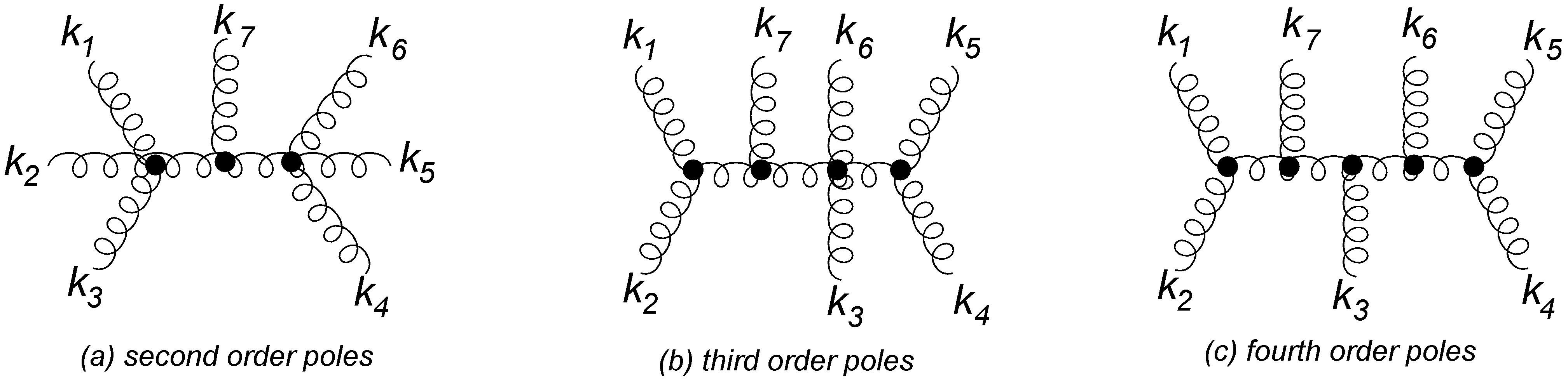}}
\caption{Tree kind of Feynman diagrams used to calculate the YM subamplitude
$A_{YM}(1,2,3,4,5,6,7)$. }
\label{feynmandiag3}
\end{figure}

\label{Case of the 7-point}
\noindent Doing the procedure for the $N=7$ case is straight forward (although very tedious) once it has been done for $N=5$ and $N=6$. We will only mention here a few details.\\
\noindent First, with respect to the poles of $A_{YM}(1,2,3,4,5,6,7)$ (and of the other twenty three subamplitudes of the $7$-point basis), it has second, third and fourth order ones. They come from the Feynman diagrams in figure \ref{feynmandiag3}. The poles can happen in any of the fourteen Mandelstam variables that we have specified in eq.(\ref{Mandelstam7}): seven $\alpha_{ij}$'s and also seven $\beta_{ijk}$'s.\\
\noindent We have (shortly) presented the poles in table (\ref{poles7p}):
\begin{equation}
\begin{tabular}{cccc}
\hline
& second order & third order & fourth order \\
& (28 terms) & (84 terms) & (42 terms) \\ \hline
& ${\beta}_{167}{\beta}_{234},$ & ${\beta}_{127}{\beta}_{345}\alpha _{12},$
& ${\beta}_{167}{\beta}_{234}\alpha _{17}\alpha _{23}$, \\
$A_{YM}(1,2,3,4,5,6,7)$ & ${\beta}_{345}\alpha _{12},$ & ${\beta}_{167}{%
\beta}_{234}\alpha _{17},$ & ${\beta}_{167}{\beta}_{234}\alpha _{17}\alpha
_{34},$ \\
& $\vdots $ & $\vdots $ & $\vdots $ \\
& ${\beta}_{567}\alpha _{34}$ & ${\beta}_{234}{\beta}_{567}\alpha _{23}$ & ${%
\beta}_{167}{\beta}_{345}\alpha _{17}\alpha _{34}$ \\ \hline
\end{tabular}
\label{poles7p}
\end{equation}%
\noindent So, $F^{\{2345\}}(\alpha')$ will be naturally defined in terms of the $N=7$ Mandelstam variables that we have defined in (\ref{Mandelstam7}):
\begin{equation}
 F^{\{2345\}}(\alpha')  = F^{\{2345\}}\left[
\begin{array}{ccccccc}
\scriptstyle   \alpha_{12}, & \scriptstyle\alpha _{23}, & \scriptstyle\alpha _{34}, & %
\scriptstyle\alpha _{56}, & \scriptstyle\alpha _{56}, & \scriptstyle\alpha
_{67}, & \scriptstyle\alpha _{17} \\
\scriptstyle {\beta}_{123}, & \scriptstyle {\beta}_{234}, & \scriptstyle {\beta}_{345}, & %
\scriptstyle {\beta}_{456}, & \scriptstyle {\beta}_{567}, & \scriptstyle {\beta}_{167}, & %
\scriptstyle {\beta}_{127}%
\end{array}%
;\alpha ^{\prime }\right]  \ .
\label{canonic}
\end{equation}%
\noindent The Mandelstam variables for the remaining twenty three $F^{\{ \sigma_7 \}}(\alpha')$ momentum factors are obtained from the ones that appear in eq.(\ref{canonic}), by considering in them exactly the $\sigma_7$ permutation of indices $\{2, 3, 4, 5\}$.\\
\noindent A computational detail that coould be of interest to the reader is that finding the coefficients of the $\alpha'$ expansions, along the same lines presented in subsection \ref{Case of the 6-point} (in four stages, in particular, demanding cyclic symmetry after unitarity has been demanded in one physical pole), becomes a heavy task if compared with solving the coefficients from purely demanding unitarity in all poles. These two methods happen to agree at low orders in $\alpha'$ (but not necessarily at high orders\footnote{We have verified this computationally.}). The advantage of the second method lies in the fact that each of its equations contain a few momentum factors (containing, therefore, a low number of unknowns) while in the first method each of the equations contains many more momentum factors. \\
\noindent In Appendix \ref{Unitarity-N7} we have written the momentum factor relations that arise when unitarity of the amplitudes is demanded with respect to their $\alpha_{12}$ and their $\beta_{123}$ poles. Our complete result is that, when any ${\alpha}_{ij} \rightarrow 0$ there arise six relations and when any  ${\beta}_{ijk} \rightarrow 0$ there arise only two relations. The first type of relations involves $7$-point with $6$-point momentum factors while the second type involves $7$-point with $4$-point and $5$-point momentum factors (see Appendix \ref{Unitarity-N7} for the detailed relations).\\
\noindent We have found the $\alpha'$ expansions up to ${\alpha'}^4$ order\footnote{As mentioned before, the reason for not going to higher orders is a purely computational one: we have had memory difficulties to deal with $N=7$ ${\alpha'}^5$ calculations on the computers . In spite of this complication, as we mentioned in subsection \ref{Using}, we believe that only using the $4$-point amplitude information our method is, in principle, able to acheive ${\alpha'}^7$ order calculations, for any number of legs.}. In the follwing we have listed these expansions only up to ${\alpha'}^3$ order\footnote{There are specific cases of momentum factors which, as remarked in \cite{Mafra3}, have an $\alpha'$ expansion which begins at ${\alpha'}^4$ order. For these cases we have written the ${\alpha'}^4$ terms.}.
\begin{eqnarray}
F^{\{2345\}}(\alpha ^{\prime }) &=&1-{\alpha ^{\prime }}^{2}\zeta (2)\left( {%
\beta}_{123}\left( {\beta}_{127}-{\beta}_{456}\right) +{\beta}_{456}{\beta}%
_{567}-{\beta}_{127}\alpha _{12}+\alpha _{12}\alpha _{17}-{\beta}%
_{567}\alpha _{56}+\alpha _{56}\alpha _{67}\right)  \notag \\
&&+{\alpha ^{\prime }}^{3}\text{$\zeta (3)\left[ {\beta}_{123}^{2}\left( {%
\beta}_{127}-{\beta}_{456}\right) +{\beta}_{456}^{2}{\beta}_{567}-{\beta}%
_{127}^{2}\alpha _{12}-{\beta}_{127}\alpha _{12}^{2}+\alpha _{12}^{2}\alpha
_{17}+\alpha _{12}\alpha _{17}^{2}\right. $}  \notag \\
&&-2{\beta}_{127}\alpha _{12}\alpha _{23}+{\beta}_{456}\left( {\beta}%
_{567}^{2}+2\alpha _{12}\left( -{\beta}_{234}+\alpha _{23}+\alpha
_{34}\right) \right) -{\beta}_{567}^{2}\alpha _{56}-2{\beta}_{167}\alpha
_{12}\alpha _{56}  \notag \\
&&+2{\beta}_{234}\alpha _{12}\alpha _{56}+2{\beta}_{345}\alpha _{12}\alpha
_{56}-2\alpha _{12}\alpha _{34}\alpha _{56}-2{\beta}_{567}\alpha _{45}\alpha
_{56}-{\beta}_{567}\alpha _{56}^{2}  \notag \\
&&+\left. {\beta}_{123}\left( {\beta}_{127}^{2}-{\beta}_{456}^{2}-2{\beta}%
_{456}\alpha _{34}+2\left( -{\beta}_{345}+\alpha _{34}+\alpha _{45}\right)
\alpha _{56}\right) +\alpha _{56}^{2}\alpha _{67}+\alpha _{56}\alpha
_{67}^{2}\right]   \nonumber \\
&& +{\cal O}({\alpha'}^4)  \ ,
\label{expF2345}
\end{eqnarray}%
\begin{eqnarray}
F^{\{2354\}}(\alpha ^{\prime }) &=&{\alpha ^{\prime }}^{2}\zeta (2)\left( -{%
\beta}_{123}+{\beta}_{467}\right) \alpha _{46}  - \notag \\
&&{\alpha ^{\prime }}^{3}\text{$\zeta (3)$}\alpha _{46}\left[ -{\beta}%
_{123}^{2}+2{\beta}_{456}{\beta}_{467}-{\beta}_{467}^{2}-2{\beta}%
_{235}\alpha _{12}+2\alpha _{12}\alpha _{23}+2\alpha _{12}\alpha _{35} - \right.
\notag \\
&&\left. {\beta}_{467}\alpha _{45}-{\beta}_{467}\alpha _{46}+{\beta}%
_{123}\left( -2{\beta}_{456}+2{\beta}_{467}-2\alpha _{35}+\alpha
_{45}+\alpha _{46}-2\alpha _{67}\right) + \right . \nonumber \\
&& \left. 2{\beta}_{467}\alpha _{67}\right]   +{\cal O}({\alpha'}^4)  \ ,
\label{expF2354}
\end{eqnarray}
\begin{eqnarray}
F^{\{2435\}}(\alpha ^{\prime }) &=&{\alpha ^{\prime }}^{2}\zeta (2)\left( {%
\beta}_{124}-\alpha _{12}\right) \left( {\beta}_{356}-\alpha _{56}\right)
\notag \\
&&+{\alpha ^{\prime }}^{3}\text{$\zeta (3)\left[ {\beta}_{124}^{2}\left( {%
\beta}_{356}-\alpha _{56}\right) +{\beta}_{124}\left( {\beta}_{356}^{2}+{%
\beta}_{356}\left( -2{\beta}_{567}-2\alpha _{12}+\alpha _{34}-2\alpha
_{56}\right) \right. \right. $}  \notag \\
&&\left. -2{\beta}_{127}\left( {\beta}_{356}-\alpha _{56}\right) +\alpha
_{56}\left( 2{\beta}_{567}+2\alpha _{12}-\alpha _{34}+2\alpha _{35}+\alpha
_{56}\right) \right)  \notag \\
&&+\alpha _{12}\left( -{\beta}_{356}^{2}+2{\beta}_{127}\left( {\beta}%
_{356}-\alpha _{56}\right) -\alpha _{56}\left( 2{\beta}_{567}+\alpha
_{12}+2\alpha _{24}-\alpha _{34}+2\alpha _{35}+\alpha _{56}\right) \right)
\notag \\
&&+\alpha _{12}{\beta}_{356}\left( 2{\beta}_{567}+\alpha _{12}+2\alpha
_{24}-\alpha _{34}+2\alpha _{56}\right) +\mathcal{O}(\alpha ^{\prime }{}^{4}) \ ,
\label{expF2435}
\end{eqnarray}%
\begin{eqnarray}
F^{\{2453\}}(\alpha ^{\prime }) &=&{\alpha ^{\prime }}^{2}\zeta (2)\left( {%
\beta}_{346}-\alpha _{34}-\alpha _{46}\right) \left( -{\beta}_{345}-{\beta}%
_{467}+\alpha _{35}+\alpha _{67}\right)  \notag \\
&&-{\alpha ^{\prime }}^{3}\text{$\zeta (3)$}\left[ 2\alpha _{12}\left( {\beta%
}_{346}-\alpha _{34}-\alpha _{46}\right) \left( -{\beta}_{167}+{\beta}_{235}-%
{\beta}_{467}+\alpha _{67}\right) \right.  \notag \\
&&+\left. \left( {\beta}_{346}-\alpha _{34}-\alpha _{46}\right) \left( {\beta%
}_{345}+{\beta}_{467}-\alpha _{35}-\alpha _{67}\right) \left( 2{\beta}%
_{125}-2{\beta}_{127}+{\beta}_{345}+{\beta}_{346}\right. \right]  \notag \\
&&\left. \left. -{\beta}_{467}-2\alpha _{34}-\alpha _{46}-\alpha
_{67}\right) \right] +\mathcal{O}(\alpha ^{\prime }{}^{4}) \ ,
\label{expF2453}
\end{eqnarray}%
\begin{eqnarray}
F^{\{2534\}}(\alpha ^{\prime }) &=&-{\alpha ^{\prime }}^{2}\zeta (2)\left( {%
\beta}_{127}-{\beta}_{345}-{\beta}_{356}+\alpha _{35}\right) \left( {\beta}%
_{124}-{\beta}_{367}+\alpha _{45}\right)  \notag \\
&&+{\alpha ^{\prime }}^{3}\text{$\zeta (3)$}\left( {\beta}_{127}-{\beta}%
_{345}-{\beta}_{356}+\alpha _{35}\right) \left[ {\beta}_{124}^{2}+{\beta}%
_{345}{\beta}_{367}-{\beta}_{356}{\beta}_{367}+{\beta}_{367}^{2}+2{\beta}%
_{245}\alpha _{12}\right.  \notag \\
&&-2\alpha _{12}\alpha _{24}+2{\beta}_{367}\alpha _{36}-{\beta}_{345}\alpha
_{45}+{\beta}_{356}\alpha _{45}-3{\beta}_{367}\alpha _{45}-2\alpha
_{12}\alpha _{45}-2\alpha _{36}\alpha _{45}  \notag \\
&&+2\alpha _{45}^{2}+{\beta}_{127}\left( -{\beta}_{367}+\alpha _{45}\right)
-2{\beta}_{367}\alpha _{67}+2\alpha _{45}\alpha _{67}  \notag \\
&&+\left. {\beta}_{124}\left( {\beta}_{127}-{\beta}_{345}+{\beta}_{356}-2{%
\beta}_{367}-2\alpha _{36}+3\alpha _{45}+2\alpha _{67}\right) \right] +%
\mathcal{O}(\alpha ^{\prime }{}^{4}) \ ,
\label{expF2534}
\end{eqnarray}%
\begin{eqnarray}
F^{\{2543\}}(\alpha ^{\prime }) &=&-{\alpha ^{\prime }}^{2}\zeta (2)\left( {%
\beta}_{125}-\alpha _{12}\right) \alpha _{36}  \notag \\
&&-{\alpha ^{\prime }}^{3}\text{$\zeta (3)$}\alpha _{36}\left[ {\beta}%
_{125}^{2}+{\beta}_{125}\left( -2{\beta}_{127}+{\beta}_{345}-2\alpha
_{12}+\alpha _{36}-2\alpha _{67}\right) \right.  \notag \\
&&\left. +\alpha _{12}\left( 2{\beta}_{127}-{\beta}_{345}+\alpha
_{12}+2\alpha _{25}-\alpha _{36}+2\alpha _{67}\right) \right] +\mathcal{O}%
(\alpha ^{\prime }{}^{4}) \ ,
\label{expF2543}
\end{eqnarray}%
\begin{eqnarray}
F^{\{3245\}}(\alpha ^{\prime }) &=&{\alpha ^{\prime }}^{2}\zeta (2)\left( {%
\beta}_{137}-{\beta}_{456}\right) \alpha _{13}  \notag \\
&&+{\alpha ^{\prime }}^{3}\text{$\zeta (3)$}\alpha _{13}\left[ {\beta}%
_{137}^{2}-2{\beta}_{123}\left( {\beta}_{137}-{\beta}_{456}\right) +{\beta}%
_{456}^{2}-{\beta}_{456}\alpha _{13}+2{\beta}_{456}\alpha _{17}-{\beta}%
_{456}\alpha _{23}\right.  \notag \\
&&\left. +{\beta}_{137}\left( -2{\beta}_{456}+\alpha _{13}-2\alpha
_{17}+\alpha _{23}\right) +2{\beta}_{456}\alpha _{24}+2{\beta}_{245}\alpha
_{56}-2\alpha _{24}\alpha _{56}-2\alpha _{45}\alpha _{56}\right]  \notag \\
&&+\mathcal{O}(\alpha ^{\prime }{}^{4}) \ ,
\label{expF3245}
\end{eqnarray}%
\begin{equation}
F^{\{3254\}}(\alpha ^{\prime })=-2{\alpha ^{\prime }}^{3}\text{$\zeta (3)$}%
\alpha _{13}\alpha _{25}\alpha _{46}+\mathcal{O}(\alpha ^{\prime }{}^{4}) \ ,
\label{expF3254}
\end{equation}
\begin{eqnarray}
F^{\{3425\}}(\alpha ^{\prime }) &=&{\alpha ^{\prime }}^{2}\zeta (2)\left( -{%
\beta}_{137}-{\beta}_{234}+\alpha _{17}+\alpha _{24}\right) \left( {\beta}%
_{134}-\alpha _{13}-\alpha _{34}\right)  \notag \\
&&-{\alpha ^{\prime }}^{3}\text{$\zeta (3)$}\left( {\beta}_{134}-\alpha
_{13}-\alpha _{34}\right) \left[ -{\beta}_{137}^{2}+{\beta}_{234}^{2}+2{\beta%
}_{234}{\beta}_{256}-2{\beta}_{234}{\beta}_{567}-{\beta}_{234}\alpha
_{13}\right.  \notag \\
&&-2{\beta}_{234}\alpha _{17}-2{\beta}_{256}\alpha _{17}+2{\beta}%
_{567}\alpha _{17}+\alpha _{13}\alpha _{17}+\alpha _{17}^{2}+{\beta}%
_{134}\left( {\beta}_{137}+{\beta}_{234}-\alpha _{17}-\alpha _{24}\right)
\notag \\
&&-{\beta}_{234}\alpha _{24}-2{\beta}_{256}\alpha _{24}+2{\beta}_{567}\alpha
_{24}+\alpha _{13}\alpha _{24}+\alpha _{17}\alpha _{24}-2{\beta}_{234}\alpha
_{34}  \notag \\
&&+2\alpha _{17}\alpha _{34}+2\alpha _{24}\alpha _{34}+{\beta}_{137}\left( 2{%
\beta}_{256}-2{\beta}_{567}-\alpha _{13}+\alpha _{24}-2\alpha _{34}-2\alpha
_{56}\right)  \notag \\
&&-2{\beta}_{167}\alpha _{56}+2{\beta}_{245}\alpha _{56}+2\alpha _{17}\alpha
_{56}+\mathcal{O}(\alpha ^{\prime }{}^{4}) \ ,
\label{expF3425}
\end{eqnarray}%
\begin{eqnarray}
F^{\{3452\}}(\alpha ^{\prime }) &=&{\alpha ^{\prime }}^{2}\zeta (2)\left( {%
\beta}_{137}-{\beta}_{245}-{\beta}_{256}+\alpha _{25}\right) \left( -{\beta}%
_{134}+{\beta}_{267}-{\beta}_{345}+\alpha _{34}\right)  \notag \\
&&+{\alpha ^{\prime }}^{3}\text{$\zeta (3)$}\left( {\beta}_{137}-{\beta}%
_{245}-{\beta}_{256}+\alpha _{25}\right) \left( {\beta}_{134}-{\beta}_{267}+{%
\beta}_{345}-\alpha _{34}\right)  \notag \\
&&\times \left( {\beta}_{134}+{\beta}_{137}-{\beta}_{245}+{\beta}_{256}-{%
\beta}_{267}+2{\beta}_{345}-2\alpha _{26}-2\alpha _{34}+2\alpha _{67}\right)
\notag \\
&&+\mathcal{O}(\alpha ^{\prime }{}^{4}) \ ,
\end{eqnarray}%
\begin{eqnarray}
F^{\{3524\}}(\alpha ^{\prime }) &=&{\alpha ^{\prime }}^{2}\zeta (2)\alpha
_{14}\alpha _{36}  \notag \\
&&-{\alpha ^{\prime }}^{3}\text{$\zeta (3)$}\alpha _{14}\alpha _{36}\left( -2%
{\beta}_{145}-{\beta}_{167}-2{\beta}_{236}+\alpha _{14}+2\alpha _{17}+\alpha
_{23}+\alpha _{36}+\alpha _{45}+2\alpha _{67}\right)  \notag \\
&&+\mathcal{O}(\alpha ^{\prime }{}^{4}) \ ,
\label{expF3524}
\end{eqnarray}%
\begin{eqnarray}
F^{\{3542\}}(\alpha ^{\prime }) &=&{\alpha ^{\prime }}^{2}\zeta (2)\left( {%
\beta}_{137}+{\beta}_{167}-{\beta}_{245}-\alpha _{17}\right) \left( {\beta}%
_{135}-\alpha _{13}-\alpha _{35}\right)  \notag \\
&&-{\alpha ^{\prime }}^{3}\text{$\zeta (3)$}\left( {\beta}_{137}+{\beta}%
_{167}-{\beta}_{245}-\alpha _{17}\right) \left( {\beta}_{135}-\alpha
_{13}-\alpha _{35}\right)  \notag \\
&&\times \left( -{\beta}_{135}+{\beta}_{137}-{\beta}_{167}+{\beta}%
_{345}+\alpha _{13}+\alpha _{17}-\alpha _{24}-2\alpha _{26}+\alpha
_{35}-\alpha _{45}+2\alpha _{67}\right)  \notag \\
&&+\mathcal{O}(\alpha ^{\prime }{}^{4}) \ ,
\label{expF3542}
\end{eqnarray}
\begin{eqnarray}
F^{\{4235\}}(\alpha ^{\prime }) &=&{\alpha ^{\prime }}^{2}\zeta (2)\left( {%
\beta}_{147}-{\beta}_{356}-\alpha _{23}\right) \left( -{\beta}_{124}-{\beta}%
_{234}+{\beta}_{567}+\alpha _{24}\right)  \notag \\
&&+{\alpha ^{\prime }}^{3}\text{$\zeta (3)$}\left( {\beta}_{124}+{\beta}%
_{234}-{\beta}_{567}-\alpha _{24}\right) \left[ -{\beta}_{147}^{2}+{\beta}%
_{234}{\beta}_{356}-{\beta}_{356}^{2}-{\beta}_{356}{\beta}_{567}\right.
\notag \\
&&+2{\beta}_{356}\alpha _{14}-2{\beta}_{356}\alpha _{17}+{\beta}_{124}\left(
{\beta}_{147}-{\beta}_{356}-\alpha _{23}\right) +{\beta}_{234}\alpha _{23}-3{%
\beta}_{356}\alpha _{23}  \notag \\
&&-{\beta}_{567}\alpha _{23}+2\alpha _{14}\alpha _{23}-2\alpha _{17}\alpha
_{23}-2\alpha _{23}^{2}-2{\beta}_{235}\alpha _{56}+2\alpha _{23}\alpha
_{56}+2\alpha _{35}\alpha _{56}  \notag \\
&&\left. +{\beta}_{147}\left( -{\beta}_{234}+2{\beta}_{356}+{\beta}%
_{567}-2\alpha _{14}+2\alpha _{17}+3\alpha _{23}\right) \right] +\mathcal{O}%
(\alpha ^{\prime }{}^{4}) \ ,
\label{expF4235}
\end{eqnarray}%
\begin{eqnarray}
F^{\{4253\}}(\alpha ^{\prime }) &=&-\frac{1}{10}{\alpha ^{\prime }}^{4}\zeta
^{2}(2)\left( {\beta}_{157}-{\beta}_{234}-{\beta}_{346}+\alpha _{34}\right)
\left( -{\beta}_{167}+{\beta}_{245}-{\beta}_{367}+\alpha _{67}\right)  \notag
\\
&&\left[ 7\alpha _{15}\alpha _{24}+17\alpha _{24}\left( {\beta}_{167}-{\beta}%
_{234}-{\beta}_{245}+\alpha _{24}\right) -3\alpha _{15}\left( {\beta}%
_{346}-\alpha _{34}-\alpha _{36}\right) \right.  \notag \\
&&\left. -10\left( {\beta}_{167}-{\beta}_{234}-{\beta}_{245}+\alpha
_{24}\right) \left( {\beta}_{346}-\alpha _{34}-\alpha _{36}\right) \right] +%
\mathcal{O}(\alpha ^{\prime }{}^{5}) \ ,
\label{expF4253}
\end{eqnarray}%
\begin{eqnarray}
F^{\{4325\}}(\alpha ^{\prime }) &=&-{\alpha ^{\prime }}^{2}\zeta (2)\alpha
_{14}\left( {\beta}_{256}-\alpha _{56}\right)  \notag \\
&&-{\alpha ^{\prime }}^{3}\text{$\zeta (3)$}\alpha _{14}\left[ {\beta}%
_{256}^{2}+{\beta}_{234}\left( {\beta}_{256}-\alpha _{56}\right) +\alpha
_{56}\left( 2{\beta}_{567}-\alpha _{14}+2\alpha _{17}+2\alpha _{25}+\alpha
_{56}\right) \right.  \notag \\
&&\left. +{\beta}_{256}\left( -2{\beta}_{567}+\alpha _{14}-2\left( \alpha
_{17}+\alpha _{56}\right) \right) \right] +\mathcal{O}(\alpha ^{\prime
}{}^{4}) \ ,
\label{expF4325}
\end{eqnarray}%
\begin{eqnarray}
F^{\{4352\}}(\alpha ^{\prime }) &=&\frac{1}{10}{\alpha ^{\prime }}^{4}\zeta
^{2}(2)\left( -{\beta}_{147}-{\beta}_{167}+{\beta}_{235}+\alpha _{17}\right)
\left( -{\beta}_{167}+{\beta}_{245}-{\beta}_{367}+\alpha _{67}\right)  \notag
\\
&&\left[ 10\alpha _{24}\left( -{\beta}_{124}-{\beta}_{245}+{\beta}%
_{367}+\alpha _{24}\right) +27\alpha _{24}\alpha _{35}\right.  \notag \\
&&+3\left( -{\beta}_{124}-{\beta}_{245}+{\beta}_{367}+\alpha _{24}\right)
\left( {\beta}_{147}-{\beta}_{235}-{\beta}_{356}+\alpha _{35}\right)  \notag
\\
&&\left. +10\alpha _{35}\left( {\beta}_{147}-{\beta}_{235}-{\beta}%
_{356}+\alpha _{35}\right) \right] +\mathcal{O}(\alpha ^{\prime }{}^{5}) \ ,
\label{expF4352}
\end{eqnarray}%
\begin{eqnarray}
F^{\{4523\}}(\alpha ^{\prime }) &=&{\alpha ^{\prime }}^{2}\zeta (2)\left( {%
\beta}_{137}+{\beta}_{167}-{\beta}_{245}-\alpha _{17}\right) \left( {\beta}%
_{167}-{\beta}_{235}+{\beta}_{467}-\alpha _{67}\right)  \notag \\
&&-{\alpha ^{\prime }}^{3}\text{$\zeta (3)$}\left( {\beta}_{137}+{\beta}%
_{167}-{\beta}_{245}-\alpha _{17}\right) \left( {\beta}_{167}-{\beta}_{235}+{%
\beta}_{467}-\alpha _{67}\right)  \notag \\
&&\times \left[ -2{\beta}_{135}+{\beta}_{137}+{\beta}_{167}-2{\beta}_{235}-2{%
\beta}_{245}-2{\beta}_{246}+{\beta}_{467}\right.  \notag \\
&&\left. +2\alpha _{13}+\alpha _{17}+3\alpha _{24}-\alpha _{25}+3\alpha
_{35}+2\alpha _{46}+\alpha _{67}\right] +\mathcal{O}(\alpha ^{\prime }{}^{4}) \ ,
\label{expF4523}
\end{eqnarray}%
\begin{eqnarray}
F^{\{4532\}}(\alpha ^{\prime }) &=&-{\alpha ^{\prime }}^{2}\zeta (2)\left( {%
\beta}_{236}-\alpha _{23}-\alpha _{26}\right) \left( {\beta}_{145}-\alpha
_{14}-\alpha _{45}\right)  \notag \\
&&-{\alpha ^{\prime }}^{3}\text{$\zeta (3)$}\left( {\beta}_{236}-\alpha
_{23}-\alpha _{26}\right) \left( {\beta}_{145}-\alpha _{14}-\alpha
_{45}\right)  \notag \\
&&\times \left( {\beta}_{145}+{\beta}_{167}+{\beta}_{236}-\alpha
_{14}-2\alpha _{17}+\alpha _{26}-\alpha _{45}-2\alpha _{67}\right) +\mathcal{%
O}(\alpha ^{\prime }{}^{4}) \ ,
\label{expF4532}
\end{eqnarray}%
\begin{eqnarray}
F^{\{5234\}}(\alpha ^{\prime }) &=&{\alpha ^{\prime }}^{2}\zeta (2)\left( -{%
\beta}_{125}-{\beta}_{235}+{\beta}_{467}+\alpha _{25}\right) \left( {\beta}%
_{157}-{\beta}_{234}-{\beta}_{346}+\alpha _{34}\right)  \notag \\
&&+{\alpha ^{\prime }}^{3}\text{$\zeta (3)$}\left( {\beta}_{125}+{\beta}%
_{235}-{\beta}_{467}-\alpha _{25}\right) \left( {\beta}_{157}-{\beta}_{234}-{%
\beta}_{346}+\alpha _{34}\right)  \notag \\
&&\times \left( {\beta}_{125}-{\beta}_{157}+2{\beta}_{234}-{\beta}_{235}+{%
\beta}_{346}+{\beta}_{467}-2\alpha _{15}+2\alpha _{17}-2\alpha _{34}\right)
\notag \\
&&+\mathcal{O}(\alpha ^{\prime }{}^{4}) \ ,
\label{expF5234}
\end{eqnarray}%
\begin{eqnarray}
F^{\{5243\}}(\alpha ^{\prime }) &=&\frac{1}{10}{\alpha ^{\prime }}^{4}\zeta
^{2}(2)\left( -{\beta}_{147}-{\beta}_{167}+{\beta}_{235}+\alpha _{17}\right)
\left( -{\beta}_{134}+{\beta}_{267}-{\beta}_{345}+\alpha _{34}\right)  \notag
\\
&&\times \left[ 3\alpha _{26}\left( {\beta}_{134}-\alpha _{14}-\alpha
_{34}\right) -7\alpha _{26}\alpha _{35}\right.  \notag \\
&&+\left. 10\left( {\beta}_{134}-\alpha _{14}-\alpha _{34}\right) \left( {%
\beta}_{167}-{\beta}_{235}-{\beta}_{345}+\alpha _{35}\right) \right. \nonumber \\
 && \left. -17\alpha_{35}\left( {\beta}_{167}-{\beta}_{235}-{\beta}_{345}+\alpha _{35}\right) %
\right]  +\mathcal{O}(\alpha ^{\prime }{}^{5}) \ ,
\label{expF5243}
\end{eqnarray}%
\begin{eqnarray}
F^{\{5324\}}(\alpha ^{\prime }) &=&{\alpha ^{\prime }}^{2}\zeta (2)\left( {%
\beta}_{246}-\alpha _{24}-\alpha _{46}\right) \left( {\beta}_{167}-{\beta}%
_{235}+{\beta}_{467}-\alpha _{67}\right)  \notag \\
&&+{\alpha ^{\prime }}^{3}\text{$\zeta (3)$}\left( {\beta}_{167}-{\beta}%
_{234}+{\beta}_{246}-{\beta}_{467}+2\alpha _{15}-2\alpha _{17}+\alpha
_{23}-\alpha _{24}+\alpha _{35}-\alpha _{46}-\alpha _{67}\right)  \notag \\
&&\times \left( {\beta}_{246}-\alpha _{24}-\alpha _{46}\right) \left( {\beta}%
_{167}-{\beta}_{235}+{\beta}_{467}-\alpha _{67}\right) +\mathcal{O}(\alpha
^{\prime }{}^{4}) \ ,
\label{expF5324}
\end{eqnarray}%
\begin{eqnarray}
F^{\{5342\}}(\alpha ^{\prime }) &=&\frac{1}{10}{\alpha ^{\prime }}^{4}\text{$%
\zeta ^{2}(2)$}\alpha _{15}\alpha _{26}\left[ -3{\beta}_{246}\alpha
_{15}+10\alpha _{15}\alpha _{24}+{\beta}_{135}\left( 3{\beta}_{246}-10\alpha
_{24}-3\alpha _{26}\right) \right.  \notag \\
&&\left. +3\alpha _{15}\alpha _{26}-10{\beta}_{246}\alpha _{35}+27\alpha
_{24}\alpha _{35}+10\alpha _{26}\alpha _{35}\right] +\mathcal{O}(\alpha
^{\prime }{}^{5}) \ ,
\label{expF5342}
\end{eqnarray}%
\begin{eqnarray}
F^{\{5423\}}(\alpha ^{\prime }) &=&-{\alpha ^{\prime }}^{2}\zeta (2)\left( {%
\beta}_{236}-\alpha _{23}-\alpha _{36}\right) \left( {\beta}_{145}-\alpha
_{15}-\alpha _{45}\right)  \notag \\
&&-{\alpha ^{\prime }}^{3}\text{$\zeta (3)$}\left( {\beta}_{236}-\alpha
_{23}-\alpha _{36}\right) \left( {\beta}_{145}-\alpha _{15}-\alpha
_{45}\right)  \notag \\
&&\times \left( {\beta}_{145}+{\beta}_{167}+{\beta}_{236}+\alpha
_{15}-2\alpha _{17}-\alpha _{23}-\alpha _{36}-2\alpha _{67}\right) +\mathcal{%
O}(\alpha ^{\prime }{}^{4}) \ ,
\label{expF5423}
\end{eqnarray}%
\begin{eqnarray}
F^{\{5432\}}(\alpha ^{\prime }) &=&{\alpha ^{\prime }}^{2}\zeta (2)\alpha
_{15}\alpha _{26} +{\alpha ^{\prime }}^{3}\text{$\zeta (3)$}\alpha _{15}\alpha _{26}\left( {%
\beta}_{167}+\alpha _{15}-2\alpha _{17}+\alpha _{26}-2\alpha _{67}\right) +%
\mathcal{O}(\alpha ^{\prime }{}^{4}) \ . \hspace{0.8cm}
\label{expF5432}
\end{eqnarray}
\noindent The corresponding expressions of the ${\alpha'}^4$ terms for the remaining terms are included in the text files that we have submitted, attached to this work, to the hep-th arXiv preprint basis.\\
\noindent We have confirmed that our results are in perfect agreement with the ones found previously in \cite{Broedel3}.

\vspace{0.5cm}

\section{Summary and conclusions}

\label{Summary}

\noindent We have successfully derived a closed formula for the tree level $N$-point amplitude of open massless superstrings (for $3 \leq N \leq 7$). This is the same formula first found by Mafra, Schlotterer and Stieberger in \cite{Mafra1}, using the Pure Spinor formalism \cite{Berkovits1}.\\
\noindent Our approach has consisted in working only within the RNS formalism, so spacetime supersymmetry has been present throughout our approach, but in a {\it non manifestly} manner. First, it has been present implicitly in the computation of the $N$-point gauge boson amplitude in a closed form (see eq.(\ref{Ab})) because, in order to arrive to it, the condition of absence of $(\zeta \cdot k)^N$ terms\footnote{This is the main observation of our revisited S-matrix method \cite{Barreiro0}.} has been used. And second, it has been used to find {\it uniquely} the amplitudes involving fermions, once the $N$-point formula for gauge bosons, eq.(\ref{Ab}), has been found.  \\
\noindent We believe that a deeper understanding of our procedure can eventually arrive to the MSS formula in eq.(\ref{MSS}), for arbitrary $N$.\\
\noindent The kinematic analysis that we have required to arrive to MSS formula, naturally leads us to a space of $N$-point gauge boson subamplitudes which is $(N-3)!$-dimensional (at least for $3 \leq N \leq 7$)\footnote{In the main body of this work we have referred to this space by ${\cal V}_N$. See section \ref{Kinematical} for more details.}. At this point is where the basis of Yang-Mills subamplitudes first proposed by Bern, Carrasco and Johansson \cite{Bern1}, plays an important role in our procedure. Once this basis has been identified as a basis for ${\cal V}_N$, then the MSS formula and the explicit BCJ relations themselves become linear algebra problems in which the components of a certain vector, with respect to a given basis, are desired to be found. We have done these calculations in section \ref{Closed} and Appendix \ref{BCJ}, respectively. \\
\noindent Following the same spirit of the revisited S-matrix approach \cite{Barreiro0}, we have found $\alpha'$ correction terms to the open superstring $N$-point amplitudes (where $N=5,6,7$) by only using the well known $4$-point amplitude $\alpha'$ expansion (see eqs.(\ref{formula1}) and (\ref{expansionGamma})) and demanding cyclic symmetry and tree level unitarity for the scattering amplitudes. We have done these calculations to at most ${\alpha'}^6$ order. It is quite remarkable that we have not needed to compute, for the calculations that we have proposed to do in this work, any coefficient as a numerical series or any integral involving polylogarithms.\\
\noindent We expect that, within our approach, only for ${\alpha'}^8$ order onwards\footnote{With the remarkable exception of the ${\alpha'}^9$ order terms, as mentioned in subsection \ref{Using}.} the $\alpha'$ expansion of the $5$-point amplitude will be required to obtain some of the coefficients of the series\footnote{These coefficients would be the ones which have dependence in the non trivial MZV's that we referred to in subsection \ref{Using}}.\\
\noindent In the following table we have summarized the existing parallel of the revisited S-matrix method in the determination of the OSLEEL and the corresponding scattering amplitude $\alpha'$ calculations.
\begin{eqnarray}
\begin{tabular}{|c|c|c|}
\hline
Step of the revisited      &  Open superstring                               &  Open superstring  \\
S-matrix method            &  low energy effective lagrangian        & scattering amplitude \\
at ${\alpha'}^p$ order  &      at ${\alpha'}^p$ order                 &    at ${\alpha'}^p$ order\\
\hline
\hline
$Step \  I:$ &  &  \\
Requirement of  absence                    &                                                       &   Finds that the $N$-point  \\
of $(\zeta \cdot k)^N$ terms  in the   &   Reduces the general basis         &  amplitude can be written \\
$N$-point amplitude                            &   to a constrained basis of terms                             &   in terms of the ${\cal B}_N$ basis \\
$(N=4, \ldots, p+2)$                          &                                                      &   (see eq.(\ref{basis}))  \\
\hline
$Step \  II:$                                        &                   &  \\
 Use of $n$-point amplitudes              &                                                   &    \\
information   ($n \ll p+2$)                 &  Determines $all$ the coefficients     &   Determines $all$ the  \\
(presumably, only  $n=4$                  & of the constrained basis            &   momentum factors \\
 and $n=5$)                                      &                                                    & \\
\hline
\end{tabular}
\label{table3}
\end{eqnarray}

\vspace{0.5cm}

\noindent In a forthcoming work we will use the revisited S-matrix method to compute the ${\alpha'}^5$ order terms of the OSLEEL (in analogy to the calculations that we did in \cite{Barreiro0}) and also to compute those terms in the $N=7$ scattering amplitude \cite{Barreiro3}.

\section*{Acknowledgements}

We would like to thank N. Berkovits, R. Boels, E. Hatefi and S. Stieberger for useful e-mail correspondence at different stages in the development of this work. This work has been partially supported by the brazilian agency FAPEMIG (research projects PPM-00617-11 and APQ-00140-12).

\appendix

\section{Conventions}

\label{conventions}

\noindent We use the following convention for the Minkowski metric:
\begin{eqnarray}
\label{metric} \eta_{\mu \nu} = \mbox{diag}(-, +, \ldots , +) \ .
\end{eqnarray}
\noindent Gauge fields are matrices in the adjoint representation of the Lie group internal space, so that
$A_{\mu} = A^{\mu}_{\ a} \lambda^a$, where
\begin{eqnarray}
(\lambda^a)^{bc} = -i f^{abc} \ .
\label{adjointm}
\end{eqnarray}
In (\ref{adjointm}) the $f^{abc}$'s are the Lie group structure constants and the $\lambda^a$'s obey the normalization condition
\begin{eqnarray}
\mbox{tr}(\lambda^a \lambda^b) = \delta^{ab} \ .\label{group1}
\end{eqnarray}
The field strength is defined by
\begin{eqnarray}
\label{group2} F_{\mu \nu} & = & \partial_{\mu} A_{\nu} -
\partial_{\nu} A_{\mu}
- i [ A_{\mu}, A_{\nu}] \ .
\end{eqnarray}

\section{Known relations for Yang-Mills subamplitudes before BCJ}

\label{Known relations}

\noindent Before the BCJ relations were found, based on the conjucture between color and kinematics \cite{Bern1}, other relations were very well  known for the Yang-Mills (tree level) subamplitudes \cite{Mangano1}:
\begin{enumerate}
\item On-shell gauge invariance (item {\it 3} in eq. (\ref{requirements})).
\item Cyclicity:
\begin{eqnarray}
\label{cyclic}
A(1,2, \ldots, N-1, N) = A(2,3, \ldots, N, 1) = \ldots= A(N,1, \ldots, N-2, N-1) \ .
\end{eqnarray}
\item Reflection:
\begin{eqnarray}
\label{reflection}
A(1,2, \ldots, N-1, N) = (-1)^N A(N-1,N-2, \ldots, 1, N) \ .
\end{eqnarray}
\item Dual Ward identity\footnote{This identity is also known as `subcylic property' or `photon decoupling identity'.}:
\begin{eqnarray}
\label{dual}
A(1,2,3, \ldots, N) + A(2,1,3, \ldots, N) + A(2,3,1,\ldots, N) + \dots + A(2,3, \ldots, 1, N) = 0 \ . \ \
\end{eqnarray}
\item Kleiss-Kuijf relations \cite{Kleiss1}\footnote{See section II of ref.\cite{Bern1}, for example, for the details about the intrinsic notation used on these relations.}:
\begin{eqnarray}
\label{Kleiss}
A(1,\{\alpha\},N,\{\beta\}) = (-1)^{n_{\beta}} \sum_{\{\sigma\}_i  \varepsilon \ {\cal OP(\{\alpha\},\{\beta^T\})}} A(1,\{\sigma\}_i,N) \ .
\end{eqnarray}
\end{enumerate}
\noindent It is known that the first three of these relations are obeyed, also, by gauge boson subamplitudes in Open String Theory \cite{Schwarz1} and that there is a String Theory version of the Kleiss-Kuijf relations in (\ref{Kleiss}) \cite{Bjerrum1, Stieberger3}.\\
The Kleiss-Kuijf relations in (\ref{Kleiss}) are extremely important since using them it was possible to find an $(N-2)!$ dimensional set of `independent' Yang-Mills subamplitudes from which all the remaining ones could be obtained \cite{Kleiss1}\footnote{After the discovery of the BCJ relations it became understood that these $(N-2)!$ were not really all `independent': only a a subset of $(N-3)!$ of them is.}.

\section{Calculations that lead to one 4-point kinematical expression}

\label{Calculations}

In this appendix we give the details of the results mentioned in subsection \ref{N4}, namely, we give the details of how starting with the general expression in (\ref{4point}) for $T(1,2,3,4)$ we arrive to (\ref{A1234-2}), after demanding on-shell gauge invariance.\\
On-shell gauge invariance of $T(1,2,3,4)$ means that it becomes zero (after demanding momentum conservation, the physical state condition (\ref{transversality}) and the mass-shell condition (\ref{mass-shell})) whenever any of the polarizations vectors $\zeta_i$ becomes $k_i$ ($i=1,2,3,4$). Let us see this:  \\

\noindent {\bf 1.} Case of $\zeta_1 \rightarrow k_1$:\\

\noindent In this case, using that
$k_1 \cdot k_2 =   s/2$ and $k_1 \cdot k_3 = - (s+t)/2$, where $s$ and $t$ are the Mandelstam variables defined in eq.(\ref{Mandelstam4}), we can arrive to
\begin{multline}
T(1,2,3,4)\Big|_{\zeta_1 = k_1}   =  \\
\begin{split}
& \left[ \frac{  }{  } (\lambda_1 + \lambda_5 + \lambda_9) (\zeta_2 \cdot k_1) (\zeta_3 \cdot k_1) (\zeta_4 \cdot k_1) +
(\lambda_2 + \lambda_6) (\zeta_2 \cdot k_1) (\zeta_3 \cdot k_1) (\zeta_4 \cdot k_2) + \right. \\
&  \hphantom{ \left[ \frac{}{} \right. }   (\lambda_3 + \lambda_{10}) (\zeta_2 \cdot k_1) (\zeta_3 \cdot k_2) (\zeta_4 \cdot k_1) +
\lambda_4 (\zeta_2 \cdot k_1) (\zeta_3 \cdot k_2) (\zeta_4 \cdot k_2) +   \\
&  \hphantom{  \   }   (\lambda_7 + \lambda_{11}) (\zeta_2 \cdot k_3) (\zeta_3 \cdot k_1) (\zeta_4 \cdot k_1) +
\lambda_8 (\zeta_2 \cdot k_3) (\zeta_3 \cdot k_1) (\zeta_4 \cdot k_2) +   \lambda_{12} (\zeta_2 \cdot k_3) (\zeta_3 \cdot k_2) (\zeta_4 \cdot k_1)  \left.     \frac{}{} \right] +   \\
&   \Bigl[ \  \Bigl\{- \frac{1}{2} (s + t) \lambda_{15} + \frac{1}{2} \lambda_{13} s + \rho_3 \Bigr \} (\zeta_2 \cdot \zeta_3) (\zeta_4 \cdot k_1) +   \Bigl\{ -\frac{1}{2} (s + t) \lambda_{16} + \frac{1}{2} \lambda_{14} s \Bigr \} (\zeta_2 \cdot \zeta_3) (\zeta_4 \cdot k_2) + \\
&   \hphantom{  \Bigl[ \  }    \Bigl\{ - \frac{1}{2} (s + t) \lambda_{19} + \frac{1}{2} \lambda_{17} s + \rho_2 \Bigr \} (\zeta_2 \cdot \zeta_4) (\zeta_3 \cdot k_1) +
 \Bigl\{ -\frac{1}{2} (s + t) \lambda_{20} + \frac{1}{2} \lambda_{18} s \Bigr \} (\zeta_2 \cdot \zeta_4) (\zeta_3 \cdot k_2) + \\
&   \hphantom{  \Bigl[ \  }    \Bigl\{ -\frac{1}{2} (s + t) \lambda_{23} + \frac{1}{2} \lambda_{21} s + \rho_1 \Bigr \} (\zeta_3 \cdot \zeta_4) (\zeta_2 \cdot k_1)   +   \Bigl\{ - \frac{1}{2} (s + t) \lambda_{24} + \frac{1}{2} \lambda_{22} s \Bigr \} (\zeta_3 \cdot \zeta_4) (\zeta_2 \cdot k_3)   \left.   \frac{}{}   \right] \ .
\end{split}
\label{4pointzeta1}
\end{multline}
In (\ref{4pointzeta1}) the fisrt square bracket contains seven $(\zeta \cdot k)^3$ terms while the second square bracket contains six $(\zeta \cdot \zeta)^1 (\zeta \cdot k)^1$ ones. Since the complete expression in (\ref{4pointzeta1}) should be zero and all its kinematic terms are linearly independent then the coefficient of each of them should be zero. This leads precisely to the linear system of equations of eq. (\ref{4point1}) in the main text.\\

\noindent {\bf 2.} Case of $\zeta_2 \rightarrow k_2$:\\

\noindent In this case, using that
$k_1 \cdot k_2 =   s/2$ and $k_2 \cdot k_3 =   t/2$, we can arrive to
\begin{multline}
T(1,2,3,4)\Big|_{\zeta_2 = k_2}   =  \\
\begin{split}
& \left[ \frac{  }{  } \lambda_1  (\zeta_1 \cdot k_2) (\zeta_3 \cdot k_1) (\zeta_4 \cdot k_1) +
(\lambda_2 + \lambda_{17}) (\zeta_1 \cdot k_2) (\zeta_3 \cdot k_1) (\zeta_4 \cdot k_2) + \right. \\
&  \hphantom{ \left[ \frac{}{} \right. }    (\lambda_3 + \lambda_{13}) (\zeta_1 \cdot k_2) (\zeta_3 \cdot k_2) (\zeta_4 \cdot k_1) +
   (\lambda_4 + \lambda_{14}+  \lambda_{18} ) (\zeta_1 \cdot k_2) (\zeta_3 \cdot k_2) (\zeta_4 \cdot k_2)     +   \\
&     \lambda_{19} (\zeta_1 \cdot k_3) (\zeta_3 \cdot k_1) (\zeta_4 \cdot k_2) +
\lambda_{15} (\zeta_1 \cdot k_3) (\zeta_3 \cdot k_2) (\zeta_4 \cdot k_1) + (\lambda_{16}+  \lambda_{20}) (\zeta_1 \cdot k_3) (\zeta_3 \cdot k_2) (\zeta_4 \cdot k_2)  \left.   \frac{}{}   \right]    \\
&   + \Bigl[ \   \frac{1}{2}  \Bigl\{ \lambda_{5}s +  \lambda_{7} t  \Bigr \} (\zeta_1 \cdot \zeta_3) (\zeta_4 \cdot k_1) +   \Bigl\{  \frac{1}{2} (\lambda_6 s + \lambda_8 t)  + \rho_2  \Bigr \} (\zeta_1 \cdot \zeta_3) (\zeta_4 \cdot k_2) + \\
&   \hphantom{  + \Bigl[ \ \ }  \frac{1}{2}  \Bigl\{ \lambda_{9}s + \lambda_{11} t  \Bigr \} (\zeta_1 \cdot \zeta_4) (\zeta_3 \cdot k_1) +   \Bigl\{  \frac{1}{2} (\lambda_{10} s + \lambda_{12} t)  + \rho_3  \Bigr \} (\zeta_1 \cdot \zeta_4) (\zeta_3 \cdot k_2) +   \\
&   \hphantom{  + \Bigl[  }  \Bigl\{  \frac{1}{2} (\lambda_{21} s + \lambda_{22} t)  + \rho_1  \Bigr \} (\zeta_3 \cdot \zeta_4) (\zeta_1 \cdot k_2)   + \frac{1}{2}  \Bigl\{ \lambda_{23}s + \lambda_{24} t  \Bigr \} (\zeta_3 \cdot \zeta_4) (\zeta_1 \cdot k_3)    \left.   \frac{}{}   \right] \ .
\end{split}
\label{4pointzeta2}
\end{multline}
In a similar way to what was argued in the previous case, the expression in (\ref{4pointzeta2}) should be zero and this leads to the following set of thirteen equations (which happens to be linearly independent):
\begin{eqnarray}
\lambda_1 = 0 \ , \ \lambda_{15}=0 \ , \ \lambda_{19}=0 \ , \
\lambda_2 + \lambda_{17}=0 \ ,\ \lambda_{3} +\lambda_{13}=0 \ , \ \lambda_{4}  +\lambda_{14}+\lambda_{18}=0 \ ,\ \lambda_{16} +\lambda_{20}=0 \ , \nonumber \\
\ 2 \rho_{1} +\lambda_{21}s+\lambda_{22}t=0 \ , \
2 \rho_{2} +\lambda_{6}s+\lambda_{8}t=0 \ , \ 2 \rho_{3} +\lambda_{10}s+\lambda_{12}t=0 \ , \ \nonumber \\
 \lambda_{5} s + \lambda_{7}t =0 \ , \ \lambda_{9}s + \lambda_{11} t =0 \ ,  \ \lambda_{23}s + \lambda_{24} t =0 \ . \hspace{1cm}
\label{4point4}
\end{eqnarray}

\noindent {\bf 3.} Case of $\zeta_3 \rightarrow k_3$:\\

\noindent In this case, besides using the expressions for products of momenta in terms of Mandelstam variables ($k_1 \cdot k_3 =  - (s+t)/2$ and $k_2 \cdot k_3 =   t/2$), we need to use the condition
\begin{eqnarray}
(\zeta_4 \cdot k_3) & = & - (\zeta_4 \cdot k_1) - (\zeta_4 \cdot k_2) \ ,
\label{zeta4k3a}
\end{eqnarray}
which comes from the gauge (or the physical state) condition in (\ref{transversality}) and momentum conservation.
Doing these substitutions, after $\zeta_3 \rightarrow k_3$, we can arrive to
\begin{multline}
T(1,2,3,4)\Big|_{\zeta_3 = k_3}   =  \\
\begin{split}
& \left[ \frac{}{}  -\lambda_{21}  (\zeta_1 \cdot k_2) (\zeta_2 \cdot k_1) (\zeta_4 \cdot k_1)  -\lambda_{21}  (\zeta_1 \cdot k_2) (\zeta_2 \cdot k_1) (\zeta_4 \cdot k_2) + \right. \\
&  \hphantom{ \left[ \frac{}{} \right. }   (\lambda_{13} - \lambda_{22}) (\zeta_1 \cdot k_2) (\zeta_2 \cdot k_3) (\zeta_4 \cdot k_1) +    (\lambda_{14} - \lambda_{22}) (\zeta_1 \cdot k_2) (\zeta_2 \cdot k_3) (\zeta_4 \cdot k_2) +   \\
&  \hphantom{ \left(  \ \right.  }   (\lambda_5 - \lambda_{23}) (\zeta_1 \cdot k_3) (\zeta_2 \cdot k_1) (\zeta_4 \cdot k_1) +    (\lambda_6 - \lambda_{23}) (\zeta_1 \cdot k_3) (\zeta_2 \cdot k_1) (\zeta_4 \cdot k_2) + \\
&  \hphantom{ \left[ \frac{}{} \right. }  (\lambda_7 +\lambda_{15}-\lambda_{24}) (\zeta_1 \cdot k_3) (\zeta_2 \cdot k_3) (\zeta_4 \cdot k_1) +   (\lambda_{8} + \lambda_{16} - \lambda_{24}) (\zeta_1 \cdot k_3) (\zeta_2 \cdot k_3) (\zeta_4 \cdot k_2)  \left.   \frac{}{}   \right] +   \\
&   \Bigl[ \  \Bigl\{ -\frac{1}{2} (s + t) \lambda_{1} + \frac{1}{2} \lambda_{3} t - \rho_1 \Bigr \} (\zeta_1 \cdot \zeta_2) (\zeta_4 \cdot k_1) +   \Bigl\{ - \frac{1}{2} (s + t) \lambda_{2} + \frac{1}{2} \lambda_{4} t - \rho_1 \Bigr \} (\zeta_1 \cdot \zeta_2) (\zeta_4 \cdot k_2) + \\
&   \hphantom{  \Bigl[ \  }    \Bigl\{ - \frac{1}{2} (s + t) \lambda_{9} + \frac{1}{2} \lambda_{10} t  \Bigr \} (\zeta_1 \cdot \zeta_4) (\zeta_2 \cdot k_1) +
 \Bigl\{ - \frac{1}{2} (s + t) \lambda_{11} + \frac{1}{2} \lambda_{12} t + \rho_3 \Bigr \} (\zeta_1 \cdot \zeta_4) (\zeta_2 \cdot k_3) + \\
&   \hphantom{  \Bigl[ \  }    \Bigl\{ - \frac{1}{2} (s + t) \lambda_{17} + \frac{1}{2} \lambda_{18} t  \Bigr \} (\zeta_2 \cdot \zeta_4) (\zeta_1 \cdot k_2)   +   \Bigl\{ - \frac{1}{2} (s + t) \lambda_{19} + \frac{1}{2} \lambda_{20} t + \rho_2 \Bigr \} (\zeta_2 \cdot \zeta_4) (\zeta_1 \cdot k_3)   \left.   \frac{}{}   \right] \ .
\end{split}
\label{4pointzeta3}
\end{multline}
This time demanding the previous expression to vanish leads to fourteen equations, but clearly, two of them are identical ($-\lambda_{21}=0$), so we just have a  linear system of thirteen equations (which, once more, happens to be linearly independent) and which is equivalent to the following one:
\begin{eqnarray}
\lambda_{21} = 0 \ , \ \lambda_{23}-\lambda_6=0 \ , \ \lambda_{22}-\lambda_{14}=0 \ , \
\lambda_{24} - \lambda_{8} - \lambda_{16}=0 \ ,\ \lambda_7 - \lambda_8 + \lambda_{15} - \lambda_{16} =0 \ , \nonumber \\
\lambda_{13} -\lambda_{14}=0 \ , \ \lambda_{5}  - \lambda_{6}=0 \ , \ 2 \rho_{1} +\lambda_{1}(s+t) - \lambda_{3}t=0 \ , \
2 \rho_{2} - \lambda_{19}(s+t) + \lambda_{20}t=0 \ , \nonumber \\
2 \rho_{3} - \lambda_{11} (s+t) + \lambda_{12}t=0 \ , \ \lambda_4 t - \lambda_3 t + \lambda_1 (s+t) - \lambda_2 (s+t)=0  \ ,
 \lambda_{10}t - \lambda_{9} (s+t) =0 \ , \nonumber \\
 \lambda_{18} t - \lambda_{17}(s+t) =0 \ .  \hspace{0.2cm}
\label{4point5}
\end{eqnarray}

\noindent {\bf 4.} Case of $\zeta_4 \rightarrow k_4$:\\

\noindent Similarly to the previous case, besides using the expressions for products of momenta in terms of Mandelstam variables ($k_1 \cdot k_4 =  t/2$ and $k_2 \cdot k_4 =  - (s+t)/2$), we need to use the conditions
\begin{eqnarray}
(\zeta_1 \cdot k_4)  =  - (\zeta_1 \cdot k_2) - (\zeta_1 \cdot k_3) \ , \
(\zeta_2 \cdot k_4)  =  - (\zeta_2 \cdot k_1) - (\zeta_2 \cdot k_3) \ , \
(\zeta_3 \cdot k_4)  =  - (\zeta_3 \cdot k_1) - (\zeta_3 \cdot k_2) \ , \nonumber \\
\label{zetak4}
\end{eqnarray}
which come from the gauge (or the physical state) condition in (\ref{transversality}) and momentum conservation.
Doing these substitutions, after $\zeta_4 \rightarrow k_4$, we can arrive to
\begin{multline}
T(1,2,3,4)\Big|_{\zeta_4 = k_4}   =  \\
\begin{split}
& - \left[ \frac{  }{  } (\lambda_9 + \lambda_{17} + \lambda_{21}) (\zeta_1 \cdot k_2) (\zeta_2 \cdot k_1) (\zeta_3 \cdot k_1) +
(\lambda_{10} + \lambda_{18} +\lambda_{21}) (\zeta_1 \cdot k_2) (\zeta_2 \cdot k_1) (\zeta_3 \cdot k_2) + \right. \\
&  \hphantom{ \left[ \frac{}{} \right. - }   (\lambda_{11} + \lambda_{17}+\lambda_{22}) (\zeta_1 \cdot k_2) (\zeta_2 \cdot k_3) (\zeta_3 \cdot k_1) +
(\lambda_{12}+\lambda_{18}+\lambda_{22}) (\zeta_1 \cdot k_2) (\zeta_2 \cdot k_3) (\zeta_3 \cdot k_2) +   \\
&  \hphantom{ \left[ \frac{}{} \right. - }   (\lambda_{9} + \lambda_{19}+\lambda_{23}) (\zeta_1 \cdot k_3) (\zeta_2 \cdot k_1) (\zeta_3 \cdot k_1) +
(\lambda_{10}+\lambda_{20}+\lambda_{23}) (\zeta_1 \cdot k_3) (\zeta_2 \cdot k_1) (\zeta_3 \cdot k_2) +   \\
&  \hphantom{ \left[ \frac{}{} \right. - }  (\lambda_{11} + \lambda_{19}+\lambda_{24}) (\zeta_1 \cdot k_3) (\zeta_2 \cdot k_3) (\zeta_3 \cdot k_1) +
(\lambda_{12}+\lambda_{20}+\lambda_{24}) (\zeta_1 \cdot k_3) (\zeta_2 \cdot k_3) (\zeta_3 \cdot k_2)    \left.   \frac{}{}   \right] +   \\
&   \Bigl[ \  \Bigl\{ - \frac{1}{2} (s + t) \lambda_{2} + \frac{1}{2} \lambda_{1} t - \rho_1 \Bigr \} (\zeta_1 \cdot \zeta_2) (\zeta_3 \cdot k_1) +   \Bigl\{ - \frac{1}{2} (s + t) \lambda_{4} + \frac{1}{2} \lambda_{3} t -\rho_1 \Bigr \} (\zeta_1 \cdot \zeta_2) (\zeta_3 \cdot k_2) + \\
&   \hphantom{  \Bigl[ \  }    \Bigl\{ - \frac{1}{2} (s + t) \lambda_{6} + \frac{1}{2} \lambda_{5} t - \rho_2 \Bigr \} (\zeta_1 \cdot \zeta_3) (\zeta_2 \cdot k_1) +
 \Bigl\{ - \frac{1}{2} (s + t) \lambda_{8} + \frac{1}{2} \lambda_{7} t - \rho_2  \Bigr \} (\zeta_1 \cdot \zeta_3) (\zeta_2 \cdot k_3) + \\
&   \hphantom{   }    \Bigl\{ - \frac{1}{2} (s + t) \lambda_{14} + \frac{1}{2} \lambda_{13} t - \rho_3 \Bigr \} (\zeta_2 \cdot \zeta_3) (\zeta_1 \cdot k_2)   +   \Bigl\{ -\frac{1}{2} (s + t) \lambda_{16} + \frac{1}{2} \lambda_{15} t   - \rho_3 \Bigr \} (\zeta_2 \cdot \zeta_3) (\zeta_1 \cdot k_3)   \left.   \frac{}{}   \right]
\end{split}
\label{4pointzeta4}
\end{multline}
As it happended in the case of $\zeta_3 \rightarrow k_3$, demanding the previous expression to vanish leads to fourteen equations, where only thirteen of them are linearly independent. The set of equations is equivalent to the following thirteen ones:
\begin{eqnarray}
\lambda_{11}-\lambda_{12}+\lambda_{19}-\lambda_{20} = 0 \ , \ \lambda_{10} + \lambda_{11} - \lambda_{12} + \lambda_{17} + \lambda_{21}=0 \ ,\ \lambda_{9} - \lambda_{10} - \lambda_{11} + \lambda_{12} = 0 \ , \nonumber \\
\lambda_{11}-\lambda_{12}+\lambda_{17}-\lambda_{18} = 0 \ , \ \lambda_{10} + \lambda_{20} + \lambda_{23} =0 \ ,\ \lambda_{12} + \lambda_{20} + \lambda_{24} = 0 \ ,  \lambda_{11} + \lambda_{17} + \lambda_{22} = 0 \ ,  \nonumber \\
\ 2 \rho_{1} - \lambda_{1} t + \lambda_{2}(s+t)=0 \ , \
2 \rho_{2} -\lambda_{5} t + \lambda_{6}(s+t)=0 \ , \ 2 \rho_{3} - \lambda_{15} t + \lambda_{16}(s+t)=0 \ , \ \nonumber \\
\lambda_{3} t + (\lambda_{2} - \lambda_4)(s+t) - \lambda_1 t =0 \ , \
\lambda_{7} t + (\lambda_{6} - \lambda_8)  (s+t) - \lambda_5 t =0 \ , \nonumber \\
\lambda_{13} t + (\lambda_{16} - \lambda_{14})  (s+t) - \lambda_{15} t =0 \ . \hspace{1cm}
\label{4point6}
\end{eqnarray}
Now, the solution of the linear system formed by equations  (\ref{4point1}),  (\ref{4point4}),  (\ref{4point5}) and  (\ref{4point6}), is given by
\begin{eqnarray}
\lambda_{1} = 0 \ , \ \lambda_{4}=0 \ ,\ \lambda_{8}= 0 \ , \ \lambda_{12} = 0 \ , \ \lambda_{15} =0 \ ,\ \lambda_{19} = 0 \ ,  \lambda_{21} = 0 \ ,  \nonumber \\
\lambda_7 = - \lambda_{11} = \lambda_{16} = - \lambda_{20} = \lambda_{24} \ , \nonumber \\
\lambda_2 =  - \lambda_5 = - \lambda_6 =  \lambda_9 = - \lambda_{17}  = - \lambda_{23} =  \lambda_{24}  \frac{ \ t \ }{s} \ , \nonumber \\
- \lambda_3 =   \lambda_{10} =  \lambda_{13} =  \lambda_{14} = - \lambda_{18}  =  \lambda_{22} =   \lambda_{24}  \frac{ s+ t }{s} \ ,
\label{4pointsolution}
\end{eqnarray}
where $\lambda_{24}$ is completely arbitrary.\\
In this way, substituing the final solution found in eq. (\ref{4pointsolution}), in the original expression for $T(1,2,3,4)$, given in eq. (\ref{4point}), this subamplitude becomes a kinematical expression which contains as a global factor $\lambda_{24}$.
Due to the arbitrariness of $\lambda_{24}$, for convinience we may rewrite it as
\begin{eqnarray}
\lambda_{24} & = & 4 g^2 \lambda /t \ .
\label{lambda24}
\end{eqnarray}
Afterwards, using appropriately the relations in (\ref{zetak4}) plus the relations
\begin{eqnarray}
\label{u}
u & = & -s - t \ , \\
\label{additional}
(\zeta_4 \cdot k_3) & = & - (\zeta_4 \cdot k_1) - (\zeta_4 \cdot k_2) \ ,
\end{eqnarray}
we can finally arrive to the symmetric formula in (\ref{A1234-1}).\\
In (\ref{A1234-1}) the expression inside the curly brackets is precisely the well known 4-point kinematic factor\cite{Schwarz1}.

\section{Finding an $N$-point basis for the scattering amplitude ($N=5,6,7$)}

\label{N-point basis}

\noindent In this appendix we give the details of the calculations that support the main claim of section \ref{Kinematical}, namely, that the ${\cal B}_N$ set of subamplitudes, given in eq.(\ref{basis}), is a basis for the space of $N$-point amplitudes ${\cal V}_N$, defined by the requirements in eq.(\ref{requirements}) (where the polarizations and the momenta obey relations (\ref{momentum}), (\ref{mass-shell}) and (\ref{transversality})). \\
\noindent The importance of this claim is that the $N$-point tree level scattering subamplitudes of gauge bosons in Open Superstring Theory (and also in Yang-Mills theory) can all be written as a linear combination of the subamplitudes in $(\zeta \cdot \zeta)^1 (\zeta \cdot k)^{N-2}$ terms of $A_{YM}(1, \ldots, N)$. At least we have succeeded in doing so for $3 \leq N \leq 7$.\\
\noindent We begin by defining, in Appendix \ref{Mandelstam variables}, the independent Mandelstam variables that appear in an $N$-point scattering process of massless particles (for $N=5,6,7$). We also write there the expression for the remaining scalar invariants, $k_i \cdot k_j$, in terms of the corresponding Mandelstam variables. Then, in Appendix \ref{zetazeta-zetak} we present, in an abbreviated fashion, the expressions of the $(\zeta \cdot \zeta)^1 (\zeta \cdot k)^{N-2}$ terms of $A_{YM}(1, \ldots, N)$. As argued at the end of subsection \ref{Finding the components}, the explicit expression of these terms will allow us to find the momentum factors in the case of open superstring subamplitudes (section \ref{Closed}) and also in the case of Yang-Mills subamplitudes (Appendix \ref{BCJ}).\\
\noindent In Appendix \ref{Linear} we use the $(\zeta \cdot \zeta)^1 (\zeta \cdot k)^{N-2}$ terms of $A_{YM}(1, \ldots, N)$, written in \ref{zetazeta-zetak}, to prove that the ${\cal B}_N$ set is linearly independent in ${\cal V}_N$.\\
\noindent We end by checking, in Appendix \ref{Dimension}, that ${\cal V}_N$, as defined by the requirements in eq.(\ref{requirements}), is an $(N-3)!$-dimensional space (at least in the case of $N=5,6,7$) and, therefore, ${\cal B}_N$ is indeed a basis for ${\cal V}_N$.\\

\subsection{Mandelstam variables}

\label{Mandelstam variables}
In general, in an $N$-point scattering process of massless particles there are $N(N-3)/2$ independent Mandelstam variables. In the cases of $N=5,6,7$ these variables can be respectively chosen as \footnote{The $s_i$ and $t_j$ variables that appear in eqs.(\ref{Mandelstam5}), (\ref{Mandelstam6}) and (\ref{Mandelstam7}), are the ones that naturally appear as poles of $A_{YM}(1, \ldots, N)$, for $N=5,6,7$, respectively.}
\begin{eqnarray}
\label{Mandelstam5}
N=5 \ : \ s_i  = & (k_i + k_{i+1})^2 & = \ 2 \ k_i \cdot k_{i+1} \ ,  \hspace{4.3cm}  (i=1, \ldots, 5) \ .\\
\label{Mandelstam6}
N=6 \ : \ s_i =  & (k_i + k_{i+1})^2 & = \ 2 \ k_i \cdot k_{i+1} \ ,  \hspace{4.3cm}   (i=1, \ldots, 6) \ . \nonumber \\
t_j = &(k_j + k_{j+1} + k_{j+2})^2 & = \ 2 \ (k_j \cdot k_{j+1} + k_j \cdot k_{j+2} + k_{j+1} \cdot k_{j+2} ) \ , \hspace{0.2cm}   (j=1, 2, 3) \ . \\
\label{Mandelstam7}
N=7 \ : \ s_i =  & (k_i + k_{i+1})^2 & = \ 2 \ k_i \cdot k_{i+1} \ , \hspace{4.3cm} (i=1, \ldots, 7) \ , \nonumber \\
t_j = &(k_j + k_{j+1} + k_{j+2})^2 & = \ 2 \ (k_j \cdot k_{j+1} + k_j \cdot k_{j+2} + k_{j+1} \cdot k_{j+2} ) \ , \hspace{0.2cm} (j=1, \ldots, 7) \ . \hspace{0.7cm}
\end{eqnarray}
In eqs.(\ref{Mandelstam5}), (\ref{Mandelstam6}) and (\ref{Mandelstam7}) it is implicit the identification of indexes $(6 \leftrightarrow 1)$, $(7 \leftrightarrow 1)$ and $(8 \leftrightarrow 1, \ 9 \leftrightarrow 2)$, respectively. For example, according to this identification of indexes, in (\ref{Mandelstam7}) we have that $t_7 = (k_7 + k_{8} + k_{9})^2= (k_7 + k_{1} + k_{2})^2= 2 \ (k_7 \cdot k_{1} + k_7 \cdot k_{2} + k_{1} \cdot k_{2} ) \ $.\\
The remaining scalar products of the momenta that appear in the corresponding $N$-point scattering process of massless particles can all be written in terms of the corresponding Mandelstam variables, in the following way\footnote{This comes from demanding momentum conservation and the mass-shell condition for the external massless states.}:
\begin{eqnarray}
\label{Mandelstam5-2}
N=5:  &  \hspace{0cm} &  2\ k_1 \cdot k_3 = s_4 - s_1 - s_2 \ , \ \ 2 \ k_1 \cdot k_4 = s_2 - s_5 - s_4 \ , \ \  2 \ k_2 \cdot k_4 = s_5 - s_2 - s_3 \ , \nonumber \\
          &  \hspace{0cm} &   2 \ k_2 \cdot k_5 = s_3 - s_1 - s_5 \ , \ \ 2 \ k_3 \cdot k_5 = s_1 - s_4 - s_3 \ ,  \\
\label{Mandelstam6-2}
N=6:  &  \hspace{0cm} &  2\ k_1 \cdot k_3 = t_1 - s_1 - s_2 \ , \ \ 2 \ k_1 \cdot k_4 = s_2 + s_5 - t_1 - t_2 \ , \ \  2 \ k_1 \cdot k_5 = t_1 - s_5 - s_6 \ , \nonumber \\
          &  \hspace{0cm} &   2 \ k_2 \cdot k_4 = t_2 - s_2 - s_3 \ , \ \ 2 \ k_2 \cdot k_5 = s_3 + s_6 - t_2 - t_3 \ , \ \  2 \ k_2 \cdot k_6 = t_3 - s_1 - s_6 \ , \nonumber \\
          &  \hspace{0cm} &   2 \ k_3 \cdot k_5 = t_3 - s_3 - s_4 \ , \ \ 2 \ k_3 \cdot k_6 = s_1 + s_4 - t_1 - t_3 \ , \ \  2 \ k_4 \cdot k_6 = t_1 - s_4 - s_5 \ , \hspace{0.6cm} \\
\label{Mandelstam7-2}
N=7:  &  \hspace{0cm} &  2\ k_1 \cdot k_3 = t_1 - s_1 - s_2 \ , \ \ 2 \ k_1 \cdot k_4 = s_2 + t_5 - t_1 - t_2 \ , \ \  2 \ k_1 \cdot k_5 = s_6 + t_2 - t_5 - t_6 \ , \nonumber \\
          &  \hspace{0cm} &   2 \ k_1 \cdot k_6 = t_6 - s_6 - s_7 \ , \ \ 2 \ k_2 \cdot k_4 = t_2 - s_2 - s_3 \ , \ \  2 \ k_2 \cdot k_5 = s_3 + t_6 - t_2 - t_3 \ , \nonumber \\
          &  \hspace{0cm} &   2 \ k_2 \cdot k_6 = s_7 + t_3 - t_6 - t_7 \ , \ \ 2 \ k_2 \cdot k_7 = t_7 - s_1 - s_7  \ , \ \  2 \ k_3 \cdot k_5 = t_3 - s_3 - s_4 \ ,
\nonumber \\
  &  \hspace{0cm} &  2\ k_3 \cdot k_6 = s_4 + t_7 - t_3 - t_4 \ , \ \ 2 \ k_3 \cdot k_7 = s_1 + t_4 - t_1 - t_7 \ , \ \  2 \ k_4 \cdot k_6 = t_4 - s_4 - s_5 \ , \nonumber \\
          &  \hspace{0cm} &   2 \ k_4 \cdot k_7 = t_1 + s_5 - t_4 - t_5 \ , \ \ 2 \ k_5 \cdot k_7 = t_5 - s_5 - s_6 \ . \
\end{eqnarray}

\vspace{0.5cm}

\subsection{$(\zeta \cdot \zeta)^1 (\zeta \cdot k)^{N-2}$ terms of the Yang-Mills $N$-point subamplitude}

\label{zetazeta-zetak}

\noindent As mentioned at the end of subsection \ref{Finding the components}, our proposal is that it will be enough to consider the $(\zeta \cdot \zeta)^1 (\zeta \cdot k)^{N-2}$ terms of the Yang-Mills $N$-point subamplitude, in order to find the momentum factors that appear in the BCJ relations (and also the momentum factors that appear in the closed expression of the open superstring scattering subamplitudes). So, in the next subsection of this appendix we will give the explicit expression of those terms in the case of $N=5$. We have obtained it by calculating the $5$-point subamplitude in Yang-Mills theory (in the Lorentz gauge) by using Feynman rules. We have also written the $(\zeta \cdot k)$ terms of them in the basis mentioned in eq. (\ref{zetakterms}). For the Mandelstam variables we are using the conventions of eq. (\ref{Mandelstam5}).\\
In the case of $N=6$ and $N=7$ we have written in subsection \ref{N67} only the general structure of the $(\zeta \cdot \zeta)^1 (\zeta \cdot k)^{4}$ and the $(\zeta \cdot \zeta)^1 (\zeta \cdot k)^{5}$ terms, respectively, using the convention for the Mandelstam variables in (\ref{Mandelstam6}) and (\ref{Mandelstam7}). Although we have used the complete expression of them in order to acheive important results like the ones in eqs.(\ref{6point}) and (\ref{7point}), it is not instructive to write down those huge expressions here\footnote{We have specified on these expressions the number of terms that they have. See equations (\ref{A6}) and (\ref{A7}).}. We have verified that those expressions satisfy basic properties of the Yang-Mills subamplitudes, like cyclicity and reflection. In any case, the interested reader can request the authors for the complete explicit expressions of the $6$ and $7$-point subamplitudes.

\subsubsection{Case of $N=5$}

\label{N5}

\begin{multline}
A_{YM}(1,2,3,4,5)\Big|_{(\zeta \cdot \zeta)^1 (\zeta \cdot k)^3}  = \\
\begin{split}
 & =8 \ g^3 \Biggl[   \ (\zeta_2 \cdot \zeta_3) \ \biggl\{   \  - \Big( \frac{1}{s_1 s_4} + \frac{1}{s_2 s_4} \Big) (\zeta_1 \cdot k_2) (\zeta_4 \cdot k_1) (\zeta_5 \cdot k_3)  -   \Big( \frac{1}{s_1 s_4} + \frac{1}{s_2 s_4}  + \frac{1}{s_2 s_5} \Big) \times \\
&  \times  (\zeta_1 \cdot k_2) (\zeta_4 \cdot k_2) (\zeta_5 \cdot k_3) +
  \Big( \frac{1}{s_1 s_3} + \frac{1}{s_2 s_4}  + \frac{1}{s_3 s_5}  + \frac{1}{s_4 s_1} + \frac{1}{s_5 s_2}  \Big) (\zeta_1 \cdot k_2) (\zeta_4 \cdot k_3) (\zeta_5 \cdot k_1)  + \\
&  \Big( \frac{1}{s_1 s_3} + \frac{1}{s_2 s_4}  + \frac{1}{s_3 s_5}  + \frac{1}{s_4 s_1} + \frac{1}{s_5 s_2}  \Big) (\zeta_1 \cdot k_2) (\zeta_4 \cdot k_3) (\zeta_5 \cdot k_2)  +   \frac{1}{s_2 s_4}   (\zeta_1 \cdot k_3) (\zeta_4 \cdot k_1) (\zeta_5 \cdot k_2) - \\
& \Big( \frac{1}{s_2 s_4} + \frac{1}{s_2 s_5} \Big) (\zeta_1 \cdot k_3) (\zeta_4 \cdot k_2) (\zeta_5 \cdot k_1) -
\Big( \frac{1}{s_2 s_4} + \frac{1}{s_2 s_5} \Big) (\zeta_1 \cdot k_3) (\zeta_4 \cdot k_2) (\zeta_5 \cdot k_3) + \\
&  \Big( \frac{1}{s_2 s_4} + \frac{1}{s_3 s_5}  + \frac{1}{s_5 s_2} \Big)  (\zeta_1 \cdot k_3) (\zeta_4 \cdot k_3) (\zeta_5 \cdot k_2)   - \frac{1}{s_2 s_5}  (\zeta_1 \cdot k_4) (\zeta_4 \cdot k_2) (\zeta_5 \cdot k_3)  +  \\
&    \Big( \frac{1}{s_2 s_5} + \frac{1}{s_3 s_5} \Big) (\zeta_1 \cdot k_4) (\zeta_4 \cdot k_3) (\zeta_5 \cdot k_2)   \   \biggr\} \  + \\
&    \hphantom{ \ = 2 \ g^3 \Biggl[   } \ (\zeta_1 \cdot \zeta_4) \ \biggl\{    \  - \frac{1}{s_1 s_4}  (\zeta_2 \cdot k_1) (\zeta_3 \cdot k_1) (\zeta_5 \cdot k_1)                  - \frac{1}{s_1 s_4}  (\zeta_2 \cdot k_1) (\zeta_3 \cdot k_1) (\zeta_5 \cdot k_2)   - \\
&  - \frac{1}{s_1 s_4}  (\zeta_2 \cdot k_1) (\zeta_3 \cdot k_1) (\zeta_5 \cdot k_3)     -   \Big( \frac{1}{s_1 s_4} + \frac{1}{s_2 s_4} \Big)  (\zeta_2 \cdot k_1) (\zeta_3 \cdot k_2) (\zeta_5 \cdot k_1)     -   \Big( \frac{1}{s_1 s_4} + \frac{1}{s_2 s_4} \Big) \times \\
&  \times   (\zeta_2 \cdot k_1) (\zeta_3 \cdot k_2) (\zeta_5 \cdot k_2)   -  \Big( \frac{1}{s_1 s_4} + \frac{1}{s_2 s_4} \Big)  (\zeta_2 \cdot k_1) (\zeta_3 \cdot k_2) (\zeta_5 \cdot k_3) +  \frac{1}{s_1 s_3} \times \\
& \times (\zeta_2 \cdot k_1) (\zeta_3 \cdot k_4) (\zeta_5 \cdot k_1)  +  \frac{1}{s_1 s_3} (\zeta_2 \cdot k_1) (\zeta_3 \cdot k_4) (\zeta_5 \cdot k_2)
+  \frac{1}{s_2 s_4} (\zeta_2 \cdot k_3) (\zeta_3 \cdot k_1) (\zeta_5 \cdot k_1)  + \\
&   \frac{1}{s_2 s_4} (\zeta_2 \cdot k_3) (\zeta_3 \cdot k_1) (\zeta_5 \cdot k_2)  +   \frac{1}{s_2 s_4} (\zeta_2 \cdot k_3) (\zeta_3 \cdot k_1) (\zeta_5 \cdot k_3) -    \Big( \frac{1}{s_2 s_5} + \frac{1}{s_5 s_3} \Big)   \times     \\
& \times  (\zeta_2 \cdot k_3) (\zeta_3 \cdot k_4) (\zeta_5 \cdot k_1)
+  \frac{1}{s_2 s_5} (\zeta_2 \cdot k_4) (\zeta_3 \cdot k_2) (\zeta_5 \cdot k_1)
-  \frac{1}{s_3 s_5} (\zeta_2 \cdot k_4) (\zeta_3 \cdot k_4) (\zeta_5 \cdot k_1)   \  \biggr\}  \Biggr] +  \\
&  \hphantom{ 2 \ g^3 \Biggl[   } + (\mbox{ \ cyclic permutations \ })
\end{split}
\label{A5}
\end{multline}

\subsubsection{Case of $N=6$ and $N=7$}

\label{N67}

\noindent The  $(\zeta \cdot \zeta)^1 (\zeta \cdot k)^{4}$ and the $(\zeta \cdot \zeta)^1 (\zeta \cdot k)^{5}$ terms of the $6$ and the $7$-point Yang-Mills subampltiudes have the following structure\footnote{In the $6$-point case, in eq. (\ref{A6}), there is no mistake in the fact that the $(\zeta_1 \cdot \zeta_4)$ term does not carry a factor `2', as every other terms does: this factor will arise when summing the cyclic permutations.}:

\begin{multline}
A_{YM}(1,2,3,4,5,6)\Big|_{(\zeta \cdot \zeta)^1 (\zeta \cdot k)^4}  = \\
\begin{split}
 & = 8 \ g^4 \Biggl[ \ 2  \ (\zeta_1 \cdot \zeta_2) \ \biggl\{    \   \frac{1}{s_1 s_5 t_1}   (\zeta_3 \cdot k_1) (\zeta_4 \cdot k_1) (\zeta_5 \cdot k_1)(\zeta_6 \cdot k_2) + (\mbox{ \ $67$ terms \ })    \  \biggr\}  \ + \\
 & \hphantom{  = \ g^4 \Biggl[ } \ 2 \ (\zeta_1 \cdot \zeta_3) \ \biggl\{    \   \frac{1}{s_1 s_5 t_1}   (\zeta_2 \cdot k_1) (\zeta_4 \cdot k_1) (\zeta_5 \cdot k_1)(\zeta_6 \cdot k_3) + (\mbox{ \ $57$ terms \ })    \  \biggr\}  \ + \\
 & \hphantom{  =2 \ g^4 \Biggl[ } \ \ (\zeta_1 \cdot \zeta_4) \ \biggl\{    \   \frac{1}{s_1 s_5 t_1}   (\zeta_2 \cdot k_1) (\zeta_3 \cdot k_1) (\zeta_5 \cdot k_1)(\zeta_6 \cdot k_4) + (\mbox{ \ $51$ terms \ })    \  \biggr\}  \   \Biggr] +   \\
&  \hphantom{ 2 \ g^3 \Biggl[   } + (\mbox{ \ cyclic permutations \ })
\end{split}
\label{A6}
\end{multline}
and
\begin{multline}
A_{YM}(1,2,3,4,5,6,7)\Big|_{(\zeta \cdot \zeta)^1 (\zeta \cdot k)^5}  = \\
\begin{split}
 & = 32 \ g^5 \Biggl[  \  (\zeta_1 \cdot \zeta_2) \ \biggl\{    \   \frac{1}{s_1 s_6 t_1 t_5}   (\zeta_3 \cdot k_1) (\zeta_4 \cdot k_1) (\zeta_5 \cdot k_1)(\zeta_6 \cdot k_1)(\zeta_7 \cdot k_2) + (\mbox{ \ $565$ terms \ })    \  \biggr\}  \ + \\
& \hphantom{ = 2 \ g^5 \Biggl[  \  \ }   (\zeta_1 \cdot \zeta_3) \ \biggl\{    \   \frac{1}{s_1 s_6 t_1 t_5}   (\zeta_2 \cdot k_1) (\zeta_4 \cdot k_1) (\zeta_5 \cdot k_1)(\zeta_6 \cdot k_1)(\zeta_7 \cdot k_3) + (\mbox{ \ $466$ terms \ })    \  \biggr\}  \ + \\
& \hphantom{ = 2 \ g^5 \Biggl[  \  \ }   (\zeta_1 \cdot \zeta_4) \ \biggl\{    \   \frac{1}{s_1 s_6 t_1 t_5}   (\zeta_2 \cdot k_1) (\zeta_3 \cdot k_1) (\zeta_5 \cdot k_1)(\zeta_6 \cdot k_1)(\zeta_7 \cdot k_4) + (\mbox{ \ $408$ terms \ })    \  \biggr\}    \  \biggr\}  \   \Biggr] +   \\
&  \hphantom{ 2 \ g^3 \Biggl[   } + (\mbox{ \ cyclic permutations \ }) \ .
\end{split}
\label{A7}
\end{multline}

\subsection{Linear independence of the set ${\cal B}_N$}

\label{Linear}

\noindent The color/kinematic duality in Yang-Mills theory \cite{Bern1} and the monodromy relations found for the open string subamplitudes \cite{Bjerrum1, Stieberger3} succeed in finding all $N$-point subamplitudes of Yang-Mills theory in terms of a linear combinations of a set of $(N-3)!$ subamplitudes (which can be chosen to be the ones in the ${\cal B}_N$ set of eq.(\ref{basis}))\footnote{In the case of the color/kinematic duality approach, the authors of \cite{Bern1} have conjectured the explicit form of BCJ relations for arbitrary $N$, and they have found evidence for their conjecture up to $N=8$, but the proof for arbitrary $N$ has not been given yet.} . The procedure of each of these approaches is self consistent in the sense that it is found a linear system for the subamplitudes which is overdetermined; then, it is found that this linear system has a $unique$ solution (which consists, precisely, in the BCJ relations for the $N$-point subamplitudes).  But this result still does not guarantee that this set ${\cal B}_N$ is indeed a basis of ${\cal V}_N$: nothing forbids that there could still be additional restrictions that relate the Yang-Mills subamplitudes in ${\cal B}_N$. \\
\noindent In fact, it is known that in D=4, the $N=5$ BCJ relations can all be found in terms of \underline{only one} Yang-Mills subamplitude, instead of two (see section 5.2 of ref.\cite{Nastase1}). This does not invalidate the BCJ relations that have been found for $N=5$: it simply states that, at least in D=4, the set ${\cal B}_5$ is not linearly independent and the BCJ relations can have an even simpler form.   \\
\noindent In the way that the BCJ relations have been found \cite{Bern1, Bjerrum1, Stieberger3}, their validity is for $any$ spacetime dimension D. So, what may happen (and the mentioned result of ref.\cite{Nastase1} is an explicit evidence for it) is that if $N$ is sufficiently high (as compared to D), there may still be further relations among the Yang-Mills subamplitudes of the ${\cal B}_N$ set in eq.(\ref{basis}).\\
\noindent The same as in \cite{Bern1} and \cite{Bjerrum1, Stieberger3}, our approach throughout this work has only considered the general case, that is, the one in which the BCJ relations are written being valid for $any$ spacetime dimension D (or equivalently, we will assume that $N$ is sufficiently small, as compared to D, such that no new extra relations between the subamplitudes arise).\\
\noindent Having stated this subtlety about the dependence of the basis of ${\cal V}_N$  in the spacetime dimension D, we will proceed with our analysis only using arguments that are valid for any spactime dimension.\\
In this section of Appendix \ref{N-point basis} we will see that in fact,  as mentioned at the end of subsection \ref{Finding the components}, considering only the $(\zeta \cdot \zeta)^1 (\zeta \cdot k)^{N-2}$ terms of the Yang-Mills $N$-point subamplitudes, there is enough information (which is even redundant when we consider $all$ those terms) to prove the linear independence of the set of subamplitudes ${\cal B}_N$, given in eq. (\ref{basis}).

\subsubsection{Case of $N=5$}

\label{N5-2-1}

\noindent In order to prove that ${\cal B}_5 = \{ A_{YM}(1,2,3,4,5) , A_{YM}(1,3,2,4,5) \}$ is a linearly independent set in ${\cal V}_5$, we need to consider the null linear combination of $A_{YM}(1,2,3,4,5)$ and $A_{YM}(1,3,2,4,5)$:
\begin{eqnarray}
\lambda^{\{23\}}  A_{YM}(1,2,3,4,5) + \lambda^{\{32\}}  A_{YM}(1,3,2,4,5) =0 \ .
\label{linear5}
\end{eqnarray}
Here we want to prove that (\ref{linear5}) happens if and only if $\lambda^{\{23\}} = \lambda^{\{32\}} = 0$.\\
Considering only the $(\zeta \cdot \zeta)^1 (\zeta \cdot k)^{3}$ terms in (\ref{linear5}), we have that
\begin{eqnarray}
\lambda^{\{23\}}  A_{YM}(1,2,3,4,5)\Big|_{(\zeta \cdot \zeta)^1 (\zeta \cdot k)^3} + \lambda^{\{32\}}  A_{YM}(1,3,2,4,5)\Big|_{(\zeta \cdot \zeta)^1 (\zeta \cdot k)^3} =0 \ ,
\label{linear5-2}
\end{eqnarray}
and using the explicit expression for the $(\zeta \cdot \zeta)^1 (\zeta \cdot k)^{3}$ terms of the 5-point subamplitude, given in (\ref{A5}), we have
\begin{multline}
8 \ g^3 \  \Biggl[  \  \biggl\{ -\lambda^{\{23\}}  \Big( \frac{1}{s_1 s_4} + \frac{1}{s_2 s_4} \Big)  + \lambda^{\{32\}}  \frac{1}{s_2 s_4}  \biggr\} (\zeta_2 \cdot \zeta_3) (\zeta_1 \cdot k_2) (\zeta_4 \cdot k_1)(\zeta_5 \cdot k_3) \ + \\
\begin{split}
 &    \biggl\{ -\lambda^{\{23\}}  \frac{1}{s_1 s_4}  - \lambda^{\{32\}}  \frac{1}{(s_4 - s_1 - s_2)  s_4}  \biggr\} (\zeta_1 \cdot \zeta_4) (\zeta_2 \cdot k_1) (\zeta_3 \cdot k_1)(\zeta_5 \cdot k_1) \ +    \big(\mbox{$140$ terms}\big) \ \Biggr] = 0 \ .
\end{split}
\label{linear5-4}
\end{multline}
Since the $(\zeta \cdot \zeta)^1 (\zeta \cdot k)^{3}$ terms in (\ref{linear5-4}) are linearly independent, then this equation is valid if and only if the coefficient of each of these terms is zero:
\begin{eqnarray}
-\lambda^{\{23\}}  \Big( \frac{1}{s_1 s_4} + \frac{1}{s_2 s_4} \Big)  + \lambda^{\{32\}}  \frac{1}{s_2 s_4}  = 0 \ , \ \  -\lambda^{\{23\}}  \frac{1}{s_1 s_4}  - \lambda^{\{32\}}  \frac{1}{(s_4 - s_1 - s_2)  s_4} = 0 \ , \ \ldots \ . \hspace{0.7cm}
\label{linear5-5}
\end{eqnarray}
(\ref{linear5-5}) is a linear homogeneous system of $142$ equations for  $\lambda^{(5)}_1$ and $\lambda^{(5)}_2$. From the first two equations, explicitly written in (\ref{linear5-5}), it is easy to see that the $unique$ solution of this system is the trivial one:
\begin{eqnarray}
\lambda^{\{23\}} = \lambda^{\{32\}} = 0 \ .
\label{linear5-6}
\end{eqnarray}

\subsubsection{Case of $N=6$ and $N=7$}

\label{N6-2-1}

\noindent Doing exactly the same procedure of the previous subsection, but for $N=6$ and $N=7$, using the corresponding expressions of the $(\zeta \cdot \zeta)^1 (\zeta \cdot k)^4$ and the $(\zeta \cdot \zeta)^1 (\zeta \cdot k)^5$ terms, mentioned in eqs.(\ref{A6}) and (\ref{A7}), respectively, we arrive to the same conclusion, namely, that the six $\lambda^{\{ \sigma_6 \}}$'s and the twentyfour $\lambda^{\{ \sigma_7\}}$'s are $all$ zero, implying that the ${\cal B}_6$ and ${\cal B}_7$ sets are linearly independent in their corresponding spaces, ${\cal V}_6$ and ${\cal V}_7$.\\
\noindent An interesting remark is that we have not needed to examine $all$ the $(\zeta \cdot \zeta)^1 (\zeta \cdot k)^{N-2}$ terms of the Yang-Mills $N$-point subamplitudes (for $N=5,6,7$), in order to arrive at the conclusion that the $\lambda^{\{ \sigma_N\}}$'s are all zero : it has been enough to just analyze the $(\zeta_1 \cdot \zeta_2)(\zeta \cdot k)^{N-2}$ kinematical structures (which are just one of the many possible structures of the $(\zeta \cdot \zeta)^1 (\zeta \cdot k)^{N-2}$ terms).\\

\subsection{Dimension of ${\cal V}_N$ and a basis for it}

\label{Dimension}

In the case of $N=3$ and $N=4$, in subsections \ref{N3} and \ref{N4}, respectively, it was checked that the dimension of ${\cal V}_N$ is, in fact,
\begin{eqnarray}
\label{dimVN}
 dim({\cal V}_N)=(N-3)! \ \ ,
\end{eqnarray}
and that the set of subamplitudes ${\cal B}_N$, given in eq.(\ref{basis}), is indeed a basis of this space.\\
In this subsection we will argue that the validity of the formula (\ref{dimVN}) and of the set ${\cal B}_N$ proposed as a basis for ${\cal V}_N$, given in eq.(\ref{basis}), indeed holds for $N=5,6,7$.\\
\noindent We will argue this in subsection \ref{N5-3} for the $N=5$ case, with some detail, and in subsection \ref{N67-2} for the $N=6$ and the $N=7$ case. In these last two cases we will just mention the final results, because the intermediate formulas are simply too big and they do not give any additional idea.\\
\noindent Before going into the details of sections \ref{N5-3} and \ref{N67-2} we warn the reader that our calculations have not lead directly to the set ${\cal B}_N$ as a basis for ${\cal V}_N$ (as explicited in eq. (\ref{basis})): what we have found for each $N$ (=5,6,7) is another set of $(N-3)!$ independent kinematical expressions\footnote{These kinematical expressions are on-shell gauge invariant.} which become a basis for the corresponding ${\cal V}_N$ space. Within our approach, it would simple be too tricky to arrive, for any $N \geq 5$, directly to the set ${\cal B}_N$.\\
\noindent So, at this point we must admit that, without knowing the BCJ result for the basis of Yang-Mills subamplitudes, given in eq.(\ref{basis}), demanding on-shell gauge invariance would have lead us to the validity of eq.(\ref{dimVN}), but we would have hardly arrived to a basis in which $all$ the elements can be directly associated to Yang-Mills theory.

\subsubsection{Case of $N=5$}

\label{N5-3}

Let $T(1,2,3,4,5) \ \varepsilon \ {\cal V}_5$. According to the table in eq. (\ref{table1}), this element of ${\cal V}_5$ can be constructed from $270$ $(\zeta \cdot \zeta)^1 (\zeta \cdot k)^3$ and from $45$ $(\zeta \cdot \zeta)^2 (\zeta \cdot k)^1$ terms, which, after being written using the $(\zeta \cdot k)$ terms  of eq. (\ref{zetakterms}), are all linearly independent. So, initially, it contains 315 kinematical terms. \\
Then, we impose the third requirement that the list in eq.(\ref{requirements}) demands for the elements of ${\cal V}_5$, namely, $T(1,2,3,4,5)$ should satisfy on-shell gauge invariance:
\begin{eqnarray}
T(1,2,3,4,5)\Big|_{\zeta_i = k_i} = 0 \ , \hspace{1cm} (i=1, 2, \ldots, 5) \ .
\label{5-point-on-shell-1}
\end{eqnarray}
Doing exactly the same procedure that we explained in subsection \ref{N4} and in Appendix \ref{Calculations} for the $N=4$ case,
in this case we have arrived to the conclusion that $T(1,2,3,4,5)$ can be written in terms of two known (and independent) kinematic expressions:
\begin{eqnarray}
T(1,2,3,4,5) & = & \rho^{(5)}_1 K^{(5)}_1 (\zeta , k) +  \rho^{(5)}_2 K^{(5)}_2 (\zeta , k) \ ,
\label{5point2}
\end{eqnarray}
where $K^{(5)}_1 (\zeta , k)$ and $K^{(5)}_2 (\zeta , k)$ are the known (but big) kinematic expressions (without poles) and $\{  \rho^{(5)}_1 ,  \rho^{(5)}_2 \}$ are the arbitrary momentum factors. So, by construction $K^{(5)}_1 (\zeta , k)$ and $K^{(5)}_2 (\zeta , k)$ are (on-shell) gauge invariant kinematical expressions.\\
$K^{(5)}_1 (\zeta , k)$ consists in $133$ $(\zeta \cdot k)^3 (\zeta \cdot \zeta)^1$ and $33$ $(\zeta \cdot k)^1 (\zeta \cdot \zeta)^2$ independent terms while
$K^{(5)}_2 (\zeta , k)$ consists in $134$ $(\zeta \cdot k)^3 (\zeta \cdot \zeta)^1$ and $31$ $(\zeta \cdot k)^1 (\zeta \cdot \zeta)^2$ independent ones.
These kinematical expressions have the following form\footnote{If it has any usefulness to the reader, the explicit expressions of $ K^{(5)}_1 (\zeta , k)$ and $ K^{(5)}_2 (\zeta , k)$ can be requested to the authors.}:
\begin{eqnarray}
\label{K51}
K^{(5)}_1 (\zeta , k) & = & g^3 \biggl[ \Big\{ (s_2 + s_3 - s_5)(s_1 s_2 + s_2 s_5 + s_3 s_4 + s_3 s_5 - s_2 s_3) (\zeta_1 \cdot k_2)(\zeta_2 \cdot k_3)(\zeta_3 \cdot k_4)(\zeta_4 \cdot \zeta_5) + \nonumber \\
&&  \hphantom{  \Big\{ }  \ 132 \ (\zeta \cdot k)^3 (\zeta \cdot \zeta)^1 \ \mbox{terms} \Big\} + \nonumber \\
&&  \Big\{   s_2 s_3 (s_1 s_2 + s_2 s_5 + s_3 s_4 - s_2 s_3 - s_4 s_5 ) (\zeta_1 \cdot k_4)(\zeta_2 \cdot \zeta_4)(\zeta_3 \cdot \zeta_5)  + \nonumber \\
&&  \hphantom{  \Big\{ }  \ 32 \ (\zeta \cdot k)^1 (\zeta \cdot \zeta)^2 \ \mbox{terms} \Big\} \biggr] \ , \\
\label{K52}
K^{(5)}_2 (\zeta , k) & = &  g^3 \biggl[  \Big\{ -( s_1 s_2 s_4 + s_1 s_3 s_4 + s_1 s_3 s_5 + s_2 s_3 s_5 + s_2 s_4 s_5 ) (\zeta_1 \cdot k_2)(\zeta_2 \cdot k_3)(\zeta_3 \cdot k_4)(\zeta_4 \cdot \zeta_5) + \nonumber \\
&&  \hphantom{  \Big\{ }  \ 133 \ (\zeta \cdot k)^3 (\zeta \cdot \zeta)^1 \ \mbox{terms} \Big\} + \nonumber \\
&&  \Big\{   s_2 (s_1 s_2 s_4 + s_1 s_3 s_4 + s_1 s_3 s_5 + s_2 s_3 s_5 + s_2 s_4 s_5 ) (\zeta_1 \cdot k_4)(\zeta_2 \cdot \zeta_4)(\zeta_3 \cdot \zeta_5)  + \nonumber \\
&&  \hphantom{  \Big\{ }  \ 30 \ (\zeta \cdot k)^1 (\zeta \cdot \zeta)^2 \ \mbox{terms} \Big\}  \biggr] \ .
\end{eqnarray}
In (\ref{K51}) and (\ref{K52}) $\{s_1, s_2, s_3, s_4, s_5 \}$ are the five independent Mandelstam variables that appear in massless 5-point scattering, and that were defined in eq. (\ref{Mandelstam5}). \\
So, our result in (\ref{5point2}) indeed verifies that $dim({\cal V}_5)=2$, in agreement with (\ref{dimVN}).\\
Since in subsection \ref{N5-2-1} we proved that the two-element set ${\cal B}_5$ is linearly independent in ${\cal V}_5$, it is clear that a change of basis can be done,  namely,
\begin{eqnarray}
\Big\{  \ K^{(5)}_1(\zeta , k) \ ,  \ K^{(5)}_2(\zeta , k) \ \Big\} \rightarrow \Big \{ \ A_{YM}(1,2,3,4,5) \ , \ A_{YM}(1,3,2,4,5) \ \Big\} \ ,
\label{5-point basis}
\end{eqnarray}
and, therefore, ${\cal B}_5$ is a possible basis of ${\cal V}_5$.\\
So, at the end, instead of (\ref{5point2}) it is possible to write
\begin{eqnarray}
T(1,2,3,4,5) & = &\lambda^{\{23\}} \cdot A_{YM}(1,2,3,4,5) + \lambda^{\{32\}} \cdot A_{YM}(1,3,2,4,5) \ ,
\label{5pointthree}
\end{eqnarray}
as claimed in eq. (\ref{5point3}) of the main text of this work.\\

\subsubsection{Case of $N=6$ and $N=7$}

\label{N67-2}

Let $T(1,2,3,4,5,6) \ \varepsilon \ {\cal V}_6$ and $T(1,2,3,4,5,6,7) \ \varepsilon \ {\cal V}_7$. Although the computational effort becomes greater as $N$ grows, we have succeeded in finding that, repeating exactly the same procedure that we have described in subsection \ref{N4} and in Appendix \ref{Calculations} for the $N=4$ case, and in subsection \ref{N5-3} for the $N=5$ case, we arrive that these two elements can be respectively written as
\begin{eqnarray}
T(1,2,3,4,5,6) & = & \rho^{(6)}_1 K^{(6)}_1 (\zeta , k) +  \rho^{(6)}_2 K^{(6)}_2 (\zeta , k) + \rho^{(6)}_3 K^{(6)}_3 (\zeta , k) +  \rho^{(6)}_4 K^{(6)}_4 (\zeta , k) + \nonumber \\
&& \rho^{(6)}_5 K^{(6)}_5 (\zeta , k) +  \rho^{(6)}_6 K^{(6)}_6 (\zeta , k)
\label{6point2}
\end{eqnarray}
and
\begin{eqnarray}
T(1,2,3,4,5,6,7) & = & \rho^{(7)}_1 K^{(7)}_1 (\zeta , k) +  \rho^{(7)}_2 K^{(7)}_2 (\zeta , k) + \rho^{(7)}_3 K^{(7)}_3 (\zeta , k) +  \rho^{(7)}_4 K^{(7)}_4 (\zeta , k) + \nonumber \\
&& \rho^{(7)}_5 K^{(7)}_5 (\zeta , k) +  \rho^{(7)}_6 K^{(7)}_6 (\zeta , k)  +  \rho^{(7)}_7 K^{(7)}_5 (\zeta , k) +  \rho^{(7)}_8 K^{(7)}_6 (\zeta , k)  +   \nonumber  \\
& & \rho^{(7)}_9 K^{(7)}_9 (\zeta , k) +  \rho^{(7)}_{10} K^{(7)}_{10} (\zeta , k) + \rho^{(7)}_{11} K^{(7)}_{11} (\zeta , k) +  \rho^{(7)}_{12} K^{(7)}_{12} (\zeta , k) + \nonumber \\
& & \rho^{(7)}_{13} K^{(7)}_{13} (\zeta , k) +  \rho^{(7)}_{14} K^{(7)}_{14} (\zeta , k) + \rho^{(7)}_{15} K^{(7)}_{15} (\zeta , k) +  \rho^{(7)}_{16} K^{(7)}_{16} (\zeta , k) + \nonumber \\
& & \rho^{(7)}_{17} K^{(7)}_{17} (\zeta , k) +  \rho^{(7)}_{18} K^{(7)}_{18} (\zeta , k) + \rho^{(7)}_{19} K^{(7)}_{19} (\zeta , k) +  \rho^{(7)}_{20} K^{(7)}_{20} (\zeta , k) + \nonumber \\
& & \rho^{(7)}_{21} K^{(7)}_{21} (\zeta , k) +  \rho^{(7)}_{22} K^{(7)}_{22} (\zeta , k) + \rho^{(7)}_{23} K^{(7)}_{23} (\zeta , k) +  \rho^{(7)}_{24} K^{(7)}_{24} (\zeta , k) \ ,
\label{7point2}
\end{eqnarray}
where the $K^{(N)}_i(\zeta , k)$'s are known kinematic expressions (without poles).\\
\noindent In the case of $N=6$ we have found our result in (\ref{6point2}) analytically, in the same way as we did in (\ref{5point2}).\\
\noindent In the case of $N=7$ we have obtained our result numerically: we have found the explicit form of the $K^{(7)}_i(\zeta , k)$'s only for fixed values of the Mandelstam variables. \\
\noindent The important result is that, independently of being an analytic or a numerical result, we can conclude that
\begin{eqnarray}
\label{dimV67}
 dim({\cal V}_6)=6 \ \ \ \ \ \mbox{and} \ \ \ \ \ dim({\cal V}_7)=24 \ ,
\end{eqnarray}
in agreement with formula (\ref{dimVN}).\\
\noindent Since in subsection \ref{N6-2-1} we proved that the six-element set ${\cal B}_6$ and the twenty four-element set ${\cal B}_7$  are linearly independent in ${\cal V}_6$ and ${\cal V}_7$, respectively, due to the matching with the dimensions found in (\ref{dimV67}), these sets can be chosen as basis of the corresponding spaces.

\section{Kinematical derivation of BCJ relations}

\label{BCJ}

BCJ relations were discovered by Bern, Carrasco and Johansson in 2008 by means of a conjectured duality between color and kinematics in Yang-Mills theories \cite{Bern1}. These relations state that the $(N-1)!$ Yang-Mills subamplitudes that appear in the $N$-point formula (see eq.(\ref{N-point})\footnote{Formula (\ref{N-point}) was written in the Introduction as valid for gauge bosons in tree level Open Superstring Theory, but it is well knwon that it is also valid for tree level Yang-Mills $N$-point amplitudes \cite{Mangano1}.}) can be written as linear combinations of a subset which contains only $(N-3)!$ of them. Although not exactly the same, from the work in \cite{Bern1} it can be easily seen that the set of subamplitudes that we propose in eq. (\ref{basis}) can be chosen as a basis for the Yang-Mills subamplitudes in the BCJ relations\footnote{The reason of our choice of basis is, of course, that we want to arrive to the MSS formula in eq.(\ref{MSS}), which is written in this basis.}. \\
\noindent A rigorous proof of the BCJ relations was found first using String Theory methods, independently, by the authors of \cite{Bjerrum1} and \cite{Stieberger3}. Later, a pure field theory proof of these relations was also found in \cite{Vaman1} (at least for $N=4,5,6$).\\
\noindent In this appendix we present an alternative derivation of BCJ relations for $N=4,5,6,7$. We will not deal with color/kinematic duality of Yang-Mills subamplitudes at any moment. We will not need to consider Kleiss-Kuijf relations \cite{Kleiss1} either. From our perspective the result is that BCJ relations arise as a natural consequence of the basis found for ${\cal V}_N$, considered in section \ref{Kinematical} of this work. The only unknowns in the BCJ relations will then be the momentum factors that participate in them as the components of a given Yang-Mills subamplitude with respect to the ${\cal B}_N$ basis.\\
\noindent As mentioned at the end of subsection \ref{Finding the components}, our proposal is that it is enough to consider the explicit expression of the $(\zeta \cdot \zeta)^1 (\zeta \cdot k)^{N-2}$ terms of the $N$-point Yang-Mills subamplitudes, in order to obtain the momentum factors that appear in the BCJ relations. These subamplitudes can be obtained using Feynman rules from the Yang-Mills lagrangian (in the Lorentz gauge) or by selecting them in the low energy limit of the corresponding subamplitude of gauge bosons in String Theory.\\
\noindent We will find the BCJ relations for a set of $(N-1)!/2-(N-3)!$ Yang-Mills subamplitudes. The first of these numbers, $(N-1)!/2$, comes from considering the number of independent subamplitiudes, from the point of view of cyclic and reflection symmetries (see eqs.(\ref{cyclic}) and (\ref{reflection}), respectively). The second one, $(N-3)!$, is simply the number of Yang-Mills subamplitudes of the ${\cal B}_N$ basis, which are taken from this set.\\
\noindent Here, we will give the details of our derivation of the BCJ relations for the case of $N=4$ and $N=5$. The procedure for $N=6$ and $N=7$ is exactly the same one, but the intermediate formulas become too big, so we will just present some examples of these relations.\\

\subsection{Case of $N=4$}

\label{N4-2}

In this case there are $(N-1)!/2=3$ subamplitudes which are not related by cyclic neither reflection symmetry. We will choose them to be $A_{YM}(1,2,3,4)$, $A_{YM}(1,3,4,2)$ and $A_{YM}(1,4,2,3)$\footnote{Notice that the first subamplitude is $A_{YM}(1,2,3,4)$, that is, the one that has been chosen as a basis for ${\cal V}_4$ in eq.(\ref{A1234-2}).}.\\
\noindent We will now apply eq.(\ref{A1234-2}) for $T(1,2,3,4)=A_{YM}(1,3,4,2)$, namely:
\begin{eqnarray}
A_{YM}(1,3,4,2) & = & \lambda^{\{2\}} \cdot A_{YM}(1,2,3,4) \ .
\label{N41342}
\end{eqnarray}
where $\lambda^{\{2\}}$ is the momentum factor that we want to determine.\\
\noindent Using the explicit expression for $A_{YM}(1,2,3,4)$ given in eq.(\ref{A1234-1}) (and the corresponding one for $A_{YM}(1,3,4,2)$), and afterwards only selecting the $(\zeta \cdot \zeta)^1 (\zeta \cdot k)^{2}$ terms on them, eq. (\ref{N41342}) becomes
\begin{multline}
- 8 g^2 \frac{1}{us}   \Biggl\{ \frac{1}{2} u \Bigl[ (\zeta_1 \cdot k_2)(\zeta_4 \cdot
  k_3)(\zeta_3 \cdot \zeta_2) + (\zeta_3 \cdot k_4)(\zeta_2 \cdot
  k_1)(\zeta_1 \cdot \zeta_4) +(\zeta_1 \cdot k_4)(\zeta_2 \cdot
  k_3)(\zeta_3 \cdot \zeta_4) + \\
\hphantom{ 8 g^2 \frac{1}{st}   \Biggl\{  + \frac{1}{2} u \Bigl[ }
  (\zeta_3 \cdot k_2)(\zeta_4 \cdot k_1)(\zeta_1 \cdot \zeta_3)
  \Bigr]  + \frac{1}{2} s \Bigl[ (\zeta_3 \cdot k_1)(\zeta_2 \cdot
  k_4)(\zeta_4 \cdot \zeta_1) + (\zeta_4 \cdot k_2)(\zeta_1 \cdot
  k_3)(\zeta_3 \cdot \zeta_2) +\\
\hphantom{ 8 g^2 \frac{1}{st}   \Biggl\{  + \frac{1}{2} u \Bigl[}
 (\zeta_3 \cdot k_2)(\zeta_1 \cdot k_4)(\zeta_4 \cdot \zeta_2) +
  (\zeta_4 \cdot k_1)(\zeta_2 \cdot k_3)(\zeta_3 \cdot \zeta_1)
  \Bigr] + \frac{1}{2} t \Bigl[ (\zeta_1 \cdot k_3)(\zeta_2 \cdot
  k_4)(\zeta_4 \cdot \zeta_3) +  \\
 (\zeta_4 \cdot k_2)(\zeta_3 \cdot k_1)(\zeta_1 \cdot \zeta_2)
+  (\zeta_1 \cdot k_2)(\zeta_3 \cdot k_4)(\zeta_4 \cdot \zeta_2) +
  (\zeta_4 \cdot k_3)(\zeta_2 \cdot k_1)(\zeta_1 \cdot \zeta_3)
  \Bigr] \ \Biggr\}= \\
\begin{split}
&-  \lambda^{\{2\}}  \ 8 g^2 \frac{1}{st}   \Biggl\{ \frac{1}{2} s \Bigl[ (\zeta_1 \cdot k_4)(\zeta_3 \cdot
  k_2)(\zeta_2 \cdot \zeta_4) + (\zeta_2 \cdot k_3)(\zeta_4 \cdot
  k_1)(\zeta_1 \cdot \zeta_3) +(\zeta_1 \cdot k_3)(\zeta_4 \cdot
  k_2)(\zeta_2 \cdot \zeta_3) + \\
& \hphantom{= \lambda^{\{2\}}  \ 8 g^2 \frac{1}{st}   \Biggl\{  }
  (\zeta_2 \cdot k_4)(\zeta_3 \cdot k_1)(\zeta_1 \cdot \zeta_4)
  \Bigr]  + \frac{1}{2} t \Bigl[ (\zeta_2 \cdot k_1)(\zeta_4 \cdot
  k_3)(\zeta_3 \cdot \zeta_1) + (\zeta_3 \cdot k_4)(\zeta_1 \cdot
  k_2)(\zeta_2 \cdot \zeta_4) +\\
&  \hphantom{= \lambda^{\{2\}}   8 g^2 \frac{1}{st}   \Biggl\{  }
 (\zeta_2 \cdot k_4)(\zeta_1 \cdot k_3)(\zeta_3 \cdot \zeta_4) +
  (\zeta_3 \cdot k_1)(\zeta_4 \cdot k_2)(\zeta_2 \cdot \zeta_1)
  \Bigr] + \frac{1}{2} u \Bigl[ (\zeta_1 \cdot k_2)(\zeta_4 \cdot
  k_3)(\zeta_3 \cdot \zeta_2) +  \\
&  \hphantom{= \lambda^{\{2\}}  \ 8 g^2 \frac{1}{st}   \Biggl\{  }
(\zeta_3 \cdot k_4)(\zeta_2 \cdot k_1)(\zeta_1 \cdot \zeta_4)
 +  (\zeta_1 \cdot k_4)(\zeta_2 \cdot k_3)(\zeta_3 \cdot \zeta_4) +
  (\zeta_3 \cdot k_2)(\zeta_4 \cdot k_1)(\zeta_1 \cdot \zeta_2)
  \Bigr] \ \Biggr\}  \ .
\end{split}
\label{N41342-2}
\end{multline}
Next, we write all $(\zeta \cdot k)$ terms in both sides of (\ref{N41342-2}) in the basis mentioned in (\ref{zetakterms}), by using the relations (\ref{restrict1}) for this case, namely,
\begin{eqnarray}
\label{zeta4k3}
(\zeta_4 \cdot k_3) & = & - (\zeta_4 \cdot k_1) - (\zeta_4 \cdot  k_2) \ , \\
\label{zeta1k4}
(\zeta_1 \cdot k_4) & = & - (\zeta_1 \cdot k_2) - (\zeta_1 \cdot  k_3) \ , \\
\label{zeta2k4}
(\zeta_2 \cdot k_4) & = & - (\zeta_2 \cdot k_1) - (\zeta_2 \cdot  k_3) \ , \\
\label{zeta3k4}
(\zeta_3 \cdot k_4) & = & - (\zeta_3 \cdot k_1) - (\zeta_3 \cdot  k_2) \ .
\end{eqnarray}
Once both sides of (\ref{N41342-2}) have been written in terms of the same basis for the $(\zeta \cdot \zeta)^1 (\zeta \cdot k)^{2}$ terms, it is possible to compare their coefficients, arriving at a set of eight linearly dependent equations for $\lambda^{\{2\}}$. After using that $u=-s-t \ $ it can be seen that there is really only one linearly independent equation, namely, $\lambda^{\{2\}} = t/u$, from which it comes that
\begin{eqnarray}
A_{YM}(1,3,4,2) & = & \frac{t}{u} A_{YM}(1,2,3,4) \ .
\label{N41342-3}
\end{eqnarray}
\noindent Doing exactly the same procedure to find $A_{YM}(1,4,2,3)$ in terms of $A_{YM}(1,2,3,4)$ we find that
\begin{eqnarray}
A_{YM}(1,4,2,3) & = & \frac{s}{u} A_{YM}(1,2,3,4) \ .
\label{N41423}
\end{eqnarray}
Eqs. (\ref{N41342-3}) and (\ref{N41423}) are the independent BCJ relations for $N=4$. Any other of the remaining twenty one 4-point subamplitudes can be written in terms of $A_{YM}(1,2,3,4)$ by means of these two relations and using cyclic or/and reflection symmetries.\\
It is clear that in this case ($N=4$) it would have been very much simple to just compare the coefficients of the $(\zeta \cdot \zeta)^2$ kinematical terms since there are only three of them in the whole subamplitude. The result would have been just the same one: $\lambda^{\{2\}} = t/u$. But, as indicated in the final paragraph of subsection \ref{Finding the components}, our proposal for the $N$-point subamplitudes, for any value of $N$, is that it will always be enough to consider the $(\zeta \cdot \zeta)^1 (\zeta \cdot k)^{N-2}$ terms and this is what we have confirmed here.\\

\subsection{Case of $N=5$}

\label{N5-2}

In this case there are $(N-1)!/2=12$ subamplitudes which are not related by cyclic neither reflection symmetry. We will choose them to be
 $A_{YM}(1,2,3,4,5)$, $A_{YM}(1,2,4,3,5)$, $A_{YM}(1,4,2,3,5)$, $A_{YM}(2,1,3,4,5)$, $A_{YM}(2,3,1,4,5)$, $A_{YM}(2,1,4,3,5)$ and six additional 5-point subamplitudes obtained from the previous ones by exchanging labels $2$ and $3$ \footnote{We have chosen the same set of twelve 5-point subamplitudes of ref. \cite{Bjerrum1}.}.\\
\noindent In this subsection we will present, in some detail, the calculations that lead to the BCJ relation for the subamplitude $A_{YM}(2,1,3,4,5)$, that is, we will find the momentum factors, $\lambda^{\{23\}}$ and $\lambda^{\{32\}}$, that allow to write $A_{YM}(2,1,3,4,5)$ in terms of $A_{YM}(1,2,3,4,5)$ and $A_{YM}(1,2,3,4,5)$, according to eq. (\ref{5point3}):
\begin{eqnarray}
\label{N521345}
A_{YM}(2,1,3,4,5) & = &\lambda^{\{23\}}  A_{YM}(1,2,3,4,5) + \lambda^{\{32\}}  A_{YM}(1,3,2,4,5) \ .
\end{eqnarray}
\noindent Using the explicit expression for the $(\zeta \cdot \zeta)^1 (\zeta \cdot k)^{3}$ terms of $A_{YM}(1,2,3,4,5)$, given in eq.(\ref{A5}), the corresponding terms of $A_{YM}(1,3,2,4,5)$ and $A_{YM}(2,1,3,4,5)$ can be obtained. Substituing all these $(\zeta \cdot \zeta)^1 (\zeta \cdot k)^{3}$ terms in (\ref{N521345}) we arrive to an equation which has the following form\footnote{See the convention for the Mandelstam variables $s_i$ in eq. (\ref{Mandelstam5}) of Appendix \ref{N-point basis}.}:
\begin{multline}
 8  g^3 \Biggl[   \ (\zeta_1 \cdot \zeta_3) \ \biggl\{   \  - \Big( \frac{1}{s_1 s_4} + \frac{1}{(s_4 - s_1 - s_2) s_4} \Big) (\zeta_2 \cdot k_1) (\zeta_4 \cdot k_2) (\zeta_5 \cdot k_3) + (\mbox{ \ $9$ terms \ })  \   \biggr\} \  + \\
 \ (\zeta_3 \cdot \zeta_4) \ \biggl\{    \  + \frac{1}{s_1 s_3}  (\zeta_1 \cdot k_2) (\zeta_2 \cdot k_4) (\zeta_5 \cdot k_3)                        + (\mbox{ \ $15$ terms \ }) \  \biggr\}  +     \hspace{3cm}     \\
 (\mbox{ \ cyclic permutations of indexes (2,1,3,4,5) \ }) \ \Biggr]   =      \hspace{1cm}     \\
\begin{split}
&=\lambda^{\{23\}} 8 g^3 \Biggl[   \ (\zeta_2 \cdot \zeta_3) \ \biggl\{   \  - \Big( \frac{1}{s_1 s_4} + \frac{1}{s_2 s_4} \Big) (\zeta_1 \cdot k_2) (\zeta_4 \cdot k_1) (\zeta_5 \cdot k_3) + (\mbox{ \ $9$ terms \ })  \   \biggr\} \  + \\
&   \hphantom{ \ = \lambda^{\{23\}} \cdot   2 \ g^3 \Biggl[   } \ (\zeta_1 \cdot \zeta_4) \ \biggl\{    \  - \frac{1}{s_1 s_4}  (\zeta_2 \cdot k_1) (\zeta_3 \cdot k_1) (\zeta_5 \cdot k_1)                        + (\mbox{ \ $13$ terms \ }) \  \biggr\} \ +  \\
&   \hphantom{ \ = \lambda^{\{23\}} \cdot   2 \ g^3 \Biggl[   } \  (\mbox{ \ cyclic permutations  of indexes (1,2,3,4,5) \ }) \   \Biggr] \ + \\
&+\lambda^{\{32\}}  8  g^3 \Biggl[   \ (\zeta_3 \cdot \zeta_2) \ \biggl\{   \  - \Big( \frac{1}{(s_4 - s_1 - s_2) s_4} + \frac{1}{s_2 s_4} \Big) (\zeta_1 \cdot k_3) (\zeta_4 \cdot k_1) (\zeta_5 \cdot k_2) + (\mbox{ \ $9$ terms \ })  \   \biggr\} \  + \\
&    \hphantom{ \ = \lambda^{\{23\}} \cdot   2 \ g^3 \Biggl[   } \ (\zeta_1 \cdot \zeta_4) \ \biggl\{    \  - \frac{1}{(s_4 - s_1 - s_2) s_4}  (\zeta_3 \cdot k_1) (\zeta_2 \cdot k_1) (\zeta_5 \cdot k_1)                        + (\mbox{ \ $13$ terms \ }) \  \biggr\}  \ +  \\
&   \hphantom{ \ = \lambda^{\{23\}} \cdot   2 \ g^3 \Biggl[   } \ (\mbox{ \ cyclic permutations  of indexes (1,3,2,4,5) \ }) \ \Biggr] \ .
\end{split}
\label{N521345-2}
\end{multline}
In this equation we have preferred to present explicitly only some of the $(\zeta \cdot \zeta)^1 (\zeta \cdot k)^{3}$ terms in order to save some space on it\footnote{It is clear, from the expression of the $(\zeta \cdot \zeta)^1 (\zeta \cdot k)^{3}$ terms of $A_{YM}(1,2,3,4,5)$, given in eq.(\ref{A5}), that we could have written explicitly $all$ those kinematic terms in eq. (\ref{N521345-2}).}.\\
The peculiar factor $(s_4 - s_1 - s_2)$ in some of the denominators of eq. (\ref{N521345-2}) simply corresponds to $k_1 \cdot k_3$, as given in eq. (\ref{Mandelstam5}).\\
Writing all $(\zeta \cdot k)$ terms in both sides of (\ref{N521345-2}) in the basis mentioned in (\ref{zetakterms}), once again, by appropriately using the relations (\ref{restrict1}), we arrive to a set of $142$ linearly dependent equations\footnote{This set comes from demanding that the same $142$ $(\zeta \cdot \zeta)^1 (\zeta \cdot k)^3$ considered in Appendix \ref{N-point basis}, in eq. (\ref{linear5-4}), are linearly independent.}, finding $\lambda^{\{23\}}(\alpha')=(s_5-s_3)/(s_3-s_1-s_5)$ and $\lambda^{\{32\}}(\alpha')=(s_5-s_2-s_3)/(s_3-s_1-s_5)$ as a consistent unique solution of it (as expected). Therefore, eq. (\ref{N521345}) becomes
\begin{eqnarray}
\label{N521345-3}
A_{YM}(2,1,3,4,5) & = & \frac{s_5-s_3}{s_3-s_1-s_5}  A_{YM}(1,2,3,4,5) +  \frac{s_5-s_2-s_3}{s_3-s_1-s_5}  A_{YM}(1,3,2,4,5) \ .
\end{eqnarray}
Doing this same procedure we arrive to the following additional relations:
 \begin{eqnarray}
\label{N512435}
A_{YM}(1,2,4,3,5) & = & \frac{s_4-s_1}{s_1-s_3-s_4}  A_{YM}(1,2,3,4,5) +  \frac{s_4-s_1-s_2}{s_1-s_3-s_4}  A_{YM}(1,3,2,4,5) \ , \\
\label{N523145}
A_{YM}(2,3,1,4,5) & = & -\frac{s_1 \ s_3}{(s_3-s_1-s_5)(s_2-s_4-s_5)}  A_{YM}(1,2,3,4,5) +  \nonumber \\
&&\frac{(s_1+s_5)(s_5-s_2-s_3)}{(s_3-s_1-s_5)(s_2-s_4-s_5)}  A_{YM}(1,3,2,4,5) \ , \\
\label{N514235}
A_{YM}(1,4,2,3,5) & = & -\frac{s_1 \ s_3}{(s_1-s_3-s_4)(s_2-s_4-s_5)}  A_{YM}(1,2,3,4,5) +  \nonumber \\
&&\frac{(s_3+s_4)(s_4-s_1-s_2)}{(s_1-s_3-s_4)(s_2-s_4-s_5)}  A_{YM}(1,3,2,4,5) \ , \\
\label{N521435}
A_{YM}(2,1,4,3,5) & = & \frac{p(s_1, s_2, s_3, s_4, s_5)}{(s_1-s_3-s_4)(s_2-s_4-s_5)(s_3-s_1-s_5)}  A_{YM}(1,2,3,4,5) -  \nonumber \\
&&\frac{(s_4+s_5)(s_4-s_1-s_2)(s_5-s_2-s_2)}{(s_1-s_3-s_4)(s_2-s_4-s_5)(s_3-s_1-s_5)}  A_{YM}(1,3,2,4,5) \ ,
\end{eqnarray}
where the polynomial appearing in (\ref{N521435}) is given by
\begin{eqnarray}
\label{p}
p(s_1, s_2, s_3, s_4, s_5) &=&s_3 s_4^2 - s_3 s_2 s_4+ s_3 s_5 s_4 - s_3 s_5 s_1 - s_3 s_4 s_1 + s_1 s_5 s_4 - s_1 s_5 s_2 + s_5 s_2 s_4 - s_5^2 s_4 + \nonumber \\
&& s_5^2 s_1 - s_5 s_4^2 \ .
\end{eqnarray}
After using the expressions for the momentum products $k_i \cdot k_j$, given in eqs. (\ref{Mandelstam5}) and (\ref{Mandelstam5-2}) of Appendix \ref{N-point basis}, it can be verified that all five relations agree with the BCJ ones found in eq.(13) of ref. \cite{Bjerrum1}\footnote{Since in ref. \cite{Bjerrum1} those authors work with Open String Theory subamplitudes, in order to see the matching of our results with theirs it should be taken the low energy limit of their relations.}.\\
The BCJ relations for the remaining subamplitudes can be obtained from eqs. (\ref{N521345-3})-(\ref{N521435}) by just exchanging labels $2$ and $3$ \footnote{In order for these relations to have the same format as the first ones, the $s_i$ Mandelstam variables should first be rewritten in terms of momentum products $k_i \cdot k_j$, then exchange labels $2$ and $3$ and then write the final expression in terms of the Mandelstam variables again.}. \\
\noindent This acheives writing ten of the twelve subamplitudes, mentioned at the beginning of this section of Appendix \ref{BCJ}, in terms of the remaining two ones.\\

\subsection{Case of $N=6$}

\label{N6-2}

In this case there are $(N-1)!/2=60$ subamplitudes which are not related by cyclic neither reflection symmetry. The outcome of our procedure is that we have succeeded in finding an expression for $54$ of them in terms of the six ones of ${\cal B}_6$, as predicted in eq.(\ref{6point}).\\
\noindent We have found that these 54 relations can be classified into two types. In the first type, $42$ subamplitudes are in fact given as a linear combination of the six ones of ${\cal B}_6$, like for example\footnote{See eq.(\ref{Mandelstam6}) for the definition of the nine independent Mandelstam variables that appear in the $6$-point scattering process: there are six $s_i$'s and three $t_j$'s.}:
\begin{eqnarray}
A_{YM}(1,4,5,3,2,6) &=& -\frac{s_{1}(-t_{1}+s_{2}+t_{3}-s_{6})}{(-s_{1}+t_{3}-s_{6})(-t_{1}+s_{2}-s_{6}+s_{4})}A_{YM}(1,2,3,4,5,6) +\nonumber \\
&& \frac{(t_{1}-s_{1}-s_{2})(s_{2}+t_{3}-s_{6})}{(-s_{1}+t_{3}-s_{6})(-t_{1}+s_{2}-s_{6}+s_{4})}A_{YM}(1,3,2,4,5,6) + \nonumber \\
&&\frac{s_{1}(t_{1}-s_{2}+s_{3}-t_{3}+s_{6})}{(-s_{1}+t_{3}-s_{6})(-t_{1}+s_{2}-s_{6}+s_{4})}A_{YM}(1,2,4,3,5,6) + \nonumber \\
&&\frac{(s_{2}-t_{2}+s_{3}-s_{1})(t_{3}-s_{6}-t_{1}+s_{2}-s_{3})}{(-s_{1}+t_{3}-s_{6})(-t_{1}+s_{2}-s_{6}+s_{4})}A_{YM}(1,3,4,2,5,6) - \nonumber \\
&& \frac{(-t_{3}-t_{2}+s_{6}+s_{3})(t_{1}-s_{1}-s_{2})}{(-s_{1}+t_{3}-s_{6})(-t_{1}+s_{2}-s_{6}+s_{4})}A_{YM}(1,4,2,3,5,6)\nonumber - \\
&& \frac{(-t_{3}-t_{2}+s_{6}+s_{3})(t_{1}-s_{1}-s_{2}+s_{3})}
{(-s_{1}+t_{3}-s_{6})(-t_{1}+s_{2}-s_{6}+s_{4})}A_{YM}(1,4,3,2,5,6) \ , \\
A_{YM}(1,5,4,3,2,6) & = & \frac{s_1(s_2+t_3-s_6-t_1 )(s_5+s_6-s_4-t_2) }  {(s_1+s_5-t_2)(s_{1}+s_{6}-t_{3})(s_6+t_{1}-s_{2}-s_{4})}A_{YM}(1,2,3,4,5,6) + \nonumber \\
 & & \frac{(s_1 +s_2-t_1) (s_6-s_2-t_3 ) ( s_5+s_6-s_4-t_2 )}
{(s_1+s_5-t_2)(s_{1}+s_{6}-t_{3})(s_6+t_{1}-s_{2}-s_{4})}A_{YM}(1,3,2,4,5,6) + \nonumber \\
 & & \frac{s_1 ( s_3 + s_4 - t_3 ) ( t_1 + t_2 - s_2 - s_5 )}
{(s_1+s_5-t_2)(s_{1}+s_{6}-t_{3})(s_6+t_{1}-s_{2}-s_{4})}A_{YM}(1,2,4,3,5,6) + \nonumber \\
 & & \frac{ ( s_1 + s_2 - t_1 )  ( s_4 + t_2 -s_5 - s_6 )  ( t_2 + t_3 - s_3 - s_6 ) }
{(s_1+s_5-t_2)(s_{1}+s_{6}-t_{3})(s_6+t_{1}-s_{2}-s_{4})}A_{YM}(1,3,4,2,5,6) + \nonumber \\
 & & \frac{ ( s_3 + s_4 - t_3  )  ( s_2 + s_5 - t_1 - t_2 )  ( s_1 + t_2 - s_2 - s_3 ) }
{(s_1+s_5-t_2)(s_{1}+s_{6}-t_{3})(s_6+t_{1}-s_{2}-s_{4})}A_{YM}(1,4,2,3,5,6) + \nonumber \\
 & & \frac{ ( s_6 + s_1 - s_3 - s_4 )  ( t_1 + t_2 - s_2 - s_5 )  ( t_2 + t_3 - s_3 - s_6 )}
{(s_1+s_5-t_2)(s_{1}+s_{6}-t_{3})(s_6+t_{1}-s_{2}-s_{4})}A_{YM}(1,4,3,2,5,6) \ . \nonumber \\
\end{eqnarray}
\noindent In the second type, $12$ subamplitudes are found to be given only in terms of three subamplitudes of ${\cal B}_6$, like for example\footnote{Notice, in eqs.(\ref{AYM123546}), (\ref{AYM124536}) and (\ref{AYM143526}), that the three subamplitudes of ${\cal B}_6$ that appear in them are not necessarilly all the same.}:
\begin{eqnarray}
\label{AYM123546}
A_{YM}(1,2,3,5,4,6) &=& \frac{t_1-s_5}{t_1-s_5-s_{4}}A_{YM}(1,2,3,4,5,6)-\frac{t_1-s_5+s_{3}}{t_1-s_5-s_{4}}A_{YM}(1,2,4,3,5,6)\nonumber \\
& \ &+\frac{s_5-t_1-t_2+s_{2}}{-s_5-s_{4}+t_1}A(1,3,4,2,5,6) \ , \\
\label{AYM124536}
A_{YM}(1,2,4,5,3,6) &=&\frac{t_{1}-s_1}{s_{1}-t_{3}+s_{4}-t_1}A_{YM}(1,2,3,4,5,6) + \frac{t_1-s_{1}-s_{2}}{s_{1}-t_{3}+s_{4}-t_1}A_{YM}(1,3,2,4,5,6)+\nonumber \\
& \ &  \frac{s_{3}+t_1-s_{1}}{s_{1}-t_{3}+s_{4}-t_1}A_{YM}(1,2,4,3,5,6) \ , \\
\label{AYM143526}
A_{YM}(1,4,3,5,2,6) &=& -\frac{s_{1}}{s_{1}-t_{3}+s_{6}}A_{YM}(1,2,4,3,5,6)-\frac{s_{1}-s_{3}+t_2}{s_{1}-t_{3}+s_{6}}A_{YM}(1,4,3,2,5,6) + \nonumber \\
& \ &\frac{s_2 + s_3 - s_1 - t_2}{s_{1}-t_{3}+s_{6}}A(1,3,4,2,5,6) \ ,
\end{eqnarray}

\noindent We have checked that all our $N=6$ results agree with the ones in ref. \cite{Hemily2} (which were obtained in detail using the BCJ color/kinematic duality).

\subsection{Case of $N=7$}

\label{N7-2}

In this case there are $(N-1)!/2=360$ subamplitudes which are not related by cyclic neither reflection symmetry. This time we have found $336$ of them in terms of the twenty four ones of ${\cal B}_7$, as predicted in eq.(\ref{7point}).\\
\noindent The $336$ BCJ relations that we have found can be classified into three types. In the first type, $216$ subamplitudes are in fact given as a linear combination of the twenty four ones of ${\cal B}_7$. In the second type, $72$ subamplitudes are given only in terms of twelve of the ones of ${\cal B}_7$. And in the third type, $48$ subamplitudes  are given only in terms of four of the ones of ${\cal B}_7$\footnote{Because of computer limitations, we have arrived to these results numerically, for the first two types of BCJ relations, and analytically for the last ones. See eqs.(\ref{AYM1234657}), (\ref{AYM1235647}) and (\ref{AYM1243657}).}.\\
\noindent A few exampes of the third type of BCJ relations are the following\footnote{See eq.(\ref{Mandelstam7}) for the definition of the fourteen independent Mandelstam variables that appear in the $7$-point scattering process: there are seven $s_i$'s and seven $t_j$'s.}:
\begin{multline}
A_{YM}(1,2,3,4,6,5,7) =\\
\begin{split}
& \frac{t_5-s_6}   {s_5 + s_6 - t_5}
A_{YM}(1,2,3,4,5,6,7)+\frac{s_4 + t_5 - s_6}    {s_5 + s_6 - t_5}
A_{YM}(1,2,3,5,4,6,7) + \\
&\frac{ t_3 + t_5 - s_3 - s_6  }    {s_5 + s_6 - t_5}
A_{YM}(1,2,4,5,3,6,7)  +\frac{ t_5 + t_6 - s_6 - t_2  }    {s_5 + s_6 - t_5}
A_{YM}(1,3,4,5,2,6,7)  \ ,
\end{split}
\label{AYM1234657}
\end{multline}
\begin{multline}
\label{AYM1235647}
A_{YM}(1,2,3,5,6,4,7) =\\
\begin{split}
& \frac{t_{5}-t_1}   {s_5 + t_1 - t_4 - t_5 }
A_{YM}(1,2,3,4,5,6,7) + \frac{ s_4 + t_5 - t_1 }     {s_5 + t_1 - t_4 - t_5 }
A_{YM}(1,2,3,5,4,6,7)+ \\
&  \frac{ t_5 - s_3 - t_1  }     {s_5 + t_1 - t_4 - t_5 }
A_{YM}(1,2,4,3,5,6,7) +\frac{ s_2 + t_5 - t_1 -t_2  }     {s_5 + t_1 - t_4 - t_5 }
A_{YM}(1,3,4,2,5,6,7)   \ , \\
\end{split}
\end{multline}
\begin{multline}
\label{AYM1243657}
A_{YM}(1,2,4,3,6,5,7) =\\
\begin{split}
& \frac{ t_5 - s_6 }   {s_5 + s_6 - t_5}
A_{YM}(1,2,4,3,5,6,7)+\frac{ t_3 + t_5 - s_3 - s_4 - s_6 }   {s_5 + s_6 - t_5}
A_{YM}(1,2,5,3,4,6,7) + \\
& \frac{ t_3 + t_5 - s_3 - s_6  }   {s_5 + s_6 - t_5}
A_{YM}(1,2,5,4,3,6,7) +\frac{ t_5 + t_6 - s_6 - t_2  }   {s_5 + s_6 - t_5}
A_{YM}(1,3,5,4,2,6,7) \ .
\end{split}
\end{multline}
\noindent Our confidence in the correctness of these (and the remaining $45$) relations is the fact that the linear system that we have found for the momentum factors in each of them, following the procedure of subsection \ref{Finding the components}, is overdetermined, and we have found a unique an consistent solution for it. \\

\section{Momentum factors in the open superstring formula}

\label{Momentum factors}

\noindent In this appendix  we will consider the $\alpha'$ expansion of the momentum factors that appear in the $N$-point open superstring formula. These are the
$F^{\{\sigma_N\}}(\alpha')$'s that appear in eq.(\ref{MSS}). \\
\noindent In section \ref{Gamma factor} we will consider the momentum factor that appears in the case of $N=4$ and in section \ref{N5 and higher} we will consider the case of $N \geq 5$.

\subsection{Gamma factor}

\label{Gamma factor}

As mentioned in \cite{DeRoo2}, using the
Taylor expansion for $\mbox{ln} \ \Gamma(1+z)$ \footnote{See
formula (10.44c) of \cite{Arfken1}, for example.},
\begin{eqnarray}
\mbox{ln} \ \Gamma(1+z) = -\gamma z + \sum_{k=2}^{\infty} (-1)^k
\frac{\zeta(k)}{k} z^k \ \ \ \ \ \ \ (-1 < z \leq 1) \ ,
\label{Taylor}
\end{eqnarray}
it may be proved that the explicit $\alpha'$ expansion for the
Gamma factor is given by
\begin{eqnarray}
F^{\{2\}}(\alpha')={\alpha'}^2 s t \ \frac{\Gamma( \alpha' s) \Gamma( \alpha' t)}
{\Gamma(1+ \alpha' s + \alpha' t)} =
\mbox{exp} \left\{ \sum_{k=2}^{\infty} \frac{\zeta(k)}{k}
{\alpha'}^k ((s+t)^k-s^k-t^k) \right \}   \ .
\label{formula1}
\end{eqnarray}
\noindent So, up to ${\cal O}({\alpha'}^6)$ terms we have that
\begin{eqnarray}
F^{\{2\}}(\alpha') & = & 1 -\frac{\
\pi^2}{6} st \ {\alpha'}^2 + \zeta(3) (s+t) st \ {\alpha'}^3  -
\frac{ \ \pi^4}{360}(4 s^2 + st + 4 t^2) st \ {\alpha'}^4 \nonumber \\
 & & - \left[ \frac{\ \pi^2}{6} \zeta(3) \ s^2 t^2 (s+t) - \zeta(5)
(s^3 + 2 s^2 t + 2 s t^2+ t^3) st \right]  {\alpha'}^5 \nonumber \\
 & & + \left[ \frac{}{} \frac{1}{2} \zeta(3)^2 s^2 t^2 (s+t)^2 - \frac{\
 \pi^6}{15120}(16 s^4 + 12 s^3 t + 23 s^2 t^2 + 12 s t^3 + 16 t^4) st
 \right] {\alpha'}^6
\nonumber \\
 & &   + \ {\cal O}({\alpha'}^7) \ .
\label{expansionGamma}
\end{eqnarray}

\subsection{$N=5$ and higher $N$-point momentum factors}

\label{N5 and higher}

\noindent In subsection \ref{Multiple} we make a short review about MZV's. We present enough material for the reader to become aware of the $non \ trivial$ MZV's that we mentioned in section \ref{Brief review} of the main body of this work ($\zeta(3,5)$, $\zeta(3,7)$, $\zeta(3,3,5)$, etc.).\\
\noindent Then, in subsection \ref{Structure} we briefly review the structure of the $\alpha'$ expansion of the $N$-point momentum factors (for $N \geq 5$)\footnote{This subsection has been taken literally (except for some differences in the notation) from part of section {\it 3.2} of ref.\cite{Schlotterer1}.}.  Here it will be seen that the $non \ trivial$ MZV's only show up for the first time at ${\alpha'}^8$ order. \\

\subsubsection{Multiple zeta values (MZV's)}

\label{Multiple}

\noindent MZV's are defined as
\begin{eqnarray}
\zeta(n_1, \ldots, n_r) := \sum_{0 < k_1 < \ldots < k_r} \ \prod_{l=1}^r \frac{1}{ \ k_l^{n_l}} \ ,
\label{multiple}
\end{eqnarray}
where all the $n_l$'s are positive integers and the last one of the list, $n_r$, should be greater than $1$.\\
\noindent In (\ref{multiple}),
\begin{eqnarray}
r \ \ \ \ \ \mbox{and} \ \ \ \ \ w  =  \sum_{l=1}^r n_l
\label{weight}
\end{eqnarray}
are called, respectively, the `depth' and the `weight' (or the transcendentality) of $\zeta(n_1, \ldots, n_r)$. \\
In the case of $r=1$, (\ref{multiple}) becomes a single zeta value, that is, Riemann's zeta function evaluated at an integer $n \geq 2$, $\zeta(n)$. In this case the weight is $w=n$.\\
The following are a few examples of (\ref{multiple}) for $r=2$ and weights $3$, $4$ and $5$, respectively:
\begin{eqnarray}
\label{zeta12}
\zeta(1,2) :=  \sum_{k_2 = 2}^{\infty}  \sum_{k_1 = 1}^{k_2 -1} \frac{1}{ \ k_1^1 \ k_2^2 \ } =  \zeta(3) \ , \hspace{1.8cm} \\
\label{zeta22}
\zeta(2,2) :=  \sum_{k_2 = 2}^{\infty}  \sum_{k_1 = 1}^{k_2 -1} \frac{1}{ \ k_1^2 \ k_2^2 \ } =  \frac{3}{10} \zeta(2)^2 \ , \hspace{1.2cm} \\
\label{zeta14}
\zeta(1,4) :=  \sum_{k_2 = 2}^{\infty}  \sum_{k_1 = 1}^{k_2 -1} \frac{1}{ \ k_1^1 \ k_2^4 \ } =  2 \zeta(5) - \zeta(2) \zeta(3) \ .
\end{eqnarray}
In order to compute these double zeta values and arrive to the expressions that we have written in these equations, in terms of single zeta ones, Harmonic sums and its algebraic property \cite{Vermaseren1} can be used, for example.  \\
\noindent An extremely important data mine of proven results for MZV's is ref. \cite{Blumlein1}. Higher order coefficients of momentum factor $\alpha'$ expansions in String Theory, for $N \geq 5$, are based on the MZV's basis of this reference (see \cite{Schlotterer1} and \cite{Boels1}, for example). \\
For $w \leq 7$ it is always possible to write any MZV as a rational linear combination of products of $\zeta(n)$'s, where $n \leq w$ (as in (\ref{zeta12}), (\ref{zeta22}) and (\ref{zeta14}), for example). But already at $w=8$ there is a first MZV (which is present in the MZV basis of ref.\cite{Blumlein1}), namely $\zeta(3,5)$, which is believed to not admit such sort of expression \cite{Brown0}. \\
\noindent Other MZV's, for higher weights, which are also believed that cannot be written as a rational linear combinations of products of $\zeta(n)$'s, and which are present in the MZV basis of ref.\cite{Blumlein1}, are $\zeta(3,7)$, $\zeta(3,3,5)$, $\zeta(3,9)$ and $\zeta(1,1,4,6)$. Section $2$ of ref.\cite{Schlotterer1} contains tables in which these type of MZV's can be found up to $w=16$. We recall that in section \ref{Brief review} of the main body of the present work we referred to these peculiar MZV's as ``$non \ trivial$ MZV's''.\\

\subsubsection{Structure of the $\alpha'$ expansion of the $N$-point momentum factors ($N \geq 5$) }

\label{Structure}

\noindent Following \cite{Schlotterer1}, we wil first consider the case $N=5$.\\
\noindent From eq. (\ref{A12345-3}) we can write that
\begin{eqnarray}
\label{A12345-2}
A_b(1,2,3,4,5) & = &  F^{\{23\}}(\alpha')  \ A_{YM}(1,2,3,4,5) +    F^{\{32\}}(\alpha')  \ A_{YM}(1,3,2,4,5) \ , \\
\label{A13245}
A_b(1,3,2,4,5) & = &  \tilde{F}^{\{23\}}(\alpha')  \ A_{YM}(1,3,2,4,5) +    \tilde{F}^{\{32\}}(\alpha')  \ A_{YM}(1,2,3,4,5) \ ,
\end{eqnarray}
where eq.(\ref{A12345-2}) is literally the same one written in (\ref{A12345-3}) and eq.(\ref{A13245}) has been obtained from the first one by interchanging indexes $2$ and $3$, therefore
\begin{eqnarray}
\tilde{F}^{\{23\}}(\alpha') =  F^{\{23\}}(\alpha')|_{2 \leftrightarrow 3} \ , \hspace{0.5cm} \tilde{F}^{\{32\}}(\alpha') =  F^{\{32\}}(\alpha')|_{2 \leftrightarrow 3} \ .
\label{interchange}
\end{eqnarray}
\noindent In matrix notation, eqs.(\ref{A12345-2}) and (\ref{A13245}) can be written as
\begin{eqnarray}
\label{relatedAs}
\vec{A} & = & \Bbb{F} \ \vec{A}_{YM} \ ,
\end{eqnarray}
where $\vec{A}$ and $\vec{A}_{YM}$ are the 2-component vectors given by
\begin{eqnarray}
\vec{A} = \left( \begin{array}{c}
                A_b(1,2,3,4,5) \\
                A_b(1,3,2,4,5)
               \end{array}
      \right) \ \ , \ \
\vec{A}_{YM} = \left( \begin{array}{c}
                A_{YM}(1,2,3,4,5) \\
                A_{YM}(1,3,2,4,5)
               \end{array}
      \right) \ \ ,
\label{As}
\end{eqnarray}
and where $\Bbb{F}$ is a $2 \times 2$ matrix given by
\begin{eqnarray}
\label{matrixF}
\Bbb{F} = {\left( \begin{array}{cc}
                                         F^{\{23\}}(\alpha') \ & \ F^{\{32\}}(\alpha') \\
                                        \tilde{F}^{\{32\}}(\alpha') \ & \ \tilde{F}^{\{23\}}(\alpha')
                                         \end{array}
                                \right)}  \ .
\end{eqnarray}
\noindent In ref.\cite{Schlotterer1} the authors propose that the $\alpha'$ expansion of $\Bbb{F}$ can be decomposed in the following way\footnote{We will not explain here what is the prescription for the ordering colons,`: ... :', in eq.(\ref{F}). The interested reader can find this detail in eq.(3.18) of ref \cite{Schlotterer1}.}:
\begin{eqnarray}
\Bbb{F} &=& \Big\{   \Bbb{I} + \sum_{n=1}^{\infty} \ \Bbb{P}_{2n} \  \zeta(2)^n {\alpha'}^{2n} \Big\} \ \Bbb{Q} \ : \mbox{exp} \Big\{ \sum_{n=1}^{\infty}  \Bbb{M}_{2n+1} \ \zeta(2n+1) {\alpha'}^{2n+1} \Big\} : \ ,
\label{F}
\end{eqnarray}
where the $\Bbb{P}_{2n}$'s, $\Bbb{Q}$ and the $\Bbb{M}_{2n+1}$'s are also $2 \times 2$ matrices\footnote{In eqs.(\ref{F}) and (\ref{matrixQ}) $\Bbb{I}$ denotes de $2 \times 2$ identity matrix.}, given by
\begin{eqnarray}
\label{matrixP}
\Bbb{P}_{2n} & = & \Bbb{F}\big|_{\zeta(2n) {\alpha'}^{2n}}  \ , \\
\label{matrixQ}
\Bbb{Q} & = & \Bbb{I} + \sum_{n=8}^{\infty} \Bbb{Q}_n {\alpha'}^n \ , \\
\label{matrixM}
\Bbb{M}_{2n+1} & = & \Bbb{F}\big|_{\zeta(2n+1) {\alpha'}^{2n+1}}  \ .
\end{eqnarray}
So, from eqs.(\ref{matrixP}) and (\ref{matrixM}) we see that the $\Bbb{P}_{2n}$'s and the $\Bbb{M}_{2n+1}$'s matrices contain the dependence of $\Bbb{F}$ in the $5$-point Mandelstam variables that we have specified in eq. (\ref{Mandelstam5}).\\
In (\ref{matrixQ}) the $\Bbb{Q}_n$'s are $2 \times 2$ matrices given in terms of $w=n$ MZV's and commutators of the $\Bbb{M}_{2r+1}$ matrices:
\begin{eqnarray}
\label{Qns}
\Bbb{Q}_8 = \frac{1}{5} \ \zeta(3,5) \ \big[ \Bbb{M}_5 , \Bbb{M}_3 \big] \ \ \ , \ \ \Bbb{Q}_9 = \Bbb{O} \ \ \ , \ \ \Bbb{Q}_{10} = \Big\{ \ \frac{3}{14} \zeta(5)^2 + \frac{1}{14} \zeta(3,7) \Big\}\ \big[ \Bbb{M}_7 , \Bbb{M}_3 \big] \ \ , \ \mbox{etc.} \hspace{1cm}
\end{eqnarray}
In section {\it 3.2} of \cite{Schlotterer1} the authors give explicit expressions for $\Bbb{Q}_{11}$, $\Bbb{Q}_{12}$, \ldots, $\Bbb{Q}_{16}$. Higher weight $non \ trivial$ MZV's (like $\zeta(3,3,5)$, $\zeta(3,9)$, $\zeta(1,1,4,6)$, and many others) show up on these $\Bbb{Q}_n$'s.\\
So we see that, in the conjectured relation in (\ref{F}), the $non \ trivial$ MZV's are $all$ contained in matrix $\Bbb{Q}$ and that, due to expression (\ref{matrixQ}), these $non \ trivial$ MZV's arise, for the first time, only at weight $w=8$.\\
\noindent The authors of \cite{Schlotterer1} have tested their conjecture in (\ref{F}) up to ${\alpha'}^{16}$ order. The authors of \cite{Broedel2} have gone further, having checked this conjecture up to ${\alpha'}^{21}$ order.\\
\noindent Now, for the case of $N>5$, the authors of \cite{Schlotterer1} conjectured that the $(N-3)! \times (N-3)!$ generalization of matrix $\Bbb{F}$ \footnote{It is clear from MSS formula in eq.(\ref{MSS}), that the matrix generalization of (\ref{relatedAs}) will contain an $(N-3)! \times (N-3)!$ matrix $\Bbb{F}$ and $(N-3)!$-component vectors $\vec{A}$ and $\vec{A}_{YM}$.} in (\ref{matrixF}), satisfies exactly \underline{the same} conjecture in eq.(\ref{F}), but for the corresponding $\Bbb{P}_{2n}$, $\Bbb{Q}$ and $\Bbb{M}_{2n+1}$ matrices. The definitions in (\ref{matrixP}), (\ref{matrixQ}) and (\ref{matrixM}) will now lead to $(N-3)! \times (N-3)!$ matrix expressions for them, and these expressions now depend on the $N$-point Mandelstam variables (because $\Bbb{F}$ does so).\\
\noindent The authors of \cite{Broedel2} have tested this last conjecture up to ${\alpha'}^{9}$ order for $N=6$ and up to ${\alpha'}^{7}$ order for $N=7$.\\

\subsection{Unitarity relations for the momentum factors}

\label{Unitarity relations}

\subsubsection{Case of $N=5$}

\label{Unitarity-N5}

\noindent Here we give the details of the proof of the relation between the $5$ and the $4$-point momentum factors, $F^{\{23\}}(\alpha')$ and $F^{\{2\}}(\alpha')$, respectively (see eq.(\ref{factorization5p1})).\\
\noindent In the $N=5$ case there are five independent Mandelstam variables (which are identified with the poles of the amplitude) and, therefore, there are five unitarity relations like the one in eq.(\ref{collinear}). We will give the details of the unitarity relation of $A_b(1,2,3,4,5)$ with respect to its pole $\alpha_{12}$.\\
\noindent According to eq.(\ref{collinear}), when $\alpha_{12} \rightarrow 0$, in this case we have that
\begin{eqnarray}
\label{collinear-5p-st}
A(  \zeta_1, k_1; \ \zeta_2, k_2; \ \ldots ; \ \zeta_5, k_5  ) \ &\sim &
\frac{1}{ \alpha_{12}}\  \ V^{\mu }_{(12)}  \ \frac{\partial }{\partial \zeta
^{\mu }}A(\zeta ,k_{1}+k_{2};\ \zeta_3,k_{3}; \ \zeta_4,k_{4}  ; \ \zeta_{5},k_{5})\ ,  \\
\label{collinear-5p-ym}
A_{YM}(  \zeta_1, k_1; \ \zeta_2, k_2; \ \ldots ; \ \zeta_5, k_5  ) &\sim &
\frac{1}{ \alpha_{12}}\  \ V^{\mu }_{(12)}  \ \frac{\partial }{\partial \zeta
^{\mu }}A_{YM}(\zeta ,k_{1}+k_{2};\ \zeta_3,k_{3}; \ \zeta_4,k_{4}  ; \ \zeta_{5},k_{5})\ ,  \hspace{0.5cm}
\end{eqnarray}
where the expression for $V^{\mu }_{(12)}$ is given in eq.(\ref{V12}). We have explicited the case of Yang-Mills subamplitudes in (\ref{collinear-5p-ym}) because we will soon need this relation.\\
\noindent It is understood that the limit $\alpha_{12} \rightarrow 0$ also is being taken in the $4$-point subamplitudes on the right hand-side of eqs.(\ref{collinear-5p-st}) and (\ref{collinear-5p-ym}).\\
\noindent The proof of (\ref{factorization5p1}) goes as follows. Using eq.(\ref{A1234-3}), we may write the $4$-point amplitude of the right hand-side of (\ref{collinear-5p-st}) as
\begin{eqnarray}
A(\zeta ,k_{1}+k_{2};\ \zeta_3,k_{3}; \ \zeta_4,k_{4}  ; \ \zeta_{5},k_{5}) & = & F^{\{2\}}[ \ 2 (k_1+k_2) \cdot k_3, \ 2 k_3 \cdot k_4; \ \alpha ^{\prime }] \ \times \nonumber \\
&& \hspace{1.5cm} A_{YM}(\zeta ,k_{1}+k_{2};\ \zeta_3,k_{3}; \ \zeta_4,k_{4}  ; \ \zeta_{5},k_{5}) \ , \nonumber \\
\label{A4pzeta}
\end{eqnarray}
where, after using eqs.(\ref{Mandelstam5}) and (\ref{Mandelstam5-2}), we may prove that $2(k_1+k_2) \cdot k_3 = 2 (k_4 \cdot k_5 - k_1 \cdot k_2)$, and since we are considering the limit $\alpha_{12} \rightarrow 0$, we have that
\begin{eqnarray}
2(k_1+k_2) \cdot k_3 \rightarrow \alpha_{45}
\label{limit}
\end{eqnarray}
and  in (\ref{A4pzeta}) we are left with
\begin{eqnarray}
A(\zeta ,k_{1}+k_{2};\ \zeta_3,k_{3}; \ \zeta_4,k_{4}  ; \ \zeta_{5},k_{5}) & = & F^{\{2\}}[ \ \alpha_{45}, \ \alpha_{34}; \ \alpha ^{\prime }] \  A_{YM}(\zeta ,k_{1}+k_{2};\ \zeta_3,k_{3}; \ \zeta_4,k_{4}  ; \ \zeta_{5},k_{5}) \ . \nonumber \\
\label{useful}
\end{eqnarray}
Substituing (\ref{useful}) in (\ref{collinear-5p-st}) we have that
\begin{eqnarray}
A(  \zeta_1, k_1; \ \zeta_2, k_2; \ \ldots ; \ \zeta_5, k_5  ) \ &\sim &  F^{\{2\}}[ \ \alpha_{45}, \ \alpha_{34}; \ \alpha ^{\prime }] \times  \\
&& \hspace{1cm} \Big\{ \frac{1}{ \alpha_{12}}\  \ V^{\mu }_{(12)}  \ \frac{\partial }{\partial \zeta
^{\mu }}A_{YM}(\zeta ,k_{1}+k_{2};\ \zeta_3,k_{3}; \ \zeta_4,k_{4}  ; \ \zeta_{5},k_{5})  \Big\}  \ , \nonumber
\label{collinear-5p-st-2a}
\end{eqnarray}
and identifying the term in the curly brackets (in the $\alpha_{12} \rightarrow 0$ limit, which is being taken) as the Yang-Mills $5$-point subamplitude (see eq.(\ref{collinear-5p-ym})) we finally arrive at
\begin{eqnarray}
A(  \zeta_1, k_1; \ \zeta_2, k_2; \ \ldots ; \ \zeta_5, k_5  ) \ &\sim &  F^{\{2\}}[ \ \alpha_{45}, \ \alpha_{34}; \ \alpha ^{\prime }] \ A_{YM}(  \zeta_1, k_1; \ \zeta_2, k_2; \ \ldots ; \ \zeta_5, k_5  )
   \ ,
\label{collinear-5p-st-2b}
\end{eqnarray}
or even using the simpler notation,
\begin{eqnarray}
A_b(1,2,3,4,5) \ &\sim &  F^{\{2\}}[ \ \alpha_{45}, \ \alpha_{34}; \ \alpha ^{\prime }] \ A_{YM}(1,2,3,4,5 ) \ .
\label{collinear-5p-st-3}
\end{eqnarray}
\noindent This relation is to be compared with the one in (\ref{A12345-7}). Taking there the $\alpha_{12} \rightarrow 0$ limit, the leading divergent term in the right hand-side comes from $F^{\{23\}}(\alpha')A_{YM}(1,2,3,4,5 )$\footnote{The other term, $F^{\{32\}}(\alpha') A_{YM}(1,3,2,4,5)$, remains finite when this limit is taken because it does not have any pole at $\alpha_{12}=0$.}. Comparing this leading term with the one in (\ref{collinear-5p-st-3}) we finally arrive at eq.(\ref{factorization5p1}).\\

\vspace{0.5cm}

\subsubsection{Case of $N=6$}

\label{Unitarity-N6}

\noindent Here we give the details of the proof of the relation between two of the $6$ and the two $5$-point momentum factors, $\{ F^{\{234\}}(\alpha')$, $ F^{\{234\}}(\alpha')\}$ and $\{F^{\{23\}}(\alpha')$, $F^{\{32\}}(\alpha')\}$, respectively (see eqs.(\ref{fact-F234}) and (\ref{fact-F243})). We also give here the details of the proof of the relation between two of the $6$ and the $4$-point momentum factors, $\{ F^{\{234\}}(\alpha')$, $ F^{\{324\}}(\alpha')\}$ and $F^{\{2\}}(\alpha')$, respectively (see eq.(\ref{fact-mixed})).\\
\noindent In the $N=6$ case there are nine independent Mandelstam variables (which are identified with the poles of the amplitude). Six of these variables (the $\alpha_{ij}$'s) have to do with the poles coming from two adjacent legs and the other three Mandelstam variables (the $\beta_{ijk}$'s) have to do with the poles that come from three adjacent legs. Therefore, the first six unitarity relations are like the one in eq.(\ref{collinear}) and the remaining three ones are like the one in eq.(\ref{unitarity2}), with $m=4$. \\
\noindent Here we will give the details of the unitarity relations of $A_b(1,2,3,4,5,6)$ with respect to its $\alpha_{12}$ pole and of the one with respect to its $\beta_{123}$ pole.\\

\noindent \underline{{\bf i)} Case of $\alpha_{12} \rightarrow 0$}:\\

\noindent According to eq.(\ref{collinear}), when $\alpha_{12} \rightarrow 0$, in this case we have that\footnote{For reasons of space in the writing, we are using two different notations in both sides eqs.(\ref{collinear-6p-st}), (\ref{collinear-6p-ym1}) and (\ref{collinear-6p-ym2}): in the left hand-side we are using $A(1,2, \ldots, N)$ to denote $A(  \zeta_1, k_1; \ \zeta_2, k_2; \ \ldots ; \ \zeta_N, k_N  )$.}
\begin{eqnarray}
\label{collinear-6p-st}
A_b(1,2,3,4,5,6) \ &\sim &
\frac{1}{ \alpha_{12}}\  \ V^{\mu }_{(12)}  \ \frac{\partial }{\partial \zeta
^{\mu }}A(\zeta ,k_{1}+k_{2};\ \zeta_3,k_{3}; \ \zeta_4,k_{4}; \ \zeta_5,k_{5}; \ \zeta_{6},k_{6})\ ,  \\
\label{collinear-6p-ym1}
A_{YM}(1,2,3,4,5,6) &\sim &
\frac{1}{ \alpha_{12}}\  \ V^{\mu }_{(12)}  \ \frac{\partial }{\partial \zeta
^{\mu }}A_{YM}(\zeta ,k_{1}+k_{2};\ \zeta_3,k_{3}; \ \zeta_4,k_{4}; \ \zeta_5,k_{5}; \ \zeta_{6},k_{6})  \ ,  \hspace{0.5cm} \\
\label{collinear-6p-ym2}
A_{YM}(1,2,4,3,5,6) &\sim &
\frac{1}{ \alpha_{12}}\  \ V^{\mu }_{(12)}  \ \frac{\partial }{\partial \zeta
^{\mu }}A_{YM}(\zeta ,k_{1}+k_{2};\ \zeta_4,k_{4}; \ \zeta_3,k_{3}; \ \zeta_5,k_{5}; \ \zeta_{6},k_{6})  \ ,  \hspace{0.5cm}
\end{eqnarray}
where the expression for $V^{\mu }_{(12)}$ is given in eq.(\ref{V12}). We have explicited the case of Yang-Mills subamplitudes, in (\ref{collinear-6p-ym1}) and (\ref{collinear-6p-ym2}), because we will soon need these relations.\\
\noindent It is understood that the limit $\alpha_{12} \rightarrow 0$ also is being taken in the $5$-point subamplitudes on the right hand-side of eqs.(\ref{collinear-6p-st}), (\ref{collinear-6p-ym1}) and  (\ref{collinear-6p-ym2}).\\
\noindent Using eq.(\ref{A12345-3}), we may write the $5$-point amplitude of the right hand-side of (\ref{collinear-6p-st}) as
\begin{multline}
A(\zeta ,k_{1}+k_{2};\ \zeta_3,k_{3}; \ \ldots  ; \ \zeta_{6},k_{6}) = \\
\begin{split}
&= F^{\{23\}}[ \ 2 (k_1+k_2) \cdot k_3, \ 2 k_3 \cdot k_4,  \ 2 k_4 \cdot k_5,  \ 2 k_5 \cdot k_6, 2 k_6 \cdot (k_1+k_2); \ \alpha ^{\prime }]  \times  \\
& \ \ \ A_{YM}(\zeta ,k_{1}+k_{2};\ \zeta_3,k_{3}; \ \zeta_4,k_{4}  ; \ \zeta_{5},k_{5};   \ \zeta_{6},k_{6} ) \ + \\
& F^{\{32\}}[ \ 2 (k_1+k_2) \cdot k_4, \ 2 k_3 \cdot k_4,  \ 2 k_3 \cdot k_5,  \ 2 k_5 \cdot k_6, \ 2 k_6 \cdot (k_1+k_2); \ \alpha ^{\prime }]  \times  \\
& \ \ \ A_{YM}(\zeta ,k_{1}+k_{2};\ \zeta_4,k_{4}; \ \zeta_3,k_{3}  ; \ \zeta_{5},k_{5};   \ \zeta_{6},k_{6} ) \ ,
\end{split}
\label{A5pzeta}
\end{multline}
where, after using eqs.(\ref{Mandelstam6}) and (\ref{Mandelstam6-2}), we may prove that $2(k_1+k_2) \cdot k_3 = \beta_{123}-\alpha_{12}$ and also $2 k_6 \cdot (k_1+k_2)= \beta_{345}-\alpha_{12}$, and since we are considering the limit $\alpha_{12} \rightarrow 0$, we have that
\begin{eqnarray}
2(k_1+k_2) \cdot k_3 \rightarrow \beta_{123} \ , \ \ \ \ \ 2 k_6 \cdot (k_1+k_2)  \rightarrow \beta_{345} \ .
\label{limit2}
\end{eqnarray}
\noindent Also using eqs.(\ref{Mandelstam6}) and (\ref{Mandelstam6-2}), we may prove that $2(k_1+k_2) \cdot k_4 = \beta_{124} - \alpha_{12} $ and, therefore, in the limit $\alpha_{12} \rightarrow 0$,
\begin{eqnarray}
2(k_1+k_2) \cdot k_4 \rightarrow \beta_{124}  \ .
\label{limit3}
\end{eqnarray}
\noindent So, substituing (\ref{limit2}) and (\ref{limit3}) in (\ref{A5pzeta}) we are left with\footnote{In passing from eq.(\ref{A5pzeta}) to eq.(\ref{useful2}) we have gone back to the notations (\ref{alphaij}) and (\ref{tijk}) for the momentum invariants.}
\begin{multline}
A(\zeta ,k_{1}+k_{2};\ \zeta_3,k_{3}; \ \ldots  ; \ \zeta_{6},k_{6}) =    \\
\begin{split}
&= F^{\{23\}}[ \ \beta_{123}, \ \alpha_{34},  \ \alpha_{45},  \ \alpha_{56}, \ \beta_{345}; \ \alpha ^{\prime }]  \ A_{YM}(\zeta ,k_{1}+k_{2};\ \zeta_3,k_{3}; \ \zeta_4,k_{4}  ; \ \zeta_{5},k_{5};   \ \zeta_{6},k_{6} ) \ + \\
& \ \ \ \ F^{\{32\}}[ \  \beta_{124}, \ \alpha_{34},  \  \alpha_{35},  \ \alpha_{56}, \ \beta_{345} ; \ \alpha ^{\prime }] \ A_{YM}(\zeta ,k_{1}+k_{2};\ \zeta_4,k_{4}; \ \zeta_3,k_{3}  ; \ \zeta_{5},k_{5};   \ \zeta_{6},k_{6} ) \ .
\end{split}
\label{useful2}
\end{multline}
\noindent Substituing (\ref{useful2}) in (\ref{collinear-6p-st}) we have that
\begin{eqnarray}
A(1,2,3,4,5,6) \ &\sim & F^{\{23\}}[ \ \beta_{123}, \ \alpha_{34},  \ \alpha_{45},  \ \alpha_{56}, \ \beta_{345}; \ \alpha ^{\prime }]  \times \nonumber  \\
&& \hspace{0cm} \Big\{ \frac{1}{ \alpha_{12}}\  \ V^{\mu }_{(12)}  \ \frac{\partial }{\partial \zeta
^{\mu }}A_{YM}(\zeta ,k_{1}+k_{2};\ \zeta_3,k_{3}; \ \zeta_4,k_{4}  ; \ \zeta_{5},k_{5} ; \ \zeta_{6},k_{6})  \Big\}  + \nonumber \\
 & &  F^{\{32\}}[ \  \beta_{124}, \ \alpha_{34},  \  \alpha_{35},  \ \alpha_{56}, \ \beta_{345} ; \ \alpha ^{\prime }] \times  \nonumber \\
&& \hspace{0cm} \Big\{ \frac{1}{ \alpha_{12}}\  \ V^{\mu }_{(12)}  \ \frac{\partial }{\partial \zeta
^{\mu }}A_{YM}(\zeta ,k_{1}+k_{2};\ \zeta_4,k_{4}; \ \zeta_3,k_{3}  ; \ \zeta_{5},k_{5} ; \ \zeta_{6},k_{6})  \Big\}  \ , \nonumber \\
\label{collinear-6p-st-2}
\end{eqnarray}
and identifying the terms in the curly brackets (in the $\alpha_{12} \rightarrow 0$ limit, which is being taken) as the Yang-Mills $6$-point subamplitudes (see eqs.(\ref{collinear-6p-ym1}) and (\ref{collinear-6p-ym2})) we finally arrive at
\begin{eqnarray}
A(1,2,3,4,5,6 ) \sim   \hspace{11.5cm}      \nonumber \\
  F^{\{23\}}[ \ \beta_{123}, \ \alpha_{34},  \ \alpha_{45},  \ \alpha_{56}, \ \beta_{345}; \ \alpha ^{\prime }] \ A_{YM}(  \zeta_1, k_1; \ \zeta_2, k_2; \ \zeta_3, k_3; \ \zeta_4, k_4;  \ \zeta_5, k_5 ;  \ \zeta_6, k_6 ) + \nonumber \\
  F^{\{32\}}[ \  \beta_{124}, \ \alpha_{34},  \  \alpha_{35},  \ \alpha_{56}, \ \beta_{345} ; \ \alpha ^{\prime }]  \ A_{YM}(  \zeta_1, k_1; \ \zeta_2, k_2; \ \zeta_4, k_4; \ \zeta_3, k_3;  \ \zeta_5, k_5;  \ \zeta_6, k_6   ) \ ,  \hspace{0.15cm}  \nonumber \\
\label{collinear-6p-st-3}
\end{eqnarray}
or even using the simpler notation,
\begin{eqnarray}
A(1,2,3,4,5,6 ) \ &\sim &  F^{\{23\}}[ \ \beta_{123}, \ \alpha_{34},  \ \alpha_{45},  \ \alpha_{56}, \ \beta_{345}; \ \alpha ^{\prime }]
\ A_{YM}(1,2,3,4,5,6) + \nonumber \\
 && F^{\{32\}}[ \  \beta_{124}, \ \alpha_{34},  \  \alpha_{35},  \ \alpha_{56}, \ \beta_{345} ; \ \alpha ^{\prime }]  \ A_{YM}( 1,2,4,3,5,6 ) \ .
\label{collinear-6p-st-4}
\end{eqnarray}
\noindent This relation is to be compared with the one in (\ref{6point-aux}). Taking there the $\alpha_{12} \rightarrow 0$ limit, there are two leading divergent terms in the right hand-side: they come from $F^{\{234\}}(\alpha')A_{YM}(1,2,3,4,5,6 )$ and $F^{\{243\}}(\alpha')A_{YM}(1,2,4,3,5,6 )$\footnote{The other four terms remain finite when this limit is taken because they do not have any pole at $\alpha_{12}=0$.}. Comparing these leading terms with the ones in (\ref{collinear-6p-st-4}) we finally arrive at eqs.(\ref{fact-F234}) and (\ref{fact-F243}).\\

\noindent \underline{{\bf ii)} Case of $\beta_{123} \rightarrow 0$}:\\

\noindent According to eq.(\ref{unitarity2}) for $m=4$, when $\beta_{123} \rightarrow 0$ in the $N=6$ case we have that\footnote{For reasons of space in the writing, we are using two different notations in both sides eqs.(\ref{collinear-6p-st}), (\ref{collinear-6p-ym1}) and (\ref{collinear-6p-ym2}): in the left hand-side we are using $A(1,2, \ldots, N)$ to denote $A(  \zeta_1, k_1; \ \zeta_2, k_2; \ \ldots ; \ \zeta_N, k_N  )$.}
\begin{eqnarray}
\label{trilinear-6p-st}
A_b(1,2,3,4,5,6) \ &\sim &
\frac{1}{ \beta_{123}} \ \frac{\partial }{\partial \zeta
_{\mu }}A(\zeta_1,k_{1}; \ \zeta_2,k_{2}; \ \zeta_3,k_{3} ;\zeta, k ) \ \frac{\partial }{\partial \zeta
^{\mu }} A(\zeta , -k ;\ \zeta_4,k_{4}; \ \zeta_5,k_{5}; \ \zeta_{6},k_{6})  \ ,  \nonumber  \\
\end{eqnarray}
\begin{eqnarray}
A_{YM}(1,2,3,4,5,6) \sim   \hspace{11.5cm}    \\
\sim \frac{1}{ \beta_{123}} \ \frac{\partial }{\partial \zeta
_{\mu }}A_{YM}(\zeta_1,k_{1}; \ \zeta_2,k_{2}; \ \zeta_3,k_{3} ;\zeta, k ) \ \frac{\partial }{\partial \zeta
^{\mu }} A_{YM}(\zeta , -k ;\ \zeta_4,k_{4}; \ \zeta_5,k_{5}; \ \zeta_{6},k_{6})  \ ,  \nonumber
\label{trilinear-6p-ym}
\end{eqnarray}
where we have explicited the case of the Yang-Mills subamplitude in (\ref{trilinear-6p-ym}), because we will soon need it.\\
\noindent It is understood that the limit $\beta_{123} \rightarrow 0$ also is being taken in the $4$-point subamplitudes on the right hand-side of eqs.(\ref{trilinear-6p-st}) and (\ref{trilinear-6p-ym}).\\
\noindent In relations (\ref{trilinear-6p-st}) and (\ref{trilinear-6p-ym}) the momentum $k^{\mu}$ is given by
\begin{eqnarray}
k^{\mu} = - (k_1 + k_2 + k_3)^{\mu} = (k_4 + k_5 + k_6)^{\mu} \ .
\label{kmu}
\end{eqnarray}
\noindent Using eq.(\ref{A1234-3}), we may write the $4$-point subamplitudes of the right hand-side of (\ref{trilinear-6p-st}) as
\begin{eqnarray}
\label{A4p-11}
A(\zeta_1,k_{1}; \ \zeta_2,k_{2}; \ \zeta_3,k_{3} ;\zeta, k ) & = & F^{\{2\}}[\alpha_{12}, \alpha_{23}; \ \alpha' ] \  A_{YM}(\zeta_1,k_{1}; \ \zeta_2,k_{2}; \ \zeta_3,k_{3} ;\zeta, k ) \ , \\
\label{A4p-12}
A(\zeta , -k ;\ \zeta_4,k_{4}; \ \zeta_5,k_{5}; \ \zeta_{6},k_{6}) & = & F^{\{2\}}[ - 2 k \cdot k_4, \alpha_{45}; \ \alpha' ] \  A_{YM}(\zeta , -k ;\ \zeta_4,k_{4}; \ \zeta_5,k_{5}; \ \zeta_{6},k_{6}) \ , \hspace{1cm}
\end{eqnarray}
where $k^{\mu}$ is given by eq.(\ref{kmu}). \\
\noindent It may be proved,  using eqs.(\ref{Mandelstam6}) and (\ref{Mandelstam6-2}), that $-2 k \cdot k_4= \alpha_{56}-\beta_{123}$, and since we are considering the limit $\beta_{123} \rightarrow 0$, we have that
\begin{eqnarray}
-2 k \cdot k_4 \rightarrow \alpha_{56} \ .
\label{limit4}
\end{eqnarray}
\noindent So, in (\ref{A4p-12}) we have that
\begin{eqnarray}
\label{A4p-12new}
A(\zeta , -k ;\ \zeta_4,k_{4}; \ \zeta_5,k_{5}; \ \zeta_{6},k_{6}) & = & F^{\{2\}}[  \alpha_{56}, \alpha_{45}; \ \alpha' ] \  A_{YM}(\zeta , -k ;\ \zeta_4,k_{4}; \ \zeta_5,k_{5}; \ \zeta_{6},k_{6}) \ . \hspace{1cm}
\end{eqnarray}
Substituing (\ref{A4p-11}) and (\ref{A4p-12new}) in (\ref{trilinear-6p-st}), it becomes
\begin{multline}
A_b(1,2,3,4,5,6) \sim  F^{\{2\}}[  \alpha_{12}, \alpha_{23}; \ \alpha' ]  \ F^{\{2\}}[  \alpha_{56}, \alpha_{45}; \ \alpha' ] \times   \\
\begin{split}
 \hspace{1cm} \biggl\{  \frac{1}{ \beta_{123}} \ \frac{\partial }{\partial \zeta
_{\mu }}A_{YM}(\zeta_1,k_{1}; \ \zeta_2,k_{2}; \ \zeta_3,k_{3} ;\zeta, k ) \ \frac{\partial }{\partial \zeta
^{\mu }} A_{YM}(\zeta , -k ;\ \zeta_4,k_{4}; \ \zeta_5,k_{5}; \ \zeta_{6},k_{6})  \biggr\}  \ ,
\end{split}
\label{trilinear-6p-st-2}
\end{multline}
and identifying the terms in the curly brackets (in the $\beta_{123} \rightarrow 0$ limit, which is being taken) as the Yang-Mills $6$-point subamplitude (see eq.(\ref{trilinear-6p-ym})) we arrive at
\begin{eqnarray}
A_b(1,2,3,4,5,6) \sim  F^{\{2\}}[  \alpha_{12}, \alpha_{23}; \ \alpha' ]  \ F^{\{2\}}[  \alpha_{56}, \alpha_{45}; \ \alpha' ] \ A_{YM}(1,2,3,4,5,6) \ .
\label{trilinear-6p-st-3}
\end{eqnarray}
\noindent On the other side, taking the $\beta_{123} \rightarrow 0$ limit in (\ref{6point-aux}), we have that
\begin{eqnarray}
A_b(1,2,3,4,5,6)  \sim  F^{\{234\}}(\alpha')  A_{YM}(1,2,3,4,5,6) +  F^{\{324\}}(\alpha')  A_{YM}(1,3,2,4,5,6)   \ ,
\label{6point-aux2}
\end{eqnarray}
because, according to the unitarity relation, eq.(\ref{trilinear-6p-ym}), when $\beta_{123} \rightarrow 0$ $A_{YM}(1,2,3,4,5,6)$ and $A_{YM}(1,3,2,4,5,6)$ are the only divergent terms in (\ref{6point-aux}). In fact, according to (\ref{unitarity2}) for $m=4$, the unitarity relation for $A_{YM}(1,3,2,4,5,6)$, when $\beta_{123} \rightarrow 0$, is given by
\begin{multline}
A_{YM}(1,3,2,4,5,6) \sim \\
\begin{split}
\sim \frac{1}{ \beta_{123}} \ \frac{\partial }{\partial \zeta
_{\mu }}A_{YM}(\zeta_1,k_{1}; \ \zeta_3,k_{3}; \ \zeta_2,k_{2} ;\zeta, k ) \ \frac{\partial }{\partial \zeta
^{\mu }} A_{YM}(\zeta , -k ;\ \zeta_4,k_{4}; \ \zeta_5,k_{5}; \ \zeta_{6},k_{6})  \ ,
\end{split}
\label{trilinear-6p-ym2}
\end{multline}
where $k^{\mu}$ is given by eq.(\ref{kmu}).\\
\noindent Now, from the $N=4$ BCJ relations we have that
\begin{eqnarray}
A_{YM}(\zeta_1,k_{1}; \ \zeta_3,k_{3}; \ \zeta_2,k_{2} ;\zeta, k ) & = & - \frac{\alpha_{12}}{\alpha_{12}+\alpha_{23}} A_{YM}(\zeta_1,k_{1}; \ \zeta_2,k_{2}; \ \zeta_3,k_{3} ;\zeta, k ) \ ,
\label{trilinear-6p-ym3}
\end{eqnarray}
\noindent so, in (\ref{trilinear-6p-ym2}), it becomes
\begin{eqnarray}
A_{YM}(1,3,2,4,5,6) \sim  \hspace{11.2cm}  \nonumber \\
 - \frac{\alpha_{12}}{\alpha_{12}+\alpha_{23}} \  \biggl\{  \frac{1}{ \beta_{123}} \ \frac{\partial }{\partial \zeta
_{\mu }}A_{YM}(\zeta_1,k_{1}; \ \zeta_2,k_{2}; \ \zeta_3,k_{3} ;\zeta, k ) \ \frac{\partial }{\partial \zeta
^{\mu }} A_{YM}(\zeta , -k ;\ \zeta_4,k_{4}; \ \zeta_5,k_{5}; \ \zeta_{6},k_{6}) \biggr\} \  . \nonumber \\
\label{trilinear-6p-ym4a}
\end{eqnarray}
\noindent Identifying the expression in the curly brackets with the leading behaviour of $A_{YM}(1,3,2,4,5,6)$ (see eq.(\ref{trilinear-6p-ym2})), we have that
\begin{eqnarray}
A_{YM}(1,3,2,4,5,6) \sim  - \frac{\alpha_{12}}{\alpha_{12}+\alpha_{23}} A_{YM}(1,2,3,4,5,6) \ .
\label{trilinear-6p-ym4}
\end{eqnarray}
\noindent Substituing (\ref{trilinear-6p-ym4}) in (\ref{6point-aux2}) leads us to
\begin{eqnarray}
A_b(1,2,3,4,5,6)  \sim  \Big[ F^{\{234\}}(\alpha')   - \frac{\alpha_{12}}{\alpha_{12}+\alpha_{23}}  F^{\{324\}}(\alpha') \Big] A_{YM}(1,2,3,4,5,6)     \ .
\label{6point-aux3a}
\end{eqnarray}
\noindent Finally, comparing expressions (\ref{trilinear-6p-st-3}) and (\ref{6point-aux3a}) we conclude, precisely, the desired relation between $F^{\{234\}}(\alpha')$, $F^{\{324\}}(\alpha')$ and $F^{\{2\}}(\alpha')$ (see eq.(\ref{fact-mixed})). \\
\noindent In order to derive conclusions for the $\alpha'$ series of the momentum factors which are involved in relation (\ref{fact-mixed}), one has to write the arguments which in one of them are different (say, $\{\alpha_{13}$, $\alpha_{24}$, $\beta_{245}\}$, in the case of $F^{\{324\}}(\alpha')$) in terms of the nine Mandelstam arguments of the other ($F^{\{324\}}(\alpha')$), taking into account the condition $\beta_{123} \rightarrow 0$. Using relations (\ref{Mandelstam6}) and (\ref{Mandelstam6-2}) it can easily be seen that the expressions for $\alpha_{13}$, $\alpha_{24}$ and $\beta_{245}$ are indeed the ones that we have written in eq.(\ref{fact-mixed}).

\vspace{0.5cm}

\subsubsection{Case of $N=7$}

\label{Unitarity-N7}

\vspace{0.5cm}

\noindent \underline{{\bf i)} Case of $\alpha_{12} \rightarrow 0$}:\\

\noindent It may be proved that demanding the unitarity relation (\ref{collinear}) to be obeyed by $A_b(1,2,3,4,5,6,7)$ implies the following six relations for the $7$-point momentum factors, $F^{\{ \sigma_7 \}}(\alpha')$, in which the $\sigma_7=\{2,{\sigma}_6'\}$, where ${\sigma}_6'$  is a permutation of indices $\{3,4,5\}$:
\begin{multline}
F^{\{2345\}}\left[
\begin{array}{ccccccc}
\scriptstyle0, & \scriptstyle\alpha _{23}, & \scriptstyle\alpha _{34}, & %
\scriptstyle\alpha _{56}, & \scriptstyle\alpha _{56}, & \scriptstyle\alpha
_{17}, & \scriptstyle\alpha _{17} \\
\scriptstyle {\beta}_{123}, & \scriptstyle {\beta}_{234}, & \scriptstyle {\beta}_{345}, & %
\scriptstyle {\beta}_{456}, & \scriptstyle {\beta}_{567}, & \scriptstyle {\beta}_{167}, & %
\scriptstyle {\beta}_{127}%
\label{Unitarity7-F2345}
\end{array}%
;\alpha ^{\prime }\right] = \\
=F^{\{234\}}[{\beta}_{123},\alpha _{34},\alpha _{45},\alpha _{56},\alpha
_{67},{\beta}_{127},{\beta}_{567},{\beta}_{345},{\beta}_{456};\alpha ^{\prime }] \ ,
\end{multline}%
\begin{multline}
F^{\{2354\}}\left[
\begin{array}{ccccccc}
\scriptstyle0, & \scriptstyle\alpha _{23}, & \scriptstyle\alpha _{35}, & %
\scriptstyle\alpha _{45}, & \scriptstyle\alpha _{46}, & \scriptstyle\alpha
_{67}, & \scriptstyle\alpha _{17} \\
\scriptstyle {\beta}_{123}, & \scriptstyle {\beta}_{235}, & \scriptstyle {\beta}_{345}, & %
\scriptstyle {\beta}_{456}, & \scriptstyle {\beta}_{467}, & \scriptstyle {\beta}_{167}, & %
\scriptstyle {\beta}_{127}%
\end{array}%
;\alpha ^{\prime }\right] = \\
=F^{\{243\}}[{\beta}_{123},\alpha _{35},\alpha _{45},\alpha _{46},\alpha
_{67},{\beta}_{127},{\beta}_{467},{\beta}_{345},{\beta}_{456};\alpha ^{\prime }] \ ,
\end{multline}%
\begin{multline}
F^{\{2435\}}\left[
\begin{array}{ccccccc}
\scriptstyle0, & \scriptstyle\alpha _{24}, & \scriptstyle\alpha _{34}, & %
\scriptstyle\alpha _{35}, & \scriptstyle\alpha _{56}, & \scriptstyle\alpha
_{67}, & \scriptstyle\alpha _{17} \\
\scriptstyle {\beta}_{124}, & \scriptstyle {\beta}_{234}, & \scriptstyle {\beta}_{345}, & %
\scriptstyle {\beta}_{356}, & \scriptstyle {\beta}_{567}, & \scriptstyle {\beta}_{167}, & %
\scriptstyle {\beta}_{127}%
\end{array}%
;\alpha ^{\prime }\right] = \\
=F^{\{324\}}[{\beta}_{124},\alpha _{34},\alpha _{35},\alpha _{56},\alpha
_{67},{\beta}_{127},{\beta}_{567},{\beta}_{345},{\beta}_{356};\alpha ^{\prime }] \ ,
\end{multline}%
\begin{multline}
F^{\{2534\}}\left[
\begin{array}{ccccccc}
\scriptstyle0, & \scriptstyle\alpha _{25}, & \scriptstyle\alpha _{35}, & %
\scriptstyle\alpha _{34}, & \scriptstyle\alpha _{46}, & \scriptstyle\alpha
_{67}, & \scriptstyle\alpha _{17} \\
\scriptstyle {\beta}_{125}, & \scriptstyle {\beta}_{235}, & \scriptstyle {\beta}_{345}, & %
\scriptstyle {\beta}_{346}, & \scriptstyle {\beta}_{467}, & \scriptstyle {\beta}_{167}, & %
\scriptstyle {\beta}_{127}%
\end{array}%
;\alpha ^{\prime }\right] = \\
=F^{\{423\}}[{\beta}_{125},\alpha _{35},\alpha _{34},\alpha _{46},\alpha
_{67},{\beta}_{127},{\beta}_{467},{\beta}_{345},{\beta}_{346};\alpha ^{\prime }] \ ,
\end{multline}%
\begin{multline}
F^{\{2453\}}\left[
\begin{array}{ccccccc}
\scriptstyle0, & \scriptstyle\alpha _{24}, & \scriptstyle\alpha _{45}, & %
\scriptstyle\alpha _{35}, & \scriptstyle\alpha _{36}, & \scriptstyle\alpha
_{67}, & \scriptstyle\alpha _{17} \\
\scriptstyle {\beta}_{124}, & \scriptstyle {\beta}_{245}, & \scriptstyle {\beta}_{345}, & %
\scriptstyle {\beta}_{356}, & \scriptstyle {\beta}_{367}, & \scriptstyle {\beta}_{167}, & %
\scriptstyle {\beta}_{127}%
\end{array}%
;\alpha ^{\prime }\right] = \\
=F^{\{342\}}[{\beta}_{124},\alpha _{45},\alpha _{35},\alpha _{36},\alpha
_{67},{\beta}_{127},{\beta}_{367},{\beta}_{345},{\beta}_{356};\alpha ^{\prime }] \ ,
\end{multline}%
\begin{multline}
F^{\{2543\}}\left[
\begin{array}{ccccccc}
\scriptstyle0, & \scriptstyle\alpha _{25}, & \scriptstyle\alpha _{45}, & %
\scriptstyle\alpha _{34}, & \scriptstyle\alpha _{36}, & \scriptstyle\alpha
_{67}, & \scriptstyle\alpha _{17} \\
\scriptstyle {\beta}_{123}, & \scriptstyle {\beta}_{245}, & \scriptstyle {\beta}_{345}, & %
\scriptstyle {\beta}_{346}, & \scriptstyle {\beta}_{367}, & \scriptstyle {\beta}_{167}, & %
\scriptstyle {\beta}_{127}%
\end{array}%
;\alpha ^{\prime }\right] = \\
=F^{\{432\}}[{\beta}_{123},\alpha _{45},\alpha _{34},\alpha _{36},\alpha
_{67},{\beta}_{127},{\beta}_{367},{\beta}_{345},{\beta}_{346};\alpha ^{\prime }] \ .
\label{Unitarity7-F2543}
\end{multline}

\noindent \underline{{\bf ii)} Case of $\beta_{123} \rightarrow 0$}:\\

\noindent It may be proved that demanding the unitarity relation (\ref{}) to be obeyed by $A_b(1,2,3,4,5,6,7)$ implies the following two relations for the $7$-point momentum factors, $\{F^{\{2345\}}(\alpha')$, $F^{\{3245\}}(\alpha')$, $F^{\{2354\}}(\alpha')$, $F^{\{3254\}}(\alpha')\}$,  and the $4$-point and $5$-point momentum factors, $\{F^{\{2\}}(\alpha')$, $F^{\{23\}}(\alpha')$, $F^{\{32\}}(\alpha')\}$:
\begin{multline}
F^{\{2345\}}\left[
\begin{array}{ccccccc}
\scriptstyle {\alpha}_{12}, & \scriptstyle\alpha _{23}, & \scriptstyle\alpha _{34},
& \scriptstyle\alpha _{45}, & \scriptstyle\alpha _{56}, & \scriptstyle\alpha
_{67}, & \scriptstyle\alpha _{17} \\
\scriptstyle0, & \scriptstyle {\beta}_{234}, & \scriptstyle {\beta}_{345}, & \scriptstyle
{\beta}_{456}, & \scriptstyle {\beta}_{567}, & \scriptstyle {\beta}_{167}, & \scriptstyle %
{\beta}_{127}%
\end{array}%
;\alpha ^{\prime }\right] - \\
-\frac{{\alpha}_{12}}{{\alpha}_{12}+\alpha _{23}}F^{\{3245\}}\left[
\begin{array}{ccccccc}
\scriptstyle\alpha _{13}, & \scriptstyle\alpha _{23}, & \scriptstyle\alpha
_{24}, & \scriptstyle\alpha _{45}, & \scriptstyle\alpha _{56}, & \scriptstyle%
\alpha _{67}, & \scriptstyle\alpha _{17} \\
\scriptstyle0, & \scriptstyle {\beta}_{234}, & \scriptstyle {\beta}_{245}, & \scriptstyle
{\beta}_{456}, & \scriptstyle {\beta}_{567}, & \scriptstyle {\beta}_{167}, & \scriptstyle %
{\beta}_{137}%
\end{array}%
;\alpha ^{\prime }\right] = \\
=F^{\{2\}}[{\alpha}_{12},\alpha _{23};\alpha ^{\prime }]F^{\{23\}}[{\beta}_{567},\alpha
_{45},\alpha _{56},\alpha _{67},{\beta}_{127};\alpha ^{\prime }] \ ,
\label{Unitarity7p-1}
\end{multline}%
\begin{multline}
F^{\{2354\}}\left[
\begin{array}{ccccccc}
\scriptstyle {\alpha}_{12}, & \scriptstyle\alpha _{23}, & \scriptstyle\alpha _{35},
& \scriptstyle\alpha _{45}, & \scriptstyle\alpha _{46}, & \scriptstyle\alpha
_{67}, & \scriptstyle\alpha _{17} \\
\scriptstyle0, & \scriptstyle {\beta}_{235}, & \scriptstyle {\beta}_{345}, & \scriptstyle
{\beta}_{456}, & \scriptstyle {\beta}_{467}, & \scriptstyle {\beta}_{167}, & \scriptstyle %
{\beta}_{127}%
\end{array}%
;\alpha ^{\prime }\right] - \\
-\frac{{\alpha}_{12}}{{\alpha}_{12}+\alpha _{23}}F^{\{3254\}}\left[
\begin{array}{ccccccc}
\scriptstyle\alpha _{13}, & \scriptstyle\alpha _{23}, & \scriptstyle\alpha
_{25}, & \scriptstyle\alpha _{45}, & \scriptstyle\alpha _{46}, & \scriptstyle%
\alpha _{67}, & \scriptstyle\alpha _{17} \\
\scriptstyle0, & \scriptstyle {\beta}_{235}, & \scriptstyle {\beta}_{245}, & \scriptstyle
{\beta}_{456}, & \scriptstyle {\beta}_{467}, & \scriptstyle {\beta}_{167}, & \scriptstyle %
{\beta}_{137}%
\end{array}%
;\alpha ^{\prime }\right] = \\
=F^{\{2\}}[{\alpha}_{12},\alpha _{23};\alpha ^{\prime }]F^{\{32\}}[{\beta}_{467},\alpha
_{45},\alpha _{46},\alpha _{67},{\beta}_{456};\alpha ^{\prime }] \ .
\label{Unitarity7p-2}
\end{multline}

\vspace{0.5cm}

\noindent Care must be taken when using relations (\ref{Unitarity7-F2345})-(\ref{Unitarity7-F2543}) and/or relations (\ref{Unitarity7p-1})-(\ref{Unitarity7p-2}) to compute the coefficients of the $\alpha'$ series of the momentum factors. It is understood implicitly in them that all Mandelstam variables are written in terms of the ones for $F^{\{2345\}}(\alpha')$, namely,  $\{\alpha_{12}$, $\alpha_{23}$,  $\alpha_{34}$, $\alpha_{45}$, $\alpha_{56}$, $\alpha_{67}$, $\alpha_{17}$, $\beta_{123}$, $\beta_{234}$,  $\beta_{345}$,  $\beta_{456}$,   $\beta_{567}$,  $\beta_{167}$,  $\beta_{127}\}$ where $\alpha_{12}=0$ or $\beta_{123}=0$, depending on the relation that is being used.

\vspace{0.5cm}

\subsection{Cyclic transformation of the  momentum factors}

\label{Cyclic transformation}

\subsubsection{Case of $N=5$}

\label{Cyclicity-N5}

\noindent Here we will prove that relations (\ref{F23cycl}) and (\ref{F32cycl}) arise when we demand cyclic symmetry to be obeyed by $A_b(1,2,3,4,5)$.\\
\noindent Cyclic invariance of $A_b(1,2,3,4,5)$ means that $A_b(1,2,3,4,5) =A_b(2,3,4,5,1)$, so in (\ref{A12345-7}) cyclic invariance implies that
\begin{eqnarray}
A_b(1,2,3,4,5) &=&F_{cycl}^{\{23\}}(\alpha')A_{YM}(2,3,4,5,1)+F_{cycl}^{\{32\}}(\alpha')A_{YM}(2,4,3,5,1)
 \ , \nonumber \\
&=&F_{cycl}^{\{23\}}(\alpha')A_{YM}(1,2,3,4,5)+F_{cycl}^{\{32\}}(\alpha')A_{YM}(1,2,4,3,5)  \ ,
\label{cyclic5}
\end{eqnarray}%
where in (\ref{cyclic5}) we have used the cyclic property of Yang-Mills subamplitudes.\\
\noindent $F_{cycl}^{\{23\}}(\alpha')$ and $F_{cycl}^{\{32\}}(\alpha')$ denote
doing $\{k_1 \rightarrow k_2$, $k_2 \rightarrow k_3$, $\ldots$, $k_5 \rightarrow k_1\}$ in
$F^{\{23\}}(\alpha')$ and $F^{\{32\}}(\alpha')$, respectively.\\
\noindent Now, the BCJ relation for $A_{YM}(1,2,4,3,5)$ (see eq.(\ref{N512435})) says that
\begin{equation}
A_{YM}(1,2,4,3,5)=\frac{\alpha _{13}+\alpha _{23}}{\alpha _{35}}%
A_{YM}(1,2,3,4,5)+\frac{\alpha _{13}}{\alpha _{35}}A_{YM}(1,3,2,4,5) \ ,
\label{A12435}
\end{equation}
where $\alpha_{13}=k_1 \cdot k_3$ and $\alpha_{35}=k_3 \cdot k_5$ are given in terms of the Mandelstam variables in eq.(\ref{Mandelstam5-2}).\\
\noindent So, substituing (\ref{A12435}) in (\ref{cyclic5}) we have that
\begin{eqnarray}
A_b(1,2,3,4,5) &=&\Big\{   F_{cycl}^{\{23\}}(\alpha ^{\prime })+\frac{\alpha _{13}+\alpha _{23}}{\alpha _{35}}F_{cycl}^{\{32\}}(\alpha ^{\prime })        \Big\} A_{YM}(1,2,3,4,5)+ \nonumber \\
&&   \Big\{      \frac{\alpha _{13}}{\alpha _{35}} F_{cycl}^{\{32\}}(\alpha')       \Big\}        A_{YM}(1,2,3,4,5)
\end{eqnarray}
and comparisson of this last relation with (\ref{A12345-7}) leads precisely to relations (\ref{F23cycl}) and (\ref{F32cycl}).\\

\subsubsection{Case of $N=6$}

\label{Cyclicity-N6}

\noindent Cyclic invariance of $A_b(1,2,3,4,5,6)$ means that $A_b(1,2,3,4,5,6) =A_b(2,3,4,5,6,1)$, so in (\ref{6point-aux}) cyclic invariance implies that

\begin{eqnarray}
A_b(1,2,3,4,5,6)  =  \hspace{11.9cm} \nonumber \\
F^{\{234\}}_{cycl}(\alpha')  A_{YM}(1,2,3,4,5,6) +  F^{\{324\}}_{cycl}(\alpha')  A_{YM}(2,4,3,5,6,1) + F^{\{243\}}_{cycl}(\alpha')  A_{YM}(2,3,5,4,6,1) +  \nonumber \\
  F^{\{342\}}_{cycl}(\alpha')  A_{YM}(2,4,5,3,6,1) +  F^{\{423\}}_{cycl}(\alpha')  A_{YM}(2,5,3,4,6,1) + F^{\{432\}}_{cycl}(\alpha')  A_{YM}(2,5,4,3,6,1)  \ , \nonumber\\
 = \hspace{14.5cm}  \nonumber \\
 F^{\{234\}}_{cycl}(\alpha')  A_{YM}(1,2,3,4,5,6) +  F^{\{324\}}_{cycl}(\alpha')  A_{YM}(1,2,4,3,5,6) + F^{\{243\}}_{cycl}(\alpha')  A_{YM}(1,2,3,5,4,6) +  \nonumber \\
 F^{\{342\}}_{cycl}(\alpha')  A_{YM}(1,2,4,5,3,6) +  F^{\{423\}}_{cycl}(\alpha')  A_{YM}(1,2,5,3,4,6) + F^{\{432\}}_{cycl}(\alpha')  A_{YM}(1,2,5,4,3,6)  \ , \nonumber\\
\label{6point-aux3}
\end{eqnarray}
\noindent where in the second equality we have used the cyclic property of Yang-Mills subamplitudes.\\
\noindent $F_{cycl}^{\{ \sigma_6 \}}(\alpha')$ denotes
doing $\{k_1 \rightarrow k_2$, $k_2 \rightarrow k_3$, $\ldots$, $k_6 \rightarrow k_1\}$ in
$F^{\{ \sigma_6\}}(\alpha')$, for all six permutations $\sigma_6$.\\
\noindent Now, in order to write the right hand-side of (\ref{6point-aux3}) in terms of the original Yang-Mills basis, we use the $N=6$ BCJ relations. In Appendix \ref{N6-2} we have found them but, in order to save space, there we have explicited only a few examples of them. In particular in eqs.(\ref{AYM123546}) and (\ref{AYM124536}) we have the BCJ relations for $A_{YM}(1,2,3,5,4,6)$ and $A_{YM}(1,2,4,5,3,6)$, respectively. Writing these equations in the conventions of the $\alpha_{ij}$'s and the $\beta_{ijk}$'s, they are given by \\
\begin{eqnarray}
\label{AYM123546-2}
A_{YM}(1,2,3,5,4,6) &=& \frac{\beta_{123}-\alpha_{56}}{\beta_{123}-\alpha_{56}-\alpha_{45}}A_{YM}(1,2,3,4,5,6)-\frac{\beta_{123}-\alpha_{56}+\alpha_{34}}{\beta_{123}-\alpha_{56}-\alpha_{45}}A_{YM}(1,2,4,3,5,6)\nonumber \\
& \ &+\frac{\alpha_{56}-\beta_{123}-\beta_{234}+\alpha_{23}}{-\alpha_{56}-\alpha_{45}+\beta_{123}}A(1,3,4,2,5,6) \ , \\
\label{AYM124536-2}
A_{YM}(1,2,4,5,3,6) &=&\frac{\beta_{123}-\alpha_{12}}{\alpha_{12}-\beta_{345}+\alpha_{45}-\beta_{123}}A_{YM}(1,2,3,4,5,6) +\nonumber \\
&& \frac{\beta_{123}-\alpha_{12}-\alpha_{23}}{\alpha_{12}-\beta_{345}+\alpha_{45}-\beta_{123}}A_{YM}(1,3,2,4,5,6)+\nonumber \\
& \ & \frac{\alpha_{34}+\beta_{123}-\alpha_{12}}{\alpha_{12}-\beta_{345}+\alpha_{45}-\beta_{123}}A_{YM}(1,2,4,3,5,6) \ .
\end{eqnarray}
Notice that there is one subamplitude in (\ref{AYM123546-2}) which is not present in (\ref{AYM124536-2}), and vice versa.\\
\noindent Substituing the expressions (\ref{AYM123546-2}), (\ref{AYM124536-2}) and the BCJ relations for $A_{YM}(1,2,4,3,5,6)$,  \ \ \ \ \ \ \ \\
\noindent $A_{YM}(1,2,5,3,4,6)$ and $A_{YM}(1,2,5,4,3,6)$ (which we have not explicited here) in the second equality in (\ref{6point-aux3}), and then comparing the coefficient of each subamplitude, $A_{YM}(1,\{2_{\sigma},3_{\sigma},4_{\sigma} \},5,6)$, with the corresponding one in eq.(\ref{6point-aux}), we arrive to the six relations for the $F^{\{ \sigma_6\}}(\alpha')$'s in terms of the $F^{\{ \sigma_6\}}_{cycl}(\alpha')$'s, given in eqs.(\ref{F234-expansion})-(\ref{F432-expansion}).

\vspace{0.5cm}

\section{Linear system for the $F^{\{ \sigma_6\}}(\alpha')$ and the $F^{\{ \sigma_7\}}(\alpha')$ momentum factors}

\label{Linear system}

\noindent We have implemented the $N$-point formula in eq.(\ref{ANfermionic}) computationally, for $4 \leq N \leq 7$, in such a way that we have done on it all the Grassmann integrations (for the $(N-2)$ $\theta_i$'s and for the $N$ $\phi_j$'s). So we have an explicit expression for the $A_b(1, \ldots, N)$ in terms of momentum factors (= $(N-3)$ dimensional integrals) and the kinematical structures that we have listed in eq.(\ref{general-form}).\\
\noindent In this appendix we give the details of how we have found the $F^{\{ \sigma_6\}}(\alpha')$ and the $F^{\{ \sigma_7\}}(\alpha')$ momentum factors of subsection \ref{N67-3}.\\

\subsection{Case of the $F^{\{ \sigma_6\}}(\alpha')$'s }

\label{F6}

\noindent Choosing $T(1,2,3,4,5,6)=A_b(1,2,3,4,5,6)$ in eq.(\ref{6point}) and considering then only the $(\zeta \cdot \zeta)^1 (\zeta \cdot k)^4$ terms, we have that:
\begin{multline}
A_b(1,2,3,4,5,6)\Big|_{(\zeta \cdot \zeta)^1 (\zeta \cdot k)^4}  = \\
\begin{split}
&=  F^{\{234\}}(\alpha')  \ A_{YM}(1,2,3,4,5,6)\Big|_{(\zeta \cdot \zeta)^1 (\zeta \cdot k)^4} +    F^{\{324\}}(\alpha')  \ A_{YM}(1,3,2,4,5,6)\Big|_{(\zeta \cdot \zeta)^1 (\zeta \cdot k)^4} \ + \\
& \ \ \ \ F^{\{243\}}(\alpha')  \ A_{YM}(1,2,4,3,5,6)\Big|_{(\zeta \cdot \zeta)^1 (\zeta \cdot k)^4} +    F^{\{342\}}(\alpha')  \ A_{YM}(1,3,4,2,5,6)\Big|_{(\zeta \cdot \zeta)^1 (\zeta \cdot k)^4} \ + \\
& \ \ \ \ F^{\{423\}}(\alpha')  \ A_{YM}(1,4,2,3,5,6)\Big|_{(\zeta \cdot \zeta)^1 (\zeta \cdot k)^4} +    F^{\{432\}}(\alpha')  \ A_{YM}(1,4,3,2,5,6)\Big|_{(\zeta \cdot \zeta)^1 (\zeta \cdot k)^4} \ .
\end{split}
\label{A123456-4}
\end{multline}
\noindent Considering the coefficient of the following six kinematical structures,
\begin{eqnarray}
\Big\{ \  (\zeta_1 \cdot \zeta_2)(\zeta_3 \cdot k_1)(\zeta_4 \cdot k_5)(\zeta_5 \cdot k_1)(\zeta_6 \cdot k_2) \ , \ (\zeta_1 \cdot \zeta_2)(\zeta_3 \cdot k_2)(\zeta_4 \cdot k_5)(\zeta_5 \cdot k_1)(\zeta_6 \cdot k_2) \ , \ \ \ \nonumber \\
\hphantom{ \Big\{ \  }  (\zeta_1 \cdot \zeta_2)(\zeta_3 \cdot k_5)(\zeta_4 \cdot k_1)(\zeta_5 \cdot k_1)(\zeta_6 \cdot k_2) \ , \ (\zeta_1 \cdot \zeta_2)(\zeta_3 \cdot k_5)(\zeta_4 \cdot k_2)(\zeta_5 \cdot k_1)(\zeta_6 \cdot k_2) \ , \ \ \ \nonumber \\
\hphantom{ \Big\{ \  }  (\zeta_1 \cdot \zeta_3)(\zeta_2 \cdot k_5)(\zeta_4 \cdot k_1)(\zeta_5 \cdot k_1)(\zeta_6 \cdot k_3) \ , \ (\zeta_1 \cdot \zeta_3)(\zeta_2 \cdot k_5)(\zeta_4 \cdot k_3)(\zeta_5 \cdot k_1)(\zeta_6 \cdot k_3) \ \Big\} \ ,
\label{six}
\end{eqnarray}
in both sides of (\ref{A123456-4}), we have arrived, respectively,  at the following six linearly independent equations for the momentum factors\footnote{On these equations we introduce the $N=6$ Mandelstam variables.  See eqs.(\ref{Mandelstam6}) and (\ref{Mandelstam6-2}).}:
\begin{eqnarray}
-    F^{\{324\}}(\alpha') s_1+   F^{\{234\}}(\alpha') (s_1+s_2-t_1)&=& s_1 s_4 (s_1+s_2-t_1) t_1 \ {\alpha'}^3 \ \Big[ \ \frac{1}{x_2 x_3 (1-x_4)} \ \Big] \ ,  \nonumber \\
-F^{\{324\}}(\alpha') s_1+F^{\{234\}}(\alpha') (s_1+s_2)&=& s_1 s_2 s_4 t_1 \ {\alpha'}^3 \  \Big[ \ \frac{1}{x_2 (x_3-x_2) (1-x_4)} \ \Big]   \ ,
\label{eqs1}
\end{eqnarray}
\begin{multline}
-F^{\{423\}}(\alpha') s_1+F^{\{243\}}(\alpha') (-s_2-s_5+t_1+t_2)=  \\
\begin{split}
&= s_1 (s_1-s_3+s_5-t_1) (s_2+s_5-t_1-t_2) (s_3+s_4-t_3) \ {\alpha'}^3 \  \Big[ \ \frac{1}{x_2 (1-x_3) x_4} \ \Big] \ , \ \
\end{split}
\\
F^{\{423\}}(\alpha') s_1+F^{\{243\}}(\alpha') (-s_1+s_2+s_3-t_2)=  \hspace{6.2cm} \\
\begin{split}
&= s_1 (s_1-s_3+s_5-t_1) (s_2+s_3-t_2) (s_3+s_4-t_3)  \ {\alpha'}^3 \ \Big[ \ \frac{1}{x_2 (1-x_3) (x_4-x_2)} \ \Big] \ , \hspace{0cm}
\end{split}
\label{eqs2}
\end{multline}
\begin{multline}
\  F^{\{432\}}(\alpha') (s_1+s_2-t_1)+F^{\{342\}}(\alpha') (s_1+s_2-s_3-t_1)= \\
\begin{split}
&=-s_3 (s_1+s_2-t_1) (-s_1+s_3+s_5-t_2) (s_3+s_6-t_2-t_3)   \ {\alpha'}^3 \ \Big[ \ \frac{1}{(1-x_2) x_3 (x_4-x_3)} \ \Big] \ , \hspace{0.8cm}
\end{split}
\\
F^{\{432\}}(\alpha') (s_1+s_2-t_1)+F^{\{342\}}(\alpha') (-s_2-s_5+t_1+t_2)=  \hspace{5.4cm} \\
\begin{split}
&=(s_1+s_2-t_1)
(s_2+s_5-t_1-t_2)(s_1-s_3-s_5+t_2) (-s_3-s_6+t_2+t_3)     {\alpha'}^3  \Big[ \ \frac{1}{(1-x_2) x_3 x_4} \ \Big]
\end{split}
\label{eqs3}
\end{multline}
\noindent where we are using the notation \cite{Hemily1}
\begin{eqnarray}
\label{notation1}
\Big[ f(x_2, x_3, x_4) \Big] &= & \int_0^1 dx_{4}  \int_0^{x_4} dx_{3}  \int_0^{x_3} dx_{2} \ {x_2}^{2 \alpha' k_2 \cdot k_1} {x_3}^{2 \alpha' k_3 \cdot k_1} {x_4}^{2 \alpha' k_4 \cdot k_1} {(x_3-x_2)}^{2 \alpha' k_3 \cdot k_2} \times \nonumber \\
&& \hphantom{  \int_0^1 dx_{4}  \int_0^{x_4} dx_{3}  \int_0^{x_3} dx_{2} \   }
 {(x_4-x_2)}^{2 \alpha' k_4 \cdot k_2} {(1-x_2)}^{2 \alpha' k_5 \cdot k_2} {(x_4-x_3)}^{2 \alpha' k_4 \cdot k_3}   \times \nonumber \\
&& \hphantom{  \int_0^1 dx_{4}  \int_0^{x_4} dx_{3}  \int_0^{x_3} dx_{2} \   }
 {(1-x_3)}^{2 \alpha' k_5 \cdot k_3}  {(1-x_4)}^{2 \alpha' k_5 \cdot k_4}  \  \times  f(x_2, x_3, x_4)   \ , \hspace{1.0cm} \\
\label{notation2}
& = & \int_0^1 dx_{4}  \int_0^{x_4} dx_{3}  \int_0^{x_3} dx_{2}   \ \biggl(  \prod_{i>j \geq 1}^{5} (x_i - x_j)^{2 \alpha' k_i \cdot k_j} \biggr)
  \times  f(x_2, x_3, x_4)  \ \ .
\end{eqnarray}
\noindent In eq.(\ref{notation2}) we have assumed that $\{ x_1=0, x_5=1\}$.\\
\noindent Notice that the six equations come in three blocks of two equations and two unknowns each.\\
\noindent Solving the first block of two equations, eq.(\ref{eqs1}), gives
\begin{eqnarray}
\label{F234-4}
F^{\{234\}}(\alpha')  & = &  {\alpha'}^3 \  \int_0^1 dx_{4}  \int_0^{x_4} dx_{3}  \int_0^{x_3} dx_{2}   \ \biggl(  \prod_{i>j \geq 1}^{5} (x_i - x_j)^{2 \alpha' k_i \cdot k_j} \biggr)  \ \times \nonumber \\
&& \hphantom{    \int_0^1 dx_{4}  \int_0^{x_4} dx_{3}  \int_0^{x_3} dx_{2}   }
\Big\{ \ \frac{s_1}{x_2}   \cdot  \frac{s_4}{1-x_4} \cdot  \Big(   \frac{t_1-s_1-s_2}{x_3} + \frac{s_2}{x_3-x_2}  \Big) \ \Big\} \ , \\
\label{F324-4}
F^{\{324\}}(\alpha')  & = & {\alpha'}^3 \  \int_0^1 dx_{4}  \int_0^{x_4} dx_{3}  \int_0^{x_3} dx_{2}   \ \biggl(  \prod_{i>j \geq 1}^{5} (x_i - x_j)^{2 \alpha' k_i \cdot k_j} \biggr)  \ \times  \nonumber \\
&& \hphantom{    \int_0^1 dx_{4}  \int_0^{x_4} dx_{3}  \int_0^{x_3} dx_{2}   }
\Big\{ \ \frac{t_1-s_1-s_2}{x_3}   \cdot  \frac{s_4}{1-x_4} \cdot  \Big(   \frac{s_1}{x_2} + \frac{s_2}{x_2-x_3}  \Big) \ \Big\} \ .
\end{eqnarray}
\noindent Substituing back the $N=6$ Mandelstam variables (see eqs.(\ref{Mandelstam6}) and (\ref{Mandelstam6-2})) in (\ref{F234-4}) and (\ref{F324-4}), it is easy to see that the expressions for $F^{\{234\}}(\alpha')$ and $F^{\{324\}}(\alpha')$ correspond exactly to the ones given in eqs.(\ref{G1}) and (\ref{G2}) of the main body of this work, respectively.\\
\noindent Solving the two blocks of equations in (\ref{eqs2}) and (\ref{eqs3}) it can be verified, also, that the expression for $F^{\{243\}}(\alpha')$,  $F^{\{423\}}(\alpha')$,  $F^{\{342\}}(\alpha')$ and  $F^{\{432\}}(\alpha')$ are in agreement with the one in (\ref{G1}) by doing the corresponding $\sigma_6$ permutation of indices $\{2,3,4\}$ {\it inside} the curly brackets of that equation.

\subsection{Case of the $F^{\{ \sigma_7\}}(\alpha')$'s }

\label{F7}

\noindent Choosing $T(1,2,3,4,5,6,7)=A_b(1,2,3,4,5,6,7)$ in eq.(\ref{7point}) and considering then only the $(\zeta \cdot \zeta)^1 (\zeta \cdot k)^5$ terms, we have that:
\begin{multline}
A_b(1,2,3,4,5,6,7)\Big|_{(\zeta \cdot \zeta)^1 (\zeta \cdot k)^5}  = \\
\begin{split}
&  F^{\{2345\}}(\alpha')  \ A_{YM}(1,2,3,4,5,6,7)\Big|_{(\zeta \cdot \zeta)^1 (\zeta \cdot k)^5} +    F^{\{2354\}}(\alpha')  \ A_{YM}(1,2,3,5,4,6,7)\Big|_{(\zeta \cdot \zeta)^1 (\zeta \cdot k)^5} \ + \\
&  F^{\{2435\}}(\alpha')  \ A_{YM}(1,2,4,3,5,6,7)\Big|_{(\zeta \cdot \zeta)^1 (\zeta \cdot k)^5} +    \ldots + F^{\{5432\}}(\alpha')  \ A_{YM}(1,5,4,3,2,6,7)\Big|_{(\zeta \cdot \zeta)^1 (\zeta \cdot k)^5}
\end{split}
\label{A1234567-4}
\end{multline}
\noindent Considering the coefficient of the following twenty four kinematical structures,
\begin{eqnarray}
\Big\{   (\zeta_1 \cdot \zeta_2)(\zeta_3 \cdot k_2)(\zeta_4 \cdot k_6)(\zeta_5 \cdot k_6)(\zeta_6 \cdot k_1)(\zeta_7 \cdot k_2)  , \ (\zeta_1 \cdot \zeta_2)(\zeta_3 \cdot k_1)(\zeta_4 \cdot k_6)(\zeta_5 \cdot k_6)(\zeta_6 \cdot k_1)(\zeta_7 \cdot k_2) \ ,  \nonumber \\
\hphantom{ \Big\{  } (\zeta_1 \cdot \zeta_2)(\zeta_3 \cdot k_2)(\zeta_4 \cdot k_5)(\zeta_5 \cdot k_6)(\zeta_6 \cdot k_1)(\zeta_7 \cdot k_3)  , \ (\zeta_1 \cdot \zeta_2)(\zeta_3 \cdot k_1)(\zeta_4 \cdot k_5)(\zeta_5 \cdot k_6)(\zeta_6 \cdot k_1)(\zeta_7 \cdot k_2) \ ,  \nonumber \\
\hphantom{ \Big\{ }  (\zeta_1 \cdot \zeta_2)(\zeta_3 \cdot k_6)(\zeta_4 \cdot k_2)(\zeta_5 \cdot k_3)(\zeta_6 \cdot k_1)(\zeta_7 \cdot k_2)  , \ (\zeta_1 \cdot \zeta_2)(\zeta_3 \cdot k_6)(\zeta_4 \cdot k_1)(\zeta_5 \cdot k_3)(\zeta_6 \cdot k_1)(\zeta_7 \cdot k_2) \ ,  \nonumber \\
\hphantom{  \Big\{  }  (\zeta_1 \cdot \zeta_2)(\zeta_3 \cdot k_2)(\zeta_4 \cdot k_5)(\zeta_5 \cdot k_6)(\zeta_6 \cdot k_1)(\zeta_7 \cdot k_3)  , \ (\zeta_1 \cdot \zeta_2)(\zeta_3 \cdot k_1)(\zeta_4 \cdot k_5)(\zeta_5 \cdot k_6)(\zeta_6 \cdot k_1)(\zeta_7 \cdot k_2) \ ,  \nonumber \\
\hphantom{  \Big\{ }  (\zeta_1 \cdot \zeta_4)(\zeta_2 \cdot k_6)(\zeta_3 \cdot k_1)(\zeta_5 \cdot k_2)(\zeta_6 \cdot k_1)(\zeta_7 \cdot k_4)  , \ (\zeta_1 \cdot \zeta_4)(\zeta_2 \cdot k_6)(\zeta_3 \cdot k_4)(\zeta_5 \cdot k_2)(\zeta_6 \cdot k_1)(\zeta_7 \cdot k_3) \ ,  \nonumber \\
\hphantom{  \Big\{ }  (\zeta_1 \cdot \zeta_4)(\zeta_2 \cdot k_6)(\zeta_3 \cdot k_1)(\zeta_5 \cdot k_6)(\zeta_6 \cdot k_1)(\zeta_7 \cdot k_4)  , \ (\zeta_1 \cdot \zeta_4)(\zeta_2 \cdot k_6)(\zeta_3 \cdot k_4)(\zeta_5 \cdot k_6)(\zeta_6 \cdot k_1)(\zeta_7 \cdot k_3) \ ,  \nonumber \\
\hphantom{  \Big\{ }  (\zeta_1 \cdot \zeta_4)(\zeta_2 \cdot k_6)(\zeta_3 \cdot k_2)(\zeta_5 \cdot k_4)(\zeta_6 \cdot k_1)(\zeta_7 \cdot k_4)  , \ (\zeta_1 \cdot \zeta_4)(\zeta_2 \cdot k_6)(\zeta_3 \cdot k_2)(\zeta_5 \cdot k_4)(\zeta_6 \cdot k_1)(\zeta_7 \cdot k_5) \ ,  \nonumber \\
\hphantom{  \Big\{ }  (\zeta_1 \cdot \zeta_4)(\zeta_2 \cdot k_6)(\zeta_3 \cdot k_6)(\zeta_5 \cdot k_4)(\zeta_6 \cdot k_1)(\zeta_7 \cdot k_4)  , \ (\zeta_1 \cdot \zeta_4)(\zeta_2 \cdot k_3)(\zeta_3 \cdot k_6)(\zeta_5 \cdot k_1)(\zeta_6 \cdot k_1)(\zeta_7 \cdot k_4) \ ,  \nonumber \\
\hphantom{  \Big\{ }  (\zeta_1 \cdot \zeta_3)(\zeta_2 \cdot k_6)(\zeta_4 \cdot k_2)(\zeta_5 \cdot k_3)(\zeta_6 \cdot k_1)(\zeta_7 \cdot k_5)  , \ (\zeta_1 \cdot \zeta_3)(\zeta_2 \cdot k_6)(\zeta_4 \cdot k_6)(\zeta_5 \cdot k_3)(\zeta_6 \cdot k_1)(\zeta_7 \cdot k_3) \ ,  \nonumber \\
\hphantom{  \Big\{ }  (\zeta_1 \cdot \zeta_3)(\zeta_2 \cdot k_4)(\zeta_4 \cdot k_6)(\zeta_5 \cdot k_3)(\zeta_6 \cdot k_1)(\zeta_7 \cdot k_3)  , \ (\zeta_1 \cdot \zeta_3)(\zeta_2 \cdot k_6)(\zeta_4 \cdot k_6)(\zeta_5 \cdot k_1)(\zeta_6 \cdot k_1)(\zeta_7 \cdot k_3) \ ,  \nonumber \\
\hphantom{  \Big\{ }  (\zeta_1 \cdot \zeta_2)(\zeta_3 \cdot k_6)(\zeta_4 \cdot k_3)(\zeta_5 \cdot k_2)(\zeta_6 \cdot k_1)(\zeta_7 \cdot k_5)  , \ (\zeta_1 \cdot \zeta_2)(\zeta_3 \cdot k_6)(\zeta_4 \cdot k_6)(\zeta_5 \cdot k_2)(\zeta_6 \cdot k_1)(\zeta_7 \cdot k_2) \ ,  \nonumber \\
\hphantom{  \Big\{ }  (\zeta_1 \cdot \zeta_2)(\zeta_3 \cdot k_6)(\zeta_4 \cdot k_6)(\zeta_5 \cdot k_1)(\zeta_6 \cdot k_1)(\zeta_7 \cdot k_2)  , \ (\zeta_1 \cdot \zeta_2)(\zeta_3 \cdot k_4)(\zeta_4 \cdot k_6)(\zeta_5 \cdot k_1)(\zeta_6 \cdot k_1)(\zeta_7 \cdot k_2)  \Big\}  \nonumber \\
\label{twentyfour}
\end{eqnarray}
in both sides of (\ref{A1234567-4}), we have arrived, respectively,  at the following twenty four linearly independent equations for the momentum factors\footnote{On these equations we have introduced the $N=7$ Mandelstam variables.  See eqs.(\ref{Mandelstam7}) and (\ref{Mandelstam7-2}).}:
\begin{multline*}
\text{ }(-\text{$F^{\{3254\}}(\alpha')s_{1}$}+\text{$F^{\{2354\}}(\alpha')$}(\text{$s_{1}$}+%
\text{$s_{2}$}-\text{$t_{1}$}))(s_5-t_4)+\nonumber \\
\text{$F^{\{3245\}}(\alpha')s_{1}$}(\text{$s_{4}$}+%
\text{$s_{5}$}-\text{$t_{4}$})-\text{$F^{\{2345\}}(\alpha')$}(\text{$s_{1}$}+\text{$%
s_{2}$}-\text{$t_{1}$})(\text{$s_{4}$}+\text{$s_{5}$}-\text{$t_{4}$})= \\
=-{\alpha'}^4\text{$s_{1}$ }(\text{$s_{1}$}+\text{$s_{2}$}-\text{$t_{1}$})\text{$t_{1}
$}(\text{$s_{4}$}+\text{$s_{5}$}-\text{$t_{4}$})\text{$t_{4}$}\Big[ \ \frac{1%
}{x_{2}x_{3}(1-x_{4})(1-x_{5})} \ \Big] \ ,
\end{multline*}%
\begin{multline*}
-\text{$s_{5}$}(\text{$F^{\{2354\}}(\alpha')$}(\text{$s_{1}$}+\text{$s_{2}$}-\text{$%
t_{1}$})+\text{$F^{\{3254\}}(\alpha')$}(-\text{$s_{1}$}+\text{$t_{1}$})) \\
-\text{$F^{\{3245\}}(\alpha')$}(\text{$s_{1}$}-\text{$t_{1}$})(\text{$s_{4}$}+\text{$%
s_{5}$}-\text{$t_{4}$})+\text{$F^{\{2345\}}(\alpha')$}(\text{$s_{1}$}+\text{$s_{2}$}-%
\text{$t_{1}$})(\text{$s_{4}$}+\text{$s_{5}$}-\text{$t_{4}$})= \\
={\alpha'}^4(\text{$s_{2}s_{5}$}(\text{$s_{1}$}+\text{$s_{2}$}-\text{$t_{1}$})\text{$%
t_{1}$}(\text{$s_{4}$}+\text{$s_{5}$}-\text{$t_{4}$})\text{$t_{4}$})\Big[ \
\frac{1}{x_{3}(x_{3}-x_{2})(1-x_{4})(1-x_{5})} \ \Big] \ ,
\end{multline*}%
\begin{multline*}
\text{$F^{\{3245\}}(\alpha')s_{1}$}(\text{$s_{4}$}+\text{$s_{5}$})+\text{$s_{5}$}(-%
\text{$F^{\{3254\}}(\alpha')s_{1}$}+\text{$F^{\{2354\}}(\alpha')$}(\text{$s_{1}$}+\text{$s_{2}$%
}-\text{$t_{1}$})) \\
-\text{$F^{\{2345\}}(\alpha')$}(\text{$s_{4}$}+\text{$s_{5}$})(\text{$s_{1}$}+\text{$%
s_{2}$}-\text{$t_{1}$})=-{\alpha'}^4\text{$s_{1}s_{4}s_{5}$}(\text{$s_{1}$}+\text{$%
s_{2}$}-\text{$t_{1}$})\text{$t_{1}t_{4}$}\Big[ \ \frac{1}{%
x_{2}x_{3}(1-x_{5})(x_{5}-x_{4})} \ \Big] \ ,
\end{multline*}%
\begin{multline}
-\text{$F^{\{3245\}}(\alpha')$}(\text{$s_{4}$}+\text{$s_{5}$})(\text{$s_{1}$}-\text{$%
t_{1}$})+\text{$F^{\{2345\}}(\alpha')$}(\text{$s_{4}$}+\text{$s_{5}$})(\text{$s_{1}$}+%
\text{$s_{2}$}-\text{$t_{1}$}) \\
-\text{$s_{5}$}(\text{$F^{\{2354\}}(\alpha')$}(\text{$s_{1}$}+\text{$s_{2}$}-\text{$%
t_{1}$})+\text{$F^{\{3254\}}(\alpha')$}(-\text{$s_{1}$}+\text{$t_{1}$}))= \\
={\alpha'}^4\text{$s_{2}s_{4}s_{5}$}(\text{$s_{1}$}+\text{$s_{2}$}-\text{$t_{1}$})%
\text{$t_{1}t_{4}$}\Big[ \ \frac{1}{x_{3}(x_{3}-x_{2})(1-x_{5})(x_{5}-x_{4})}%
\Big] \ ,
\label{eqs1-2}
\end{multline}
\begin{multline*}
-\text{$s_{5}$}(\text{$F^{\{4253\}}(\alpha')s_{1}$}+\text{$F^{\{2453\}}(\alpha')$}(\text{$s_{2}
$}-\text{$t_{1}$}-\text{$t_{2}$}+\text{$t_{5}$}))+\text{$F^{\{4235\}}(\alpha')s_{1}$}%
(-\text{$s_{4}$}+\text{$t_{3}$}+\text{$t_{4}$}-\text{$t_{7}$}) \\
-\text{$F^{\{2435\}}(\alpha')$}(\text{$s_{2}$}-\text{$t_{1}$}-\text{$t_{2}$}+\text{$%
t_{5}$})(\text{$s_{4}$}-\text{$t_{3}$}-\text{$t_{4}$}+\text{$t_{7}$})= \\
=-{\alpha'}^4\text{$s_{1}s_{5}$}(\text{$s_{1}$}-\text{$s_{3}$}-\text{$t_{1}$}+\text{$%
t_{5}$})(\text{$s_{2}$}-\text{$t_{1}$}-\text{$t_{2}$}+\text{$t_{5}$}) \\
\times (-\text{$s_{3}$}+\text{$s_{5}$}-\text{$t_{4}$}+\text{$t_{7}$})(\text{$%
s_{4}$}-\text{$t_{3}$}-\text{$t_{4}$}+\text{$t_{7}$})\Big[ \ \frac{1}{%
x_{2}(1-x_{3})x_{4}(1-x_{5})} \ \Big] \ ,
\end{multline*}%
\begin{multline*}
\text{$F^{\{4235\}}(\alpha')s_{1}$}(\text{$s_{3}$}+\text{$s_{4}$}-\text{$s_{5}$}-%
\text{$t_{3}$})+\text{$F^{\{2435\}}(\alpha')$}(\text{$s_{3}$}+\text{$s_{4}$}-\text{$%
s_{5}$}-\text{$t_{3}$})(\text{$s_{2}$}-\text{$t_{1}$}-\text{$t_{2}$}+\text{$%
t_{5}$}) \\
+\text{$s_{5}$}(\text{$F^{\{4253\}}(\alpha')s_{1}$}+\text{$F^{\{2453\}}(\alpha')$}(\text{$s_{2}
$}-\text{$t_{1}$}-\text{$t_{2}$}+\text{$t_{5}$}))= \\
=-{\alpha'}^4\text{$s_{1}s_{5}$}(\text{$s_{3}$}+\text{$s_{4}$}-\text{$t_{3}$})(-\text{%
$s_{2}$}+\text{$t_{1}$}+\text{$t_{2}$}-\text{$t_{5}$}) \\
\times (\text{$s_{1}$}-\text{$s_{3}$}-\text{$t_{1}$}+\text{$t_{5}$})(-\text{$%
s_{3}$}+\text{$s_{5}$}-\text{$t_{4}$}+\text{$t_{7}$})\Big[ \ \frac{1}{%
x_{2}x_{4}(1-x_{5})(x_{5}-x_{3})} \ \Big] \ ,
\end{multline*}%
\begin{multline*}
(\text{$F^{\{4235\}}(\alpha')$}(\text{$s_{3}$}+\text{$s_{4}$}-\text{$s_{5}$}-\text{$%
t_{3}$})(\text{$s_{3}$}+\text{$t_{1}$}-\text{$t_{5}$})+\text{$F^{\{2435\}}(\alpha')$}(%
\text{$s_{3}$}+\text{$s_{4}$}-\text{$s_{5}$}-\text{$t_{3}$})(\text{$s_{2}$}-%
\text{$t_{1}$}-\text{$t_{2}$}+\text{$t_{5}$}) \\
+\text{$s_{5}$}(\text{$F^{\{4253\}}(\alpha')$}(\text{$s_{3}$}+\text{$t_{1}$}-\text{$%
t_{5}$})+\text{$F^{\{2453\}}(\alpha')$}(\text{$s_{2}$}-\text{$t_{1}$}-\text{$t_{2}$}+%
\text{$t_{5}$})))= \\
={\alpha'}^4(\text{$s_{5}$}(\text{$s_{2}$}+\text{$s_{3}$}-\text{$t_{2}$})(\text{$s_{3}
$}+\text{$s_{4}$}-\text{$t_{3}$})(-\text{$s_{1}$}+\text{$s_{3}$}+\text{$t_{1}
$}-\text{$t_{5}$}) \\
\times (\text{$s_{2}$}-\text{$t_{1}$}-\text{$t_{2}$}+\text{$t_{5}$})(-\text{$%
s_{3}$}+\text{$s_{5}$}-\text{$t_{4}$}+\text{$t_{7}$}))\Big[ \ \frac{1}{%
x_{4}(x_{4}-x_{2})(1-x_{5})(x_{5}-x_{3})} \ \Big] \ ,
\end{multline*}%
\begin{multline}
\text{$s_{5}$}(\text{$F^{\{4253\}}(\alpha')$}(\text{$s_{3}$}+\text{$t_{1}$}-\text{$%
t_{5}$})+\text{$F^{\{2453\}}(\alpha')$}(\text{$s_{2}$}-\text{$t_{1}$}-\text{$t_{2}$}+%
\text{$t_{5}$}))+\text{$F^{\{4235\}}(\alpha')$}(\text{$s_{3}$}+\text{$t_{1}$}-\text{$%
t_{5}$})(\text{$s_{4}$}-\text{$t_{3}$}-\text{$t_{4}$}+\text{$t_{7}$}) \\
+\text{$F^{\{2435\}}(\alpha')$}(\text{$s_{2}$}-\text{$t_{1}$}-\text{$t_{2}$}+\text{$%
t_{5}$})(\text{$s_{4}$}-\text{$t_{3}$}-\text{$t_{4}$}+\text{$t_{7}$})= \\
=-{\alpha'}^4(\text{$s_{5}$}(\text{$s_{2}$}+\text{$s_{3}$}-\text{$t_{2}$})(-\text{$%
s_{1}$}+\text{$s_{3}$}+\text{$t_{1}$}-\text{$t_{5}$})(\text{$s_{2}$}-\text{$%
t_{1}$}-\text{$t_{2}$}+\text{$t_{5}$}) \\
\times (-\text{$s_{4}$}+\text{$t_{3}$}+\text{$t_{4}$}-\text{$t_{7}$})(-\text{%
$s_{3}$}+\text{$s_{5}$}-\text{$t_{4}$}+\text{$t_{7}$}))\Big[ \ \frac{1}{%
(1-x_{3})x_{4}(x_{4}-x_{2})(1-x_{5})} \ \Big] \ ,
\label{eqs2-2}
\end{multline}
\begin{multline*}
\text{$F^{\{3452\}}(\alpha')s_{5}$}(\text{$s_{2}$}-\text{$t_{1}$}-\text{$t_{2}$}+%
\text{$t_{5}$})-\text{$F^{\{4352\}}(\alpha')s_{5}$}(\text{$s_{2}$}+\text{$s_{3}$}-%
\text{$t_{1}$}-\text{$t_{2}$}+\text{$t_{5}$})-(\text{$F^{\{3425\}}(\alpha')$}(-\text{$%
s_{2}$}+\text{$t_{1}$}+\text{$t_{2}$}-\text{$t_{5}$}) \\
+\text{$F^{\{4325\}}(\alpha')$}(\text{$s_{2}$}+\text{$s_{3}$}-\text{$t_{1}$}-\text{$%
t_{2}$}+\text{$t_{5}$}))(\text{$s_{7}$}+\text{$t_{3}$}-\text{$t_{6}$}-\text{$%
t_{7}$})= \\
={\alpha'}^4(\text{$s_{3}s_{5}$}(-\text{$s_{1}$}+\text{$s_{3}$}-\text{$t_{2}$}+\text{$%
t_{5}$})(\text{$s_{2}$}-\text{$t_{1}$}-\text{$t_{2}$}+\text{$t_{5}$}) \\
\times (\text{$s_{3}$}+\text{$s_{5}$}+\text{$s_{7}$}-\text{$t_{2}$}-\text{$%
t_{7}$})(\text{$s_{7}$}+\text{$t_{3}$}-\text{$t_{6}$}-\text{$t_{7}$}))\Big[ \
\frac{1}{(1-x_{2})x_{4}(x_{4}-x_{3})(1-x_{5})} \ \Big] \ ,
\end{multline*}%
\begin{multline*}
F^{\{4352\}}(\alpha')s_{5}(s_{1}+s_{2}-t_{1})+F^{\{3452%
\}}(\alpha')s_{5}(-s_{2}+t_{1}+t_{2}-t_{5})+(F^{\{4325\}}(\alpha')(s_{1}+s_{2}-t_{1}) \\
+F^{\{3425\}}(\alpha')(-s_{2}+t_{1}+t_{2}-t_{5}))(s_{7}+t_{3}-t_{6}-t_{7})= \\
=-{\alpha'}^4s_{5}(s_{1}+s_{2}-t_{1})(s_{2}-t_{1}-t_{2}+t_{5})(s_{7}+t_{3}-t_{6}-t_{7})
\\
\times (\text{$s_{1}$}-\text{$s_{3}$}+\text{$t_{2}$}-\text{$t_{5}$})(\text{$%
s_{3}$}+\text{$s_{5}$}+\text{$s_{7}$}-\text{$t_{2}$}-\text{$t_{7}$})\Big[ \
\frac{1}{(1-x_{2})x_{3}x_{4}(1-x_{5})} \ \Big] \ ,
\end{multline*}%
\begin{multline*}
-(\text{$F^{\{4325\}}(\alpha')$}(\text{$s_{1}$}+\text{$s_{2}$}-\text{$t_{1}$})+\text{$%
F^{\{3425\}}(\alpha')$}(-\text{$s_{2}$}+\text{$t_{1}$}+\text{$t_{2}$}-\text{$t_{5}$}%
))(\text{$s_{7}$}+\text{$t_{3}$}-\text{$t_{6}$}-\text{$t_{7}$}) \\
\text{$F^{\{4352\}}(\alpha')$}(\text{$s_{1}$}+\text{$s_{2}$}-\text{$t_{1}$})(\text{$%
s_{3}$}+\text{$s_{7}$}-\text{$t_{2}$}-\text{$t_{7}$})+\text{$F^{\{3452\}}(\alpha')$}(%
\text{$s_{2}$}-\text{$t_{1}$}-\text{$t_{2}$}+\text{$t_{5}$})(-\text{$s_{3}$}-%
\text{$s_{7}$}+\text{$t_{2}$}+\text{$t_{7}$})= \\
={\alpha'}^4((\text{$s_{1}$}+\text{$s_{2}$}-\text{$t_{1}$})(\text{$s_{1}$}-\text{$%
s_{3}$}+\text{$t_{2}$}-\text{$t_{5}$})(\text{$s_{2}$}-\text{$t_{1}$}-\text{$%
t_{2}$}+\text{$t_{5}$})(\text{$s_{3}$}-\text{$t_{2}$}-\text{$t_{3}$}+\text{$%
t_{6}$}) \\
\times (\text{$s_{3}$}+\text{$s_{5}$}+\text{$s_{7}$}-\text{$t_{2}$}-\text{$%
t_{7}$})(\text{$s_{7}$}+\text{$t_{3}$}-\text{$t_{6}$}-\text{$t_{7}$}))\Big[ \
\frac{1}{(1-x_{2})x_{3}x_{4}(x_{5}-x_{2})} \ \Big] \ ,
\end{multline*}%
\begin{multline}
\text{$F^{\{3452\}}(\alpha')$}(\text{$s_{2}$}-\text{$t_{1}$}-\text{$t_{2}$}+\text{$%
t_{5}$})(\text{$s_{3}$}+\text{$s_{7}$}-\text{$t_{2}$}-\text{$t_{7}$})-\text{$%
F^{\{4352\}}(\alpha')$}(\text{$s_{2}$}+\text{$s_{3}$}-\text{$t_{1}$}-\text{$t_{2}$}+%
\text{$t_{5}$})(\text{$s_{3}$}+\text{$s_{7}$}-\text{$t_{2}$}-\text{$t_{7}$})
\\
+(\text{$F^{\{3425\}}(\alpha')$}(-\text{$s_{2}$}+\text{$t_{1}$}+\text{$t_{2}$}-\text{$%
t_{5}$})+\text{$F^{\{4325\}}(\alpha')$}(\text{$s_{2}$}+\text{$s_{3}$}-\text{$t_{1}$}-%
\text{$t_{2}$}+\text{$t_{5}$}))(\text{$s_{7}$}+\text{$t_{3}$}-\text{$t_{6}$}-%
\text{$t_{7}$})= \\
=-{\alpha'}^4(\text{$s_{3}$}(-\text{$s_{1}$}+\text{$s_{3}$}-\text{$t_{2}$}+\text{$%
t_{5}$})(\text{$s_{2}$}-\text{$t_{1}$}-\text{$t_{2}$}+\text{$t_{5}$})(\text{$%
s_{3}$}-\text{$t_{2}$}-\text{$t_{3}$}+\text{$t_{6}$}) \\
\times (\text{$s_{3}$}+\text{$s_{5}$}+\text{$s_{7}$}-\text{$t_{2}$}-\text{$%
t_{7}$})(\text{$s_{7}$}+\text{$t_{3}$}-\text{$t_{6}$}-\text{$t_{7}$}))\Big[ \
\frac{1}{(1-x_{2})x_{4}(x_{4}-x_{3})(1-x_{5})} \ \Big] \ ,
\label{eqs3-2}
\end{multline}

\begin{multline*}
-(\text{$F^{\{5423\}}(\alpha')$}(-\text{$s_{2}$}+\text{$t_{1}$}+\text{$t_{2}$}-\text{$%
t_{5}$})+\text{$F^{\{4523\}}(\alpha')$}(\text{$s_{2}$}+\text{$s_{4}$}-\text{$t_{1}$}-%
\text{$t_{2}$}+\text{$t_{5}$}))(\text{$s_{7}$}+\text{$t_{3}$}-\text{$t_{6}$}-%
\text{$t_{7}$}) \\
+\text{$F^{\{5432\}}(\alpha')$}(\text{$s_{2}$}-\text{$t_{1}$}-\text{$t_{2}$}+\text{$%
t_{5}$})(\text{$s_{4}$}-\text{$t_{3}$}-\text{$t_{4}$}+\text{$t_{7}$})+\text{$%
F^{\{4532\}}(\alpha')$}(\text{$s_{2}$}+\text{$s_{4}$}-\text{$t_{1}$}-\text{$t_{2}$}+%
\text{$t_{5}$})(\text{$s_{4}$}-\text{$t_{3}$}-\text{$t_{4}$}+\text{$t_{7}$})=
\\
=-{\alpha'}^4\text{$s_{4}$}(\text{$s_{2}$}-\text{$t_{1}$}-\text{$t_{2}$}+\text{$t_{5}$%
})(\text{$s_{4}$}-\text{$t_{3}$}-\text{$t_{4}$}+\text{$t_{7}$})(\text{$s_{2}$%
}+\text{$s_{4}$}+\text{$s_{6}$}-\text{$t_{1}$}-\text{$t_{6}$}) \\
\times (\text{$s_{2}$}+\text{$s_{4}$}+\text{$s_{7}$}-\text{$t_{4}$}-\text{$%
t_{6}$})(\text{$s_{7}$}+\text{$t_{3}$}-\text{$t_{6}$}-\text{$t_{7}$})\Big[ \
\frac{1}{(1-x_{2})(1-x_{3})x_{4}(x_{5}-x_{4})} \ \Big] \ ,
\end{multline*}%
\begin{multline*}
(\text{$F^{\{5432\}}(\alpha')$}(-\text{$s_{2}$}+\text{$t_{1}$}+\text{$t_{2}$}-\text{$%
t_{5}$})+\text{$F^{\{4532\}}(\alpha')$}(\text{$s_{2}$}+\text{$s_{4}$}-\text{$t_{1}$}-%
\text{$t_{2}$}+\text{$t_{5}$}))(\text{$s_{4}$}-\text{$t_{3}$}-\text{$t_{4}$}+%
\text{$t_{7}$}) \\
+\text{$F^{\{5423\}}(\alpha')$}(\text{$s_{2}$}-\text{$t_{1}$}-\text{$t_{2}$}+\text{$%
t_{5}$})(\text{$s_{2}$}+\text{$s_{4}$}-\text{$t_{3}$}-\text{$t_{4}$}+\text{$%
t_{7}$})-\text{$F^{\{4523\}}(\alpha')$}(\text{$s_{2}$}+\text{$s_{4}$}-\text{$t_{1}$}-%
\text{$t_{2}$}+\text{$t_{5}$})(\text{$s_{2}$}+\text{$s_{4}$}-\text{$t_{3}$}-%
\text{$t_{4}$}+\text{$t_{7}$}) \\
=-{\alpha'}^4(\text{$s_{2}s_{4}$}(\text{$s_{2}$}-\text{$t_{1}$}-\text{$t_{2}$}+\text{$%
t_{5}$})(\text{$s_{2}$}+\text{$s_{4}$}+\text{$s_{6}$}-\text{$t_{1}$}-\text{$%
t_{6}$}) \\
\times (\text{$s_{2}$}+\text{$s_{4}$}+\text{$s_{7}$}-\text{$t_{4}$}-\text{$%
t_{6}$})(\text{$s_{4}$}-\text{$t_{3}$}-\text{$t_{4}$}+\text{$t_{7}$}))\Big[ \
\frac{1}{(1-x_{3})(x_{3}-x_{2})x_{4}(x_{5}-x_{4})} \ \Big] \ ,
\end{multline*}%
\begin{multline*}
(F^{\{5423\}}(\alpha')(s_{2}-t_{1}-t_{2}+t_{5})+F^{\{4523%
\}}(\alpha')(s_{6}+t_{2}-t_{5}-t_{6}))(s_{7}+t_{3}-t_{6}-t_{7}) \\
+F^{\{5432\}}(\alpha')(s_{2}-t_{1}-t_{2}+t_{5})(s_{4}-t_{3}-t_{4}+t_{7})+F^{\{4532%
\}}(\alpha')(s_{6}+t_{2}-t_{5}-t_{6})(s_{4}-t_{3}-t_{4}+t_{7})= \\
=-{\alpha'}^4(s_{2}-t_{1}-t_{2}+t_{5})(s_{6}+t_{2}-t_{5}-t_{6})(s_{4}-t_{3}-t_{4}+t_{7})
\\
\times (-s_{7}-(\text{$s_{2}$}+\text{$s_{4}$}+\text{$s_{6}$}-\text{$t_{1}$}-%
\text{$t_{6}$})(\text{$s_{2}$}+\text{$s_{4}$}+\text{$s_{7}$}-\text{$t_{4}$}-%
\text{$t_{6}$})\Big[ \ \frac{1}{(1-x_{2})(1-x_{3})x_{4}x_{5}} \ \Big] \ ,
\end{multline*}%
\begin{multline}
F^{\{5432\}}(\alpha')(s_{2}-t_{1}-t_{2}+t_{5})(s_{4}-t_{3}-t_{4}+t_{7})-F^{\{5423%
\}}(\alpha')(s_{2}-t_{1}-t_{2}+t_{5})(s_{2}+s_{4}-t_{3}-t_{4}+t_{7}) \\
-(s_{6}+t_{2}-t_{5}-t_{6})(F^{\{4532\}}(\alpha')(-s_{4}+t_{3}+t_{4}-t_{7})+F^{\{4523%
\}}(\alpha')(s_{2}+s_{4}-t_{3}-t_{4}+t_{7}))= \\
=-{\alpha'}^4\text{$s_{2}$}%
(s_{2}-t_{1}-t_{2}+t_{5})(s_{6}+t_{2}-t_{5}-t_{6})(s_{4}-t_{3}-t_{4}+t_{7})
\\
\times (\text{$s_{2}$}+\text{$s_{4}$}+\text{$s_{6}$}-\text{$t_{1}$}-\text{$%
t_{6}$})(\text{$s_{2}$}+\text{$s_{4}$}+\text{$s_{7}$}-\text{$t_{4}$}-\text{$%
t_{6}$})\Big[ \ \frac{1}{(1-x_{3})(x_{3}-x_{2})x_{4}x_{5}} \ \Big] \ ,
\label{eqs4-2}
\end{multline}
\begin{multline*}
(-\text{$F^{\{5324\}}(\alpha')$}(\text{$s_{1}$}+\text{$s_{2}$}-\text{$t_{1}$})+\text{$%
F^{\{3524\}}(\alpha')$}(\text{$s_{1}$}+\text{$s_{2}$}+\text{$s_{3}$}+\text{$s_{4}$}-%
\text{$t_{1}$}-\text{$t_{3}$}))(\text{$s_{7}$}+\text{$t_{3}$}-\text{$t_{6}$}-%
\text{$t_{7}$}) \\
-\text{$F^{\{5342\}}(\alpha')$}(\text{$s_{1}$}+\text{$s_{2}$}-\text{$t_{1}$})(\text{$%
s_{2}$}+\text{$s_{3}$}-\text{$s_{7}$}-\text{$t_{2}$}-\text{$t_{3}$}+\text{$%
t_{6}$}+\text{$t_{7}$}) \\
+\text{$F^{\{3542\}}(\alpha')$}(\text{$s_{1}$}+\text{$s_{2}$}+\text{$s_{3}$}+\text{$%
s_{4}$}-\text{$t_{1}$}-\text{$t_{3}$})(\text{$s_{2}$}+\text{$s_{3}$}-\text{$%
s_{7}$}-\text{$t_{2}$}-\text{$t_{3}$}+\text{$t_{6}$}+\text{$t_{7}$})= \\
={\alpha'}^4((\text{$s_{1}$}+\text{$s_{2}$}-\text{$t_{1}$})(\text{$s_{2}$}+\text{$%
s_{3}$}-\text{$t_{2}$})(\text{$s_{3}$}+\text{$s_{4}$}-\text{$t_{3}$})(\text{$%
s_{1}$}+\text{$s_{2}$}+\text{$s_{3}$}+\text{$s_{4}$}-\text{$s_{6}$}-\text{$%
t_{1}$}-\text{$t_{2}$}-\text{$t_{3}$}+\text{$t_{5}$}+\text{$t_{6}$}) \\
\times (\text{$s_{7}$}+\text{$t_{3}$}-\text{$t_{6}$}-\text{$t_{7}$})(\text{$%
s_{2}$}+\text{$s_{3}$}+\text{$s_{4}$}+\text{$s_{5}$}-\text{$s_{7}$}-\text{$%
t_{2}$}-\text{$t_{3}$}-\text{$t_{4}$}+\text{$t_{6}$}+\text{$t_{7}$}))\Big[ \
\frac{1}{(1-x_{2})x_{3}(x_{4}-x_{2})(x_{5}-x_{3})} \ \Big] \ ,
\end{multline*}%
\begin{multline*}
-\text{$F^{\{5342\}}(\alpha')$}(\text{$s_{1}$}+\text{$s_{2}$}-\text{$t_{1}$})(\text{$%
s_{4}$}+\text{$s_{5}$}-\text{$t_{4}$})+\text{$F^{\{3542\}}(\alpha')$}(\text{$s_{1}$}+%
\text{$s_{2}$}+\text{$s_{3}$}+\text{$s_{4}$}-\text{$t_{1}$}-\text{$t_{3}$})(%
\text{$s_{4}$}+\text{$s_{5}$}-\text{$t_{4}$}) \\
-(-\text{$F^{\{5324\}}(\alpha')$}(\text{$s_{1}$}+\text{$s_{2}$}-\text{$t_{1}$})+\text{%
$F^{\{3524\}}(\alpha')$}(\text{$s_{1}$}+\text{$s_{2}$}+\text{$s_{3}$}+\text{$s_{4}$}-%
\text{$t_{1}$}-\text{$t_{3}$}))(\text{$s_{7}$}+\text{$t_{3}$}-\text{$t_{6}$}-%
\text{$t_{7}$})= \\
=-{\alpha'}^4((\text{$s_{1}$}+\text{$s_{2}$}-\text{$t_{1}$})(\text{$s_{3}$}+\text{$%
s_{4}$}-\text{$t_{3}$})(\text{$s_{4}$}+\text{$s_{5}$}-\text{$t_{4}$})(\text{$%
s_{1}$}+\text{$s_{2}$}+\text{$s_{3}$}+\text{$s_{4}$}-\text{$s_{6}$}-\text{$%
t_{1}$}-\text{$t_{2}$}-\text{$t_{3}$}+\text{$t_{5}$}+\text{$t_{6}$}) \\
\times (\text{$s_{7}$}+\text{$t_{3}$}-\text{$t_{6}$}-\text{$t_{7}$})(\text{$%
s_{2}$}+\text{$s_{3}$}+\text{$s_{4}$}+\text{$s_{5}$}-\text{$s_{7}$}-\text{$%
t_{2}$}-\text{$t_{3}$}-\text{$t_{4}$}+\text{$t_{6}$}+\text{$t_{7}$}))\Big[ \
\frac{1}{(1-x_{2})x_{3}(1-x_{4})(x_{5}-x_{3})} \ \Big] \ ,
\end{multline*}%
\begin{multline*}
(-\text{$F^{\{5342\}}(\alpha')$}(\text{$s_{1}$}+\text{$s_{2}$}-\text{$t_{1}$})+\text{$%
F^{\{3542\}}(\alpha')$}(\text{$s_{1}$}+\text{$s_{2}$}+\text{$s_{3}$}+\text{$s_{4}$}-%
\text{$t_{1}$}-\text{$t_{3}$}))(\text{$s_{4}$}+\text{$s_{5}$}-\text{$t_{4}$})
\\
+\text{$F^{\{5324\}}(\alpha')$}(\text{$s_{1}$}+\text{$s_{2}$}-\text{$t_{1}$})(\text{$%
s_{2}$}+\text{$s_{3}$}+\text{$s_{4}$}+\text{$s_{5}$}-\text{$t_{2}$}-\text{$%
t_{4}$}) \\
-\text{$F^{\{3524\}}(\alpha')$}(\text{$s_{1}$}+\text{$s_{2}$}+\text{$s_{3}$}+\text{$%
s_{4}$}-\text{$t_{1}$}-\text{$t_{3}$})(\text{$s_{2}$}+\text{$s_{3}$}+\text{$%
s_{4}$}+\text{$s_{5}$}-\text{$t_{2}$}-\text{$t_{4}$})= \\
=-{\alpha'}^4((\text{$s_{1}$}+\text{$s_{2}$}-\text{$t_{1}$})(\text{$s_{2}$}+\text{$%
s_{3}$}-\text{$t_{2}$})(\text{$s_{3}$}+\text{$s_{4}$}-\text{$t_{3}$})(\text{$%
s_{4}$}+\text{$s_{5}$}-\text{$t_{4}$})(\text{$s_{1}$}+\text{$s_{2}$}+\text{$%
s_{3}$}+\text{$s_{4}$}-\text{$s_{6}$}-\text{$t_{1}$}-\text{$t_{2}$}-\text{$%
t_{3}$}+\text{$t_{5}$}+\text{$t_{6}$}) \\
\times (\text{$s_{2}$}+\text{$s_{3}$}+\text{$s_{4}$}+\text{$s_{5}$}-\text{$%
s_{7}$}-\text{$t_{2}$}-\text{$t_{3}$}-\text{$t_{4}$}+\text{$t_{6}$}+\text{$%
t_{7}$}))\Big[ \ \frac{1}{x_{3}(1-x_{4})(x_{4}-x_{2})(x_{5}-x_{3})} \ \Big] \ ,
\end{multline*}%
\begin{multline}
F^{\{5342\}}(\alpha')(s_{1}+s_{2}-t_{1})(s_{4}+s_{5}-t_{4})-F^{\{3542%
\}}(\alpha')(s_{4}+s_{5}-t_{4})(s_{6}+t_{2}-t_{5}-t_{6}) \\
+(F^{\{5324\}}(\alpha')(s_{1}+s_{2}-t_{1})+F^{\{3524%
\}}(\alpha')(-s_{6}-t_{2}+t_{5}+t_{6}))(-s_{7}-t_{3}+t_{6}+t_{7})= \\
=-{\alpha'}^4(s_{1}+s_{2}-t_{1})(s_{4}+s_{5}-t_{4})(s_{6}+t_{2}-t_{5}-t_{6})(-s_{7}-t_{3}+t_{6}+t_{7})
\\
\times (\text{$s_{1}$}+\text{$s_{2}$}+\text{$s_{3}$}+\text{$s_{4}$}-\text{$%
s_{6}$}-\text{$t_{1}$}-\text{$t_{2}$}-\text{$t_{3}$}+\text{$t_{5}$}+\text{$%
t_{6}$}) \\
\times (\text{$s_{2}$}+\text{$s_{3}$}+\text{$s_{4}$}+\text{$s_{5}$}-\text{$%
s_{7}$}-\text{$t_{2}$}-\text{$t_{3}$}-\text{$t_{4}$}+\text{$t_{6}$}+\text{$%
t_{7}$})\Big[ \ \frac{1}{(1-x_{2})x_{3}(1-x_{4})x_{5}} \ \Big] \ ,
\label{eqs5-2}
\end{multline}
\begin{multline*}
(\text{$F^{\{5234\}}(\alpha')s_{1}$}-\text{$F^{\{2534\}}(\alpha')$}(\text{$s_{1}$}+\text{$s_{3}
$}-\text{$t_{2}$}-\text{$t_{3}$}+\text{$t_{6}$}))(\text{$s_{4}$}-\text{$t_{3}
$}-\text{$t_{4}$}+\text{$t_{7}$}) \\
-\text{$F^{\{5243\}}(\alpha')s_{1}$}(\text{$s_{3}$}+\text{$s_{4}$}-\text{$t_{3}$}-%
\text{$t_{4}$}+\text{$t_{7}$})+\text{$F^{\{2543\}}(\alpha')$}(\text{$s_{1}$}+\text{$%
s_{3}$}-\text{$t_{2}$}-\text{$t_{3}$}+\text{$t_{6}$})(\text{$s_{3}$}+\text{$%
s_{4}$}-\text{$t_{3}$}-\text{$t_{4}$}+\text{$t_{7}$})= \\
={\alpha'}^4(\text{$s_{1}s_{3}$}(\text{$s_{1}$}+\text{$s_{3}$}+\text{$s_{6}$}-\text{$%
t_{3}$}-\text{$t_{5}$})(\text{$s_{3}$}-\text{$t_{2}$}-\text{$t_{3}$}+\text{$%
t_{6}$}) \\
\times (-\text{$s_{4}$}+\text{$t_{3}$}+\text{$t_{4}$}-\text{$t_{7}$})(\text{$%
s_{3}$}-\text{$s_{5}$}-\text{$t_{3}$}+\text{$t_{7}$}))\Big[ \ \frac{1}{%
x_{2}(1-x_{3})(x_{4}-x_{3})(x_{5}-x_{3})} \ \Big] \ ,
\end{multline*}%
\begin{multline*}
-\text{$F^{\{5243\}}(\alpha')s_{1}$}(\text{$s_{4}$}+\text{$s_{5}$}-\text{$t_{4}$})+(%
\text{$s_{1}$}+\text{$s_{3}$}-\text{$t_{2}$}-\text{$t_{3}$}+\text{$t_{6}$})(%
\text{$F^{\{2543\}}(\alpha')$}(\text{$s_{4}$}+\text{$s_{5}$}-\text{$t_{4}$}) \\
+\text{$F^{\{2534\}}(\alpha')$}(-\text{$s_{4}$}+\text{$t_{3}$}+\text{$t_{4}$}-\text{$%
t_{7}$}))+\text{$F^{\{5234\}}(\alpha')s_{1}$}(\text{$s_{4}$}-\text{$t_{3}$}-\text{$%
t_{4}$}+\text{$t_{7}$})= \\
=-{\alpha'}^4\text{$s_{1}$}(\text{$s_{4}$}+\text{$s_{5}$}-\text{$t_{4}$})(\text{$s_{1}
$}+\text{$s_{3}$}+\text{$s_{6}$}-\text{$t_{3}$}-\text{$t_{5}$})(\text{$s_{3}$%
}-\text{$t_{2}$}-\text{$t_{3}$}+\text{$t_{6}$}) \\
\times (-\text{$s_{3}$}+\text{$s_{5}$}+\text{$t_{3}$}-\text{$t_{7}$})(\text{$%
s_{4}$}-\text{$t_{3}$}-\text{$t_{4}$}+\text{$t_{7}$})\Big[ \ \frac{1}{%
x_{2}(1-x_{3})(1-x_{4})(x_{5}-x_{2})} \ \Big] \ ,
\end{multline*}%
\begin{multline*}
-F^{\{5243\}}(\alpha')s_{1}(s_{4}+s_{5}-t_{4})+F^{\{5234%
\}}(\alpha')s_{1}(s_{4}-t_{3}-t_{4}+t_{7}) \\
+(s_{6}+t_{2}-t_{5}-t_{6})(-F^{\{2543\}}(\alpha')(s_{4}+s_{5}-t_{4})+F^{\{2534%
\}}(\alpha')(s_{4}-t_{3}-t_{4}+t_{7}))= \\
=-{\alpha'}^4s_{1}(s_{4}+s_{5}-t_{4})(s_{6}+t_{2}-t_{5}-t_{6})(s_{4}-t_{3}-t_{4}+t_{7})
\\
\times (\text{$s_{1}$}+\text{$s_{3}$}+\text{$s_{6}$}-\text{$t_{3}$}-\text{$%
t_{5}$})(\text{$s_{3}$}-\text{$s_{5}$}-\text{$t_{3}$}+\text{$t_{7}$})\Big[ \
\frac{1}{x_{2}(1-x_{3})(1-x_{4})x_{5}} \ \Big] \ ,
\end{multline*}%
\begin{multline}
\text{$F^{\{5243\}}(\alpha')s_{1}$}(\text{$s_{4}$}+\text{$s_{5}$}-\text{$t_{4}$})+%
\text{$F^{\{5234\}}(\alpha')s_{1}$}(\text{$s_{3}$}-\text{$s_{4}$}-\text{$s_{5}$}+%
\text{$t_{4}$}) \\
+(\text{$F^{\{2543\}}(\alpha')$}(\text{$s_{4}$}+\text{$s_{5}$}-\text{$t_{4}$})+\text{$%
F^{\{2534\}}(\alpha')$}(\text{$s_{3}$}-\text{$s_{4}$}-\text{$s_{5}$}+\text{$t_{4}$}))(%
\text{$s_{6}$}+\text{$t_{2}$}-\text{$t_{5}$}-\text{$t_{6}$})= \\
=-{\alpha'}^4\text{$s_{1}s_{3}$}(\text{$s_{4}$}+\text{$s_{5}$}-\text{$t_{4}$})(\text{$%
s_{1}$}+\text{$s_{3}$}+\text{$s_{6}$}-\text{$t_{3}$}-\text{$t_{5}$}) \\
\times (\text{$s_{6}$}+\text{$t_{2}$}-\text{$t_{5}$}-\text{$t_{6}$})(\text{$%
s_{3}$}-\text{$s_{5}$}-\text{$t_{3}$}+\text{$t_{7}$})\Big[ \ \frac{1}{%
x_{2}(1-x_{4})(x_{4}-x_{3})x_{5}} \ \Big] \ .
\label{eqs6-2}
\end{multline}

\noindent where, in this case, we are using the notation
\begin{eqnarray}
\label{notation1-2}
\Big[ g(x_2, x_3, x_4, x_5) \Big] &= & \int_0^{1} dx_{5} \int_0^{x_5} dx_{4} \int_0^{x_4} dx_{3} \int_0^{x_3} dx_{2} \
{x_2}^{2 \alpha' k_2 \cdot k_1} {x_3}^{2 \alpha' k_3 \cdot k_1} {x_4}^{2 \alpha' k_4 \cdot k_1} {x_5}^{2 \alpha' k_5 \cdot k_1} \ \times \nonumber \\
&& \hphantom{ \int_0^{1} dx_{5} \int_0^{x_5} dx_{4} \int_0^{x_4} dx_{3} } {(x_3-x_2)}^{2 \alpha' k_3 \cdot k_2} {(x_4-x_2)}^{2 \alpha' k_4 \cdot k_2} {(x_5-x_2)}^{2 \alpha' k_5 \cdot k_2} \times \nonumber \\
&& \hphantom{ \int_0^1 dx_{4} \int_0^{x_4} dx_{3} \int_0^{x_3} dx_{2} \ }
{(1-x_2)}^{2 \alpha' k_6 \cdot k_2} {(x_4-x_3)}^{2 \alpha' k_4 \cdot k_3} {(x_5-x_3)}^{2 \alpha' k_5 \cdot k_3} \times \nonumber \\
&& \hphantom{ \int_0^1 dx_{4} \int_0^{x_4} dx_{3} \int_0^{x_3} dx_{2} \ }
{(1-x_3)}^{2 \alpha' k_6 \cdot k_3} {(x_5-x_4)}^{2 \alpha' k_5 \cdot k_4} {(1-x_4)}^{2 \alpha' k_6 \cdot k_4} \ \times \nonumber \\
&& \hphantom{ \int_0^1 dx_{4} \int_0^{x_4} dx_{3} \int_0^{x_3} dx_{2} \ }
{(1-x_5)}^{2 \alpha' k_6 \cdot k_5} \ \times g(x_2, x_3, x_4, x_5) \ \ , \nonumber \\
&& \\
\label{notation2-2}
& = & \int_0^{1} dx_{5} \int_0^{x_5} dx_{4} \int_0^{x_4} dx_{3} \int_0^{x_3} dx_{2} \ \biggl( \prod_{i>j \geq 1}^{6} (x_i - x_j)^{2 \alpha' k_i \cdot k_j} \biggr)
\times g(x_2, x_3, x_4, x_5) \nonumber \\
\end{eqnarray}
\noindent In eq.(\ref{notation2-2}) we have assumed that $\{ x_1=0, x_6=1\}$.\\
\noindent Notice that the twenty four equations come in six blocks of four equations and four unknowns each.\\
\noindent Solving the first block of four equations, eq.(\ref{eqs1-2}), gives
\begin{eqnarray}
\label{F2345-4}
F^{\{2345\}}(\alpha') & = & {\alpha'}^4 \ \int_0^{1} dx_{5} \int_0^{x_5} dx_{4} \int_0^{x_4} dx_{3} \int_0^{x_3} dx_{2} \ \biggl( \prod_{i>j \geq 1}^{6} (x_i - x_j)^{2 \alpha' k_i \cdot k_j} \biggr) \ \times \nonumber \\
&& \hphantom{ \int0^{1} }
\Big\{ \ \frac{s_1}{x_2} \cdot \frac{s_5}{1-x_5} \cdot \Big( \frac{t_1-s_1-s_2}{x_3} + \frac{s_2}{x_3-x_2} \Big) \cdot \Big( \frac{s_4}{x_5-x_4} + \frac{t_4-s_4-s_5}{1-x_4} \Big) \ \Big\} \ , \hspace{1cm} \\
\label{F2354-4}
F^{\{2354\}}(\alpha') & = & {\alpha'}^4 \ \int_0^{1} dx_{5} \int_0^{x_5} dx_{4} \int_0^{x_4} dx_{3} \int_0^{x_3} dx_{2} \ \biggl( \prod_{i>j \geq 1}^{6} (x_i - x_j)^{2 \alpha' k_i \cdot k_j} \biggr) \ \times \nonumber \\
&& \hphantom{ \int0^{1} }
\Big\{ \ \frac{s_1}{x_2} \cdot \frac{t_4-s_4-s_5}{1-x_4} \cdot \Big( \frac{t_1-s_1-s_2}{x_3} + \frac{s_2}{x_3-x_2} \Big) \cdot \Big( \frac{s_4}{x_4-x_5} + \frac{s_5}{1-x_5} \Big) \ \Big\} \ , \hspace{1cm} \\
\label{F3245-4}
F^{\{3245\}}(\alpha') & = & {\alpha'}^4 \ \int_0^{1} dx_{5} \int_0^{x_5} dx_{4} \int_0^{x_4} dx_{3} \int_0^{x_3} dx_{2} \ \biggl( \prod_{i>j \geq 1}^{6} (x_i - x_j)^{2 \alpha' k_i \cdot k_j} \biggr) \ \times \nonumber \\
&& \hphantom{ \int0^{1} }
\Big\{ \ \frac{t_1-s_1-s_2}{x_3} \cdot \frac{s_5}{1-x_5} \cdot \Big( \frac{s_1}{x_2} + \frac{s_2}{x_2-x_3} \Big) \cdot \Big( \frac{s_4}{x_5-x_4} + \frac{t_4-s_4-s_5}{1-x_4} \Big) \ \Big\} \ , \hspace{1cm} \\
\label{F3254-4}
F^{\{3254\}}(\alpha') & = & {\alpha'}^4 \ \int_0^{1} dx_{5} \int_0^{x_5} dx_{4} \int_0^{x_4} dx_{3} \int_0^{x_3} dx_{2} \ \biggl( \prod_{i>j \geq 1}^{6} (x_i - x_j)^{2 \alpha' k_i \cdot k_j} \biggr) \ \times \nonumber \\
&& \hphantom{ \int0^{1} }
\Big\{ \ \frac{t_1-s_1-s_2}{x_3} \cdot \frac{t_4-s_4-s_5}{1-x_4} \cdot \Big( \frac{s_1}{x_2} + \frac{s_2}{x_2-x_3} \Big) \cdot \Big( \frac{s_4}{x_4-x_5} + \frac{s_5}{1-x_5} \Big) \ \Big\} \ , \hspace{1cm} \ .
\end{eqnarray}
\noindent Substituing back the $N=7$ Mandelstam variables (see eqs.(\ref{Mandelstam7}) and (\ref{Mandelstam7-2})) in (\ref{F2345-4}) and (\ref{F3245-4}), it is easy to see that the expressions for $F^{\{2345\}}(\alpha')$ and $F^{\{3245\}}(\alpha')$ correspond exactly to the ones given in eqs.(\ref{H1}) and (\ref{H7}) of the main body of this work, respectively.\\
\noindent Solving the five blocks of equations in (\ref{eqs2-2})-(\ref{eqs6-2}) it can be verified, also, that the expression for the remaining twenty $F^{\{ \sigma_7\}}(\alpha')$'s are in agreement with the one in (\ref{G1}) by doing the corresponding $\sigma_7$ permutation of indices $\{2,3,4,5\}$ {\it inside} the curly brackets of that equation.

\end{document}